\documentclass[12pt]{article}
\usepackage{amsmath}
\usepackage{adjustbox}
\usepackage{graphicx,psfrag,epsf}
\usepackage{algorithm}
\usepackage{algpseudocode}
\usepackage{float}
\usepackage{amsfonts}
\usepackage{amsmath}
\usepackage{amssymb}
\usepackage{natbib}

\usepackage{enumerate}
\usepackage{url}
\usepackage{xcolor}
\usepackage{booktabs}

\usepackage{amsthm}

\theoremstyle{plain}

\newtheorem{proposition}{Proposition}[section]

\newtheorem{lemma}{Lemma}[section]
\newtheorem{corollary}{Corollary}[section]
\newtheorem{definition}{Definition}[section]

\theoremstyle{definition}
\newtheorem{remark}{Remark}[section]

\newtheoremstyle{boldproofstyle}
  {10pt}
  {10pt}
  {\normalfont}
  {}
  {\bfseries}
  {.}
  {5pt plus 1pt minus 1pt}
  {}

\theoremstyle{boldproofstyle}

\newcommand{\customproof}[1]{\begin{proof}[{\textbf{#1}}]}

\newcommand{\bs}[1]{\boldsymbol{#1}}

\newcommand{\reals}{\mathbb{R}}

\newcommand{\ev}{\mathbb{E}}

\begin{document}

\def\spacingset#1{\renewcommand{\baselinestretch}%
{#1}\small\normalsize} \spacingset{1}

\title{{Evidence-Based Policy Choice in Macroeconomics}{\Large \thanks{%
We thank Giovanni Caggiano, Eva Janssens, Sean McCrary, Geert Mesters, Mikkel Plagborg-M{\o}ller, Kirill Ponomarev, Silvia Sarpietro, and Moto Shintani for valuable discussions and comments. We also thank the conference and seminar participants at Brown, EIEF, 2026 MEG, the University of Tokyo, and WEEE in Perugia for beneficial comments. The authors gratefully acknowledge the visiting scholarship at CIRJE (University of Tokyo) and the financial support from NSF (FAIN 234361).
}}}
\author{Toru Kitagawa\thanks{ Department of Economics, Brown University. Email: toru\_kitagawa@brown.edu} and Simone Martinalli\thanks{Department of Economics, Brown University. Email: simone\_martinalli@brown.edu}}
\date{October 15, 2026}
\maketitle
\thispagestyle{empty}
\begin{abstract}
This paper studies data-driven macroeconomic policy choice in a structural VAR framework. We formulate the policymaker's problem of optimizing the future path of a policy instrument as a statistical decision problem under two alternative policy regimes: a ``contemporaneous'' regime, where the chosen path is implemented through unexpected policy shocks each period, and a ``forward guidance'' regime, where the expected path is announced in advance and enforced through anticipated policy shocks. We derive oracle, Bayes-optimal, and robust Bayes decision rules, the last for the case where structural impulse responses are only set-identified, and show that the decision problem depends on the structural rotation matrix only through the span of the columns associated with the policy shocks. We argue that the plug-in rule  is uniformly dominated by a James--Stein-type shrinkage rule in a local asymptotic Gaussian limit experiment, and is therefore inadmissible therein. Numerical analysis calibrated to monetary SVARs compares the risks of the plug-in and a benchmark Bayes rule in Monte Carlo Gaussian simulations and finds that the Bayes rule attains lower risk at 95.1--99.9 percent of the posterior draws in the two regimes. We then study the relationship between the one-shot decision for a policy path and sequential policymaking, and show that, in the contemporaneous regime, the one-shot oracle path coincides with the expectation of the sequentially optimal oracle plan; we document that this representativeness property fails under Bayesian decision-making with parameter uncertainty. In an application to U.S. monetary policy in 2006--2007, we find that the plug-in and Bayes rules can differ substantially: the Bayes-optimal rule prescribes an initial tightening followed by partial easing across different specifications, while the plug-in changes across regimes.
\end{abstract}

\noindent%
\vfill

\newpage
\setcounter{page}{1}
\spacingset{1.45}

\section{Introduction}\label{sec:intro} 
Central banks and, more generally, macroeconomic policy authorities routinely choose the path of a policy instrument on the basis of projections of how the economy would respond under alternative counterfactual scenarios. Standard practice in both academic research \citep{antolin2021structural, crump2025large} and institutional settings constructs such projections from impulse responses estimated with SVAR models or local projections. A prominent example is the Summary of Economic Projections of the Federal Open Market Committee: four times a year, each participant submits projections for output, unemployment and inflation conditional on his or her own assessment of the appropriate path of the policy rate. Each participant thus implicitly solves a policy choice problem under uncertain counterfactuals. The role of econometrics in this process is usually confined, as advocated by \citet{hildreth1963}, to reporting estimates of the impulse responses and of the implied counterfactuals, together with uncertainty quantifications for them. How the decision maker should then incorporate identification and estimation uncertainty into the choice of the policy path is a question on which the macroeconometric literature offers little guidance.

Going beyond the conventional tasks of scientific reporting, this paper studies how the econometricians should engage with the policymakers to help their decision making. 
We develop a framework of macroeconomic policy choice using statistical decision theory of \citet{wald_statistical_1950} and study how to derive policy rules with decision theoretic optimality.

We consider a policymaker that aims at minimizing a quadratic loss representing the expected sum, over a finite horizon, of weighted squared deviations of the counterfactual paths of the relevant macroeconomic variables from a known target. Within this framework, we study two distinct policy regimes that differ in how the policy path is implemented. In the contemporaneous regime, following \citet{sims2006does}, the optimal path is delivered through a sequence of policy shocks that are unanticipated by the private sector, period by period. In the forward guidance regime, following \citet{mckay2023can} and \citet{barnichon2023sufficient}, the expected policy path is instead announced ex ante and implemented by manipulating anticipated news shocks. The distinction is central, as it determines what the possible policy actions are under the two regimes. In the contemporaneous case, an action consists in implementing a certain path of the policy instrument. In the forward guidance regime, the possible actions are intended as announcements or as a credible commitment of the policy authority, as they directly manipulate expectations about the future path of the instrument.
For both regimes we derive closed-form oracle rules by exploiting the certainty equivalence induced by the linear-quadratic structure. Under parameter uncertainty we obtain the Bayes-optimal rule, also in analytical form in the quadratic case, and, for the set-identified case, we consider the posterior Gamma-minimax rule in the spirit of \citet{giacomini2021robust} and \citet{Giacomini_Kitagawa_Read_2026}, which minimizes the worst-case posterior expected loss over the set of rotation matrices consistent with reduced-form evidence, and can be numerically approximated with the algorithms discussed in the latter paper. We further consider a regret-based variant, the Robust Bayes regret rule, which minimizes the posterior expectation of worst-case regret. 

Along the way, we characterize the identification requirements of the  decision problem: in the contemporaneous regime, the loss  depends on the structural rotation only through the column associated with the policy shock, invariantly to its sign; in the forward guidance regime, it depends on the columns associated with the policy shocks only through their span, so that any orthonormal basis of the same space of policy shocks delivers the same decision problem. Point identification of the policy problem is accordingly weaker than identification of the individual shocks. The same holds for the empirically relevant case of the forward guidance regime in which we can only identify a limited number of policy news shocks, where  the action of the policymaker only pins down the expected path at certain maturities. The feasible plug-in and Bayes forward guidance rules that we obtain under this setup are the ones that we compute in the numerical analysis of Section \ref{sec:numerical-risk} and in the empirical application to monetary policy. 

We then turn to the comparison between our framework and the standard empirical practice. In the SVAR literature, the parameters of the policy choice problems are commonly treated as known when the policy exercise is carried out, and point estimates are plugged in for the unknown parameters when the method is brought to the data.  We argue that the plug-in rule that implements this approach in our framework is uniformly dominated by a James--Stein-type shrinkage rule in a local asymptotic Gaussian limit experiment, and is therefore inadmissible therein. We additionally provide a quantitative assessment of its performance in numerical simulations calibrated to the  monetary policy exercise studied in the empirical section of the paper. We find that using a Gaussian approximation for the distribution of the point estimator of the SVAR parameters, a benchmark Bayes rule attains lower risk than the plug-in at 95.1--99.9 percent of the retained posterior draws, with the plug-in exhibiting median expected regret that is almost 8 times that of the benchmark Bayes  in the contemporaneous regime, and 3 times the regret of the Bayes rule in the forward guidance one. 

Having characterized the one-shot optimal decisions for policy paths, we consider its relationship with sequential policymaking, where we allow the policymakers to revise their actions over time, as new  shocks perturb the economy and more information on the unknown parameters accrues. 
In the contemporaneous regime we establish a representativeness property of the oracle: the one-shot optimal path coincides with the time-zero expectation of the plan of a policymaker who re-optimizes every period as shocks realize, which we show to be the solution of an associated linear-quadratic dynamic program. Under Bayesian decision-making the property of representativeness fails: the coefficients of the sequential rule become functionals of the posterior that moves with the realized data, and the time-zero expectation of the sequential plan --- a mean of ratios of posterior moments --- generically differs from the ex-ante plan --- a ratio of means. The sequential problem, moreover, carries an experimentation motive: the intervention itself changes the posterior at future dates, a channel that neither the ex-ante plan nor period-by-period re-optimization internalizes.

Finally, we take the evidence-based decision framework to the data with a one-shot decision problem faced by a policymaker at the beginning of 2006: choosing an eight-quarter path of the federal funds rate to stabilize the output gap and inflation, on the basis of a Bayesian SVAR estimated on data available at the end of 2005. In the contemporaneous regime, identified with the narrative instrument of \citet{romer2004new}, both the Bayes and the plug-in rules prescribe a substantially tighter stance than the observed one, but the Bayes path is smoother and more moderate. The policy surprises needed to enforce both the plug-in and Bayes-optimal path are large in magnitude for consecutive periods, which makes the assumption that the private sector be surprised each period less credible. In the forward guidance regime, where the announcement is enforced through $m=3$ identified shocks, the Bayes announcement prescribes a moderate tightening followed by easing, whereas the plug-in announcement changes shape and sign compared to its prescription in the contemporaneous regime.

Our paper contributes to four strands of literature. The first is the
macroeconomic and econometric work of the 1960s and 1970s on control
under parameter uncertainty. 
\citet{brainard1967uncertainty} studies a scalar control problem in a linear
structural equation, where the policy coefficients are random.  He considers the case where the source of uncertainty for the policy coefficient is sampling variability.
\citet{zellner1971introduction} develops the Bayesian treatment of
the control problem when the structural equation is  a multivariate linear regression model with several controls.
\citet{zaman1981complete} and \citet{berger1982general} cast
this linear control problem as a
statistical decision.
\citet{berger1982general} give general conditions for the
admissibility and inadmissibility of generalized Bayes rules in the same control problem.
\citet{chow1976control} extends the Bayesian treatment to general
nonlinear dynamic systems, taking a maximum likelihood estimate of the
structural parameters as given and working with the large-sample
 posterior under a flat prior.
The section on sequential decision-making in our paper  relates to the classical work of \citet{simon1956dynamic} and
\citet{theil1957note}, establishing certainty equivalence for
linear-quadratic dynamic programming and  the representativeness property of the optimal strategy, respectively.
We show this result applies to SVAR counterfactuals with known
parameters in Proposition~\ref{prop:representativeness}. \citet{prescott1972multiperiod} extends the scalar
control over many periods, analyzing the properties of the sequential
solution and the experimentation motive that separates the adaptive
solution from the sequentially updated one.

Our contribution to this strand of the literature is to integrate the SVAR policy counterfactuals for contemporaneous and forward-guidance regimes into
statistical decision theory for generalized control problems. The regression models of Brainard, Zellner and
Prescott require the policy instrument to be as good as randomized in the data, and abstract from the dynamics of the macroeconomy. The SVAR allows for dynamic and endogenous dependence between the policy instrument and the targets, delivering the notion of causality based on the exogenous structural shocks \citep{sims1980macroeconomics}. 

Second, our paper contributes to the SVAR literature on counterfactual policy analysis. The use of
semi-structural time-series models to evaluate policy interventions dates back
to \citet{sims1982policy}. Contemporaneous regime counterfactuals have been pioneered by  
\citet{bernanke1997systematic},
\citet{leeper2003modest}  and \citet{sims2006does}. \citet{antolin2021structural}
 study conditional forecasts to scenarios
driven by a chosen subset of structural shocks and provide a general
framework in which our counterfactuals can be cast. Recent contributions
by \citet{mckay2023can} and \citet{barnichon2023sufficient} construct
Lucas-critique-robust counterfactuals for a large class of linear macroeconomic models 
under the assumption that the researcher can identify  the impulse responses
to multiple distinct policy shocks, and show that these counterfactuals are sufficient to identify optimal policy when the parameters are known. 
Our main contribution to this strand of the literature is  to provide a unified framework for policy choice based on statistical decision theory that characterizes optimal policy decisions when the parameters are unknown, and the policymaker uses past data to learn their values. In the DSGE literature, optimal monetary policy under parameter uncertainty has been studied from a Bayesian perspective (e.g., \citet{cogley2011bayesian}) and from a robust-control perspective (e.g., \citet{hansen2011robustness}). These approaches do integrate parameter uncertainty into the policy problem but they neither address the set-identification that is central to modern SVAR analysis nor cast the policy problem within the Waldian framework we adopt here.

Third, our paper contributes to the growing literature of statistical treatment choice initiated by \citet{manski2004statistical} and \citet{dehejia2005program}. The common setting in this literature is static; cross-sectional data of a randomized controlled trial is used to inform a treatment assignment policy, assuming external validity of the experimental data for the target population. See, e.g., \citet{HiranoPorter2009}, \citet{Stoye2009,Stoye2012}, \citet{Tetenov2012}, \citet{BhattacharyaDupas2012}, \citet{kitagawa2018should, KT21}, \citet{MT17} and \citet{AW17}. In contrast, the current paper studies statistical policy choice in a dynamic environment where the available data is time-series data and planner's action is a policy path over multiple periods. The recent literature on dynamic policy choice includes the empirical welfare maximization approach of \citep{kitagawa2022policy} that studies estimation of an optimal policy in the framework of potential outcome time-series \citep{bojinov2019time}. Their approach covers the contemporaneous regime without any functional form specification, while it does not cover the forward guidance regime. \citet{Liao_etal2022_AnnStat} studies estimation of an optimal policy in stationary Markov decision process. 

Finally, our paper contributes to the literature on robust decision making under set-identified welfare criterion including \citet{giacomini2019ambiguity,Giacomini_Kitagawa_Read_2026, christensen2026optimal}. In the static treatment choice under set-identifed welfare, \citet{Manski2000, Manski2007} apply the minimax regret criterion of \citet{Savage51}. See, e.g., \citet{ishihara2021}, \citet{kitagawa2023treatment}, \citet{montiel2023decision} and \citet{yata2021optimal} for recent advances.    

The remainder of the paper is organized as follows. Section~\ref{sec:framework} lays out the policy choice framework and the welfare criteria, and Section~\ref{sec:svar} the SVAR model and the treatment of set identification. Section~\ref{sec:counterfactuals} derives the counterfactual paths and the oracle rules under the two regimes. Section~\ref{sec:decision-theory} develops the statistical decision theory, the feasible forward guidance rules, and the suboptimality results for plug-in rules. Section~\ref{sec:sequential} studies sequential policymaking, and Section~\ref{sec:empirical} discusses the empirical application and the Monte Carlo simulations. Proofs and additional results are collected in the appendices.

\section{Statistical policy choice framework}\label{sec:framework}
We consider a policymaker (PM) that needs to take an action $\textbf{a}$ in some set $\mathcal{A}$ --- that will determine the future path of a policy instrument --- to optimize the value of the loss function
\begin{equation}\label{lossfun1}
L(\mathbf{a}, \boldsymbol{\theta}, \boldsymbol{\chi}_0) = \mathbb{E}_{\boldsymbol{\varepsilon}} \left[ \sum_{t=1}^h \left( \mathbf{y}_t(\mathbf{a}, \boldsymbol{\theta}, \boldsymbol{\chi}_0, \boldsymbol{\varepsilon}) - \mathbf{y}^* \right)' W_t \left( \mathbf{y}_t(\mathbf{a}, \boldsymbol{\theta}, \boldsymbol{\chi}_0, \boldsymbol{\varepsilon}) - \mathbf{y}^* \right) \right]
\end{equation}
which is the sum of expected squared deviations of future counterfactual paths of  macroeconomic variables $\{\mathbf{y}_t(\mathbf{a}, \cdot)\}_{t=1}^h$ from a known target vector $\textbf{y}^*$ weighted by the the known  matrix $W_t$. The canonical weighting matrix studied in the related literature \citep{barnichon2023sufficient, mckay2023can} and the one we adopt in our empirical applications is
\[
W_t = \beta^t \Lambda \qquad \Lambda = \text{diag}(\lambda_1, \ldots, \lambda_{n})
\]
with $\Lambda$ collecting the relative weights on the non-policy target variables and $\beta$ being the discount factor.  Note that $\textbf{y}^*$ is a vector including a scalar target for each variable $1,
\ldots, n$ and each future horizon $1,\ldots, h$. In particular, it is allowed to change over future horizons. We assume that the PM only cares about the realization of the macroeconomic variables up to a finite policy-relevant horizon $h$.
The expectation in the expression for the loss function is taken over a known or normalized distribution of a vector of future structural macroeconomic shocks $\boldsymbol{\varepsilon}$. The initial state $\bs{\chi}_0$ captures the relevant economic initial conditions at time $t=0$ where we assume that the policy decisions need to be made, and in the finite-order structural VAR framework we consider, this coincides with the realizations of the endogenous variables in periods $t = 1-p, \ldots, t=0$. The finite-dimensional parameter $\bs{\theta}$ collects both the reduced-form parameters --- namely the autoregressive coefficients and the entries of the variance-covariance matrix of the Wold innovations --- and the structural parameters that determine the on-impact response coefficients of the endogenous variables to the structural shocks.

The quadratic loss is a tractable benchmark rather than a methodological requirement: the general approach to evidence-based policy decisions we develop applies to any well-defined loss function, and the policymaker remains free to specify a welfare criterion tailored to their preferences. The quadratic choice is convenient because it delivers closed-form oracle, Bayes, and regret rules and a fast numerical approximation of the robust Bayes rules introduced in Section~\ref{sec:bayes}. Additionally, the analytical and numerical risk analysis for the plug-in rest on the quadratic structure, and the certainty equivalence property implied by the linear-quadratic framework is also essential for the sequential analysis in Section \ref{sec:sequential}. Finally, quadratic approximations are  standard in optimal-policy analysis, where they arise from a second-order approximation of household welfare \citep{gali2015monetary}. We do not take a stance on what the target $\mathbf{y}^*$ should be:
like the policy weights $W_t$, we treat it as a known primitive of the
decision problem, describing the policymaker's preferences. In our
empirical application to monetary policy we hold $\mathbf{y}^*$
constant---closing the output gap and stabilizing inflation at a
constant level, with the two objectives entering the loss with equal
weight---mirroring the Fed's publicly announced dual mandate.
Hierarchical mandates such as the ECB's, which prioritize price
stability, are accommodated within the same specification by shifting
the weights in $W_t$ toward inflation.  

In the structural VAR specialization developed below, $\mathbf{y}_t$ collects all endogenous variables, including the policy instrument; in our base case the loss weights only the non-policy block $\mathbf{x}_t$ --- the macroeconomic objectives or target variables --- and treats the instrument purely as a lever, in line with \citet{barnichon2023sufficient} and \citet{mckay2023can}. The framework nonetheless accommodates a richer loss that also penalizes the instrument path itself --- for instance an interest-rate-smoothing term --- at the cost of the clean separation between targets and instrument. We consider an extension with a  smoothing penalty in the empirical application of Section \ref{sec:emp_contemp}, while formal treatment of the decision problem under  generalized loss function is deferred to Appendix \ref{app:genloss}.

\subsection{Two policy regimes}\label{sec:regimes}

We study two distinct regimes that differ in how the policy path is implemented and, correspondingly, in the nature of the action $\mathbf{a}$ that the policymaker is choosing and of the resulting policy counterfactuals. The formal derivation of the counterfactual paths $\textbf{y}_t(\mathbf{a}, \cdot)$ under each regime is postponed to Section~\ref{sec:counterfactuals}, after the SVAR framework has been laid out in Section~\ref{sec:svar}. We describe the two regimes here at a conceptual level.

In the \textit{contemporaneous regime}, following \citet{sims2006does}, the action $\mathbf{a}$ is the realized future path of the policy instrument $\textbf{i}_{1:h}=(i_1, \ldots, i_h)'$. The counterfactual paths are obtained through the manipulation of the sequence of contemporaneous policy shocks that are unanticipated by the private sector in each period: the private sector continues to expect the historical policy rule embedded in the SVAR and is surprised period by period by the policymaker setting the value of the instrument to what is prescribed by the action $\textbf{a}$. This methodology is well known to be exposed to Lucas-critique-type concerns when the magnitude or the horizon of the policy intervention is large, since forward-looking rational agents would eventually learn that the shocks are systematically deviating from their unconditional distribution and adjust their behavior accordingly. As \citet{sims2006does} and \citet{antolin2021structural} argue, the contemporaneous regime remains sensible for modest policy interventions over short horizons.

In the \textit{forward guidance regime}, following \citet{mckay2023can} and \citet{barnichon2023sufficient}, we consider a policymaker that can manipulate the path of the policy instrument through different channels (multi-dimensional policymaking). The most relevant case we consider is the possibility that the policymaker can directly manipulate expectations about the future path of the instrument through announcements about future policy, which is formalized by introducing ``news'' shocks in the SVAR model. In this second regime, then, the action $\mathbf{a}$ is the announced \textit{expected} future path $\textbf{i}^e_{1:h} = (i^e_1, \dots, i^e_h)'$ of the policy instrument, which is implemented by manipulation of a set of policy shocks that all materialize at time $t=1$,  rather than a sequence of contemporaneous surprises. The realized path then equals the announced one plus the historical response of the policymaker to future policy and non-policy shocks. In Section \ref{sec:counterfactuals} we show that the loss is a quadratic function of the action and that, despite the constant, linear and quadratic coefficients have a different structure in the two regimes, they share the same decision-theoretic properties and can be analyzed in a unified framework.

\subsection{Parameter uncertainty and the limits of as-if optimization}\label{sec:asif}

While the initial state $  \bs{\chi}_0  $ is typically observed at the time of the policy decision, the structural parameter vector $  \bs{\theta}  $ is not. Policymakers must therefore rely on estimated impulse responses and counterfactual projections as inputs to their decisions. 
In the SVAR literature on policy choice, parameter uncertainty has not been formally integrated into the decision problem. Instead, the literature typically characterizes the optimal policy as a function of the unknown structural parameter and then replaces it with a point estimate to obtain an empirical counterpart of the theoretical solution. A formal reconstruction of this implicit procedure is the following two-step approach: first, optimize the policy decision for all possible values of the parameter in the so-called \textit{oracle} problem, to obtain the oracle decision rule
\[
\textbf{a}^*(\bs{\theta}) = \underset{\textbf{a}\in \mathcal{A}}{\text{argmin}} \, L(\mathbf{a}, \boldsymbol{\theta}).
\]
Second, plug some point estimate $\hat{\bs{\theta}}$ into the expression for the oracle decision rule
\[
\hat{\textbf{a}}_{plug-in} = \textbf{a}^*(\hat{\bs{\theta}})
\]
and implement the action $\hat{\textbf{a}}_{plug-in}$.

This approach to the problem of optimal policy choice under parameter uncertainty is what \citet{manski2021econometrics} calls \textit{as-if optimization}, since it can be seen as finding the optimal action that could be taken in a state of the world where $\hat{\bs{\theta}}$ is the true value of the parameter --- in our case, the true parameter vector in the underlying SVAR model. Manski points out that the as-if approach is inference-based, as it obtains a statistical decision function based on arguments that justify good properties of the point estimates that are plugged in. 
The most common such justifications are asymptotic arguments that derive consistency or asymptotic normality of the point estimates. However, asymptotic arguments do not prove that as-if optimization performs well in finite samples. This is particularly relevant in multivariate time series analysis, where sample sizes tend to be relatively small compared to the number of parameters that need to be estimated. This calls for a theory of optimal decisions that is valid in finite samples.

Beyond the points raised by Manski, there are two further limitations of the plug-in method. First, even when the structural parameter $\bs{\theta}$ is point-identified, the plug-in approach does not explicitly account for the probabilistic uncertainty that is associated with the point estimate $\hat{\bs{\theta}}$.
While the as-if approach could be appealing for its simplicity when there is little uncertainty around the point estimate, it is undesirable to treat the point estimate as the true parameter when the former is estimated with low precision. 
In addition, when the structural parameter is only set-identified by the data, the as-if optimizing PM needs to choose a single point estimate for the structural parameter among all the ones that are consistent with reduced-form evidence, which carries some degree of arbitrariness from a frequentist perspective.

\subsection{Welfare criteria under parameter uncertainty}\label{sec:welfare}

Before specializing to the SVAR framework, we lay out in general terms the expected welfare criteria we will use to evaluate decision rules in the presence of parameter uncertainty. Throughout, we maintain the notation of a generic loss function $L(\mathbf{a}, \bs{\theta})$ depending on an action $\mathbf{a} \in \mathcal{A}$ and an unknown structural parameter $\bs{\theta} \in \Theta$, suppressing the dependence on the initial state $\bs{\chi}_0$ to lighten notation whenever this is inconsequential. We assume the policymaker has access to a sample 
\[
\textbf{y} \equiv  Y^T_0 = ( \textbf{y}'_0, \ldots,  \textbf{y}'_{-T+1})' \in \mathcal{Y}^T 
\]
of observations of macroeconomic data over $T$ periods preceding the moment at which the decision needs to be made. The structural parameter $\bs{\theta}$ induces the sampling
distribution of the data from the parametric family
\[
\textbf{y}|\bs{\theta} \sim p_T(\textbf{y} \mid \bs{\theta})
\]
Anticipating the structure of the SVAR framework introduced in Section~\ref{sec:svar}, we will partition the parameter as $\bs{\theta} = (\bs{\phi}, \mathbf{q})$, where $\bs{\phi}$ is a point-identified reduced-form component  and $\mathbf{q}$ is a structural component that may only be set-identified--- the sampling distribution $p_T $ depending on $\bs{\theta}$ only through $\bs{\phi}$.   

We let $\hat{\mathbf{a}}: \mathcal{Y}^T \to \mathcal{A}$ denote a statistical decision function mapping the data into an action and
following \citet{wald_statistical_1950} we define its frequentist risk as the expected loss under repeated sampling from $p_T(\cdot \mid \bs{\theta})$:
\begin{equation}\label{freq_risk}
R(\hat{\mathbf{a}}, \bs{\theta}) = \mathbb{E}_{\mathbf{y} \sim p_T(\cdot \mid \bs{\theta})} \left[ L(\hat{\mathbf{a}}(\mathbf{y}), \bs{\theta}) \right]
\end{equation}
Because of the peculiar nature of the macroeconomic decisions we are considering here, we additionally introduce the following notion of \textit{conditional risk}. Assuming that $T\gg p$, we consider evaluating decision rules upon their average sampling performance \textit{conditional on the state} $\bs{\chi}_0$. Note that the initial state for the decision problem $\bs{\chi}_0$ coincides with the last $p$ observations of the sample path $Y_0^T$. We then define the conditional risk by averaging over the distribution of the \textit{truncated sample path} 
\[
\check{\textbf{y}} \equiv Y^{T-p}_{0}= (\textbf{y}'_{-p}, \ldots, \textbf{y}'_{-T+1})' 
\]
conditional on the last $p$ observations agreeing with $\bs{\chi}_0$. Letting this distribution be 
\[
\check{\textbf{y}} \mid \bs{\theta}, \bs{\chi}_0 \sim p_{T-p}(\check{\textbf{y}} \mid \bs{\theta}, \bs{\chi}_0)
\]
we define conditional risk as\footnote{The motivation for conditional risk is that a policymaker is typically
interested in leaning against a specific initial condition, shock, or
perturbation realized in the recent past---precisely what the
policy-problem initial state $\bs{\chi}_0$ captures. It is then natural
to evaluate a statistical decision function ex ante only across
situations in which the policymaker would actually contemplate using
it, i.e.\ holding the initial conditions fixed. The traditional,
unconditional notion of risk does not capture this aspect: averaging
over the entire distribution of the data, it evaluates a decision
function across all initial states $\bs{\chi}_0$ the economy could be
facing. Which criterion is preferable depends on the kind of policy
exercise the policymaker is after, conditional or unconditional.} 
\[
R_{\bs{\chi}_0}(\hat{\textbf{a}}, \bs{\theta}) = \ev_{\check{\textbf{y}} \sim p_{T-p}(\cdot \mid \bs{\theta}, \bs{\chi}_0)} \left[ L\big(\hat{\textbf{a}}(\check{\textbf{y}}, \bs{\chi}_0), \bs{\theta}, \bs{\chi}_0\big)\right],
\]

where $\hat{\textbf{a}}(\check{\textbf{y}}, \bs{\chi}_0)$ denotes the decision function evaluated at the full sample $\textbf{y} = (\bs{\chi}_0', \check{\textbf{y}}')'$ assembled from the truncated path and the fixed state.
Taking $R(\cdot)$ to be either risk function considered above, and for a fixed initial state $\bs{\chi}_0$, frequentist risk is a function of the unknown parameter. Since a decision rule that uniformly minimizes $R(\hat{\mathbf{a}}, \bs{\theta})$ across $\Theta$ rarely exists, additional criteria are needed to rank decision functions.

If the policymaker is willing to specify a prior distribution $\pi(\bs{\theta})$ over the parameter space, the Bayes risk of $\hat{\mathbf{a}}$ is the prior-weighted average of the frequentist risk:
\begin{equation}\label{bayes_risk}
\rho(\hat{\mathbf{a}}, \pi) = \mathbb{E}_{\pi} \left[ R(\hat{\mathbf{a}}, \bs{\theta}) \right]
\end{equation}
A standard result in decision theory \citep{ferguson1967} guarantees that a rule minimizing Bayes risk (a Bayes decision rule) can be obtained by minimizing the posterior expected loss:\footnote{Under the conditional criterion $R_{\bs{\chi}_0}$ the posterior that delivers the Bayes rule is that of the conditional experiment, $\pi(\bs{\theta} \mid \check{\textbf{y}}, \bs{\chi}_0) \propto \pi(\bs{\theta})\, p_{T-p}(\check{\textbf{y}} \mid \bs{\theta}, \bs{\chi}_0)$, for which the characterization \eqref{bayes_rule} holds verbatim. In our empirical exercise, we use throughout the posterior based on the full-sample likelihood, $\pi(\bs{\theta} \mid \textbf{y}) \propto \pi(\bs{\theta})\, p_T(\textbf{y} \mid \bs{\theta})$. Since $p_T(\textbf{y} \mid \bs{\theta}) = p_p(\bs{\chi}_0 \mid \bs{\theta})\, p_{T-p}(\check{\textbf{y}} \mid \bs{\theta}, \bs{\chi}_0)$, where $p_p(\cdot \mid \bs{\theta})$ is the stationary distribution of $p$ consecutive observations, the two posteriors differ by the likelihood of $\bs{\chi}_0$ alone,  a term that does not depend on $T$. Under the maintained assumption $T \gg p$ we ignore it, in the same way as the conventional VAR likelihood ignores the marginal of the first $p$ observations. 
}
\begin{equation}\label{bayes_rule}
\hat{\mathbf{a}}^{Bayes}(\mathbf{y}) = \underset{\mathbf{a} \in \mathcal{A}}{\text{argmin}} \, \mathbb{E}_{\bs{\theta} \mid \mathbf{y}} \left[ L(\mathbf{a}, \bs{\theta}) \right]
\end{equation}

An alternative way to evaluate decisions under uncertainty is through the notion of \textit{regret}, defined pointwise as the difference between the loss incurred by a chosen action and the loss that would have been incurred by an oracle with perfect knowledge of $\bs{\theta}$:
\begin{equation}\label{regret_def}
Re(\mathbf{a}, \bs{\theta}) = L(\mathbf{a}, \bs{\theta}) - \underset{\mathbf{a}' \in \mathcal{A}}{\inf} \, L(\mathbf{a}', \bs{\theta})
\end{equation}
Regret-based criteria put more weight on states of the world in which decisions can make a difference, since they subtract off the unavoidable component of the loss. The frequentist counterpart of regret is the expected regret $\mathbb{E}_{\mathbf{y} \sim p(\cdot \mid \bs{\theta})} \left[ Re(\hat{\mathbf{a}}(\mathbf{y}), \bs{\theta}) \right]$, and analogous Bayesian counterparts can be defined by replacing the loss with regret in (\ref{bayes_risk}) and (\ref{bayes_rule}).

When the structural component $\mathbf{q}$ is only set-identified, the policymaker may want to remain agnostic about its unidentified component. The Robust Bayesian approach of \citet{giacomini2021robust} combines a prior on the point-identified reduced-form parameter $\bs{\phi}$ with a worst-case treatment of $\mathbf{q}$ over its identified set $\mathcal{I}_{\mathbf{q}}(\bs{\phi})$, which we formally define in Section~\ref{sec:svar}. The posterior Robust Bayes rule is then defined as:
\begin{equation}\label{rb_rule}
\hat{\mathbf{a}}^{RB}(\mathbf{y}) = \underset{\mathbf{a} \in \mathcal{A}}{\text{argmin}} \, \mathbb{E}_{\bs{\phi} \mid \mathbf{y}} \left[ \underset{\mathbf{q} \in \mathcal{I}_{\mathbf{q}}(\bs{\phi})}{\sup} \, L(\mathbf{a}, \bs{\phi}, \mathbf{q}) \right]
\end{equation}
A regret-based variant, the Robust Bayes Regret rule, is obtained by replacing $L$ with $Re$ in (\ref{rb_rule}). We will use these criteria as the basic building blocks of our analysis in the following sections, and discuss in detail their properties and computational implementation in the SVAR setting in Section~\ref{sec:decision-theory}.

\section{The SVAR framework}\label{sec:svar}

In this section we lay out the SVAR model that underlies both policy regimes introduced in Section~\ref{sec:regimes}, the parametrization of the structural and reduced-form components, and the formal treatment of set-identification.

\subsection{The SVAR model and reduced-form parametrization}\label{sec:svar-model}

We assume that, before making the decision, the policymaker observes a time series of $T$ observations on past macroeconomic variables, $Y^T_0$, that are drawn from a  Structural VAR
\begin{equation}\label{SVAR}
 \textbf{y}_t = \Pi \textbf{w}_t + B \bs{\varepsilon}_t
\end{equation}
where $\bs{\varepsilon}_t$ are $i.i.d.$ with zero mean and identity variance-covariance matrix. We assume that $B$ is invertible. The $n \times 1$ vector of endogenous variables is partitioned
\[
\textbf{y}_t = \begin{pmatrix}
    i_t \\
    \textbf{x}_t
\end{pmatrix}
\]
where $i_t$ is a scalar policy instrument and $\textbf{x}_t$ are the remaining macroeconomic variables in the square system. The macroeconomic structural shocks are collected in the $n \times 1$ vector $\bs{\varepsilon}_t$. The $k := np + 1$ vector $\textbf{w}_t$ contains $p$ lagged values of each of the endogenous variables as well as a constant
\[
\textbf{w}_t = \begin{pmatrix}
    1 & \textbf{y}'_{t-1} & \ldots & \textbf{y}'_{t-p}
\end{pmatrix}'
\]
We define
\[
\textbf{u}_t = B\bs{\varepsilon}_t
\]
to be the Wold innovations with positive-definite covariance matrix
\[
 \Sigma_u = \text{Var}(\textbf{u}_t) = BB'
\]
As is common in the literature \citep[e.g.,][]{uhlig2005effects}, we parametrize the variance-covariance matrix $\Sigma_u$ and the on-impact coefficients matrix $B$ through the non-zero entries of a lower triangular matrix and a rotation matrix $Q$ as follows. We let
\[
\Sigma_l = \text{chol}(\Sigma_u)
\]
be the unique lower-triangular Cholesky factor of $\Sigma_u$ with strictly positive diagonal and collect its non-zero entries in
\[
\bs{\sigma}_l = \text{vech}^0(\Sigma_l)
\]
Then for every full-rank $B$ there exists a unique orthogonal matrix $Q \in \mathcal{O}(n)$ such that
\[
B = \Sigma_l Q
\]
and we let $\textbf{q} = \text{vec}(Q)$ be its vectorized form.

We can then partition the unknown parameter vector
\[
\bs{\theta} = \begin{pmatrix}
    \bs{\phi} \\
    \textbf{q}
\end{pmatrix}
\]
into a reduced-form subvector $\bs{\phi}$ and a structural subvector $\textbf{q}$, where $\bs{\phi}$ collects the autoregressive coefficients and the entries of the Cholesky factor, i.e., for $\bs{\pi} = \text{vec}(\Pi)$ we define
\[
\bs{\phi} = \begin{pmatrix}
    \bs{\pi} \\
    \bs{\sigma}_l
\end{pmatrix}
\]
We assume that the policymaker knows the distribution $p_T$ of the data $Y^T_0$ up to knowledge of the parameter $\bs{\theta}$, while as in \citet{giacomini2021robust}, the likelihood of the data only depends on the unknown parameter vector through the reduced form:
\begin{equation}\label{likelihood}
Y^T_0 \mid \bs{\theta} \sim p_T(Y_0^T \mid \bs{\phi})
\end{equation}
The parameter space $\Phi$ for the reduced form will, in our framework, be the set of autoregressive and variance-covariance parameters from a reduced-form VAR. The canonical case and the one we consider in all our empirical and simulation results is when $p_T$ is a multivariate Gaussian probability density function, obtained from the assumption that the structural shocks $\bs{\varepsilon}_t$ are jointly normal and independent. The general framework of the decision problem introduced below can accommodate different parametric distributions or nonparametric distributions; for the latter, we can view the Gaussian likelihood function as a quasi-likelihood exploiting the first and second moments of the data generating process. 

\subsection{Structural parametrization and set-identification}\label{sec:set-id}

As is well known in the literature \citep{kilian2017structural}, and as is apparent from Equation (\ref{likelihood}), the SVAR model in (\ref{SVAR}) is only identified up to the orthogonal matrix $Q$ that determines the mapping from the reduced-form parameter $\bs{\sigma}_l$ to the on-impact coefficients matrix $B$. In principle, \textit{any} orthogonal matrix $Q \in \mathcal{O}(n)$ is observationally equivalent. In most applications, however, economic theory suggests restrictions that narrow down the set of rotation matrices consistent with reduced-form evidence. Such restrictions can be characterized in general by constructing an \textit{identified set} for the vectorized rotation matrix,
\[
\mathcal{I}_{\textbf{q}}(\bs{\phi}) \subseteq \text{vec}\bigl(\mathcal{O}(n)\bigr),
\]
that collects all orthonormal matrices consistent with the reduced-form parameter vector $\bs{\phi}$ and with a set of economically meaningful restrictions that the policymaker is willing to impose.

Three types of restrictions have been widely studied in the SVAR
literature are short-run zero restrictions
\citep{ christiano1999monetary}, long-run zero
restrictions \citep{blanchard1989dynamic}, and sign restrictions
\citep{faust1998robustness,  uhlig2005effects}.
Each is obtained by constraining the on-impact or cumulative impulse response
coefficients implied by a given rotation matrix to be zero or to lie on the
positive or negative real line; the identified set collects all vectorized
rotation matrices whose implied IRFs satisfy the restrictions. 

Following \citet{giacomini2021robust}, we formalize zero and sign restrictions through a set of linear constraints on the columns $\textbf{q}_i$ of $Q$, stacked as
\[
R(\bs{\phi}, \textbf{q}) = \begin{pmatrix}
    F(\bs{\phi}, \textbf{q}) \\
    S(\bs{\phi}, \textbf{q})
\end{pmatrix},
\]
where
\[
F(\bs{\phi}, \textbf{q}) = \begin{pmatrix}
    F_1(\bs{\phi})\,\textbf{q}_1 \\
    F_2(\bs{\phi})\,\textbf{q}_2 \\
    \vdots \\
    F_n(\bs{\phi})\,\textbf{q}_n
\end{pmatrix}, \qquad
S(\bs{\phi}, \textbf{q}) = \begin{pmatrix}
    S_1(\bs{\phi})\,\textbf{q}_1 \\
    S_2(\bs{\phi})\,\textbf{q}_2 \\
    \vdots \\
    S_n(\bs{\phi})\,\textbf{q}_n
\end{pmatrix},
\]
and $F_i(\bs{\phi})$ and $S_i(\bs{\phi})$ are $f_i \times n$ and $s_i \times n$ matrices, respectively. Linear constraints of this form can encode zero or sign restrictions on the instantaneous or cumulative structural response of any variable in the system to any shock.\footnote{See \citet{giacomini2021robust} for analytic expressions for the matrices $F_i(\cdot)$ and $S_i(\cdot)$ corresponding to specific restrictions.} This linear structure is particularly convenient for the numerical implementation of the Robust Bayes rules discussed in Section~\ref{sec:decision-theory}. The identified set for $\textbf{q}$ is then
\[
\mathcal{I}_{\textbf{q}}(\bs{\phi}) = \left\{ \textbf{q} = \text{vec}(Q),\; Q \in \mathcal{O}(n) \,\bigm|\, F(\bs{\phi}, \textbf{q}) = \textbf{0} \;\text{and}\; S(\bs{\phi}, \textbf{q}) \geq \textbf{0} \right\}.
\]
Point identification is then the special case in which $\mathcal{I}_{\textbf{q}}(\bs{\phi})$ is a singleton for every $\bs{\phi} \in \Phi$. Beyond zero and sign restrictions, our framework can accommodate more general set-identification schemes that have been studied in the SVAR literature, such as narrative restrictions \citep{giacomini2022narrative} with set-identification conditional on narrative events or Proxy SVARs \citep{giacomini2022robust} with set-identifying exclusion restrictions. 

\subsection{Future trajectories}\label{sec:trajectories}

We now describe a general decomposition of the future paths of the endogenous variables that will be used in Section~\ref{sec:counterfactuals} to derive the counterfactuals under the two policy regimes. Without loss of generality, we assume that the statistical decision procedure is carried out at time $t=0$ and we denote the stacked future paths of the endogenous variables from period $t=1$ to period $t=h$ by $\textbf{y}_{1:h}$. In particular, we let
\[
\textbf{i}_{1:h} = \begin{pmatrix}
    i_{1} & \ldots & i_h
\end{pmatrix}'
\]
and for every $k = 1, \ldots, n-1$
\[
\textbf{x}_{k,1:h} = \begin{pmatrix}
    x_{k,1} & \ldots & x_{k,h}
\end{pmatrix}',
\]
so that for
\[
Y_h = \begin{pmatrix}
    \textbf{y}_1 & \textbf{y}_2 & \ldots & \textbf{y}_h
\end{pmatrix}',
\]
we have
\[
\textbf{y}_{1:h} = \begin{pmatrix} \textbf{i}'_{1:h} & \textbf{x}'_{1:h} \end{pmatrix}' = \text{vec}(Y_h)
\]
Following \citet{antolin2021structural}, for some appropriately defined matrix $M(\bs{\theta})$, we can decompose the future stacked paths of the variables as
\begin{equation}\label{trajdecomp}
\textbf{y}_{1:h} = \textbf{b}_{1:h}(\bs{\theta}, \bs{\chi}_0) + M(\bs{\theta})\bs{\varepsilon}_{1:h}
\end{equation}
where $\bs{\varepsilon}_{1:h}$ stacks the structural shocks in the same variable-major ordering as the stacked paths $\textbf{y}_{1:h}$. The decomposition above expresses the stacked paths as the sum of a linear response to future shocks and a \textit{predetermined} component $\textbf{b}_{1:h}$ that only depends on the parameters and the initial state for the decision problem, which in the SVAR framework can now be given a concrete expression
\[
\bs{\chi}_0 = \begin{pmatrix}
    \textbf{y}_{0}' & \ldots & \textbf{y}_{1-p}'
\end{pmatrix}'
\]
Since in the base case the loss weights only the non-policy block (Section~\ref{sec:framework}), it is enough to characterize the counterfactual paths of the target variables. In the next section we show that, for each of the two policy regimes considered in Section~\ref{sec:regimes}, the implementation of the corresponding policy action $\textbf{a} \in \mathcal{A}$ yields counterfactual paths for the non-policy variables $\textbf{x}_{1:h}(\textbf{a}, \cdot)$ that admit the representation
\begin{equation}\label{cfactexpress}
\textbf{x}_{1:h}(\textbf{a}, \bs{\theta}, \bs{\chi}_0, \bs{\varepsilon}_{1:h}) = D_b(\bs{\theta})\, \textbf{b}_{1:h}(\bs{\theta}, \bs{\chi}_0) + D_a(\bs{\theta})\,\textbf{a} + D_\varepsilon(\bs{\theta})\,\bs{\varepsilon}_{1:h},
\end{equation}
where the predetermined component $\textbf{b}_{1:h}$ retains its full ($nh$-dimensional) form from (\ref{trajdecomp}) and the multipliers $D_b, D_a, D_\varepsilon$ now map into the $(n-1)h$-dimensional space of the non-policy target paths. We then plug $\textbf{x}_{1:h}(\textbf{a},\cdot)$ into the loss function (\ref{lossfun1}). Section~\ref{sec:counterfactuals} derives the explicit form of the multiplier matrices for the two regimes, formulates the oracle problem and derives the oracle rule in each case, highlighting the main differences and similarities.

\section{Counterfactual paths and oracle rules}\label{sec:counterfactuals}
Building on the SVAR framework introduced in Section~\ref{sec:svar}, we now derive the counterfactual paths $\textbf{x}_{1:h}(\textbf{a}, \cdot)$ of the non-policy target variables and the oracle rules under each of the two policy regimes introduced in Section~\ref{sec:regimes}. 

\subsection{Contemporaneous regime: counterfactuals and oracle rule}\label{sec:contemp}

In the contemporaneous regime we follow the \citet{sims2006does} methodology to obtain counterfactual paths when the action $\mathbf{a}$ is the realized future path $\bar{\textbf{i}}_{1:h} = (i_1, \dots, i_h)'$ of the policy instrument, which is enforced through a sequence of unanticipated policy shocks.\footnote{We use the bar notation for the action $\bar{\textbf{i}}_{1:h}$ to keep it conceptually different from the future projection of the SVAR interest rate path $\textbf{i}_{1:h}$ under a general sequence of future shocks $\bs{\varepsilon}_{1:h}$. The two clearly coincide when the shocks are chosen so as to enforce the action $\bar{\textbf{i}}_{1:h}$.}
We partition the structural shock vector accordingly as
\[
\bs{\varepsilon}_t = \begin{pmatrix} v_t \\ \bs{\eta}^c_t \end{pmatrix}
\]
where $v_t$ is a scalar contemporaneous monetary policy shock and $\bs{\eta}^c_t$ is the $(n-1) \times 1$ vector of non-policy structural shocks,  which collects all remaining structural shocks in the system. Correspondingly, we stack future shocks in the variable-major order of Section~\ref{sec:trajectories}: $\textbf{v}_{1:h} = (v_1, \dots, v_h)'$ and
\[
\bs{\eta}^c_{1:h} := \operatorname{vec}\big(E^c_h\big),
\qquad
E^c_h := \big(\bs{\eta}^c_1, \dots, \bs{\eta}^c_h\big)' \in \reals^{h \times (n-1)}.
\]

Before specializing to the policy partition, it is worth recording the block structure of the impulse response matrix $M(\bs{\theta})$ from Section~\ref{sec:trajectories}, which has a transparent economic reading. Recall that in the variable-major ordering $\textbf{y}_{1:h}=\operatorname{vec}(Y_h)$ the stacked path lists all $h$ horizons of the first variable, then all $h$ horizons of the second, and so on; the same ordering applies column-wise to the stacked shocks $\bs{\varepsilon}_{1:h}$. Accordingly $M(\bs{\theta})$ partitions into $h\times h$ blocks, one for each variable--shock pair: writing $[M]_{jk}$ for the block relating variable $j$ to shock $k$, its $(s,t)$ entry $[M]_{jk}^{s,t}$ measures the impact of the shock indexed by $k$, occurring in period $t$, on the $s$-th horizon realization of variable $j$. Here $j$ indexes the $n$ macroeconomic variables and $k$ the $n$ structural shocks, while $s,t\in\{1,\dots,h\}$ index, respectively, the horizon at which the effect is evaluated and the period in which the shock occurs.

For the contemporaneous regime it is convenient to rename these blocks so that their economic content is explicit, splitting the variable index into the policy instrument $i$ and the non-policy variables $x_j$, and the shock index into the policy shock $v$ and the non-policy shocks $\eta_j$. With this convention the multiplier partitions as
\[
M(\bs{\theta}) = \begin{pmatrix} \overset{nh \times h}{M_{v,c}} & \overset{nh \times (n-1)h}{M_{\eta,c}} \end{pmatrix} = \begin{pmatrix}
    \overset{h\times h}{M_{v,c}^i} & \overset{h \times (n-1)h}{M_{\eta,c}^i} \\
    \overset{(n-1)h \times h}{M_{v,c}^x} & \overset{(n-1)h\times (n-1)h}{M_{\eta,c}^x}
\end{pmatrix}  = 
\begin{pmatrix}
    \overset{h \times nh}{M^i_c} \\
    \overset{(n-1)h\times nh}{M^x_c}
\end{pmatrix},
\]
whose $h\times h$ sub-blocks $M_{v,c}^{i},\,M_{v,c}^{x_j},\,M_{\eta_k,c}^{i},\,M_{\eta_k,c}^{x_j}$ ($j,k=1,\dots,n-1$) collect, respectively, the response of the interest rate and of each target variable to the policy and non-policy shocks. Proposition~\ref{prop:toeplitz} in the Appendix shows that each such block is lower-triangular and Toeplitz. 

We want to obtain the counterfactual distribution of the path of the non-policy variables under the restriction that the path of the policy instrument is enforced to follow some pre-specified path $\bar{\textbf{i}}_{1:h}$ by imposing that
\begin{equation}\label{instrconstr_c}
 C_i \textbf{y}_{1:h} = \bar{\textbf{i}}_{1:h}
\end{equation}
and the restriction that the non-policy shocks follow their unconditional distribution
\[
C_{\varepsilon} \bs{\varepsilon}_{1:h} = \bs{\eta}^c_{1:h} \sim N(0, I_{(n-1)h})
\]
where the selection matrices $C_i$ and $C_\varepsilon$ are
\[
\overset{h \times nh}{C_i} = \begin{pmatrix}
    I_{h} & \vdots & O_{h \times (n-1)h}
\end{pmatrix} \quad ; \quad \overset{(n-1)h \times nh}{C_\varepsilon} = \begin{pmatrix}
    O_{(n-1)h \times h} & \vdots & I_{(n-1)h}
\end{pmatrix}
\]
We can then solve for the path of the policy shocks $\textbf{v}_{1:h}(\bar{\textbf{i}}_{1:h}, \bs{\eta}^c_{1:h}, \bs{\theta}, \bs{\chi}_0)$ needed to ensure that the policy path $\bar{\textbf{i}}_{1:h}$ is enforced. Substituting (\ref{instrconstr_c}) into (\ref{trajdecomp}) and noting that $C_i M_{v,c} = M_{v,c}^i$, we obtain
\begin{align}
\textbf{v}_{1:h}(\bar{\textbf{i}}_{1:h},  \bs{\eta}^c_{1:h}, \bs{\theta}, \bs{\chi}_0) &= (M_{v,c}^i)^{-1} \left[\bar{\textbf{i}}_{1:h} - \left(\textbf{b}_{1:h}^i + M_{\eta,c}^i \bs{\eta}^c_{1:h}\right)\right] \label{vpaths_c}
\end{align}
where we have partitioned $\textbf{b}_{1:h} = ((\textbf{b}_{1:h}^i)', (\textbf{b}_{1:h}^x)')'$. Although Equation (\ref{vpaths_c}) is written in stacked form, the lower-triangular Toeplitz structure of $M_{v,c}^i$ and the block structure of  $M_{\eta,c}^i$ ensure that the policy shock $v_t$ depends only on the non-policy shocks $\bs{\eta}^c_s$ realized up to period $t$, as shown by Lemma~\ref{lem:sequential} in the Appendix. The contemporaneous regime is thus consistent with a sequential implementation in which the policymaker observes the realization of $\bs{\eta}^c_t$ at the beginning of period $t$ and then enforces the period-$t$ surprise $v_t$. Note that invertibility of $M_{v,c}^i$ is a mild assumption that we make throughout the paper for the contemporaneous regime: since $M_{v,c}^i$ is lower-triangular with constant diagonal term $[M^i_{v,c}]_{j,j} = B_{iv} = B_{1,1}$, it is invertible as soon as the policy shock has a nonzero on-impact effect on the interest rate, $B_{i,v}\neq 0$.

Substituting the expression for the policy shocks back into (\ref{trajdecomp}), the path of the non-policy variables under the contemporaneous regime can be expressed as
\begin{align}\label{contcfact}
\textbf{x}_{1:h}^{c}(\bar{\textbf{i}}_{1:h},  \bs{\eta}^c_{1:h}, \bs{\theta}, \bs{\chi}_0)
  &\equiv D_{i,c} \bar{\textbf{i}}_{1:h} + D_{b,c} \textbf{b}_{1:h} + D_{\eta,c} \bs{\eta}^c_{1:h} 
\end{align}
where
\begin{align}
\overset{(n-1)h \times h}{D_{i,c}} &= M_{v,c}^x (M_{v,c}^i)^{-1} \label{Di_c} \\
\overset{(n-1)h \times nh}{D_{b,c}} &= \begin{pmatrix} -D_{i,c} & I_{(n-1)h} \end{pmatrix} \label{Db_c} \\
\overset{(n-1)h \times (n-1)h}{D_{\eta,c}} &= M_{\eta,c}^x - D_{i,c} M_{\eta,c}^i \label{Deta_c}
\end{align}
are nonlinear transformations of the matrix $M(\bs{\theta})$ and allow us to decompose the paths $\textbf{x}_{1:h}^c$ into policy, predetermined and non-policy components.

The multipliers $D_{i,c},D_{b,c},D_{\eta,c}$ inherit two properties from the block-Toeplitz structure of $M(\bs{\theta})$ that will be central to the comparison with the forward-guidance regime. First, decompose the action multiplier into the stack of $h\times h$ blocks associated with each non-policy variable,
\[
D_{i,c}=\begin{pmatrix} D_{i,c}^{x_1} \\ \vdots \\ D_{i,c}^{x_{n-1}} \end{pmatrix},
\qquad D_{i,c}^{x_j}\in\mathbb{R}^{h\times h}.
\]
Each block $D_{i,c}^{x_j}$ is lower-triangular and Toeplitz, and the two properties carry distinct economic content. Lower-triangularity means that the choice of the interest rate at period $t$ leaves the non-policy variable $x_j$ at earlier horizons $s<t$ unaffected. The Toeplitz property adds time-invariance: the effect of the enforced rate at period $t$ on the target at horizon $s$ depends only on the lag $s-t$. Together they are precisely what distinguishes the mathematical structure of the contemporaneous regime from the forward guidance one, where both properties fail, as detailed in Section~\ref{sec:fg}. Second, the multipliers mapping the action and the predetermined component into the target paths, $D_{i,c}$ and $D_{b,c}$, depend on the structural rotation parameter $\textbf{q}$ only through the column of $Q$ associated with the policy shock, whereas $D_{\eta,c}$ depends on the whole rotation vector. Additionally, $D_{i,c}$ and $D_{b,c}$ are invariant to the sign of $\textbf{q}_1$. The following proposition formally states these properties.

\begin{proposition}[Structure of the contemporaneous multipliers]\label{prop:Dstructure}
Suppose that $M_{v,c}^{i}$ is invertible. Then the multipliers \eqref{Di_c}--\eqref{Deta_c} satisfy:
\begin{enumerate}
\item[(i)] \emph{(Lower-triangularity of the action multiplier.)} For every $j=1,\dots,n-1$, the block $D_{i,c}^{x_j}=M_{v,c}^{x_j}(M_{v,c}^{i})^{-1}$ is lower-triangular and Toeplitz. Consequently the non-policy variable $x_j$ at horizon $s$ depends on the enforced path $\bar{\textbf{i}}_{1:h}$ only through $(\bar i_1,\dots,\bar i_s)$.
\item[(ii)] \emph{(Rotation dependence.)} $D_{i,c}$ and $D_{b,c}$ depend on $\bs{\theta}$ only through $\bs{\phi}$ and the first column $\textbf{q}_1$ of $Q$, the column associated with the policy shock under the ordering $\bs{\varepsilon}_t=(v_t,(\bs{\eta}^c_t)')'$. The multiplier $D_{\eta,c}$ depends on $\bs{\phi}$ and the entire rotation $\textbf{q}$.

\item[(iii)] \emph{(Sign invariance.)} $D_{i,c}$ and $D_{b,c}$ are invariant to the sign of $\textbf{q}_1$.

\end{enumerate}
\end{proposition}

The structural interpretation of the counterfactual paths $\textbf{x}^c(\bar{\textbf{i}}_{1:h}, \cdot)$ crucially depends on the fact that we identify $v_t$ as a contemporaneous monetary policy surprise. The policy exercise resulting from the manipulation of the shock path $\textbf{v}_{1:h}$ to enforce $\bar{\textbf{i}}_{1:h}$, then, corresponds to an ex-post reaction of the policymaker that is not anticipated by the private sector of the economy. In other words, the counterfactual distribution of the paths $\textbf{x}_{1:h}^c$ that one can obtain using the proposed methodology will correctly identify the counterfactual evolution of the variables in the economy under the assumption that the policymaker privately decides at time $t=0$ to enforce the path $\bar{\textbf{i}}_{1:h}$ for the future and adjusts the value of the policy instrument on a period-by-period basis to offset the systematic policy reaction to the non-policy shocks $\bs{\eta}^c_{1:h}$, in a way that each time surprises the private sector.

We now turn to solving the \textit{oracle} problem for the contemporaneous regime, and for the remainder of this section we thus treat $\bs{\theta} = (\bs{\phi}, Q)$ as known to the policymaker. We rewrite the loss function in (\ref{lossfun1}) specialized for the contemporaneous case\footnote{Actions are the possible realized paths $\bar{\textbf{i}}_{1:h} \in \reals^h$ with associated counterfactuals (\ref{contcfact}). Note that since the future contemporaneous policy shocks are chosen to enforce a specific path for the policy instrument, they are concentrated out of the economy, as they are now a deterministic function of future non-policy shocks and model parameters. The expectation is with respect to the normalized distribution of the remaining future shocks $\bs{\eta}^c_{1:h}$.} in compact notation as 
 \begin{equation}\label{loss_c}
L_c(\bar{\textbf{i}}_{1:h}, \bs{\theta}) := \ev_{\bs{\eta}^c_{1:h}} \left[ (\textbf{x}_{1:h}^c - \textbf{x}^*)' W (\textbf{x}_{1:h}^c - \textbf{x}^*) \right]
\end{equation}
where we take the weight matrix $W$ to be a symmetric positive semidefinite matrix, the canonical case being the time-separable Kroenecker structure as in \eqref{lossfun1}. 

To facilitate the decision-theoretic analysis, we define the \textit{predetermined target gap} as
\[
\tilde{\textbf{b}}_{1:h}^c \equiv \tilde{\textbf{b}}_{1:h}^c(\bs{\theta}, \bs{\chi}_0) = D_{b,c} \textbf{b}_{1:h}(\bs{\phi}, \bs{\chi}_0) - \textbf{x}^*,
\]
the gap between the predetermined component of the paths $\textbf{x}_{1:h}^c$ and the desired target. Substituting the counterfactual paths \eqref{contcfact} into \eqref{loss_c} yields a quadratic form in the action $\bar{\textbf{i}}_{1:h}$ whose coefficients enjoy two properties that are central to what follows.

\begin{proposition}\label{prop:loss_c}
Suppose that $M_{v,c}^{i}$ is invertible and let $W$ be symmetric positive semidefinite. The loss under the contemporaneous regime is the quadratic form
\begin{equation}\label{loss_c_quadratic}
L_c(\bar{\textbf{i}}_{1:h}, \bs{\phi}, Q, \bs{\chi}_0)
= c_c + \bar{\textbf{i}}_{1:h}' H_c\, \bar{\textbf{i}}_{1:h} + 2\,\bar{\textbf{i}}_{1:h}'\, \textbf{g}_c,
\end{equation}
where
\begin{align*}
c_c &= c_c(\bs{\phi},\textbf{q}_1,\bs{\chi}_0) := (\tilde{\textbf{b}}_{1:h}^c)' W \tilde{\textbf{b}}_{1:h}^c + \text{trace}(D_{\eta,c}' W D_{\eta,c}), \\
H_c &= H_c(\bs{\phi},\textbf{q}_1) := D_{i,c}' W D_{i,c}, \\
\textbf{g}_c &= \textbf{g}_c(\bs{\phi},\textbf{q}_1,\bs{\chi}_0) := D_{i,c}' W \tilde{\textbf{b}}_{1:h}^c .
\end{align*}
In particular, the loss depends on $Q$ only through $\textbf{q}_1$ and is invariant to its sign. 
Moreover, if $H_c$ is positive definite, the loss \eqref{loss_c_quadratic} admits the unique minimizer
\begin{equation}\label{istar_c}
\bar{\textbf{i}}^{*,c}_{1:h} (\bs{\phi}, \textbf{q}_1, \bs{\chi}_0)
= -H_c^{-1}\textbf{g}_c
= - (D_{i,c}' W D_{i,c})^{-1} D_{i,c}' W \tilde{\textbf{b}}_{1:h}^c .
\end{equation}
\end{proposition}

The representation has two implications. First, by \emph{certainty equivalence}, the second moments of the future non-policy shocks $\bs{\eta}^c_{1:h}$ enter only the constant $c_c$: the curvature $H_c$ and the linear term $\textbf{g}_c$, and hence the oracle optimal path, are invariant to the volatility of $\bs{\eta}^c_{1:h}$ and to the residual multiplier $D_{\eta,c}$. Second, since the loss $L_c$ depends on the structural rotation $Q$ only through its first column $\textbf{q}_1$, so does the oracle, which also inherits the sign-invariance property.

Since the policymaker observes the reduced-form parameter $\bs{\phi}$ and the structural rotation $Q$ without noise and without ambiguity, the maximum achievable expected welfare is attained by the oracle path \eqref{istar_c} --- the negative (weighted) projection of the predetermined target gap onto the policy multiplier $D_{i,c}$. By Proposition~\ref{prop:loss_c}, the oracle inherits the dependence of $H_c$ and $\textbf{g}_c$ on the structural parameters, and thus depends on $\bs{\theta}$ only through $(\bs{\phi},\textbf{q}_1)$, invariantly to the sign of $\textbf{q}_1$.

The assumption that $H_c$ is positive definite has a transparent reading in terms of policy transmission delays. The matrix $H_c=D_{i,c}'WD_{i,c}$ is singular if some direction of the enforced path leaves the weighted target paths unchanged, i.e., if $\text{rank}(D_{i,c})<h$. In the contemporaneous regime, this can happen if there is transmission delay from the policy shock to the non-policy target variables: suppose the policy shock takes $d\ge1$ periods to affect any variable $x_j$. Then rates set in the final $d$ periods cannot affect any target before the horizon ends, and the loss is flat along those directions whatever the weight matrix $W$. Corollary~\ref{cor:rank}(i) in the Appendix formalizes this: delayed transmission to all target variables implies that $H_c$ is singular, and the undetermined directions are exactly the last $d$ entries of the path. The converse depends on which variables enter the loss. Under the canonical time-separable weights of \eqref{lossfun1}, the policymaker may assign a zero weight to some target variables, so that only a subset of them matters for the loss. Corollary~\ref{cor:rank}(ii) shows that in this case $H_c$ is positive definite if and only if at least one variable with positive weight responds to the policy shock on impact. Positive definiteness fails exactly when transmission is delayed for every variable that enters the loss function. In particular the assumption is violated under a recursive identification scheme in which the policy rate is ordered after all target variables, since then no target responds on impact regardless of the weights.

\begin{remark}[Extension: weighting the full endogenous vector]\label{rem:full_loss}
The base case weights only the non-policy targets, as in \citet{barnichon2023sufficient}. The framework extends naturally to a loss that also penalizes the policy instrument, by weighting the full endogenous vector $\textbf{y}^c_{1:h}$ against a target $\textbf{y}^*$ with a symmetric positive semidefinite matrix $W$. This leaves the structure of the problem intact --- the loss remains a quadratic form in the action and the counterfactual paths are still governed by multipliers that map the action, the predetermined component, and the non-policy shocks into the endogenous paths --- but the multipliers now carry an additional block for the instrument, and the curvature and linear term of the loss acquire deterministic pieces involving the policy weights. The leading application for monetary policy is \emph{interest-rate smoothing}: penalizing the period-to-period change of the instrument, $\sum_s(\bar i_s-\bar i_{s-1})^2$ with $i_0$ set to either the value in the data or a historical average. A direct cost on the instrument regularizes the oracle --- the curvature is invertible even when the policy multiplier is rank-deficient --- and makes it less aggressive, trading off the reduction in the target gap against the cost of moving the instrument. Crucially, both properties of Proposition~\ref{prop:loss_c} are preserved: the added terms are deterministic and only depend on $Q$ through $D_{i,c}$ and $D_{b,c}$, so certainty equivalence continues to hold and the loss still depends on $Q$ only through the span of $\textbf{q}_1$. Appendix~\ref{app:genloss} develops the general case, defining the full-vector multipliers $\tilde{D}_{a,c},\tilde{D}_{b,c},\tilde{D}_{\eta,c}$ and carrying out the smoothing derivation in closed form.
\end{remark}

\begin{remark}[On the validity of contemporaneous  counterfactuals]
As acknowledged in the original work of \citet{sims2006does}, the methodology underlying the policy counterfactuals obtained in the contemporaneous regime does not account for the Lucas critique, and since our decision procedure relies on the identification of such policy counterfactuals, it is subject to the same limitation. Lucas-critique concerns arise because forward-looking rational agents would notice that the policy surprises $\textbf{v}_{1:h}(\cdot)$, set by the policymaker to enforce the path $\bar{\textbf{i}}_{1:h}$, deviate systematically from their unconditional distribution. They would then incorporate this information into their beliefs about the future conduct of policy and adjust their behavior. The impulse responses estimated from past data would no longer characterize the private sector's response to the shocks in the system, so that the SVAR-implied counterfactuals --- which are built from those impulse responses --- would no longer be a valid representation of the economy's future trajectories under the given policy-instrument-path scenarios. 

That said, \citet{leeper2003modest} and \citet{sims2006does} provide an argument for the validity of the contemporaneous regime when the magnitude of policy interventions is small enough, such that ``\textit{\ldots the effects of the policy change are so modest that it seems likely that the public would take a long time to learn that policy had in fact changed.}'' The same logic motivates \citet{antolin2021structural} to justify the structural interpretation of their forecasts when the norm of the implied policy deviations is small. We read these arguments as suggesting that the surprise-based exercise remains of interest when the deviations $\textbf{v}_{1:h}$ are small in size and the horizon $h$ is not too large.
    
\end{remark}

\subsection{Forward guidance regime: counterfactuals and oracle rule}\label{sec:fg}

The forward guidance regime  overcomes one of the major limitations of the contemporaneous one: the expected path is enforced through news shocks released at $t=1$, and it is sufficient to assume that the private treats the $t=1$ policy shocks as unanticipated for the validity of  the counterfactuals.  Their constructions builds on a key observation that we borrow from \citet{mckay2023can}: if one  can identify the impulse responses to the distinct policy shocks, then a combination of shocks released at $t=1$ can enforce a given \textit{expected} path for the instrument \textit{ex-ante} and deliver counterfactual paths for the remaining variables under that expected path. We apply the same logic in the context of a Structural VAR model, with a few crucial differences: first, we do not model the policy rule explicitly. Rather than counterfactuals under alternative policy \textit{rules}, as in \citet{mckay2023can}, we work directly with counterfactuals under alternative \textit{expected paths} for the instrument, in the same spirit as \citet{barnichon2023sufficient}. Second, we take a finite-order invertible SVAR as our fundamental representation and we work with a truncated, finite-horizon problem. This is particularly convenient for decision-theoretic purposes, as it implies a finite-dimensional action space and allows us to write all counterfactuals as functions of a single finite-dimensional parameter vector. 

We keep the same underlying SVAR model \eqref{SVAR}, but we consider a different partition of the shock vector. Specifically, we partition $\bs{\varepsilon}_t$ as
\[
 \bs{\varepsilon}_t = \begin{pmatrix}
     \textbf{e}_t \\  
     \bs{\eta}^{fg}_t
 \end{pmatrix}
\]
where $\textbf{e}_t$ is an $h \times 1$ vector of policy shocks and $\bs{\eta}^{fg}_t$ is the $(n-h) \times 1$ vector of additional structural shocks.\footnote{As anticipated before, we take the first entry of the vector $\textbf{e}_t$ to be the contemporaneous policy shock~$v_t$.} For now, we assume that the number of distinct policy shocks is exactly equal to the number of horizons of the policy problem, and that the number of variables in the system is larger, i.e. $\text{dim}(\textbf{e}_t) = h < n$. We  relax this assumption in Section \ref{sec:feasible-fg}, where we consider the empirically relevant case where can only identify $m=\text{dim}(\textbf{e}_t)<h$ shocks.  The forward guidance partition differs from the contemporaneous one in that the additional $h-1$ news shocks $(e_{2,t}, \ldots, e_{h,t})$ that were lumped into $\bs{\eta}^c_t$ in Section~\ref{sec:contemp} are now separated out as policy shocks. The relationship between the two partitions is therefore
\[
\bs{\eta}^c_t = \begin{pmatrix} e_{2,t} \\ \vdots \\ e_{h,t} \\ \bs{\eta}^{fg}_t \end{pmatrix}.
\]

We maintain the notation of Section~\ref{sec:trajectories} and write the $h$-period-ahead stacked paths as in \eqref{trajdecomp}. For the forward guidance construction, however, it is more convenient to stack the variables and the shocks in two different orderings. We keep the \emph{paths} $\textbf{y}_{1:h}$ in variable-major order, but we now arrange the \emph{shocks} in period-major order. In practice, stacking future shocks in period-major order we write
\[
\tilde{\bs{\varepsilon}}_{1:h} = \begin{pmatrix}
    \textbf{e}_1 \\ \bs{\eta}^{fg}_1 \\ \tilde{\bs{\varepsilon}}_{2:h}
\end{pmatrix}, \qquad \text{where} \qquad \tilde{\bs{\varepsilon}}_{2:h} = \begin{pmatrix} \bs{\varepsilon}_2 \\ \bs{\varepsilon}_3 \\ \vdots \\ \bs{\varepsilon}_h \end{pmatrix},
\]
where the tilde denotes period-major rearranging, i.e. $\tilde{\bs{\varepsilon}}_{1:h}=\textbf{K}'\bs{\varepsilon}_{1:h}$, where $\textbf{K}$ is the commutation matrix of Lemma~\ref{lem:companion}. Substituting $\bs{\varepsilon}_{1:h}=\textbf{K}\tilde{\bs{\varepsilon}}_{1:h}$ into the decomposition $\textbf{y}_{1:h}=\textbf{b}_{1:h}+M(\bs{\theta})\bs{\varepsilon}_{1:h}$ leaves the paths in variable-major order and permutes only the columns of the multiplier. We correspondingly define $M_{fg}:=M\textbf{K}$ and partition it column-wise conformably to the three shock blocks
\[
M_{fg} = \begin{pmatrix}
    \overset{nh \times h}{M_{1,fg}} & \overset{nh \times (n-h)}{M_{\eta,fg}} & \overset{nh \times n(h-1)}{M_{2:h,fg}}
\end{pmatrix}
\]
Specifically, $M_{1,fg}$ collects the impulse responses to the time-$1$ policy shocks $\textbf{e}_1$, $M_{\eta,fg}$ the responses to the time-$1$ non-policy shocks $\bs{\eta}^{fg}_1$, and $M_{2:h,fg}$ the responses to all future shocks $\tilde{\bs{\varepsilon}}_{2:h}$. We further partition each block by rows into an $h\times \cdot$ instrument block (top block) and an $(n-1)h\times \cdot$ non-instrument component (bottom block), denoted by superscripts $i$ and $x$ respectively:
\[
M_{1,fg} = \begin{pmatrix} \overset{h\times h}{M_{1,fg}^i} \\ \overset{(n-1)h\times h}{M_{1,fg}^x} \end{pmatrix}, \quad
M_{\eta,fg} = \begin{pmatrix} M_{\eta,fg}^i \\ M_{\eta,fg}^x \end{pmatrix}, \quad
M_{2:h,fg} = \begin{pmatrix} M_{2:h,fg}^i \\ M_{2:h,fg}^x \end{pmatrix}.
\]
We can then write the stacked future paths as
\begin{equation}\label{yh_fg}
\textbf{y}_{1:h} = \textbf{b}_{1:h} + M_{1,fg} \textbf{e}_1 + M_{\eta,fg} \bs{\eta}^{fg}_1 + M_{2:h,fg} \tilde{\bs{\varepsilon}}_{2:h}.
\end{equation}

As anticipated in Section~\ref{sec:fg}, actions in the forward
guidance regime are choices of the \textit{expected policy path},
the expectation of the SVAR-implied path  over the horizon $1{:}h$ given
the state of the economy at $t=1$, which includes the policy news
shocks released in that period. Policy news shock and announcements about the future conduct of policy affect earlier periods' outcomes  through the changes in agents' expectations, and it is thus natural to take the expected path as the action in the forward guidance regime.\footnote{The policy invariance assumption maintained in \citet{mckay2023can, barnichon2023sufficient} precisely states that the policy news shocks \emph{only} affect target variables through the changes in expectations about the future policy path. For the validity of counterfactuals based on structural forecast in the SVAR, the policy invariance assumption is not necessary. However, the assumption is required to interpret the SVAR in a fully structural sense, as clarified in Footnote \ref{fullstruct}.} Note that, by the linear structure of the economy, the expected path
can also be interpreted as the path that would prevail over the
policy-relevant horizon if no structural shocks were to hit the
economy in periods $2,\ldots,h$. In order to interpret the first entry of the shock vector $\textbf{e}_1$ as the time $t=1$ contemporaneous shock, and to parallel the sequential implementation of the contemporaneous regime, we assume that the first entry of $\bar{\textbf{i}}^e$ corresponds to the implemented policy rate at time $t=1$. In other words, we assume that at time $t=1$, after observing the realization of $\bs{\eta}^{fg}_1$, the policy news shocks $\textbf{e}_1$ are computed to enforce\footnote{Under a \emph{fully ex-ante} convention, the policy shocks themselves are chosen at $t=0$ and cannot react to $\bs{\eta}^{fg}_1$: the constraint is imposed with respect to $I_0$, with the chosen shocks satisfying $\textbf{b}^i_{1:h}+M^i_{1,fg}\textbf{e}_1=\bar{\textbf{i}}^e_{1:h}$. The realized time-$1$ rate is then itself stochastic, and $\bar{\textbf{i}}^e_{1:h}$ is a genuine time-$0$ expectation of the entire path. Another alternative is a ``hybrid" approach, where one allows $e_{11}\equiv v_1$, the contemporaneous shock, to adjust to $\bs{\eta}_{1}^{fg}$, while $e_{12}, \ldots, e_{1h}$ are enforced ex-ante. This convention matches what happens for central banks: after an FOMC meeting, announcements are not revised  but the FFR is kept at the announced rate until the next meeting.  The conventions differ only through the trace terms capturing the contribution of the variance of future shocks, which enter the loss additively and do not depend on the action. The oracle, plug-in and Bayes rules are therefore unchanged: action-independent terms drop out of the minimization. The robust Bayes rule over the loss is sensitive to the convention: the inner supremum acts on the full loss, and the convention-dependent terms can shift the worst case. Under worst-case regret they difference away against the oracle loss, and the action-independent part that survives is the same across conventions: the regret-based robust rule is again invariant.\label{fn:exante}}
\begin{equation}\label{lcrrestr_fg}
C_i \ev[\textbf{y}_{1:h} \mid \mathcal{I}_1] = \bar{\textbf{i}}^e_{1:h}
\end{equation}
where $\mathcal{I}_1$ is the information set as of time $t=1$ and $C_i$ is the same selection matrix as in Section~\ref{sec:contemp}.
 The remaining future shocks $\tilde{\bs{\varepsilon}}_{2:h}$ are required to follow their unconditional distribution
\[
\tilde{\bs{\varepsilon}}_{2:h} \mid \mathcal{I}_1 \sim N(0, I_{n(h-1)}).
\]
Substituting \eqref{yh_fg} into \eqref{lcrrestr_fg},  noting that $\ev[\tilde{\bs{\varepsilon}}_{2:h} \mid \mathcal{I}_1] = \textbf{0}$, and assuming that $M^i_{1, fg}$ is invertible,  we can solve for the time-$1$ policy shocks
\begin{equation}\label{e1_solution}
\textbf{e}_1(\bar{\textbf{i}}^e_{1:h}, \bs{\eta}^{fg}_1, \bs{\theta}, \bs{\chi}_0) = (M_{1,fg}^i)^{-1}\bigl[\bar{\textbf{i}}^e_{1:h} - (\textbf{b}^i_{1:h} + M_{\eta,fg}^i \bs{\eta}^{fg}_1)\bigr]
\end{equation}
which mirrors the structure of \eqref{vpaths_c} in the contemporaneous regime, with $\textbf{e}_1$ replacing $\textbf{v}_{1:h}$ and the observed time-$1$ non-policy shock $\bs{\eta}^{fg}_1$ replacing the stacked vector $\bs{\eta}^c_{1:h}$. There is, however, a fundamental difference from the contemporaneous regime. The policy multiplier $M_{1,fg}^i$, which maps the policy news shocks to deviations of the expected path from its baseline, is no longer lower triangular. The shocks $\textbf{e}_1$ are all released at time $t=1$, and each of them affects the policy instrument at every horizon --- consistent with the ``news shock'' interpretation of $\textbf{e}_1$, whereby an announcement made at $t=1$ moves the expected rate at all future dates. 

Substituting \eqref{e1_solution} back into \eqref{yh_fg} and retaining the non-policy block, we obtain the counterfactual paths of the target variables under the forward guidance regime:
\begin{equation}\label{xpaths_fg}
\textbf{x}_{1:h}^{fg}(\bar{\textbf{i}}^e_{1:h}, \bs{\eta}^{fg}_1, \tilde{\bs{\varepsilon}}_{2:h}, \bs{\theta}, \bs{\chi}_0) = D_{i,fg} \bar{\textbf{i}}^e_{1:h} + D_{b,fg} \textbf{b}_{1:h} + D_{\eta,fg} \bs{\eta}^{fg}_1 + D_{2:h,fg} \tilde{\bs{\varepsilon}}_{2:h}
\end{equation}
where
\begin{align}\label{Dmatexpr}
\overset{(n-1)h \times h}{D_{i,fg}}    &= M_{1,fg}^x (M_{1,fg}^i)^{-1} \\ \nonumber
\overset{(n-1)h \times nh}{D_{b,fg}}    &= \begin{pmatrix} -D_{i,fg} & I_{(n-1)h} \end{pmatrix} \\ \nonumber
\overset{(n-1)h \times (n-h)}{D_{\eta,fg}} &= M_{\eta,fg}^x - D_{i,fg} M_{\eta,fg}^i \\ \nonumber
\overset{(n-1)h \times n(h-1)}{D_{2:h,fg}}  &= M_{2:h,fg}^x
\end{align}

It is useful to state what the counterfactual paths \eqref{xpaths_fg} are
within the model. They are structural conditional forecasts: the paths
implied by the SVAR conditional on the release of the identified policy
shocks $\textbf{e}_1$ at $t=1$, with all remaining shocks following their
unconditional distribution, in the tradition of
\citet{waggoner1999conditional} and \citet{antolin2021structural}.
The conditioning is nevertheless of a specific kind: it acts on the
\emph{expected} rate path through \emph{anticipated} shocks, so that agents
are surprised once, at the announcement date, rather than period by period
as in scenario analyses that condition on realized paths. This timing is
what the construction of \citet{mckay2023can} delivers, and it is the
sense in which the counterfactuals are robust to the Lucas critique. We do not establish the conditions under which these
conditional forecasts coincide with counterfactuals of an underlying
structural model, and take the conditional-forecast reading, internal to
the SVAR, as the primitive interpretation.\footnote{ \label{fullstruct} We refer to the alternative interpretation, in
which the SVAR is the semi-structural representation of a sequence-space
economy satisfying the policy-invariance assumption of
\citet{mckay2023can}, as the \emph{fully structural}
reading; we return to the distinction when discussing
Proposition~\ref{prop:Dstructure_fg}.}

A few comments are in order. 
First, the invertibility of $M_{1,fg}^i$ required to solve for the policy shocks in \eqref{e1_solution} is a substantially stronger condition than its contemporaneous counterpart, where it amounts to assuming that the policy shock moves the interest rate on impact. In the forward-guidance regime, $M_{1,fg}^i$ is the full $h\times h$ matrix of impulse responses of the interest-rate path to the $h$ policy shocks $\textbf{e}_1$, and its invertibility requires these responses to be linearly independent. This is the precise sense in which the forward-guidance regime needs $h$ \emph{distinct} policy shocks: $h$ shocks whose population impulse responses on the interest-rate path span an $h$-dimensional space. The condition mirrors the informational requirements of \citet{mckay2023can, barnichon2023sufficient}, and is the finite-horizon counterpart of their assumption of an infinite sequence of distinct policy shocks. It carries two implicit assumptions: the system must contain at least as many distinct policy shocks as horizons relevant to the decision problem, and the number of variables must strictly exceed the number of such horizons, $n>h$, so that non-policy structural shocks $\bs{\eta}^{fg}$ remain in the system after $h$ of them are relabeled as policy shocks. On the theoretical side, the two assumptions are less demanding than it may appear: whenever shock series capturing distinct sources of variation in the expected rate path are available, the SVAR can simply be augmented with these series, and the impulse responses to the policy shocks identified through an internal IV procedure \citep{plagborg2021local, plagborg2022instrumental}. The binding constraint is instead empirical: in practice, researchers are typically able to identify only a small number $m<h$ of shocks with genuinely distinct identifying variation. In the theoretical analysis of this section we do not take a stance on this issue and assume the existence of the $h$ shocks; Section~\ref{sec:feasible-fg} develops the feasible decision rules for the case $m<h$. Section~\ref{sec:empirical} implements them with $m=3$.

Second, the structure of \eqref{xpaths_fg} is the direct analog of \eqref{contcfact} in the contemporaneous regime, with one difference in the shape of the multipliers. Because the policy multiplier $M^i_{1,fg}$ is not necessarily lower triangular, the blocks of the policy-path multiplier $D_{i,fg}$ need not be either lower triangular or Toeplitz: both properties of Proposition~\ref{prop:Dstructure}(i) will generically fail. Since the expected path is enforced through the anticipated shocks $\textbf{e}_1$, expectations of the future interest-rate path affect earlier-period outcomes --- the forward-looking response of the agents is internalized, and lower-triangularity is lost. And whereas in the contemporaneous regime the columns of each block are the responses to the same policy shock released at successive dates, here they are the responses to $h$ distinct shocks, all released at $t=1$: the columns are no longer translates of one another, and the effect of the expected rate at period $t$ is no longer a function of the lag alone.

Finally, the dependence of the multipliers on the rotation changes in a way that parallels the  contemporaneous regime. The multipliers $D_{i,fg}$ and $D_{b,fg}$ depend on $Q$ only through the block $Q_{1:h}=(\textbf{q}_1\ \cdots\ \textbf{q}_h)$ of columns associated with the policy shocks $\textbf{e}_1$, while $D_{\eta,fg}$ and $D_{2:h,fg}$ involve the entire rotation. The sign invariance of Proposition~\ref{prop:Dstructure}(iii) extends in a substantially stronger form: the policy multipliers depend on $\textbf{q}_1,\dots,\textbf{q}_h$ only through their \emph{span}, so that any orthonormal basis of the same $h$-dimensional space of policy shocks delivers the same multipliers, and hence the same decision problem. The following proposition states these properties formally.

\begin{proposition}[Rotation dependence of the forward-guidance multipliers]\label{prop:Dstructure_fg}
Suppose $M_{1,fg}^{i}$ is invertible. Write $Q_{1:h}:=(\textbf{q}_1\ \cdots\ \textbf{q}_h)$ for the block of columns of $Q$ associated with the policy shocks $\textbf{e}_1$, and $\textbf{q}_{1:h}:=\operatorname{vec}(Q_{1:h})$ for its vectorized form.
\begin{enumerate}
\item[(i)] \emph{(Dependence on the rotation vector.)} $D_{i,fg}$ and $D_{b,fg}$ depend on $\bs{\theta}$ only through $\bs{\phi}$ and $\textbf{q}_{1:h}$. 
\item[(ii)] The matrices $D_{i,fg}$ and $D_{b,fg}$ are invariant to rotations of $\{\textbf{q}_1, \ldots, \textbf{q}_h\}$
\end{enumerate}
\end{proposition}

The dependence on the span of $\textbf{q}_1,\dots,\textbf{q}_h$ established in Proposition~\ref{prop:Dstructure_fg}(ii) has a meaningful implication about the interpretation of the structural counterfactuals \eqref{xpaths_fg}: in the SVAR model we postulate and given the enforcement mechanism implied by \eqref{lcrrestr_fg}, in principle the paths need to be understood as the outcome of implementing the expected path $\bar{\textbf{i}}^e_{1:h}$ through the release of the specific shocks identified by the columns $\textbf{q}_1,\dots,\textbf{q}_h$. Proposition~\ref{prop:Dstructure_fg}(ii) relaxes this reading. The multipliers $D_{i,fg}$ and $D_{b,fg}$, and with them the counterfactual distribution of the target paths and the expected loss, are invariant to any orthogonal transformation of the policy columns; realized paths are invariant as well once the coordinates of the future policy shocks in $\tilde{\bs{\varepsilon}}_{2:h}$ are transformed accordingly. The identification of the individual shocks is thus relaxed to the identification of their span, and the implementation of the expected path can accordingly be understood in a broader sense. To give a practical example, in the monetary policy application, for the counterfactuals obtained through the procedure described in this Section to be valid, the central bank needs not enforce the expected path by releasing the specific shock $\textbf{e}_1$ recovered in \eqref{e1_solution} under a particular identification scheme; any combination of the identified news shocks that steers the expected rate path to $\bar{\textbf{i}}^e_{1:h}$ delivers the same counterfactual. The invariance holds, however, within the identified span and not across spans: a different set of announcement channels may define a different policy subspace and generally different multipliers and counterfactuals.\footnote{Invariance across spans is the content of the policy-invariance assumption of the fully structural reading introduced above, under which the counterfactuals depend only on the expected path, irrespective of whether this is enforced through the endogenous response or through anticipated policy shocks. Within the SVAR it is not implied by the model but implies additional restrictions on $\bs{\theta}$, namely that the multipliers coincide across policy subspaces; we do not impose it. Under that assumption our result parallels the argument of \citet{mckay2023can}, for whom identification of the Lucas-critique robust counterfactuals requires the impulse responses to a sufficient number of linear combinations of policy shocks, with no reference to a particular subspace.} The implication for empirical practice is direct: point identification of the policy multipliers (and of the loss, as clarified below) only requires restrictions that pin down the span of $\textbf{q}_1,\dots,\textbf{q}_h$, not the individual columns, which remain free up to an orthogonal rotation of the block.
The map from shocks to impulse responses is linear and injective so spans of $\textbf{q}_1,\dots,\textbf{q}_h$ and spans of the corresponding impulse responses are in one-to-one correspondence: identifying an $h$-dimensional space of policy variation in the impulse responses identifies the span of $\textbf{q}_1,\dots,\textbf{q}_h$, and vice-versa. 

Based on the counterfactual paths derived in \eqref{xpaths_fg}, we can now formalize the optimal policy choice problem under the forward guidance regime. Similarly to the contemporaneous case, the policymaker chooses the target expected path $\bar{\textbf{i}}^e_{1:h}$ ex ante and evaluates expected welfare using the unconditional quadratic loss function
\begin{equation}\label{loss_fg}
L_{fg}(\bar{\textbf{i}}^e_{1:h}, \bs{\phi}, Q, \bs{\chi}_0) := \ev_{\bs{\eta}^{fg}_1, \tilde{\bs{\varepsilon}}_{2:h}} \left[ (\textbf{x}_{1:h}^{fg} - \textbf{x}^*)' W (\textbf{x}_{1:h}^{fg} - \textbf{x}^*) \right]
\end{equation}
where $W$ is symmetric positive semidefinite, as in the contemporaneous regime. By defining the forward-guidance predetermined target gap as
\begin{equation}\label{fgPreGap}
\tilde{\textbf{b}}^{fg}_{1:h}(\bs{\theta}, \bs{\chi}_0) \equiv D_{b,fg} \textbf{b}_{1:h} - \textbf{x}^*    
\end{equation}
we can substitute the expression for $\textbf{x}_{1:h}^{fg}(\cdot)$ into \eqref{loss_fg} and exploit certainty equivalence to obtain the following representation of the loss function.

\begin{proposition}[Forward guidance loss and oracle]\label{prop:loss_fg}
Suppose $M_{1,fg}^{i}$ is invertible and let $W$ be symmetric positive semidefinite. The loss under the forward-guidance regime is the quadratic form
\begin{equation}\label{loss_fg_quadratic}
L_{fg}(\bar{\textbf{i}}^e_{1:h}, \bs{\phi}, Q, \bs{\chi}_0)
= c_{fg} + (\bar{\textbf{i}}^e_{1:h})' H_{fg}\, \bar{\textbf{i}}^e_{1:h} + 2\,(\bar{\textbf{i}}^e_{1:h})'\, \textbf{g}_{fg},
\end{equation}
where
\begin{align*}
c_{fg} &= c_{fg}(\bs{\phi},\textbf{q}_{1:h},\bs{\chi}_0) := (\tilde{\textbf{b}}^{fg}_{1:h})' W \tilde{\textbf{b}}^{fg}_{1:h} + \operatorname{trace}(D_{\eta,fg}' W D_{\eta,fg}) + \operatorname{trace}(D_{2:h,fg}' W D_{2:h,fg}), \\
H_{fg} &= H_{fg}(\bs{\phi},\textbf{q}_{1:h}) := D_{i,fg}' W D_{i,fg}, \\
\textbf{g}_{fg} &= \textbf{g}_{fg}(\bs{\phi},\textbf{q}_{1:h},\bs{\chi}_0) := D_{i,fg}' W \tilde{\textbf{b}}^{fg}_{1:h} .
\end{align*}
In particular, the loss depends on $Q$ only through $\textbf{q}_{1:h}$ and, by Proposition~\ref{prop:Dstructure_fg}(ii), only through the column span of $Q_{1:h}$. \\ Moreover, if $H_{fg}$ is positive definite, the loss \eqref{loss_fg_quadratic} admits the unique minimizer
\begin{equation}\label{istar_fg}
\bar{\textbf{i}}^{*,fg}_{1:h}(\bs{\phi},\textbf{q}_{1:h},\bs{\chi}_0)
= -H_{fg}^{-1}\textbf{g}_{fg}
= -(D_{i,fg}'WD_{i,fg})^{-1}D_{i,fg}'W\tilde{\textbf{b}}^{fg}_{1:h}.
\end{equation}
\end{proposition}

As in the contemporaneous regime, the representation has two implications. First, \emph{certainty equivalence} holds: the second moments of the future shocks --- the time-$1$ non-policy shocks $\bs{\eta}^{fg}_1$ and all later shocks $\tilde{\bs{\varepsilon}}_{2:h}$ --- enter only the constant $c_{fg}$, the oracle path being invariant to their volatility. Second, each of $c_{fg}$, $H_{fg}$, and $\textbf{g}_{fg}$ --- and therefore the loss $L_{fg}$ itself --- depends on the structural rotation $Q$ only through its first $h$ columns $\textbf{q}_1,\dots,\textbf{q}_h$ and, in particular, only through their span.

Positive definiteness of $H_{fg}$ is a genuine additional assumption beyond the invertibility of $M_{1,fg}^{i}$. The latter condition implies the news shocks can be solved uniquely for any expected path, but doesn't guarantee positive definiteness of $H_{fg}$. Linear independence of the target responses to the enforced expected path, i.e. $\text{rank}(D_{i,fg})=h$, is necessary for this, and sufficient when $W$ is positive definite. Under the time-separable weights of \eqref{lossfun1} $H_{fg}$ is positive definite if and only if the responses of the variables with positive weight are linearly independent, as in the contemporaneous regime. What does not carry over is the sharp characterization of Corollary~\ref{cor:rank}. Since the blocks of $D_{i,fg}$ are neither lower triangular nor Toeplitz, rank deficiency is no longer confined to delayed transmission. Announcing today the rate for a distant period may move the targets on impact through anticipation, so the final entries of the path are not automatically undetermined when transmission is slow. Conversely, the responses to news shocks at different horizons may be collinear even when every target responds on impact. The condition must therefore be assessed case by case. Both conditions are nevertheless consistent with the dependence of the policy problem on the span of $\textbf{q}_1,\dots,\textbf{q}_h$: they are properties of the span and not of a specific basis, since under a change of basis $Q_{1:h}\mapsto Q_{1:h}R$ both blocks are postmultiplied by the orthogonal $R$ (equation~\eqref{eq:Mcol_factor_fg}), which alters neither invertibility nor rank.

Finally, the extension of the loss to the full endogenous vector of Remark~\ref{rem:full_loss} and Appendix~\ref{app:genloss} carries over to the forward-guidance regime, with one modification. Since only the expected path is enforced, the realized instrument path is stochastic, $\textbf{i}^{fg}_{1:h}=\bar{\textbf{i}}^e_{1:h}+M^i_{2:h,fg}\tilde{\bs{\varepsilon}}_{2:h}$, so the future-shock multiplier acquires a nonzero instrument block; its contribution enters only the constant of the loss and is reduced-form, leaving certainty equivalence and the span dependence unaffected. The curvature and linear term take the same form as in \eqref{eq:Hc_full}--\eqref{eq:gc_full}, and interest-rate smoothing --- now a penalty on the period-to-period change of the \emph{expected} path --- regularizes the oracle exactly as in the contemporaneous case, restoring uniqueness when $M^{x}_{1,fg}$ is rank deficient.

\section{Statistical decision theory for SVAR policy choice}\label{sec:decision-theory}

Optimizing the criteria introduced in Section~\ref{sec:welfare} to evaluate decision rules under parameter uncertainty, we obtain optimal decision rules under the SVAR environment of Section~\ref{sec:svar} and the counterfactuals derived in Section~\ref{sec:counterfactuals}.

Under both policy regimes the loss admits the common quadratic representation
\begin{equation}\label{loss_quad}
L(\bar{\textbf{i}}_{1:h}, \bs{\phi}, \textbf{q}) = c(\bs{\phi}, \textbf{q}) + \bar{\textbf{i}}_{1:h}'\, H(\bs{\phi}, \textbf{q})\, \bar{\textbf{i}}_{1:h} + 2\,\bar{\textbf{i}}_{1:h}'\, \textbf{g}(\bs{\phi}, \textbf{q}),
\end{equation}
where the baseline loss $c$, the policy-sensitivity matrix $H$, and the policy-effectiveness vector $\textbf{g}$ are the regime-specific objects constructed in Section~\ref{sec:counterfactuals} from the multipliers $D_i, D_b, D_\eta$ and, in the forward guidance regime, $D_{2:h}$. By $\bar{\textbf{i}}_{1:h}$ we denote the corresponding action: the  realized path $\bar{\textbf{i}}_{1:h}$ in the contemporaneous regime, the expected path $\bar{\textbf{i}}^e_{1:h}$ in the forward guidance one. Since the decision problem depends on the regime only through $(c, H, \textbf{g})$, every rule and every result derived in this section holds verbatim for both regimes; we accordingly suppress the regime subscript.  We additionally leave the dependence of the loss on the relevant subset of $\textbf{q}$ implicit (Propositions \ref{prop:loss_c}-\ref{prop:loss_fg}).

We maintain the partition of the parameter into a point-identified reduced-form component and a set-identified structural component, $\bs{\theta} = (\bs{\phi}, \textbf{q})$, where $\textbf{q}$ is the vectorized structural parameter introduced in Section~\ref{sec:svar}, with identified set $\mathcal{I}_{\textbf{q}}(\bs{\phi})$ defined in Section~\ref{sec:set-id}.\footnote{Similarly to the loss, we write the identified set as a function of the entire vector $\textbf{q}$ so as to encompass both regimes,    keeping in mind how the loss may depend only on a function of a subset of them.} The policymaker observes the sample $\textbf{y} \equiv Y^T_0$ of Section~\ref{sec:welfare} (or the corresponding truncated sample for the conditional decision problem) and chooses a statistical decision rule $\hat{\textbf{i}}_{1:h} : \mathcal{Y} \to \reals^h$ mapping the data into an action. We let $R(\hat{\textbf{i}}_{1:h}, \bs{\phi}, \textbf{q})$ denote either the unconditional risk or the conditional risk $R_{\bs{\chi}_0}$ of Section~\ref{sec:welfare}; all the results below hold for either choice. To lighten the notation, throughout this section we suppress the dependence on the observed initial state $\bs{\chi}_0$.

\subsection{The SVAR decision problem and the plug-in benchmark}\label{sec:plugin}

Under the Gaussian structural shocks maintained in Section~\ref{sec:svar}, the sampling distribution of the data is
\begin{equation}\label{svar_likelihood}
\textbf{y} \mid \bs{\phi}, \textbf{q} \sim N\bigl(\bs{\mu}_y(\bs{\phi}),\, \Sigma_y(\bs{\phi})\bigr),
\end{equation}
which depends on the structural parameter $\bs{\theta}$ only through the reduced-form parameter $\bs{\phi}$. Structural parameters belonging to the identified set for a given reduced-form parameter are observationally equivalent: the likelihood is flat over $\mathcal{I}_{\textbf{q}}(\bs{\phi})$,
\begin{equation}\label{flat_likelihood}
p(\textbf{y} \mid \bs{\phi}, \textbf{q}) = p(\textbf{y} \mid \bs{\phi}, \textbf{q}') \qquad \text{for all } \textbf{q}, \textbf{q}' \in \mathcal{I}_{\textbf{q}}(\bs{\phi}), \quad \bs{\phi} \in \Phi.
\end{equation}

Before turning to decision-theoretically founded rules, we record the benchmark most commonly used in practice, the \textit{plug-in rule}, which treats a point estimate $\hat{\bs{\phi}}$ as if it were the true reduced-form parameter. When $\textbf{q}$ is point-identified, i.e., $\mathcal{I}_{\textbf{q}}(\bs{\phi})$ is a singleton and hence $\hat{q}:=\mathcal{I}_{\textbf{q}}(\hat{\bs{\phi}})$, the plug-in rule simply evaluates the oracle rule of Section~\ref{sec:counterfactuals} at the estimate,
\begin{equation}\label{plugin_pointid}
\hat{\textbf{i}}^{\,\text{plug}}_{1:h} = \bar{\textbf{i}}^{*}_{1:h}(\hat{\bs{\phi}}, \hat{\textbf{q}}) = - H(\hat{\bs{\phi}}, \hat{\textbf{q}})^{-1}\, \textbf{g}(\hat{\bs{\phi}}, \hat{\textbf{q}}).
\end{equation}
Under set identification, the policymaker must take a stance on $\textbf{q}$ over the identified set. If they are willing to specify a conditional prior $\pi_{\textbf{q}\mid\bs{\phi}}$ on $\textbf{q}$ given $\bs{\phi}$,\footnote{We assume the support of the conditional prior is the identified set $\mathcal{I}_{\textbf{q}}(\bs{\phi})$, so that if the identified set is a singleton the conditional prior collapses to a Dirac mass at the point-identified $\textbf{q}$.} the conditional expected loss
\begin{equation}\label{condexploss}
\bar{L}_{\textbf{q}\mid\bs{\phi}}(\bar{\textbf{i}}_{1:h}, \bs{\phi}) := \ev_{\pi_{\textbf{q}\mid\bs{\phi}}}\bigl[L(\bar{\textbf{i}}_{1:h}, \bs{\phi}, \textbf{q})\bigr]
\end{equation}
is again quadratic in $\bar{\textbf{i}}_{1:h}$, and the \textit{conditional plug-in rule} minimizes it at $\hat{\bs{\phi}}$:
\begin{equation}\label{plugin_cond}
\hat{\textbf{i}}^{\,\text{plug}}_{\textbf{q}\mid\bs{\phi}} = \underset{\bar{\textbf{i}}_{1:h}\in\reals^h}{\arg\min}\; \bar{L}_{\textbf{q}\mid\bs{\phi}}(\bar{\textbf{i}}_{1:h}, \hat{\bs{\phi}}) = - \bigl(\ev_{\pi_{\textbf{q}\mid\bs{\phi}}}[H(\hat{\bs{\phi}}, \textbf{q})]\bigr)^{-1} \ev_{\pi_{\textbf{q}\mid\bs{\phi}}}[\textbf{g}(\hat{\bs{\phi}}, \textbf{q})].
\end{equation}
If instead they wish to remain agnostic about $\textbf{q}$, they may use the worst-case loss over the identified set,
\begin{equation}\label{wcloss}
L_{WC}(\bar{\textbf{i}}_{1:h}, \bs{\phi}) := \underset{\textbf{q}\in\mathcal{I}_{\textbf{q}}(\bs{\phi})}{\sup}\; L(\bar{\textbf{i}}_{1:h}, \bs{\phi}, \textbf{q}),
\end{equation}
and the \textit{worst-case plug-in rule} minimizes it at $\hat{\bs{\phi}}$, 
\begin{equation} \label{wcplugin}
    \hat{\textbf{i}}^{\,\text{plug}}_{WC} = \arg\min_{\bar{\textbf{i}}_{1:h}\in\reals^h} L_{WC}(\bar{\textbf{i}}_{1:h}, \hat{\bs{\phi}}). 
\end{equation}In all cases, plug-in rules   treat $\hat{\bs{\phi}}$ as the truth and ignore its sampling variability; Section~\ref{sec:suboptimality} discusses suboptimality of the plug-in rules in detail.

\subsection{Bayes-optimal rules and regret}\label{sec:bayes}
We now specialize the Bayes and regret criteria of Section~\ref{sec:welfare} to the quadratic problem~(\ref{loss_quad}). Throughout we maintain that $H(\bs{\phi}, \textbf{q})$ is positive definite, so that the oracle rule
\[
\bar{\textbf{i}}^{*}_{1:h}(\bs{\phi}, \textbf{q}) = - H(\bs{\phi}, \textbf{q})^{-1}\, \textbf{g}(\bs{\phi}, \textbf{q})
\]
and the rules introduced below are well-defined. The quadratic loss function delivers three simplifications. First, the Bayes rule admits a closed form and satisfies certainty equivalence: it is invariant to the second-order moments of the structural shocks. Second, the regret of any action also admits a closed form expression. Third, minimizing Bayes risk and minimizing Bayes expected regret yield the same rule, so the two criteria coincide in this setting. The following two propositions collect these results.

\begin{proposition}\label{prop_bayes_quad}
Under the quadratic loss~(\ref{loss_quad}), with joint prior $\pi_{\bs{\phi}, \textbf{q}}$:
\begin{enumerate}
\item the Bayes rule is
\begin{equation}\label{bayes_quad}
\hat{\textbf{i}}^{Bayes}_{1:h}(\textbf{y}) = -\bigl(\ev_{\bs{\phi},\textbf{q}\mid\textbf{y}}[H(\bs{\phi},\textbf{q})]\bigr)^{-1}\, \ev_{\bs{\phi},\textbf{q}\mid\textbf{y}}[\textbf{g}(\bs{\phi},\textbf{q})];
\end{equation}
\item the Bayes rule~(\ref{bayes_quad}) satisfies certainty equivalence: it is invariant to the second-order moments of the structural shocks and coincides with the rule obtained by setting the future shocks to their expectation.
\end{enumerate}
\end{proposition}

\begin{proposition}\label{prop_regret_ce}
Under the quadratic loss~(\ref{loss_quad}), with joint prior $\pi_{\bs{\phi}, \textbf{q}}$:
\begin{enumerate}
\item the regret of an action $\bar{\textbf{i}}_{1:h}$ relative to the oracle is
\begin{equation}\label{regret_quad}
Re(\bar{\textbf{i}}_{1:h}, \bs{\phi}, \textbf{q}) = \bigl(\bar{\textbf{i}}_{1:h} - \bar{\textbf{i}}^{*}_{1:h}(\bs{\phi}, \textbf{q})\bigr)'\, H(\bs{\phi}, \textbf{q})\, \bigl(\bar{\textbf{i}}_{1:h} - \bar{\textbf{i}}^{*}_{1:h}(\bs{\phi}, \textbf{q})\bigr)
\end{equation}
\item the rule minimizing the Bayes risk coincides with the rule minimizing the Bayes expected regret,
\begin{equation}\label{risk_eq_regret}
\underset{\hat{\textbf{i}}_{1:h}}{\arg\min}\; \ev_{\pi_{\bs{\phi},\textbf{q}}}\!\bigl[R(\hat{\textbf{i}}_{1:h}, \bs{\phi}, \textbf{q})\bigr] = \underset{\hat{\textbf{i}}_{1:h}}{\arg\min}\; \ev_{\pi_{\bs{\phi},\textbf{q}}}\!\Bigl[\ev_{\textbf{y}\mid\bs{\phi},\textbf{q}}\bigl[Re(\hat{\textbf{i}}_{1:h}(\textbf{y}), \bs{\phi}, \textbf{q})\bigr]\Bigr].
\end{equation}
\end{enumerate}
\end{proposition}

The certainty-equivalence property of Proposition~\ref{prop_bayes_quad} extends to the regret criterion as an immediate corollary. By part~(2) of Proposition~\ref{prop_regret_ce} the rule minimizing the Bayes expected regret coincides with the Bayes rule~(\ref{bayes_quad}), which is invariant to the shock second moments; hence the regret-minimizing rule is invariant to them as well.

\subsection{Robust Bayes under set identification}\label{sec:robust-bayes}

When the structural component $\textbf{q}$ is set-identified, we consider two alternative ways in which the policymaker can proceed. The first is to remain fully Bayesian: specify a conditional prior $\pi_{\textbf{q}\mid\bs{\phi}}$ supported on $\mathcal{I}_{\textbf{q}}(\bs{\phi})$, combine it with a prior $\pi_{\bs{\phi}}$ on the reduced form, and apply the Bayes rule~(\ref{bayes_rule}).
As noted in Section~\ref{sec:plugin}, however, the likelihood is flat over the identified set, so the data never updates the conditional prior \citep{moon2012bayesian}: the resulting rule, then, depends on the unrevisable conditional prior even asymptotically.

The second route we consider specifically tackles this drawback by replacing the single conditional prior with the entire class of priors consistent with the identifying restrictions --- all priors supported on $\mathcal{I}_{\textbf{q}}(\bs{\phi})$ --- and take a robust Bayes approach with respect to this class, as put forward by \citet{giacomini2021robust}. Specifically, we consider the \textit{posterior (or conditional) Gamma-minimax} approach: For a given and fixed prior $\pi_{\bs{\phi}}$ we collect the conditional priors that place probability one on $\mathcal{I}_{\textbf{q}}(\bs{\phi})$,
\begin{equation}\label{cond_priors}
\Pi_{\textbf{q}\mid\bs{\phi}} \equiv \bigl\{\, \pi_{\textbf{q}\mid\bs{\phi}} : \pi_{\textbf{q}\mid\bs{\phi}}\bigl(\mathcal{I}_{\textbf{q}}(\bs{\phi})\bigr)=1,\ \pi_{\bs{\phi}}\text{-a.s.}\,\bigr\},
\end{equation}
and let $\Pi_{\bs{\phi}, \textbf{q} \mid\textbf{y}}$ denote the corresponding set of posteriors obtained by Bayesian updating of the joint implied by $\pi_{\bs{\phi}}$ and each prior in~(\ref{cond_priors}). The conditional gamma minimax rule is then obtained by minimizing the worst-case posterior expectation over the class $\Pi_{\bs{\phi}, \textbf{q}\mid \textbf{y}}$. As discussed in \citet{Giacomini_Kitagawa_Read_2026}, the worst-case posterior expected loss over this set of priors coincides with the posterior expectation of the worst-case loss:
\begin{equation}\label{rb_rule_choquet}
\underset{\pi_{\bs{\phi}, \textbf{q}\mid \textbf{y}}\in\Pi_{ \bs{\phi},\textbf{q}\mid\textbf{y}}}{\sup}\; \ev_{\bs{\phi},\textbf{q}\mid\textbf{y}}\bigl[\, L(\mathbf{a}, \bs{\phi}, \textbf{q}) \bigr] = \ev_{\bs{\phi}\mid\textbf{y}}\Bigl[\, \underset{\textbf{q}\in\mathcal{I}_{\textbf{q}}(\bs{\phi})}{\sup}\; L(\mathbf{a}, \bs{\phi}, \textbf{q}) \Bigr].
\end{equation}
It follows that the conditional Gamma-minimax rule coincides with the robust Bayes rule~(\ref{rb_rule}) already introduced in Section~\ref{sec:welfare}.
The most prominent feature that makes the robust Bayes rule~(\ref{rb_rule}) attractive in our setting is that, by using the full-ambiguity class of priors, it removes the dependence of any Bayesian rule on the conditional prior that the data cannot revise.\footnote{This is also the prior component responsible for the asymptotic disagreement between Bayesian and frequentist inference in set-identified models, which the multiple-prior approach reconciles \citep{giacomini2021robust}.} 
A second attractive feature is computational tractability. By the equivalence~(\ref{rb_rule_choquet}), the worst-case posterior expected loss reduces to a posterior expectation over $\bs{\phi}$ of the inner worst case loss over the identified set; the rule can therefore be approximated by Monte Carlo following the algorithm in \citet{Giacomini_Kitagawa_Read_2026}, drawing $\bs{\phi}$ from its posterior and minimizing the resulting sample average.\footnote{Rule~(\ref{rb_rule}) is distinct from the unconditional gamma-minimax
rule $\hat{\textbf{i}}^{\,U\Gamma}_{1:h} = \arg\min_{\hat{\textbf{i}}_{1:h}}\,
\sup_{\pi\in\Pi_{\bs{\phi},\textbf{q}}}\,
\ev_{\pi}\bigl[R(\hat{\textbf{i}}_{1:h},\bs{\phi},\textbf{q})\bigr]$, where
$\Pi_{\bs{\phi},\textbf{q}} = \{ \pi_{\bs{\phi}} \cdot \pi_{\textbf{q}\mid\bs{\phi}}
: \pi_{\textbf{q}\mid\bs{\phi}} \in \Pi_{\textbf{q}\mid\bs{\phi}} \}$. The two
do not coincide in general, and the conditional rule is generally not Bayes for
any prior in the class \citep{Giacomini_Kitagawa_Read_2026}. The
unconditional rule is admissible under the regularity conditions of the minimax
theorem, being Bayes under a least-favorable prior, but is typically
harder to numerically approximate. We adopt the conditional criterion at the cost of this guarantee.}

Robust Bayesian rules based on the loss share a drawback with fully minimax
rules:\footnote{The fully minimax rule minimizes the worst-case risk over the
entire parameter space, $\hat{\textbf{i}}^{\,\text{mm}}_{1:h} =
\arg\min_{\hat{\textbf{i}}_{1:h}}\, \sup_{(\bs{\phi},\textbf{q})}\,
R(\hat{\textbf{i}}_{1:h},\bs{\phi},\textbf{q})$, dispensing with the prior on
$\bs{\phi}$ as well.} they are sensitive to the constant term
$c(\bs{\phi}, \textbf{q})$. The loss and the regret of action $\textbf{a}$ at
$(\bs{\theta}, \bs{\chi}_0)$ are
\begin{equation*}
L(\textbf{a}, \bs{\theta}, \bs{\chi}_0)
= c(\bs{\theta}, \bs{\chi}_0) + \textbf{a}'H\textbf{a} + 2\textbf{a}'\textbf{g},
\qquad
Re(\textbf{a}, \bs{\theta}, \bs{\chi}_0)
= \textbf{a}'H\textbf{a} + 2\textbf{a}'\textbf{g} + \textbf{g}'H^{-1}\textbf{g}.
\end{equation*}
The constant $c$ collects the variance of future shocks and the predetermined
drift of the economy: components no action can change. A worst case over the
loss is attracted by states where $c$ is large, states of distress in which
every action performs poorly. Regret subtracts the state-by-state minimum
$L^{*} = c - \textbf{g}'H^{-1}\textbf{g}$, so $c$ drops out. The worst case over regret is thus determined by the value of policy
and by the action's distance from that state's oracle, not by the level of unavoidable
distress. In addition, the robust regret criterion is invariant to the
conventions described in Footnote~\ref{fn:exante}, while the robust Bayes
based on loss is sensitive to them. This motivates the \textit{robust Bayes
regret} rule
\begin{equation}\label{rbre_rule}
\hat{\textbf{i}}^{RBRe}_{1:h}(\textbf{y}) =
\underset{\bar{\textbf{i}}_{1:h}\in\reals^h}{\arg\min}\;
\ev_{\bs{\phi}\mid\textbf{y}}\Bigl[\,
\underset{\textbf{q}\in\mathcal{I}_{\textbf{q}}(\bs{\phi})}{\sup}\;
Re(\bar{\textbf{i}}_{1:h}, \bs{\phi}, \textbf{q}) \Bigr],
\end{equation}
which retains robustness to identification uncertainty while directing the
worst case to the states where policy choice matters.

The certainty-equivalence property of Proposition~\ref{prop_regret_ce} carries over to the robust Bayes regret rule~(\ref{rbre_rule}), but not to the loss-based robust Bayes rule~(\ref{rb_rule}). Since the regret~(\ref{regret_quad}) is free of $c$, the inner worst case and the posterior expectation in~(\ref{rbre_rule}) act only on $(H, \textbf{g})$, which do not involve the shock second moments; the rule is therefore invariant to them. For the loss-based rule, by contrast, the inner worst case ranges over $c(\bs{\phi}, \textbf{q})$ as well, so the shock second moments --- which enter $c$ --- influence the selection of the worst-case $\textbf{q}$ and hence the optimal action: certainty equivalence fails.

\subsection{Feasible forward guidance decision rules 
}\label{sec:feasible-fg}

The forward guidance regime of Section~\ref{sec:fg} and the implementation
of the plug-in and Bayes rules under parameter uncertainty introduced in
Sections~\ref{sec:plugin}--\ref{sec:bayes} enforce an $h$-quarter expected
path with $h$ announcement shocks: \eqref{e1_solution} solves for
$\textbf{e}_1$ by inverting the $h\times h$ matrix $M^i_{1,fg}$. In practice
the number $m$ of announcement shocks that can be identified is often
smaller than the horizon, so that $M_{1,fg}$ has $m<h$ columns,
$M^i_{1,fg}$ is $h\times m$, and \eqref{e1_solution} has no solution: a
generic expected path $\bar{\textbf{i}}^e_{1:h}$ is not implementable. In this section we take the following approach: we redefine the forward guidance policy actions, so that the policymaker now announces the expected rate
at $m$ selected maturities, and implements the $m$ shocks that enforce these values exactly. The expected path at the remaining $h-m$ maturities is pinned down by the baseline response to the $m$-dimensional shocks that enforce the action at those horizons. We show how to derive and compute the rules of
Sections~\ref{sec:plugin}--\ref{sec:bayes} in this case.  The construction followed here is the one we implement in the numerical simulations of
Section~\ref{sec:numerical-risk} and in the empirical application of
Section~\ref{sec:empirical}, where $m=3$ and $h=8$. 

Fix maturities $1 \leq s_1 < \dots < s_m \leq h$ and let
$S_{\textbf{s}} \in \reals^{m\times h}$ select the corresponding rows. The
action is $\bar{\textbf{i}}^e_{\textbf{s}} \in \reals^m$, and the
announcement shocks are selected to enforce the $m$ constraints
\begin{equation}\label{lcrrestr_nodes}
S_{\textbf{s}}\, C_i\, \ev[\textbf{y}_{1:h} \mid \mathcal{I}_1]
= \bar{\textbf{i}}^e_{\textbf{s}},
\end{equation}
the restriction of \eqref{lcrrestr_fg} to the selected maturities. Letting
$\tilde M^i_{1,fg} := S_{\textbf{s}} M^i_{1,fg} \in \reals^{m\times m}$ and
assuming it invertible, we can solve for the unique vector of policy news
shocks that enforce the constraint:
\begin{equation}\label{em_solution}
\textbf{e}^m_1(\bar{\textbf{i}}^e_{\textbf{s}}, \bs{\eta}^{fg}_1, \bs{\theta}, \bs{\chi}_0)
= (\tilde M^i_{1,fg})^{-1}\bigl[\bar{\textbf{i}}^e_{\textbf{s}}
- S_{\textbf{s}}(\textbf{b}^i_{1:h}
+ M^i_{\eta,fg}\bs{\eta}^{fg}_1)\bigr].
\end{equation}
When $m=h$ and $\textbf{s}=(1,\dots,h)$, \eqref{em_solution} reduces to
\eqref{e1_solution}. Substituting \eqref{em_solution} into \eqref{yh_fg} and
retaining the non-policy block, the counterfactual target paths are
\begin{equation}\label{xpaths_nodes}
\tilde{\textbf{x}}^{fg}_{1:h} = \tilde D_{i,fg}\,\bar{\textbf{i}}^e_{\textbf{s}}
+ \tilde D_{b,fg}\,\textbf{b}_{1:h}
+ \tilde D_{\eta,fg}\,\bs{\eta}^{fg}_1 + D_{2:h,fg}\,\tilde{\bs{\varepsilon}}_{2:h},
\end{equation}
with
\[
\overset{((n-1)h\times m)}{\tilde D_{i,fg}} = M^x_{1,fg}(\tilde M^i_{1,fg})^{-1},\qquad
\tilde D_{b,fg} =
\begin{pmatrix} -\tilde D_{i,fg}S_{\textbf{s}} & I_{(n-1)h}\end{pmatrix},\qquad
\tilde D_{\eta,fg} =
M^x_{\eta,fg} - \tilde D_{i,fg}S_{\textbf{s}}M^i_{\eta,fg}.
\]

The following proposition shows that all properties that we proved in the forward guidance regime carry over to this feasible version. 

\begin{proposition}[Feasible forward-guidance loss and oracle]\label{prop:loss_nodes}
Suppose $\tilde M^i_{1,fg} = S_{\textbf{s}}M^i_{1,fg}$ is invertible and $W$
is symmetric positive semidefinite. Write
$Q_{1:m}:=(\textbf{q}_1\ \cdots\ \textbf{q}_m)$ and
$\textbf{q}_{1:m}:=\operatorname{vec}(Q_{1:m})$.
\begin{enumerate}
\item[(i)] $\tilde D_{i,fg}$ and $\tilde D_{b,fg}$ depend on $\bs{\theta}$
only through $(\bs{\phi}, \textbf{q}_{1:m})$ and are invariant to rotations
of $\{\textbf{q}_1,\dots,\textbf{q}_m\}$.
\item[(ii)] The loss is the quadratic form
\begin{equation}\label{loss_nodes_quadratic}
L_{fg}(\bar{\textbf{i}}^e_{\textbf{s}}, \bs{\phi}, Q, \bs{\chi}_0)
= \tilde c_{fg} + (\bar{\textbf{i}}^e_{\textbf{s}})'\tilde H_{fg}\,
\bar{\textbf{i}}^e_{\textbf{s}}
+ 2\,(\bar{\textbf{i}}^e_{\textbf{s}})'\tilde{\textbf{g}}_{fg},
\end{equation}
where, with
$\tilde{\textbf{b}}^{fg}_{1:h} := \tilde D_{b,fg}\textbf{b}_{1:h}-\textbf{x}^*$,
\begin{align*}
\tilde c_{fg} &:= (\tilde{\textbf{b}}^{fg}_{1:h})' W \tilde{\textbf{b}}^{fg}_{1:h}
+ \operatorname{trace}(\tilde D_{\eta,fg}' W \tilde D_{\eta,fg})
+ \operatorname{trace}(D_{2:h,fg}' W D_{2:h,fg}), \\
\tilde H_{fg} &:= \tilde D_{i,fg}' W \tilde D_{i,fg}, \qquad
\tilde{\textbf{g}}_{fg} := \tilde D_{i,fg}' W \tilde{\textbf{b}}^{fg}_{1:h}.
\end{align*}
In particular, the loss depends on $Q$ only through the column span of
$Q_{1:m}$.
\item[(iii)] Assuming $\tilde H_{fg}$ is positive definite, the loss admits the
unique minimizer
\begin{equation}\label{istar_nodes}
\bar{\textbf{i}}^{*,fg}_{\textbf{s}}(\bs{\phi},\textbf{q}_{1:m},\bs{\chi}_0)
= -\tilde H_{fg}^{-1}\tilde{\textbf{g}}_{fg}.
\end{equation}
\end{enumerate}
\end{proposition}

Note that while the action fixes the expected rate at the $m$ selected
maturities only, the announcement determines, at each $\bs{\theta}$, the
expected rate path over the entire horizon. Taking expectations with
respect to $\mathcal{I}_0$ in \eqref{yh_fg} and substituting
\eqref{em_solution}, 
\begin{equation}\label{impliedpath_nodes}
\ev\bigl[\,\textbf{i}_{1:h} \mid \mathcal{I}_0\bigr]
= \textbf{b}^i_{1:h}
+ M^i_{1,fg}\,\ev\bigl[\textbf{e}^m_1 \mid \mathcal{I}_0\bigr]
= \textbf{b}^i_{1:h}
+ M^i_{1,fg}(\tilde M^i_{1,fg})^{-1}
\bigl[\bar{\textbf{i}}^e_{\textbf{s}} - S_{\textbf{s}}\textbf{b}^i_{1:h}\bigr].
\end{equation}
This marks the
difference from the $m=h$ case: there, the entire expected path is chosen
by the policymaker and enforced at every value of the parameter. Here the
policymaker chooses $m$ coordinates, and the expected rate at the remaining
horizons is determined by the predetermined path $\textbf{b}^i_{1:h}$ and
by the impulse responses to the shocks that, under each $\bs{\theta}$,
enforce the action.

We also remark that the choice of maturities $\textbf{s}$ is part of the decision
problem:  the
decision rules of Sections~\ref{sec:plugin}--\ref{sec:bayes} and their risks
depend on the selection. This opens up a question on how to optimally select
the maturities $\textbf{s}$ for the decision problem. For now, we assume
they are exogenously given and leave the selection for future work. The rules of Sections~\ref{sec:plugin}--\ref{sec:bayes}, then,  apply without
modification: the plug-in evaluates \eqref{istar_nodes} at the point
estimate, and the Bayes rule is
$\hat{\textbf{i}}^{B}_{\textbf{s}} = -\,\ev[\tilde H_{fg}\mid \textbf{y}]^{-1}
\ev[\tilde{\textbf{g}}_{fg}\mid \textbf{y}]$.

\subsection{Suboptimality of plug-in rules}\label{sec:suboptimality}

In Section~\ref{sec:welfare} we have already discussed why plug-in
rules, despite their simplicity and first-order asymptotic convergence
to the oracle, are generally not appealing statistical decision
procedures for the finite-sample decision problem that we consider. This Section outlines a local asymptotic
argument: along $1/\sqrt{T}$ parameter sequences, the decision problem
with estimated multipliers converges to a Gaussian limit experiment in
the space of policy paths, within which the plug-in rule corresponds to the identity estimator of Gaussian signals unbiased for the oracle optimal path and is uniformly dominated by a
James--Stein-type shrinkage rule whenever the dimensionality of the action is at least $k\geq 3$. 
The full exposition of the regularity conditions turning this section into a formal statement
about feasible rules are ongoing work.
Section~\ref{sec:numerical-risk} complements the analysis by comparing the risks of the plug-in and of a simple Bayes rule numerically, in a Gaussian experiment calibrated on real data, for both a contemporaneous and a forward guidance exercises meant to replicate the setting of the empirical application in Section~\ref{sec:empirical}. 
We assume throughout the impulse responses to policy shocks are point-identified. We remark, however, that the results obtained by \citet{christensen2026optimal} for set-identified settings can be applied to our framework once one discretizes the action space of the policymaker.

\paragraph{The limit experiment.} Assume that $H(\bs{\theta}) \succ 0$ in a neighborhood of some $\bs{\theta}_0$ and write $\textbf{i}^{or}(\bs{\theta}) =
-H(\bs{\theta})^{-1}\textbf{g}(\bs{\theta})$ for the oracle path and
$\hat{\textbf{i}} := \textbf{i}^{or}(\hat{\bs{\theta}})$ for the
plug-in, where $\hat{\bs{\theta}}$ is an estimator of the
point-identified parameter whose properties are specified below. Let
$\bs{\theta}_0$ be stationary and in the interior of the parameter space, and consider sequences of data-generating parameters
drifting toward it at the $1/\sqrt{T}$ rate,
$\bs{\theta}_T = \bs{\theta}_0 + \bs{\tau}/\sqrt{T}$ with local
parameter $\bs{\tau}$. Let $\hat{\bs{\theta}}$ be a regular,
asymptotically linear estimator with
$\sqrt{T}(\hat{\bs{\theta}} - \bs{\theta}_T) \Rightarrow
N(\textbf{0}, V_0)$ along such sequences, as delivered by the local
asymptotic normality of a stationary Gaussian VAR together with the
regularity of the (quasi-)maximum likelihood estimator, or of the
posterior mean under Bernstein--von Mises conditions. Two observations
reduce the decision problem to a canonical form. First,
$\bs{\theta} \mapsto \textbf{i}^{or}(\bs{\theta})$ is continuously
differentiable at $\bs{\theta}_0$ with Jacobian $J_0$, assumed of full
row rank; the delta method then gives, under $\bs{\theta}_T$,
\begin{equation}\label{eq:limit_signal}
  X_T := \sqrt{T}\bigl(\hat{\textbf{i}} -
  \textbf{i}^{or}(\bs{\theta}_0)\bigr)
  \;\Rightarrow\; X \sim N\bigl(\bs{\mu},\, \Sigma_0\bigr),
  \qquad \bs{\mu} := J_0\bs{\tau}, \quad \Sigma_0 := J_0 V_0 J_0'
  \succ 0,
\end{equation}
so that, to first order, the plug-in path is an unbiased Gaussian
signal for the local oracle path, and the local mean $\bs{\mu}$ ranges
over all of $\reals^h$ as $\bs{\tau}$ varies. Second, since the loss
is quadratic, the regret of any path satisfies the exact identity
\begin{equation}\label{eq:regret_identity}
  \ell(\textbf{i};\bs{\theta}) -
  \ell(\textbf{i}^{or}(\bs{\theta});\bs{\theta})
  = \bigl(\textbf{i} - \textbf{i}^{or}(\bs{\theta})\bigr)'
    H(\bs{\theta})
    \bigl(\textbf{i} - \textbf{i}^{or}(\bs{\theta})\bigr),
\end{equation}
and for rules whose actions concentrate at the $1/\sqrt{T}$ rate
around the oracle---the rate at which the plug-in itself
concentrates---the rescaled regret $T[\ell - \ell^{or}]$ converges to
the same quadratic form with the weight given by
$H_0 := H(\bs{\theta}_0)$, since $H(\bs{\theta}_T) - H(\bs{\theta}_0)
= O(1/\sqrt{T})$ contributes only at lower order. The local asymptotic
decision problem is therefore the estimation of the mean $\bs{\mu}$ of
a single Gaussian observation $X \sim N(\bs{\mu}, \Sigma_0)$ under the
quadratic loss $(\textbf{d} - \bs{\mu})'H_0(\textbf{d} - \bs{\mu})$,
with the plug-in rule in the role of the identity estimator
$X \mapsto X$; we take both $\Sigma_0$ and $H_0$ as known constants of the
limit experiment, fixed by the localization point. The criterion of comparison is the \emph{local asymptotic risk}, or, equivalently, its regret counterpart: the
limit of the rescaled expected regret
$T\,\ev_{\bs{\theta}_T}[\ell - \ell^{or}]$ along the drifting
sequences. Convergence of risks requires more than the convergence in
distribution established above. The additional
requirement is uniform integrability of the rescaled regret, that can be achieved by replacing the plug-in with a $T$-dependent truncated version of it that shares the same limiting distribution with the untruncated one. 

\paragraph{Shrinkage dominates the identity rule.} Within the 
canonical limit experiment, domination of the identity estimator by shrinkage
is classical for identity covariance and weight
\citep{james1961estimation}; the extension to arbitrary
positive-definite weights and covariance is due to
\citet{berger1976minimax}. For the case we need, we provide a short
self-contained proof based on Stein's unbiased risk estimate
\citep{stein1981estimation} and the canonical-form reduction of
\citet{strawderman2003minimax}; unlike the general bounds in
\citet{berger1976minimax}, this yields the risk difference in
closed form.

\begin{lemma}[James--Stein domination under general quadratic
loss]\label{lemma_js}
Let $X \sim N(\bs{\mu}, \Sigma)$ with $\Sigma \succ 0$ known, let the
loss be $(\bs{\delta} - \bs{\mu})'H(\bs{\delta} - \bs{\mu})$ with
$H \succ 0$ known, and let $k = \dim(X) \geq 3$. For $t > 0$ consider
the James--Stein-type estimator of \citet{berger1976minimax},
\begin{equation}\label{js_rule}
\bs{\delta}^{JS}_t(X) = X -
\frac{t}{X'\Sigma^{-1}H^{-1}\Sigma^{-1}X}\, H^{-1}\Sigma^{-1} X.
\end{equation}
Its risk relative to the identity estimator $\bs{\delta}_0(X) = X$
satisfies
\begin{equation}\label{js_riskgain}
R(\bs{\delta}^{JS}_t, \bs{\mu}) - R(\bs{\delta}_0, \bs{\mu}) =
- \,t\,\bigl(2(k-2) - t\bigr)\,
\ev\!\left[\frac{1}{X'\Sigma^{-1}H^{-1}\Sigma^{-1}X}\right],
\end{equation}
which is strictly negative for every $\bs{\mu}$ when
$0 < t < 2(k-2)$ and is minimized at $t = k-2$. In particular,
$\bs{\delta}_0$ is inadmissible.
\end{lemma}

Applying Lemma~\ref{lemma_js} with $\Sigma = \Sigma_0$, $H = H_0$, and
$k = h$: for $h \geq 3$ and $0 < t < 2(h-2)$, the shrinkage rule
$\bs{\delta}^{JS}_t(X)$ dominates the identity rule uniformly in
$\bs{\mu}$, with risk gain given in closed form by
\eqref{js_riskgain}---within the limit experiment, the plug-in path is
inadmissible.

The feasible counterpart of the dominating rule is obtained by finding a finite sample analogue and replacing both known constants of the limit
experiment with consistent estimates. Let
$\hat{S} := \hat{J}\hat{V}\hat{J}'/T$ denote the estimated sampling
covariance of the plug-in path, and let $\textbf{i}_c$ be a fixed reference
path---one plausible choice being the status-quo path that holds
the policy rate at its current level. Writing
$\textbf{Z} := \hat{\textbf{i}} - \textbf{i}_c$, the feasible
James--Stein path is
\begin{equation}\label{eq:shrunk_plugin_feasible}
  \tilde{\textbf{i}}_t
  = \hat{\textbf{i}}
  - \frac{t}{\textbf{Z}'\hat{S}^{-1}\hat{H}^{-1}\hat{S}^{-1}
      \textbf{Z}}\;
    \hat{H}^{-1}\hat{S}^{-1}\textbf{Z},
\end{equation}
with $\hat{H} = H(\hat{\bs{\theta}})$: the plug-in path, shrunk toward
the reference path by a data-determined amount. The heuristic argument
then delivers \emph{weak dominance with strict local gains}. At every
localization point $\bs{\theta}_0$ satisfying our assumptions and
every local parameter $\bs{\tau}$, the local asymptotic risk of
\eqref{eq:shrunk_plugin_feasible} never exceeds that of the plug-in.
The gain is strict---uniformly over compact sets of $\bs{\tau}$, with
the closed form \eqref{js_riskgain}---exactly at those
$\bs{\theta}_0$ whose oracle path coincides with the reference path; at localization
points a fixed distance away, the 
two rules are first-order equivalent, so the improvement vanishes in
the limit. Since the local parameter ranges over 
$1/\sqrt{T}$-neighborhoods of each localization point, the
finite-sample content is that the strict gains accrue whenever the
true parameter's oracle path lies within sampling uncertainty of the
reference path and vanish with that
distance. 
The choice of reference path  determines where the gains concentrate, not whether dominance holds: choosing the status-quo path as a reference point locates them at parameter values whose oracle prescribes small departures from the current stance, and gives the dominating rule the interpretation of principled gradualism under multiplier uncertainty.

\section{From One-Shot to Sequential Policymaking}\label{sec:sequential}

The analysis developed so far treats the policy choice as a one-shot problem: at the beginning of the horizon, the policymaker selects a plan for the entire path and never revisits it. Actual policymaking is sequential: as time unfolds and new shocks realize, the policymaker can revise the decisions concerning the remaining periods. This section asks what happens to our framework when this possibility is taken seriously.

\subsection{Sequential Properties of the Oracle Solution}\label{sec:seq-oracle}
In the contemporaneous regime the transition from one-shot to sequential
policymaking does not require a new decision object. Consider a policymaker
who repeats the exercise of Section~\ref{sec:contemp} period after period, while the policy-relevant horizon is kept fixed at $t=h$. At
each date the optimal path is recomputed from the state the economy has
reached and only its first entry is implemented. The remaining entries are
revised at the following date once new non-policy shocks have realized. Each
of these \textit{shifted} problems is the one-shot oracle problem of
Section~\ref{sec:contemp} translated ahead in time over a shorter residual
horizon. We call the instrument path generated by this procedure the
\textit{re-optimizing} path. The question we answer in this subsection is
whether the ex-ante plan provides a reasonable summary of the re-optimizing
policy. For the oracle case the answer is positive. The ex-ante plan coincides
with the expectation of the re-optimizing path, and that path is also
sequentially optimal in the dynamic programming sense.

\paragraph{Notation for the sequential problem.} The terminal date $h$ is
fixed throughout. At each decision time $t = 0, 1, \dots, h-1$ the
policymaker observes the history $\mathcal{I}_t = \{\textbf{y}_{-T}, \ldots,
\textbf{y}_t\}$ and chooses the instrument value $i_{t+1}$ for the following
period. The chosen value is implemented through a contemporaneous surprise
$v_{t+1}$ that adjusts the response of the economy to the non-policy shocks
$\bs{\eta}_{t+1}$, consistently with the sequential implementation of
Lemma~\ref{lem:sequential}. We write $\ev_t[\cdot] := \ev[\,\cdot \mid
\mathcal{I}_t]$ and extend the conditional-expectation notation of
Section~\ref{sec:trajectories} to stacked residual paths. For $0 \leq s \leq t
\leq h$ we set $\textbf{b}_{t:h|s} := \ev_s[\textbf{y}_{t:h}]$ and for
individual entries $\textbf{b}_{t|s} := \ev_s[\textbf{y}_{t}]$. The
superscript partition $\textbf{b}^i_{\cdot|t}, \textbf{b}^x_{\cdot|t}$ carries
over unchanged. The state of the economy at time $t$ is $\bs{\chi}_t =
(\textbf{y}_t', \dots, \textbf{y}_{t+1-p}')'$. Throughout this section we set
the constant to zero, so that the variables are expressed in deviation from
the unconditional mean. With slight abuse of notation $\Pi$ then denotes the
lag-coefficient block and $\textbf{b}_{t|t-1} = \Pi \bs{\chi}_{t-1}$. Along
the sequential problem outcomes, states and forecasts are all generated by
the enforced path. We introduce no dedicated counterfactual notation, the
distinction being clear from the structure of the problem. Finally, we maintain throughout that the weight matrix $W$ is
\emph{time-separable}, so that the stacked loss \eqref{loss_c} can be written
as a sum of period losses,
\begin{equation}\label{eq:perloss}
L_c(\bar{\textbf{i}}_{1:h}, \bs{\theta})
= \sum_{t=1}^{h} \ev_{\bs{\eta}_{1:h}}\bigl[\ell_t(\bs{\chi}_t)\bigr],
\qquad
\ell_t(\bs{\chi}_t) := (\textbf{x}_t - \textbf{x}^*_t)'\, W_t\, (\textbf{x}_t - \textbf{x}^*_t),
\end{equation}
where $\textbf{x}^*_t$ denotes the period-$t$ block of $\textbf{x}^*$. The
Kronecker form of Section~\ref{sec:framework} is the leading case, with $W_t =
\beta^t \Lambda$.

The \textit{shifted} problem at decision time $t$ is the choice of the entire
path $\textbf{i}_{t+1:h}$ on the basis of $\mathcal{I}_t$. Its objective and
solution are
\begin{equation}\label{eq:shiftedproblem}
q_t(\textbf{i}_{t+1:h}; \bs{\chi}_t) := \sum_{s=t+1}^{h} \ev_t\bigl[\ell_s(\bs{\chi}_s)\bigr],
\qquad
\bar{\textbf{i}}^{*(t)}_{t+1:h}(\bs{\chi}_t) := \underset{\textbf{i}_{t+1:h} \in \reals^{h-t}}{\arg\min}\; q_t(\textbf{i}_{t+1:h}; \bs{\chi}_t)
\end{equation}
 where the expectation integrates the non-policy shocks $\bs{\eta}_{t+1:h}$
over their unconditional distribution, exactly as in \eqref{eq:perloss}. We further denote the minimal value of the shifted problem by 
\[
V_t(\bs{\chi}_t) := \min_{\textbf{i}_{t+1:h}} q_t(\textbf{i}_{t+1:h}; \bs{\chi}_t).
\]
For
every $t = 0, \ldots, h-1$, \eqref{eq:shiftedproblem} is the one-shot
contemporaneous oracle problem of Section~\ref{sec:contemp} shifted ahead in
time, with residual horizon $h-t$. Objects pertaining to the shifted problem
at decision time $t$ carry a parenthesized superscript ($D^{(t)}_{i}$,
$H^{(t)}$, and so on), with $D^{(0)}_{i} = D_{i,c}$. We write $W^{(t)}$ for
the weight matrix of the shifted problem, defined as the unique symmetric
matrix satisfying
\begin{equation}\label{eq:Wshifted}
\bigl(\textbf{x}_{t+1:h} - \textbf{x}^*_{t+1:h}\bigr)'\,W^{(t)}\,
\bigl(\textbf{x}_{t+1:h} - \textbf{x}^*_{t+1:h}\bigr)
= \sum_{s=t+1}^{h} \bigl(\textbf{x}_s - \textbf{x}^*_s\bigr)'\,W_s\,
\bigl(\textbf{x}_s - \textbf{x}^*_s\bigr),
\end{equation}
where, here and throughout, the residual paths $\textbf{x}_{t+1:h}$ and
$\textbf{x}^*_{t+1:h}$ are stacked in variable-major order, as in
Section~\ref{sec:trajectories}. In particular $W^{(0)} = W$ and, by
time-separability, $W^{(t)}$ is the restriction of $W$ to the horizons
$s > t$. Whenever $H^{(t)}$ is invertible the shifted problem has the unique
solution of Proposition~\ref{prop:loss_c} applied to the residual horizon,
\begin{equation}\label{eq:resoracle}
\bar{\textbf{i}}^{*(t)}_{t+1:h}
= -\bigl(H^{(t)}\bigr)^{-1}\bigl(D^{(t)}_{i}\bigr)'W^{(t)}\,
\Bigl( D^{(t)}_{b}\,\textbf{b}_{t+1:h|t} - \textbf{x}^*_{t+1:h} \Bigr).
\end{equation}
Lemma~\ref{lem:timeinv} in Appendix~\ref{app:sequential} shows that
invertibility of $H_c$ transfers to every $H^{(t)}$. Since the regime is
fixed within this subsection, we suppress the subscript $c$ where no
confusion arises.

The re-optimizing path introduced above is defined recursively. Let
$\bs{\chi}^*_0 := \bs{\chi}_0$ and, for $t = 0, \dots, h-1$, let
\[
i^{seq}_{t+1} := \mathbf{e}_1'\,\bar{\textbf{i}}^{*(t)}_{t+1:h}(\bs{\chi}^*_t),
\]
with $\bs{\chi}^*_{t+1}$ the state reached from $\bs{\chi}^*_t$ by
implementing $i^{seq}_{t+1}$ and after $\bs{\eta}_{t+1}$ materializes. The
re-optimizing path is $\textbf{i}^{seq}_{1:h} := (i^{seq}_1, \dots,
i^{seq}_h)'$, a random vector through the states $\bs{\chi}^*_t$.

\begin{definition}[Representativeness]\label{def:repr}
An ex-ante plan $\bar{\textbf{i}}_{1:h}$ is \emph{representative} of the
re-optimizing plan if
\begin{equation}\label{eq:repr}
\bar{\textbf{i}}_{1:h} = \ev_0\bigl[\textbf{i}^{seq}_{1:h}\bigr].
\end{equation}
\end{definition}
A representative plan coincides with the expectation, taken at the beginning
of the horizon, of the plan of a policymaker who re-optimizes as shocks
realize. The one-shot solution then correctly summarizes what such a
policymaker is expected to do. When $\bs{\theta}$ is known the expectation in
\eqref{eq:repr} is over the non-policy shocks $\bs{\eta}_{1:h}$ alone.
Section~\ref{sec:seq-bayes} extends the definition to unknown $\bs{\theta}$.

\begin{remark}[Representativeness and time inconsistency]\label{rmk:timeinc}
Representativeness must be kept distinct from \emph{time inconsistency} in
the sense of \citet{kydland1977rules}, where the ex-ante optimal plan
ceases to be optimal at later decision dates once private-sector expectations
have responded to it. A re-optimizing policymaker then deviates from the
announced plan ex post. In the contemporaneous regime the model is backward-looking and
the policy path is enforced through contemporaneous surprises. No
announcement shapes private-sector expectations, so the question of an
ex-post deviation from a communicated plan does not arise: time inconsistency
is ruled out by design. In the forward guidance regime the announcement \emph{is} the policy
action. The time-consistency problem arises there prior to and independently
of any consideration of representativeness. This section develops the
contemporaneous regime, while the forward guidance case is briefly discussed in the Conclusion and mostly left for future work.
\end{remark}

We can now state the main result of this subsection.

\begin{proposition}[Representativeness of the contemporaneous oracle]\label{prop:representativeness}
Suppose $W$ is time-separable with $W_s \succ 0$ for every $s$, and that
$H_c$ is invertible. Then the ex-ante oracle path of
Proposition~\ref{prop:loss_c} is representative of the re-optimizing path,
\[
\bar{\textbf{i}}^{*,c}_{1:h}(\bs{\chi}_0) = \ev_0\bigl[\textbf{i}^{seq}_{1:h}\bigr],
\]
and the first entries coincide exactly, $i^{seq}_1 = \bar{i}^{*,c}_1$.
\end{proposition}

The equality of the first
entries holds by construction, since the  problem at $t = 0$ is the
ex-ante problem and the re-optimizer opens with the ex-ante entry. The content
of the proposition is in the remaining entries and states that the ex-ante oracle path can be read
as the expected path of a policymaker who revises the plan as new non-policy shocks materialize.
Section~\ref{sec:seq-bayes} shows that this reading is lost once
$\bs{\theta}$ is unknown.

\begin{remark}[The value of re-optimization]\label{rmk:hierarchy}
The ex-ante plan fixes
$\textbf{i}_{1:h}$ at time zero, whereas the re-optimizer fixes the residual
path at every date and revises all but its first entry once the next shock is
observed. Fixing the path at
$t$ and re-solving at $t+1$ implements the same current entry
$\bar{i}^{*(t)}_{t+1}$ and differs only in the continuation, which is the tail
$\bar{\textbf{i}}^{*(t)}_{t+2:h}$ in the first case and $\bar{\textbf{i}}^{*(t+1)}_{t+2:h}$ in the second. Since the first is a feasible path in the shifted
problem at $t+1$, the expected one-step gain from re-solving is nonnegative,
\[
\kappa_t(\bs{\chi}^*_t) := V_t(\bs{\chi}^*_t)
- \ev_t\bigl[\ell_{t+1}(\bs{\chi}_{t+1}^*) + V_{t+1}(\bs{\chi}_{t+1}^*)\bigr]
\;\geq\; 0, \qquad t = 0, \dots, h-2 .
\]
At $t = h-1$ nothing remains to be revised and the gain is exactly zero. By iterated expectations, the overall gain of the
re-optimizing path over the ex-ante plan as of time zero is the expectation of the sum of the one-step gains
along the path,
\[
V_0(\bs{\chi}_0) - \ev_0\Bigl[\,\sum_{t=1}^{h} \ell_t(\bs{\chi}^*_t)\Bigr]
= \sum_{t=0}^{h-2} \ev_0\bigl[\kappa_t(\bs{\chi}^*_t)\bigr] \;\geq\; 0
\]
 where each $\kappa_t$ is the value of observing
$\bs{\eta}_{t+1}$ before the entries $\textbf{i}_{t+2:h}$ are chosen.
\end{remark}

\begin{remark}[Delayed transmission]\label{rmk:repr_delay}
Proposition~\ref{prop:representativeness} requires $H_c \succ 0$. By
Corollary~\ref{cor:rank} this fails under delayed transmission, such as in the 
empirical specification of Section~\ref{sec:empirical} for the contemporaneous case. Under the time-separable Kronecker weights of
\eqref{lossfun1}, if the variables entering the loss respond to the policy
shock with a delay of $d \geq 1$ periods, the re-optimizing path is then well defined in its first $h-d$
entries and the ex-ante oracle path is representative of it over those
entries, $\bar{\textbf{i}}^{*,c}_{1:h-d} = \ev_0[\textbf{i}^{seq}_{1:h-d}]$.
The last $d$ entries carry no decision content in either problem. Fixing them
at an arbitrary value casts the problem into the framework of this section
with $h-d$ decision dates and a positive definite Hessian, so that
Lemma~\ref{lem:timeinv} and Proposition~\ref{prop:representativeness} apply
verbatim. Corollary~\ref{cor:repr_delay} in Appendix~\ref{app:sequential}
states the result formally.
\end{remark}

\paragraph{Outline of the proof.} We now sketch the argument behind the
proof of Proposition~\ref{prop:representativeness}, which proceeds in three
steps. The first is the theorem of \citet{theil1957note}: in a linear
model with quadratic welfare and additive uncertainty, the strategy that
minimizes the expected loss over all strategies adapted to the information
has, as its expectation, the plan that minimizes the loss with the shocks
at their mean. The second is that this optimal strategy is the solution of
a finite-horizon linear-quadratic dynamic program. The third is that the
re-optimizing path coincides with that solution. The formal statements are
left to Appendix~\ref{app:sequential}.

Theil considers instruments $x$ and noncontrolled variables $y = Rx + s$,
where $R$ is known and $y$ in each period does not depend on the instruments
of later periods, the random vector $s$ has a law that does not depend on
$x$, and the welfare function is quadratic in $(x, y)$. In our problem the
instruments are the path $\bar{\textbf{i}}_{1:h}$ and the noncontrolled
variables are the targets $\textbf{x}_{1:h}$. The trajectory decomposition
of Section~\ref{sec:contemp} gives
\[
\textbf{x}_{1:h} = D_{i,c}\,\bar{\textbf{i}}_{1:h} + \textbf{s},
\qquad
\textbf{s} := D_{b,c}\,\textbf{b}_{1:h} + D_{\eta,c}\,\bs{\eta}_{1:h},
\]
so that $R = D_{i,c}$ and $s = \textbf{s}$. The blocks of $D_{i,c}$ are
lower triangular (Proposition~\ref{prop:Dstructure}), which is Theil's
timing restriction, and under enforcement the non-policy shocks retain
their distribution whatever the strategy (Lemma~\ref{lem:sequential}),
which is his independence requirement. The welfare function is minus the
loss, whose curvature in the instruments is $H_c$ by
Proposition~\ref{prop:loss_c}, so Theil's condition for a maximum, a
negative definite curvature of welfare, is $H_c \succ 0$. His theorem then
states that if $\textbf{f}$ minimizes the expected loss among the strategies
adapted to the information, $\ev_0[\textbf{f}]$ is the minimizer of the
loss with the shocks at their mean, which by Proposition~\ref{prop:loss_c}
is the ex-ante oracle path. As the first entry is non-stochastic, it coincides exactly in the ex-ante and dynamically optimal strategies. Strategies are understood throughout as in Theil: instrument paths
$\textbf{i} = (i_1, \dots, i_h)'$ whose entry $i_t$ is a function of the
past non-controlled components $\textbf{s}_1, \dots, \textbf{s}_{t-1}$,
with $\textbf{s}_t$ the period-$t$ entries of $\textbf{s}$, and with
$\ev_0[\|\textbf{i}\|^2] < \infty$. This is the class in which his
perturbation argument is formulated. When the parameters are known, the
history up to period $t$ is a function of $\bs{\chi}_0$, of
$\textbf{s}_1, \dots, \textbf{s}_t$ and of the strategy, and conversely
$\textbf{s}_t = \textbf{x}_t - (D_{i,c}\textbf{i})_t$ is a function of the
history, so Theil's class coincides with the class of strategies adapted to
the information $\mathcal{I}_{t-1}$ of our policymaker.

It remains to see that the re-optimizing path coincides with the optimal strategy.

Recall from \eqref{SVAR} that $B_{iv}$ is the on-impact response of the
policy rate to the contemporaneous policy shock, and let $B_{xv}$, $B_{i\eta}$
and $B_{x\eta}$ be the remaining blocks of the impact matrix. Suppose
$B_{iv} \neq 0$ and define
\[
\mathcal{D}_i := B_{xv}B_{iv}^{-1}, \qquad
\mathcal{D}_\eta := B_{x\eta} - \mathcal{D}_i B_{i\eta}, \qquad
\mathcal{D}_\chi := \begin{pmatrix} -\mathcal{D}_i & I_{n-1} \end{pmatrix}\Pi .
\]
Enforcing $i_t = \bar{i}_t$ pins down the surprise $v_t$, and substitution
into the target rows of the SVAR gives
\begin{equation}\label{eq:markov}
\textbf{x}_t = \mathcal{D}_i\, \bar{i}_t + \mathcal{D}_\chi\, \bs{\chi}_{t-1} + \mathcal{D}_\eta\, \bs{\eta}_t,
\qquad t = 1, \dots, h ,
\end{equation}
so the past affects the current outcome only through the state
(Lemma~\ref{lem:perperiod}). Stacking \eqref{eq:markov} with $i_t = \bar{i}_t$
and with the lagged entries of the state gives the linear transition
\begin{equation}\label{eq:standardform}
\bs{\chi}_{t} = \textbf{A}\,\bs{\chi}_{t-1} + \textbf{B}\, \bar{i}_t + \textbf{w}_{t},
\qquad t = 1, \dots, h,
\end{equation}
with $\textbf{A}$, $\textbf{B}$ and $\textbf{w}_t$ built from the multipliers
and from $\bs{\eta}_t$, as detailed in Appendix~\ref{app:sequential}
Each period loss $\ell_t(\bs{\chi}_t)$ in \eqref{eq:perloss} is quadratic
in the state. Choosing the instrument date by date from the current state
is therefore a finite-horizon linear-quadratic control problem with additive
disturbances \citep[Section~4.1]{bertsekas1995dynamic}, and its dynamic
program delivers the strategy that minimizes the expected loss over all
strategies adapted to the information
\citep[Section~1.3]{bertsekas1995dynamic}. Why the control of the dynamic program is also the first entry of the
shifted solution rests on the one-step gains $\kappa_t$ of
Remark~\ref{rmk:hierarchy}.
Expanding $q_{t+1}$
in \eqref{eq:shiftedproblem}, the expected loss over periods $t+2, \dots, h$
is quadratic in the tail $\textbf{i}_{t+2:h}$, with constant Hessian
$H^{(t+1)}$ and a linear coefficient 
\[
\textbf{g}_{t+1}(\bs{\chi}_{t+1}) :=
(D^{(t+1)}_{i})'W^{(t+1)}(D^{(t+1)}_{b}\,\textbf{b}_{t+2:h|t+1} -
\textbf{x}^*_{t+2:h})
\]
affine in the state. Appendix~\ref{app:sequential} shows that
\[
\kappa_t = \operatorname{trace}\Bigl(\bigl(H^{(t+1)}\bigr)^{-1}
\operatorname{Var}_t\bigl(\textbf{g}_{t+1}(\bs{\chi}_{t+1})\bigr)\Bigr)
\]
The action $\bar{i}_{t+1}$ moves the mean of $\bs{\chi}_{t+1}$ and leaves its
variance unchanged, so $\kappa_t$ is constant with respect to the action. The
stage objectives in $\bar{i}_{t+1}$ of the shifted problem and of the dynamic
program---the latter up to a constant
\[
\ev_t\bigl[\ell_{t+1}(\bs{\chi}_{t+1})\bigr]
+ \min_{\textbf{i}_{t+2:h}} \ev_t\bigl[q_{t+1}\bigr]
\qquad\text{and}\qquad
\ev_t\bigl[\ell_{t+1}(\bs{\chi}_{t+1})\bigr]
+ \ev_t\Bigl[\min_{\textbf{i}_{t+2:h}} q_{t+1}\Bigr],
\]
differ by
$\kappa_t$ and, therefore, share the minimizer $i^{seq}_{t+1}$. The re-optimizing path is therefore the path of the dynamic program, hence
the optimal strategy, and Theil's theorem applies to it: its time-zero
expectation is the ex-ante oracle path, which is the statement of
Proposition~\ref{prop:representativeness}.

\subsection{Sequential Bayesian Policymaking and the Failure of Representativeness}\label{sec:seq-bayes}

The representativeness result of Section~\ref{sec:seq-oracle} rests on the
affinity of the re-optimizing rule in the state, which in turn rests on the
curvature and the multipliers of the shifted problems being known constants.
Under parameter uncertainty both conditions fail: the coefficients of the rule
become functionals of the current posterior, and the posterior itself moves
with the realized data.

Rather than developing the general theory, this
subsection shows that even in the simplest sequential environment, $h =
2$ with a general lag order $p$, representativeness needs not hold, and will generically fail. We
first compare the sequential re-optimizer, who repeats the one-shot Bayes
exercise at each date, with the ex-ante Bayes plan. The two share the first
entry by construction, and the comparison of the second entries reveals the
failure of representativeness. We then compare both with the solution of the
Bayesian dynamic program, which values the effect of the intervention on
future beliefs and thereby carries an experimentation motive absent from the
other two problems. To see that this latter solution differs from the ex-ante and re-optimizing plans, it is sufficient to look at the first implemented entry of the path.

The policymaker enters the problem with the
posterior of Section~\ref{sec:decision-theory}, $\pi_0(\bs{\theta}) :=
\pi(\bs{\theta} \mid Y_0)$, where $Y_0$ is the observed sample from $t=-T$ to
$t=0$. At decision date $t = 0$ the policymaker observes $\bs{\chi}_0$ and
chooses $\bar{i}_1$, and period $1$ realizes under enforcement. At decision
date $t = 1$ the policymaker observes $\textbf{x}_1$, the instrument entry of
$\textbf{y}_1$ being the known enforced value, updates beliefs and chooses
$\bar{i}_2$. Period $2$ realizes and the loss is evaluated as in
\eqref{eq:perloss} with weights $W_1, W_2 \succ 0$ and targets
$\textbf{x}^*_1, \textbf{x}^*_2$.

\paragraph{Updating along the counterfactual.} Posterior updating along the
counterfactual is not the same as updating according to the SVAR dynamics: with the instrument path enforced period-by-period, the observed data no longer follow the
historical SVAR. By \eqref{eq:markov}, the period-$1$
observation is generated by
\[
\textbf{y}_1 = \begin{pmatrix} \bar{i}_1 \\ \textbf{x}_1 \end{pmatrix},
\qquad
\textbf{x}_1 \mid \bs{\chi}_0, \bs{\theta}\,;\, \bar{i}_1
\;\sim\; N\bigl(\mathcal{D}_i\,\bar{i}_1 + \mathcal{D}_\chi\,\bs{\chi}_0,\;
\mathcal{D}_\eta\mathcal{D}_\eta'\bigr),
\]
so the likelihood contribution of the counterfactual period factorizes into a
point mass at the enforced instrument, which carries no
information about $\bs{\theta}$, and the Gaussian density of the target
block. The posterior at time $t = 1$ conditions on the
historical sample, on the counterfactual observation, and on the enforced
instrument value. The latter is separated by a semicolon to mark that it
indexes the law generating $\textbf{x}_1$ rather than being itself a random
observation:\footnote{In the language of causal inference, the conditioning
is on the intervention $do(i_1 = \bar{i}_1)$ rather than on the observation
of $i_1$.}
\begin{equation}\label{eq:postupdate}
\pi_1(\bs{\theta}) := \pi\bigl(\bs{\theta} \mid Y_0, \textbf{x}_1\,;\,
\bar{i}_1\bigr)
\;\propto\; \pi_0(\bs{\theta})\cdot
\phi\Bigl(\textbf{x}_1;\;
\mathcal{D}_i(\bs{\theta})\,\bar{i}_1 + \mathcal{D}_\chi(\bs{\theta})\,\bs{\chi}_0,\;
\mathcal{D}_\eta(\bs{\theta})\mathcal{D}_\eta(\bs{\theta})'\Bigr),
\end{equation}
where $\phi(\cdot\,; \textbf{m}, V)$ denotes the multivariate normal
density. The historical sample enters through the observational likelihood of
Section~\ref{sec:decision-theory}, the counterfactual period through the
density implied by the enforced dynamics.

\begin{remark}[Conditioning versus intervening]\label{rmk:doVScond}
The second factor in \eqref{eq:postupdate} is \emph{not} the conditional
likelihood of $\textbf{x}_1 \mid i_1 = \bar{i}_1$ under the SVAR dynamics. From the proof of Lemma~\ref{lem:perperiod},  substituting the surprise consistent with
$i_1 = \bar{i}_1$ into the target rows yields, as an identity in
$\bs{\eta}_1$
\begin{equation}\label{eq:pathwise}
\textbf{x}_1 = \textbf{b}^x_{1|0}
+ \mathcal{D}_i\bigl(\bar{i}_1 - b^i_{1|0}\bigr)
+ \mathcal{D}_\eta\,\bs{\eta}_1 .
\end{equation}
The identity holds under both the historical SVAR and along the counterfactual, so the enforced density and the conditional density of $\textbf{x}_1$
given $i_1 = \bar{i}_1$ under the historical SVAR differ exclusively in the
distribution they assign to $\bs{\eta}_1$. Under enforcement the event
$\{i_1 = \bar{i}_1\}$ has probability one, since the surprise adjusts to
deliver $\bar{i}_1$ for every realization of $\bs{\eta}_1$
(Lemma~\ref{lem:sequential}). It therefore carries no information and
$\bs{\eta}_1$ retains its unconditional distribution, $N(\mathbf{0},
I_{n-1})$. Under the historical rule the instrument is stochastic and moves
with the non-policy shocks through $B_{i\eta}$. Its observed value is a noisy
signal about $\bs{\eta}_1$, and Gaussian conditioning extracts it,
\[
\bs{\eta}_1 \mid i_1 = \bar{i}_1
\;\sim\; N\Bigl(B_{i\eta}'\Sigma_{ii}^{-1}\bigl(\bar{i}_1 - b^i_{1|0}\bigr),\;
I_{n-1} - B_{i\eta}'B_{i\eta}\,\Sigma_{ii}^{-1}\Bigr),
\qquad \Sigma := BB' .
\]
Feeding the two conditional laws of $\bs{\eta}_1$ through \eqref{eq:pathwise} makes the
comparison exact. The means differ by
\[
\ev\bigl[\textbf{x}_1 \mid \bs{\chi}_0, i_1 = \bar{i}_1, \bs{\theta}\bigr]
- \ev\bigl[\textbf{x}_1 \mid \bs{\chi}_0, \bs{\theta}\,;\, \bar{i}_1\bigr]
= \mathcal{D}_\eta\,\ev\bigl[\bs{\eta}_1 \mid i_1 = \bar{i}_1\bigr]
= \mathcal{D}_\eta B_{i\eta}'\,\Sigma_{ii}^{-1}\bigl(\bar{i}_1 - b^i_{1|0}\bigr),
\]
while for the variances it holds
\begin{align*}
\text{Var}(\textbf{x}_1\mid \bs{\chi}_0, i_1=\bar{i}_1, \bs{\theta}) - \text{Var}(\textbf{x}_1 \mid \bs{\chi}_0, \bs{\theta}; \bar{i}_1)&=  \bigl(\Sigma_{xx} - \Sigma_{xi}\Sigma_{ii}^{-1}\Sigma_{ix}\bigr) - \mathcal{D}_\eta\mathcal{D}_\eta'
 \\ 
& 
= -\bigl(\mathcal{D}_\eta B_{i\eta}'\bigr)
\bigl(\mathcal{D}_\eta B_{i\eta}'\bigr)'\,\Sigma_{ii}^{-1}
\;\preceq\; 0 
\end{align*}
conditioning extracts information about $\bs{\eta}_1$ and is accordingly less
dispersed. Both differences vanish if and only if
$\mathcal{D}_\eta B_{i\eta}' = \mathbf{0}$, hence for nonsingular
$\mathcal{D}_\eta$ if and only if $B_{i\eta} = \mathbf{0}$. If that's not the case, updating with the
observations conditional on $i_1 = \bar{i}_1$ under the historical SVAR, in
place of \eqref{eq:postupdate}, would yield wrong first and second moments for the posterior.
\end{remark}

\begin{remark}[Identification of structural parameters along the counterfactual path]\label{rmk:identD}
The interventional density depends on $\bs{\theta}$ only through
$(\bs{\phi}, \textbf{q}_1)$, invariantly to the sign of $\textbf{q}_1$. For
$\mathcal{D}_i$ and $\textbf{b}^x_{1|0}$ this is immediate, and for the
variance it follows from the identity
\[
\mathcal{D}_\eta\mathcal{D}_\eta'
= \Sigma_{xx} - \mathcal{D}_i\,\Sigma_{ix} - \Sigma_{xi}\,\mathcal{D}_i'
+ \mathcal{D}_i\,\Sigma_{ii}\,\mathcal{D}_i'
= \bigl(\Sigma_{xx} - \Sigma_{xi}\Sigma_{ii}^{-1}\Sigma_{ix}\bigr)
+ \Sigma_{ii}\,(\mathcal{D}_i - \varrho)(\mathcal{D}_i - \varrho)'
\]
with $\varrho := \Sigma_{xi}\Sigma_{ii}^{-1}$. The counterfactual observation
is therefore informative about $\textbf{q}_1$ in a way the historical sample
is not. Under the historical regime $\textbf{y}_1 \mid \bs{\chi}_0 \sim
N(\textbf{b}_{1|0}, \Sigma)$, the rotation entering only through $BB' =
\Sigma$, which is why $\textbf{q}_1$ is set-identified. Enforcement breaks
this invariance by treating the columns of $B$ asymmetrically.

If the enforced value deviates from the
historical forecast, $\delta := \bar{i}_1 - b^i_{1|0} \neq 0$, the mean of the
counterfactual observation is $\textbf{b}^x_{1|0} + \mathcal{D}_i\,\delta$ by
\eqref{eq:pathwise}. It is linear in $\mathcal{D}_i$ with slope $\delta$, so
distinct values of the impact multiplier induce distinct means. Second, even
if the historical forecast is enforced, so that $\delta=0$, the variance of
$\textbf{x}_1$ depends on $\mathcal{D}_i$. We can then distinguish between two
cases:
\[
\text{(i)}\;\; \delta \neq 0: \;\; \mathcal{D}_i \text{ point identified};
\qquad
\text{(ii)}\;\; \delta = 0: \;\; \mathcal{D}_i \text{ identified up to} \,\, \mathcal{D}_i \mapsto 2\varrho - \mathcal{D}_i .
\]
To see this, consider the factorization of the joint likelihood
\[
p\bigl(Y_0, \textbf{x}_1 \mid \bs{\theta}\,;\, \bar{i}_1\bigr)
= p\bigl(Y_0 \mid \bs{\phi}\bigr)\cdot
\phi\Bigl(\textbf{x}_1;\; \textbf{b}^x_{1|0} + \mathcal{D}_i\,\delta,\;
\mathcal{D}_\eta\mathcal{D}_\eta'\Bigr),
\qquad \delta := \bar{i}_1 - b^i_{1|0}.
\]
Let $\bs{\theta}$ and $\tilde{\bs{\theta}}$ induce the same law. By
integrating $\textbf{x}_1$ out, equality of the densities implies equality of
the marginals of $Y_0$, which gives $\bs{\phi} = \tilde{\bs{\phi}}$, since
the reduced form is identified by the historical sample. The two parameters
therefore share $\textbf{b}_{1|0}$, $\Sigma$ and $\delta$, and equality of
the conditionals reduces to equality of the second term. Equality of the means reads $\mathcal{D}_i\,\delta
= \tilde{\mathcal{D}}_i\,\delta$, so for $\delta \neq 0$ we obtain
$\mathcal{D}_i = \tilde{\mathcal{D}}_i$. Equality of the variances reads
$(\mathcal{D}_i - \varrho)(\mathcal{D}_i - \varrho)' = (\tilde{\mathcal{D}}_i - \varrho)
(\tilde{\mathcal{D}}_i - \varrho)'$, that is $\mathcal{D}_i - \varrho =
\pm(\tilde{\mathcal{D}}_i - \varrho)$. 
In case (i), $\textbf{q}_1$ is in turn identified up to sign, since
$(1, \mathcal{D}_i')'$ is proportional to $\Sigma_l\textbf{q}_1$ and
$\|\textbf{q}_1\| = 1$ fixes it up to reflection.
\end{remark}

\paragraph{The re-optimizing  and ex-ante solutions}  We can now study the policymaker's optimal action at the second decision date. The
state $\bs{\chi}_1$ has realized and is observed, and beliefs are given by the
posterior $\pi_1$ of \eqref{eq:postupdate}, obtained by updating $\pi_0$ with
the interventional density. The shifted problem
is the one-shot Bayes problem at horizon one. Using \eqref{eq:markov} for the
period-$2$ outcome, the re-optimizing policymaker solves
\[
\min_{i \in \reals}\;
\ev_{\pi_1}\ev_{\bs{\eta}_2}\Bigl[
\bigl(\mathcal{D}_i\,i + \mathcal{D}_\chi\bs{\chi}_1 - \textbf{x}^*_2
+ \mathcal{D}_\eta\bs{\eta}_2\bigr)'W_2
\bigl(\mathcal{D}_i\,i + \mathcal{D}_\chi\bs{\chi}_1 - \textbf{x}^*_2
+ \mathcal{D}_\eta\bs{\eta}_2\bigr)\Bigr].
\]
Write, as in Section~\ref{sec:seq-oracle} at horizon one and with the
dependence on $\bs{\theta}$ made explicit,
\[
H^{(1)}(\bs{\theta}) := \mathcal{D}_i'W_2\mathcal{D}_i,
\qquad
g(\bs{\theta}, \bs{\chi}_1) := \mathcal{D}_i'W_2
\bigl(\mathcal{D}_\chi\bs{\chi}_1 - \textbf{x}^*_2\bigr),
\]
both scalars, the first strictly positive under the conditions of
Proposition~\ref{prop:representativeness} by Lemma~\ref{lem:timeinv}(i).
Since $\bs{\eta}_2$ has zero mean and is independent of $(\bs{\theta},
\bs{\chi}_1)$, it contributes an additive constant, and the first-order
condition gives :
\begin{equation}\label{eq:bayesseq}
i^{B,seq}_2\bigl(\bs{\chi}_1; \bar{i}_1\bigr)
= -\,\frac{\ev_{\pi_1}\bigl[\,g(\bs{\theta}, \bs{\chi}_1)\,\bigr]}
{\ev_{\pi_1}\bigl[\,H^{(1)}(\bs{\theta})\,\bigr]} ,
\end{equation}
a ratio of posterior moments. Through $\pi_1$ they are functions
of the counterfactual observation $\textbf{x}_1$, hence
$\mathcal{I}_1$-measurable random variables.

We now consider, instead, the decision taken by a policymaker ex-ante. At $t = 0$ they commit
to the pair $(\bar{i}_1, \bar{i}_2)$, evaluating the two-period loss under
$\pi_0$. We derive the second entry in two steps.

\emph{Step 1: the period-$2$ term.} By time-separability the objective splits
into the period-$1$ term, which does not involve $\bar{i}_2$, and
\[
\ev_{\pi_0}\,\ev_{\bs{\eta}_1}\,\ev_{\bs{\eta}_2}\Bigl[
\bigl(\mathcal{D}_i\,\bar{i}_2 + \mathcal{D}_\chi\bs{\chi}_1 - \textbf{x}^*_2
+ \mathcal{D}_\eta\bs{\eta}_2\bigr)'W_2\bigl(\cdot\bigr)\Bigr],
\qquad
\bs{\chi}_1 = \textbf{A}\bs{\chi}_0 + \textbf{B}\,\bar{i}_1 + \textbf{w}_1 ,
\]
where the outer expectation is over $\bs{\theta} \sim \pi_0$ and the inner
ones over the two non-policy shocks. The state entering the period-$2$ loss is
itself random through $\bs{\eta}_1$, and depends on the first entry
$\bar{i}_1$.

\emph{Step 2: expanding and taking first-order conditions.} Expanding the
quadratic, $\bs{\eta}_2$ contributes the additive constant
$\operatorname{trace}(\mathcal{D}_\eta'W_2\mathcal{D}_\eta)$ and the objective
becomes, up to terms independent of $\bar{i}_2$,
\[
\ev_{\pi_0}\ev_{\bs{\eta}_1}\Bigl[
H^{(1)}(\bs{\theta})\,\bar{i}_2^{\,2}
+ 2\,\bar{i}_2\; g\bigl(\bs{\theta}, \bs{\chi}_1\bigr)\Bigr].
\]
Since $g$ is affine in the state, integrating $\bs{\eta}_1$ replaces
$\bs{\chi}_1$ by its conditional mean given $\bs{\theta}$,
\[
\ev_{\bs{\eta}_1}\bigl[g(\bs{\theta}, \bs{\chi}_1)\bigr]
= g\bigl(\bs{\theta},\, \bs{\chi}^{det}_1(\bar{i}_1)\bigr),
\qquad
\bs{\chi}^{det}_1(\bar{i}_1)
:= \textbf{A}\,\bs{\chi}_0 + \textbf{B}\,\bar{i}_1 ,
\]
and the first-order condition in $\bar{i}_2$ yields
\begin{equation}\label{eq:bayesexante}
\bar{i}^{B}_2\bigl(\bar{i}_1\bigr)
= -\,\frac{\ev_{\pi_0}\bigl[\,
g\bigl(\bs{\theta}, \bs{\chi}^{det}_1(\bar{i}_1)\bigr)\,\bigr]}
{\ev_{\pi_0}\bigl[\,H^{(1)}(\bs{\theta})\,\bigr]} ,
\end{equation}
formally the same ratio as \eqref{eq:bayesseq}, with the time $t=0$ posterior $\pi_0$ in place of $\pi_1$ and the
 the deterministic state in place of the stochastic one.

\paragraph{Failure of representativeness and  comparison with the Bayesian dynamic program.}  We are now in a position to discuss representativeness for the Bayesian re-optimizing policymaker. Definition~\ref{def:repr}
extends to the Bayesian problem by replacing $\ev_0$ with the time-zero
predictive expectation. We say that the ex-ante Bayes plan is \emph{representative} if
\begin{equation}\label{eq:reprbayes}
\bar{\textbf{i}}^{B}_{1:2} = \ev_{\pi_0}\ev_{\bs{\eta}_{1}}
\bigl[\,\textbf{i}^{B,seq}_{1:2}\,\bigr],
\end{equation}
the expectation being over $\bs{\theta} \sim \pi_0$ and $\bs{\eta}_1 \sim
N(\mathbf{0}, I_{n-1})$ independently, that is over the predictive law of the
counterfactual observation $\textbf{x}_1$. 

We show that \eqref{eq:reprbayes} fails on the second entry.\footnote{It holds
trivially on the first: the shifted problem at $t = 0$ \emph{is} the ex-ante
problem, nothing having been observed yet, so the re-optimizer opens with the
ex-ante entry, $i^{B,seq}_1 = \bar{i}^{B}_1$, by construction.} for an \emph{arbitrary} common $\bar{i}_1$,
which strengthens the statement and covers $\bar{i}_1 = \bar{i}^{B}_1$ as the
relevant case. By the law of iterated expectations under the predictive law,
and since $\bs{\chi}_1$ is known at the second decision date,
\[
\ev_{\pi_0}\ev_{\bs{\eta}_1}\bigl[\,g(\bs{\theta}, \bs{\chi}_1)\,\bigr]
= \ev_{\pi_0}\ev_{\bs{\eta}_1}\Bigl[\,
\ev_{\pi_1}\bigl[g(\bs{\theta}, \bs{\chi}_1)\bigr]\Bigr],
\qquad
\ev_{\pi_0}\bigl[\,H^{(1)}(\bs{\theta})\,\bigr]
= \ev_{\pi_0}\ev_{\bs{\eta}_1}\Bigl[\,
\ev_{\pi_1}\bigl[H^{(1)}(\bs{\theta})\bigr]\Bigr],
\]
the first equality using in addition Step~1 above. Substituting into \eqref{eq:bayesexante} and \eqref{eq:bayesseq}, the two
sides of \eqref{eq:reprbayes} read
\begin{equation*}
\bar{i}^{B}_2(\bar{i}_1)
= -\,\frac{\ev_{\pi_0}\ev_{\bs{\eta}_1}\bigl[\ev_{\pi_1}[\,g\,]\bigr]}
{\ev_{\pi_0}\ev_{\bs{\eta}_1}\bigl[\ev_{\pi_1}\bigl[H^{(1)}\bigr]\bigr]} ,
\qquad\qquad 
\ev_{\pi_0}\ev_{\bs{\eta}_1}\bigl[\,i^{B,seq}_2(\bs{\chi}_1; \bar{i}_1)\,\bigr]
= -\,\ev_{\pi_0}\ev_{\bs{\eta}_1}\!\left[
\frac{\ev_{\pi_1}[\,g\,]}{\ev_{\pi_1}\bigl[H^{(1)}\bigr]}\right],
\end{equation*}
a \emph{ratio of means} against the \emph{mean of a ratio}. The two coincide
when the posterior curvature $\ev_{\pi_1}\bigl[H^{(1)}\bigr]$ does not vary
with the realization $\textbf{x}_1$. It then factors out of the outer
expectation and cancels, as when the transmission mechanism is effectively
known and only the remaining parameters are uncertain. However, in the general case this is not guaranteed: if such condition fails, representativeness does not hold.

The second result from section \ref{sec:seq-oracle} fails as well: under parameter uncertainty, the
re-optimizer no longer solves the dynamically optimal problem, and the divergence
appears already in the first entry. To see this, we use the result of the
sequential problem at $t = 1$ to obtain the continuation value in closed
form. The residual problem being quadratic, substituting
\eqref{eq:bayesseq} back into the objective gives
\[
V_1(\bs{\chi}_1, \pi_1)
= \ev_{\pi_1}\bigl[\,c(\bs{\theta}, \bs{\chi}_1)\,\bigr]
- \frac{\bigl(\ev_{\pi_1}[\,g\,]\bigr)^2}{\ev_{\pi_1}\bigl[H^{(1)}\bigr]}
\]
\[
c(\bs{\theta}, \bs{\chi}_1) :=
\bigl(\mathcal{D}_\chi\bs{\chi}_1 - \textbf{x}^*_2\bigr)'W_2
\bigl(\mathcal{D}_\chi\bs{\chi}_1 - \textbf{x}^*_2\bigr)
+ \operatorname{trace}\bigl(\mathcal{D}_\eta'W_2\mathcal{D}_\eta\bigr).
\]
The sequential problem at $t = 0$ chooses $\bar{i}_1$ against the period-$1$
term plus the expected continuation
$\ev_{\pi_0}\ev_{\bs{\eta}_1}\bigl[V_1(\bs{\chi}_1, \pi_1)\bigr]$. The
ex-ante problem, with its second entry concentrated out at
\eqref{eq:bayesexante}, chooses $\bar{i}_1$ against the same period-$1$ term
plus $\ev_{\pi_0}\ev_{\bs{\eta}_1}[\,c\,] -
\bigl(\ev_{\pi_0}\ev_{\bs{\eta}_1}[\,g\,]\bigr)^2 /
\ev_{\pi_0}\bigl[H^{(1)}\bigr]$. The $c$-terms coincide by the law of
iterated expectations, so the two objectives differ by
\begin{equation}\label{eq:Psi}
\Psi(\bar{i}_1)
:= \ev_{\pi_0}\ev_{\bs{\eta}_1}\!\left[
\frac{\bigl(\ev_{\pi_1}[\,g\,]\bigr)^2}{\ev_{\pi_1}\bigl[H^{(1)}\bigr]}
\right]
- \frac{\bigl(\ev_{\pi_0}\ev_{\bs{\eta}_1}[\,g\,]\bigr)^2}
{\ev_{\pi_0}\bigl[H^{(1)}\bigr]} .
\end{equation}
The term is the value of adapting the second entry to the counterfactual
observation, and, differently from the gain $\kappa_0$ in Section \ref{sec:seq-oracle}, nothing guarantees it is constant in the intervention $\bar{i}_1$: $\Psi$ is a
functional of the law of $\pi_1$, which, as qualified by
Remark~\ref{rmk:identD}, depends on $\bar{i}_1$ through the deviation
$\delta = \bar{i}_1 - b^i_{1|0}$ governing how much the counterfactual
observation discriminates across parameter values. The sequential problem
therefore carries an \emph{experimentation motive}
\citep{ prescott1972multiperiod, wieland2000monetary}:
its objective contains, through $\Psi(\bar{i}_1)$, a channel by which the
intervention shapes what will be known at the second decision date, a channel
neither the ex-ante problem nor the re-optimizer has any counterpart for.
Whether and how strongly the motive operates at the optimum is a question we
do not pursue here.\footnote{\cite{zellner1971introduction} characterizes sequentially Bayes optimal rules for the two-period setting with linear structural equations in which the policy variables in the sample are as good as randomized.} However, the mere fact
that the difference in continuation values depends on the first-period
action allows us to see that the sequentially optimal path and the re-optimizing path
are generically different, and that  the divergence appears already in the
first entry.

\section{Empirical application: U.S. monetary policy in 2006--2007}\label{sec:empirical}

We apply the methodology to a one-shot decision taken at the beginning of 2006. The policymaker seeks an eight-quarter path of the federal funds rate that keeps the output gap closed and inflation at two percent, using a Bayesian SVAR estimated on data available at the end of 2005. We report the plug-in and the Bayes rules of Section~\ref{sec:bayes} for the contemporaneous and the forward guidance regimes and compare them with the path actually followed in the historical data. Robust Bayes and robust Bayes regret rules are obtained for the contemporaneous regime. The numerical risk comparisons of Section~\ref{sec:numerical-risk} are calibrated to the empirical specifications.

\subsection{Data, identification and estimation}\label{sec:emp_setup}

\paragraph{Data and loss.}
The macroeconomic block contains the output gap (HP-filtered real GDP), annualized GDP-deflator inflation, annualized commodity-price inflation and the federal funds rate, at quarterly frequency from 1969Q1 to 2007Q4. The loss is the discounted dual mandate
\[
  \mathcal{L}(\bar\imath)=\sum_{t=1}^{h}\beta^{t}\Big[(y_t-y^\ast)^2+(\pi_t-\pi^\ast)^2\Big],
  \qquad y^\ast=0,\ \pi^\ast=2,\ \beta=0.99,\ h=8,
\]
with $t=1$ corresponding to 2006Q1 and $i_0=3.98$ the rate observed in 2005Q4.

\paragraph{Identification.}
In our main empirical application we consider point identified settings for both the contemporaneous and forward guidance cases: the structural parameter $\bs{\theta}$ is then in one-to-one correspondence with the reduced-form parameter $\bs\phi$.\footnote{Throughout this section we reparametrize $\bs{\phi}$ by replacing the diagonal entries of $\Sigma_l$ with their logarithms. The Gaussian prior introduced below is specified on the same parametrization.} In the contemporaneous regime the policy shock is identified with the narrative series of \citet{romer2004new} ordered second to last and used as an internal instrument \citep{plagborg2021local, plagborg2022instrumental}.\footnote{Ordering the instrument after the macroeconomic block adds an exogeneity insurance relative to the plain internal IV scheme with the series ordered first \citep{ramey2016macroeconomic, mckay2023can}.} Note that under such identification scheme, the non-policy variables entering the loss function do not respond to the policy shock on impact: this is a case of delayed transmission discussed in Section \ref{sec:contemp} ($d=1$). As outlined in Appendix \ref{app:proofs}, normalizing $\bar i_h$ to an arbitrary value one can then solve the resulting $(h-1)$~-dimensional problem, which has the same structure as the canonical problem discussed in Section \ref{sec:contemp}, and is the approach that we follow.  
We adopt the feasible forward guidance approach of Section \ref{sec:feasible-fg}, and the $m=3$-dimensional action is  enforced through the impulse responses to the \citet{romer2004new} narrative series, the high-frequency monetary surprise of \citet{gertler2015monetary}, and  surprises to the 10-year on-the-run Treasury yield  from \citet{barnichon2023sufficient}. The two high-frequency series are ordered first and the Romer--Romer series second to last, so that each shock is again a Cholesky response to its instrument. One of the limitations of using pre-ZLB data is that pure forward guidance was rarely used over the sample: our third shock series yields impulse responses that are measured with large uncertainty and that in a non-negligible share of posterior draws have a small cumulative effect on the interest rate path, making the smallest singular value of the matrix $\tilde{M}^i_{1,fg}$ of Section \ref{sec:feasible-fg} close to zero.  We regularize the problem by restricting the parameter space, discarding the draws whose smallest singular value falls below a threshold, as detailed in Appendix \ref{app:specification}. 

\paragraph{Estimation.}
Each system is estimated with four lags on 1970Q1--2005Q4 (the 1969 observations serving as initial conditions for the lags) under an independent Normal--inverse-Wishart prior with block exogeneity for the instrument block. Appendix \ref{app:specification} discusses the prior selection and posterior sampling. The plug-in rule is computed at the posterior mean of the parameters, the Bayes rule by averaging the loss over the draws. Counterfactual paths are reported as posterior means, with 90\% point-wise  equi-tailed credible regions around the Bayes counterfactual. Note that in each of the following figures, the ``observed'' paths for the target variables are the counterfactual paths obtained by setting $\bar\imath$ equal to the path actually followed, not the realized data.\footnote{ The realized target paths responded to non-policy shocks over 2006--2007 and are not an ex-ante object. We then choose to report the posterior mean of the counterfactuals at the observed instrument path to allow for a meaningful comparison between the counterfactual paths implied by the optimal decision rules and the ones that are implied by the observed path.} 

\subsection{The contemporaneous regime}\label{sec:emp_contemp}

\begin{figure}[tbp]
    \centering
    \includegraphics[width=1.1\linewidth]{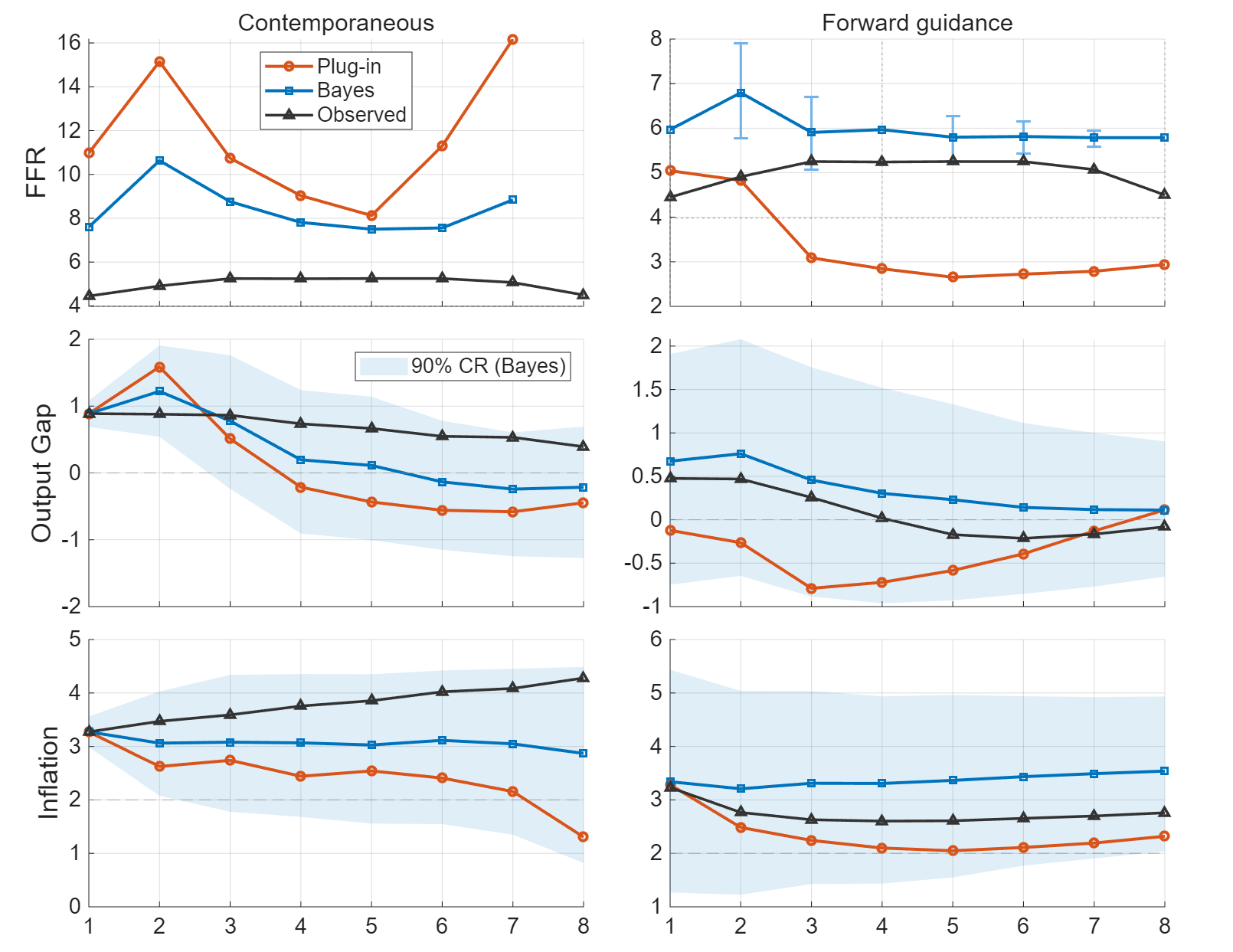}
    \caption{\small Baseline (dual-mandate) exercise, $h=8$; $t=1$
    corresponds to 2006Q1. Left column: contemporaneous regime; right
    column: feasible forward guidance regime, $m=3$. Top row: federal funds rate paths
    under the plug-in rule, the Bayes rule and the observed path; in the
    contemporaneous regime the last action is normalized
    and the rules are drawn up to $h-1$. Middle and bottom rows: posterior
    means of the counterfactual output gap and annualized inflation under
    each path; shaded areas: 90\% posterior credible regions around the
    Bayes counterfactual.}
    \label{fig:regimes_baseline}
\end{figure}

The left column of Figure~\ref{fig:regimes_baseline} reports the instrument paths and the posterior means of the implied counterfactuals from the application of the decision procedures outlined in Section~\ref{sec:bayes} for the contemporaneous regime; the right column, discussed in Section~\ref{sec:emp_fg}, reports the same objects for the feasible forward guidance regime.

Both rules prescribe a similar qualitative response: an immediate increase over the observed path and a subsequent partial easing. The Bayes path closes the output gap by mid-2007 and holds inflation around 3 percent, while the plug-in reduces inflation further, to 1.3 percent by the end of the window, but pushes the output gap below zero from 2006Q4. The plug-in is more erratic than Bayes: it reaches 15.2 percent in the second quarter, declines over the next three quarters, and rises sharply again to 16.2 percent in the seventh. The irregular shape of the plug-in path is the one the theory leads us to expect. The plug-in treats the point estimate as the truth and exploits any direction in which the estimated multipliers show leverage on the targets, however imprecisely that leverage is measured. Along directions of $\hat{H}$ with small singular values---corresponding to small estimated policy effectiveness at certain horizons---the plug-in path requires large deviations to steer the targets. The Bayes rule, instead, discounts the directions in which policy transmission is uncertain, delivering a smoother and more moderate path, which is, however, still more irregular than the observed one. The posterior expected loss of the Bayes path is 55.3, against 66.2 for the plug-in and 67.9 for the observed path. 

Both rules prescribe a markedly higher federal funds rate than the observed one, to bring back toward the target an economy with inflation between 2.9 and 3.7 percent during 2005 and a positive output gap. The policy surprises needed to enforce these paths are large and, more importantly, repeated. A sequence of sizable surprises is unlikely to keep being treated as unanticipated by the private sector.\footnote{Large surprises may also require extrapolating the SVAR model beyond the range of policy shocks observed in the estimation sample. The effects of such surprises are computed under the maintained assumptions that the model is linear and invariant to the intervention, which we keep throughout. We do not investigate this issue further here;  \citet{leeper2003modest} propose a measure of whether a policy intervention is modest enough not to induce private agents to revise their beliefs about the policy rule, and which applies directly to the contemporaneous regime.} Two features of our framework address this concern. First, adding a smoothing penalty, as discussed in the next paragraph, reduces the size of the surprises needed to implement the more gradual rules. Second, the forward guidance regime is designed to avoid a sequence of surprises altogether, since the announced path is enforced through a single set of shocks at $t=1$. 

\paragraph{Smoothing in the contemporaneous regime}

In practice policymakers may have a  preference for smooth instrument
paths. We can accommodate such preference
within our framework by adding to the loss the quadratic term
$\rho\sum_{t=1}^{h}\beta^{t}(i_t-i_{t-1})^2$, anchored at the last observed
rate $i_0=3.98$, as discussed in Appendix \ref{app:genloss} . Figure~\ref{fig:smoothing_rules} reports the
optimal paths in the contemporaneous regime at two values of $\rho$:  a moderate penalty, $\rho=0.5$ and a calibrated weight $\rho^*$ that corresponds to the smallest penalty at which
the Bayes path is no rougher --- in the discounted sum of squared first differences ---
than the observed path. 
\begin{figure}[tbp]
    \centering
    \includegraphics[width=0.8\linewidth]{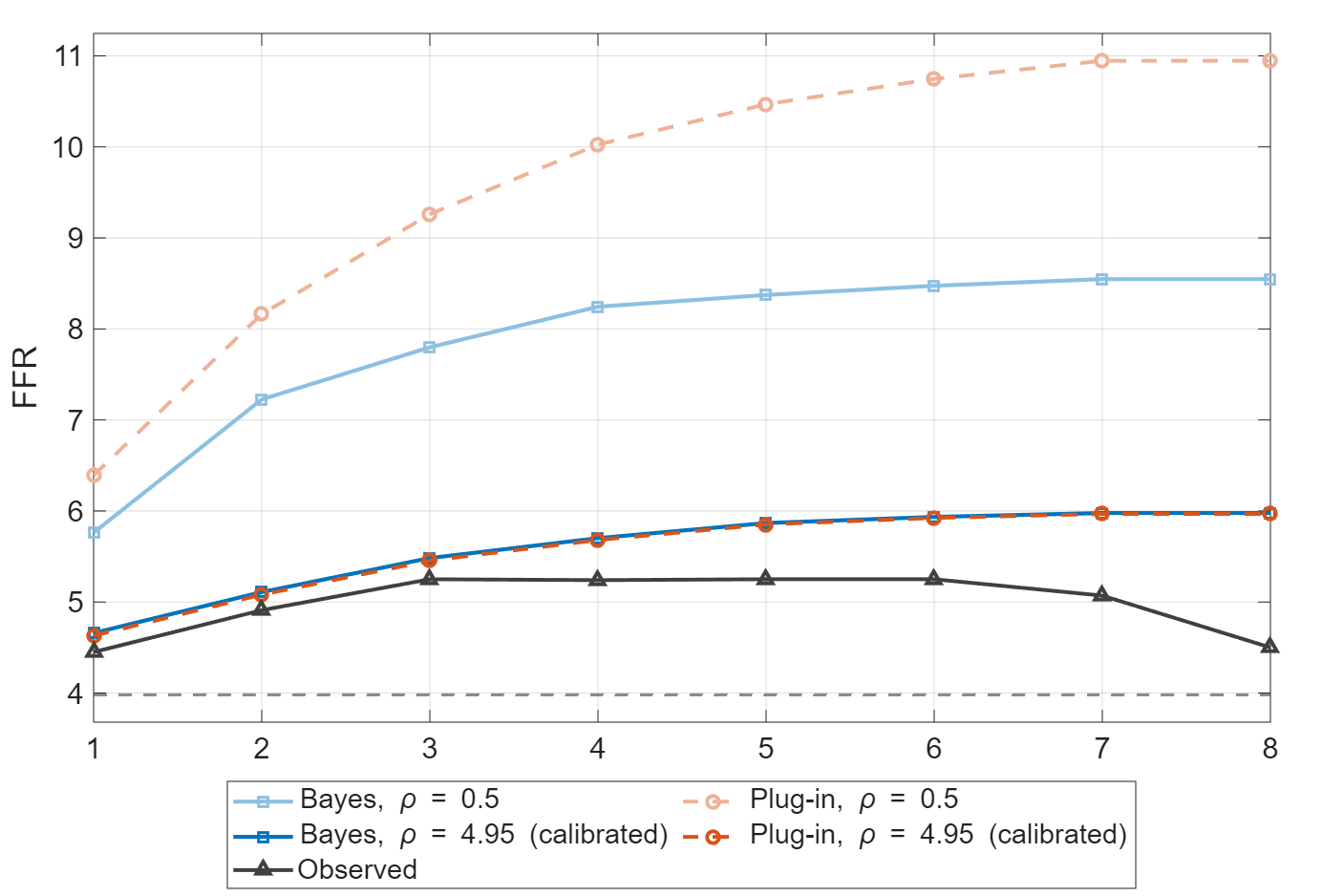}
    \caption{\small Optimal instrument paths under the interest-rate smoothing penalty, contemporaneous regime: Bayes (solid) and plug-in (dashed) rules at a moderate penalty (light) and at the calibrated weight (dark). Black: observed path; last observed rate $i_0=3.98$. $t=1$ corresponds to 2006Q1.}
    \label{fig:smoothing_rules}
\end{figure}
The smoothing penalty pulls the plug-in and the Bayes rule toward each
other and, jointly, toward the observed path: at $\rho^*$ the two rules are
nearly indistinguishable, above the observed path. At $\rho=0.5$, both prescribe a gradual tightening: for Bayes, this flattens out around the fourth horizon; the plug-in rate path is above the Bayes one at all horizons, with the difference between the two growing over time.

\subsection{The forward guidance regime}\label{sec:emp_fg}

\begin{figure}[tbp]
    \centering
    \includegraphics[width=0.7\linewidth]{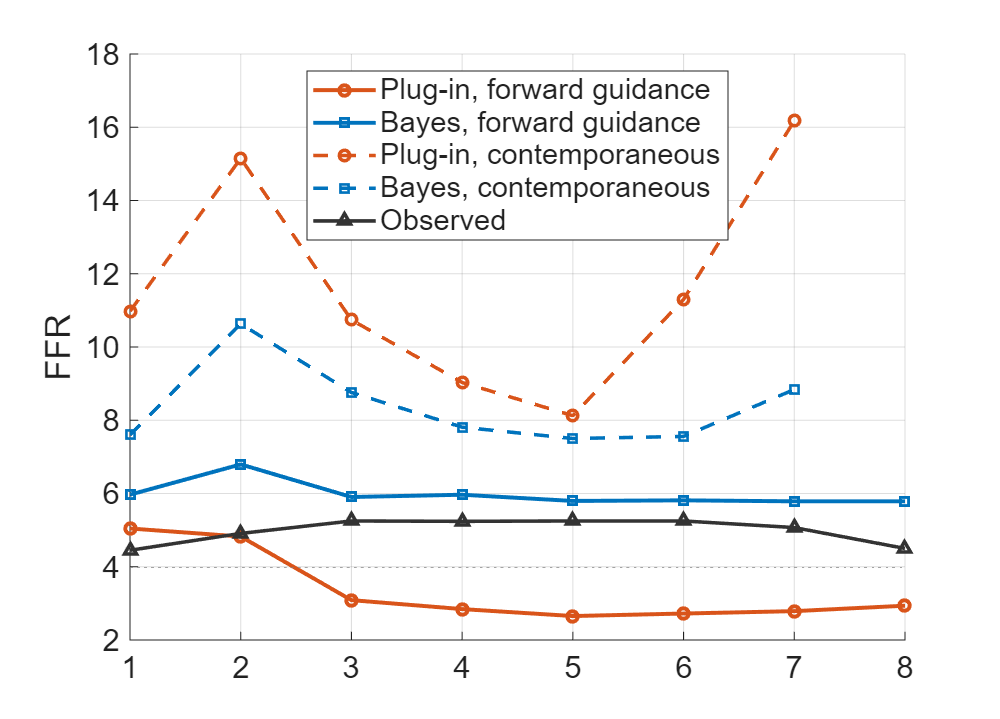}
    \caption{ \small Instrument paths under the two regimes: solid lines forward
    guidance, dashed lines contemporaneous (drawn up to $h-1$), black the
    observed path. $t=1$ corresponds to 2006Q1.}
    \label{fig:paths_regimes}
\end{figure}

The right column of Figure~\ref{fig:regimes_baseline} reports the announced paths and the counterfactuals for the feasible forward guidance regime, with the expected path enforced exactly at the maturities $t=1,4,8$.\footnote{For the remaining maturities we report the posterior mean of the implied paths and, for the Bayes rule, 90\% credible intervals. The implied paths are as in \eqref{impliedpath_nodes}.}
The forward guidance Bayes-optimal announcement and its implied mean path prescribe a tightening relative to the observed path in all eight quarters, a movement in the same direction as, but less aggressive than, its contemporaneous counterpart. The plug-in announcement is qualitatively different: after a rise in the first quarter, it brings the rate below 3\% from the fourth quarter on, an easing even relative to the observed path. At the point estimate, it is enforced mostly through the high-frequency surprise of \citet{gertler2015monetary}, and inherits its unorthodox responses in our sample: the easing lowers the mean output gap, which remains negative from the first to the seventh quarter, while mean inflation approaches the 2\% target by the fifth quarter. Under the Bayes announcement the mean output gap remains positive but shrinks, and mean inflation stays between 3.2\% and 3.5\%.
In terms of the mean counterfactual targets, the Bayes announcement looks worse than the observed one. The posterior expected loss, however, also accounts for the dispersion of the counterfactuals: the implementation of the observed and plug-in announcements requires larger shocks, whose impulse responses are estimated imprecisely, so that their posterior expected losses (150.2 and 179.5) are more than twice that of the Bayes announcement (68.6).
Figure~\ref{fig:paths_regimes} compares the Bayes and plug-in paths across the two regimes: the Bayes paths share sign and shape, while the plug-in paths differ vastly.

\subsection{Robust Bayes Rules}
We now turn to a set-identified application in the contemporaneous regime. The
system is an SVAR(4) in the same endogenous variables as the point-identified case,
excluding the \citet{romer2004new} series. Instead of an internal-IV scheme, the
contemporaneous monetary policy shock is set-identified by sign restrictions and a
relevance condition. We impose
restrictions on the policy shock only: following the construction of Section \ref{sec:set-id}, for every reduced-form parameter $\bs{\phi}$,
the identified set $\mathcal{I}_{q_1}(\bs{\phi})$ collects the $4\times 1$ unit vectors
$q$ such that
\begin{itemize}
    \item on impact and for the following three quarters, the implied shock raises
    the federal funds rate and lowers the output gap and inflation;
    \item the implied shock accounts for at least a share $\omega = 50\%$ of the
    forecast-error variance of the federal funds rate on impact.
\end{itemize}
The sign restrictions require the response to a monetary policy shock to move the
economy in the direction implied by standard New Keynesian theory. The relevance
restriction excludes from the identified set rotations that would capture a
\emph{masquerading shock} in the sense of \citet{wolf2022can}: a linear
combination of other structural shocks that satisfies the sign restrictions. Such
rotations barely move the interest rate while having sizable effects on output and
inflation, so that the multipliers $D^c_i = M^x_v (M^i_v)^{-1}$ explode as
$M^i_v$ approaches zero: without the relevance restriction the worst-case loss is
attained at these rotations and can be unbounded.

We approximate the worst-case plug-in rule \eqref{wcplugin}, the robust Bayes rule
\eqref{rb_rule} and the robust Bayes regret rule \eqref{rbre_rule} numerically,  as detailed in Appendix \ref{app:specification}.

\begin{figure}[tbp]
    \centering
 \includegraphics[width=\textwidth]{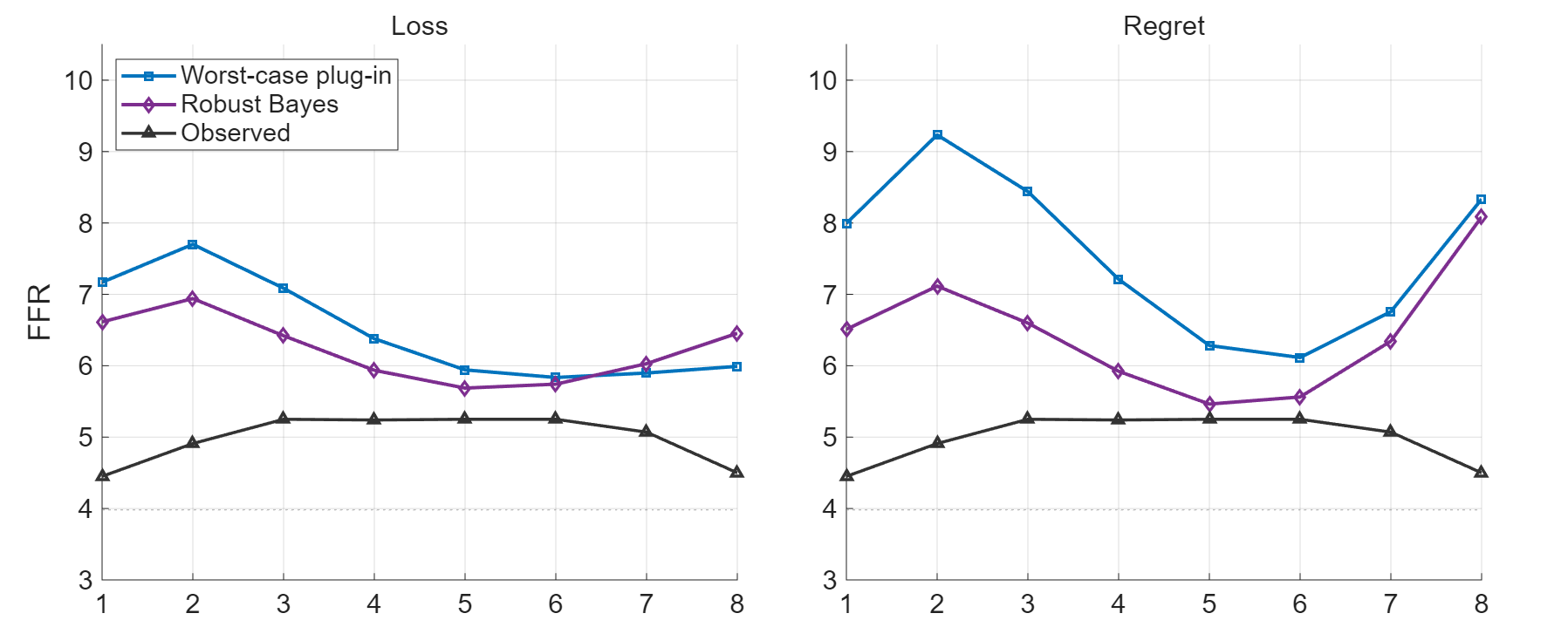}
    \caption{{\small Set-identified policy shock: rules for the federal funds rate over
    2006Q1--2007Q4 ($t = 1$ is 2006Q1). Left panel: loss criterion; right panel:
    regret criterion.  Posterior of the
    SVAR(4) estimated on the sample ending in 2005Q4, $G = 9{,}998$ draws, $R =
    5{,}000$ rotations per draw, relevance share $\omega = 0.5$.}}
    \label{fig:setid_paths}
\end{figure}

Figure~\ref{fig:setid_paths} plots, for each criterion, the worst-case plug-in and the robust Bayes rules together with the observed path. Under both criteria the two rules prescribe an initial tightening followed by a gradual easing, and a renewed increase toward the end of the window that is much more marked under the regret criterion. The robust Bayes rules are less aggressive than the worst-case plug-in rules in the first year, by 0.4--0.8 percentage points under the loss criterion and by 1.3--2.1 percentage points under the regret criterion. The gains are sizable under the regret criterion: the posterior mean of the worst-case regret is 26.5 for the robust Bayes regret rule, against 49.5 for the worst-case plug-in and 48.9 for the observed path. Under the loss criterion the differences are small (101.0 against 103.8 for the worst-case plug-in), and the reason lies in the selection of the worst case. Under the loss criterion it is attained at the relevance bound, the least relevant admissible rotation, for 98 percent of the draws, whereas under regret it is interior for 75 percent of the draws. The constant of the loss includes the variance of the targets under the policy that enforces the path. This term grows with the multipliers $D^c_i$ and is therefore largest at the least relevant admissible rotation, whatever the path, which compresses the differences between rules; the regret criterion nets it out.

\subsection{Numerical risk analysis of the plug-in in calibrated experiments}\label{sec:numerical-risk}

We compare the risk of the plug-in and of a benchmark Bayes rule numerically in a Gaussian experiment calibrated to the point-identified monetary SVARs of the empirical application (Section~\ref{sec:empirical}), one for each policy regime. The two systems share the specification of Section~\ref{sec:emp_setup} --- variables, lags, prior, and identification. The initial condition $\bs{\chi}_0$ is the same as in the empirical application.

We collect $M=10{,}000$ Gibbs draws from the posterior of $\bs{\phi}\mid Y^T_0$ and retain those that imply a stationary VAR and an oracle path of norm below 300; in the forward guidance regime the draws are also restricted to the truncated parameter space of Section~\ref{sec:emp_setup}. We conduct the Monte Carlo simulations at $I=5{,}000$ parameter points $\{\bs\phi_i\}$ drawn at random among the retained draws. At each $\bs\phi_i$ we approximate the finite-sample distribution of $\hat{\bs{\phi}}$ by a Gaussian distribution $\hat{\bs\phi}\mid\bs\phi_i\sim N(\bs\phi_i,\hat\Sigma_{0})$, $T \hat{\Sigma}_0$ being a consistent estimate of the asymptotic covariance matrix. We then compare the plug-in against a benchmark Bayes rule that uses a prior comparable to the one used for estimation.\footnote{The benchmark prior admits a closed-form posterior in the limit experiment and is described in Appendix~\ref{app:specification}. The estimation prior does not, and Gibbs sampling within the Monte Carlo loop would be computationally prohibitive.}

Without any regularization, at a fraction of the signals drawn in the contemporaneous regime $\hat H$ is nearly singular, the implied actions are large in norm, and a single signal can dominate the Monte Carlo average of the loss. We evaluate both rules on a truncated signal distribution: signals at which either rule prescribes a path with Euclidean norm above a threshold $c$ are discarded for both rules, so that the two risks are computed under the same signal distribution. Setting $c=500$ regularizes the loss distribution at all parameter points. The cap binds rarely: it removes $1.1\%$ of the signals in the contemporaneous regime and none in the forward guidance one. At almost all discarded signals ($99.8\%$), the loss of the plug-in exceeds that of the Bayes rule.\footnote{At most of the remaining $0.2\%$ of the discarded signals, it is the Bayes rule that exceeds the cap: one of the posterior draws used to compute it implies a nearly zero impact response of the federal funds rate to the monetary policy shock, which dominates the expected loss weight. This is the same issue that motivates the truncation of the parameter space in the feasible forward guidance regime. Were these signals retained, the ranking of the two risks would reverse at 31 of the 5{,}000 parameter points ($0.6\%$).} Then, for each $\bs\phi_i$, $i=1,\ldots,I$:
\begin{enumerate}
    \item we draw $S=400$ signals $\hat{\bs\phi}_s\sim N(\bs\phi_i,\hat\Sigma_{0})$,
    discarding draws that violate the norm cap;
    \item at each retained signal we compute the plug-in action
    $\bar{\textbf{i}}^*(\hat{\bs\phi}_s)$ and the Bayes action
    $\hat{\textbf{i}}_{Bayes}(\hat{\bs\phi}_s)$,\footnote{We use the
    closed-form solution for Bayes actions as in
    Proposition~\ref{prop_bayes_quad}, approximating the required posterior
    expectations by Monte Carlo with 200 draws.} and evaluate the loss at
    $\bs\phi_i$ at the two;
    \item we approximate the risks by the Monte Carlo averages of the
    realized losses over the retained signals,
    \[
    R_{PI}(\bs\phi_i)=\frac1{S_i}\sum_{s=1}^{S_i} L\big(\bar{\textbf{i}}^*(\hat{\bs\phi}_s),\bs\phi_i\big),
    \qquad
    R_B(\bs\phi_i)=\frac1{S_i}\sum_{s=1}^{S_i} L\big(\hat{\textbf{i}}_{Bayes}(\hat{\bs\phi}_s),\bs\phi_i\big),
    \]
    where $S_i\le S$ is the number of signals retained at $\bs\phi_i$;
    \item we compute the regret ratio
    \[
    RR_i := \frac{R_{PI}(\bs{\phi}_i) - L^*(\bs{\phi}_i)}{R_{B}(\bs{\phi}_i) - L^*(\bs{\phi}_i)},
    \qquad L^*(\bs\phi_i) = \min_{\textbf{i}} L(\textbf{i}, \bs{\phi}_i).
    \]
\end{enumerate}

\begin{figure}[tbp]
    \centering
    \includegraphics[width=\linewidth]{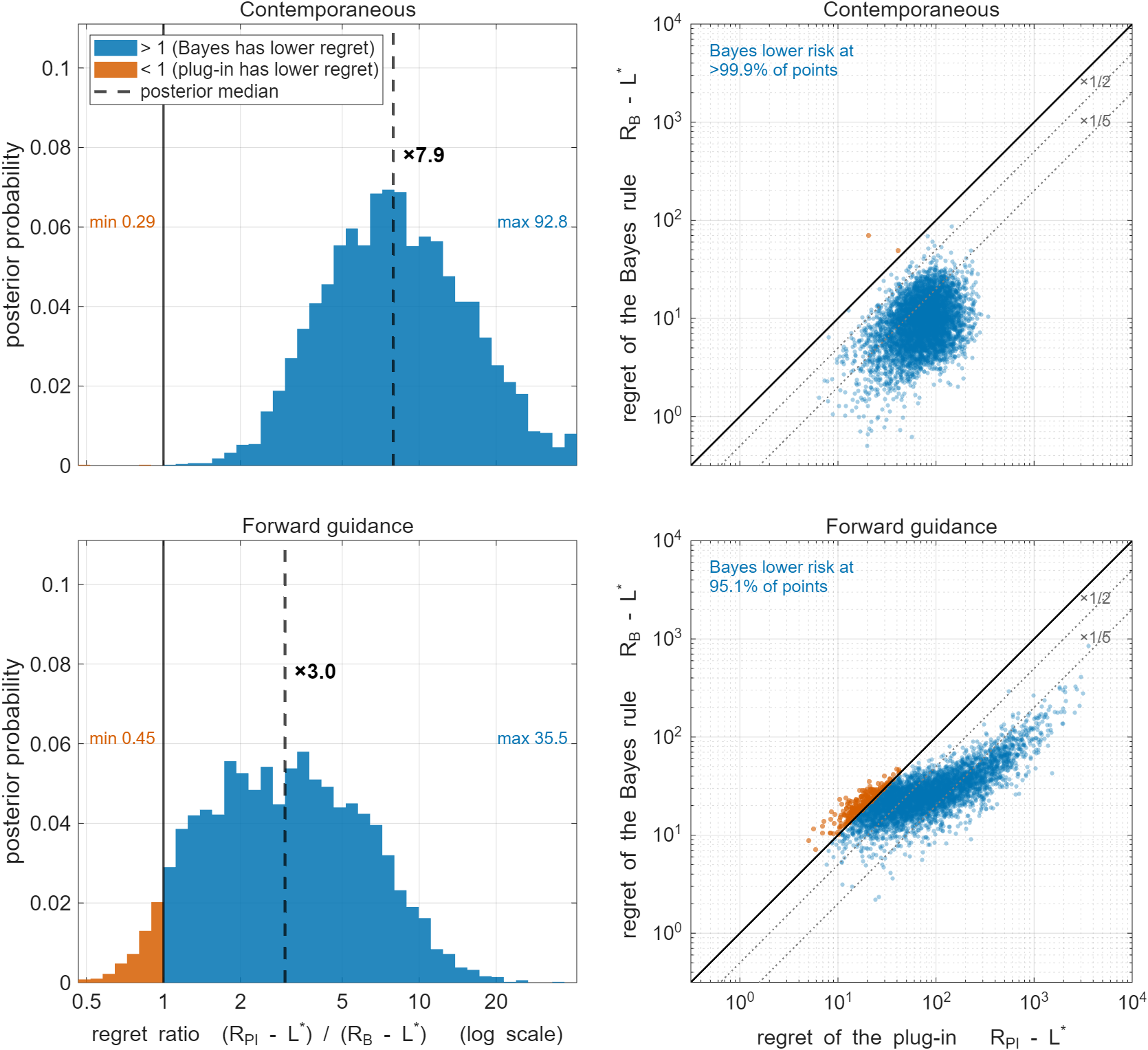}
    \caption{\small Plug-in against Bayes at the retained posterior draws: contemporaneous regime (top row) and forward guidance regime (bottom row); $h=8$, cap $c=500$. Left: histogram of the regret ratio $(R_{PI}-L^*)/(R_B-L^*)$ in log scale; blue: the Bayes rule has lower risk; orange: the plug-in has lower risk; dashed: median. Right: regret of the plug-in against regret of the Bayes rule, log-log scale; dotted lines: Bayes regret equal to $1/2$ and $1/5$ of the plug-in regret.}
    \label{fig:hist_rr}
\end{figure}

The first column of Figure~\ref{fig:hist_rr} reports the histograms of the regret ratios for the contemporaneous (upper panel) and feasible forward guidance (lower panel) regimes. Since the parameter points are posterior draws, the histograms approximate the posterior distribution of the regret ratio.
In the contemporaneous regime the
plug-in rule has higher risk than the Bayes rule in more than $99.9\%$ of the retained
parameter points, and its median regret is 8 times that of the Bayes rule. In the forward guidance regime the plug-in performs slightly better: it has lower risk than Bayes in  4.9\% of the draws and its median regret is 3 times that of Bayes in the feasible forward guidance case. The scatter plot in the bottom-right panel shows that the parameter points where the plug-in outperforms the Bayes rule are points where the regret of both rules is relatively contained. At points where the regret is higher for both rules, the Bayes rule performs markedly better than the plug-in.

In addition to comparing the plug-in rule to the benchmark Bayes rule, we also consider the feasible James--Stein (JS) rule proposed in Section~\ref{sec:suboptimality}. Table~\ref{tab:num_rules_ratio} summarizes the comparison among the three rules, where the JS rule is computed at the classical Stein constant $t=k-2$, $k$ being the dimension of the action space, and at a lower value $t\approx(k-2)/2$ implying less aggressive shrinkage. In the contemporaneous regime both JS rules improve on the plug-in at almost every point, but their gains are much smaller than those of the Bayes rule (a median regret ratio of 1.32 and 1.16, against 7.92), and the Bayes rule has lower risk than either of them at $99.9\%$ of the points. In the feasible forward guidance regime, where the action space has only three dimensions, the Stein constant $t=1$ improves on the plug-in at $72.8\%$ of the points only, while the less aggressive rule ($t=0.5$) does so at $95.9\%$, slightly more often than the Bayes rule ($95.1\%$); the Bayes rule still delivers much larger gains (a median regret ratio of 2.99, against 1.14) and has lower risk than the JS rule at $92.6\%$ of the points.

\begin{table}[tbp]
\centering\small
\label{tab:num_rules_ratio}
\begin{tabular}{lrrrr}
\toprule
 & \multicolumn{2}{c}{Regret ratio} & \multicolumn{2}{c}{Share of points (\%)} \\
\cmidrule(lr){2-3}\cmidrule(lr){4-5}
Rule & Median & Geom.\ mean & Rule beats plug-in & Bayes beats rule \\
\midrule
\multicolumn{5}{l}{\emph{Contemporaneous}} \\
Bayes & 7.92 & 8.09 & $>$99.9 & -- \\
James--Stein, $t=5$ (Stein) & 1.32 & 1.31 & 98.9 & 99.9 \\
James--Stein, $t=2$ & 1.16 & 1.16 & 99.8 & 99.9 \\
\midrule
\multicolumn{5}{l}{\emph{Forward guidance}} \\
Bayes & 2.99 & 2.99 & 95.1 & -- \\
James--Stein, $t=1$ (Stein) & 1.09 & 1.05 & 72.8 & 94.5 \\
James--Stein, $t=0.5$ & 1.14 & 1.12 & 95.9 & 92.6 \\
\bottomrule
\end{tabular}
\caption{ \small Comparison of plug-in, Bayes and James--Stein: the last two columns are the shares of points at which the rule attains a lower risk than the plug-in, and at which the Bayes rule attains a lower risk than the James--Stein rule (on the same simulated signals). The James--Stein rules shrink toward the status-quo path at $i_0$.}
\end{table}

\section{Conclusion}\label{sec:conclusion}

This paper formulates macroeconomic policy choice as a statistical decision
problem. The policymaker chooses the future path --- or the expected future path, in the forward guidance regime --- of a policy instrument under a
quadratic welfare criterion. The counterfactual paths of the target variables that enter the loss function  are obtained from structural conditional forecasts in an SVAR. In the
contemporaneous regime the path is enforced through unanticipated
surprises, while in the forward guidance regime the expected path is
announced and enforced through anticipated shocks. Both deliver a loss
that is quadratic in the action, so the decision problem under parameter
uncertainty can be analyzed in a common mathematical framework and the
oracle, Bayes and robust Bayes rules are derived once and apply to each
regime.

We show that point identification of the loss requires weaker
conditions than point identification of the structural impulse responses.
The loss depends on the structural rotation matrix only through the columns associated with the policy
shocks and is invariant to orthogonal rotations of those columns. 

We argue that the plug-in rule commonly used in the SVAR literature is
locally asymptotically dominated in our framework, and we obtain a
feasible James--Stein rule whose limit attains lower risk uniformly over
compact sets of the local parameter at at least one localization center,
while at all other centers the rule is asymptotically equivalent to the
plug-in. We complement this theoretical finding with numerical
simulations calibrated to monetary SVARs, where a benchmark Bayes rule
attains lower risk than the plug-in at 95.1 to 99.9 percent of the
posterior draws, with median regret ratios of 7.9 in the contemporaneous
regime and 3 in the forward guidance one. We bring our methodology to
to U.S. monetary policy over 2006-2007, and find that plug-in and Bayes rule prescriptions can differ substantially in both regimes. The plug-in path is qualitatively different across regimes, while the  Bayes rule prescribes an initial tightening followed by partial easing across different specifications.

\newpage

\bibliographystyle{ecta}
\bibliography{bibliography/bibliography}

@article{antolin2021structural,
  title   = {Structural scenario analysis with {SVARs}},
  author  = {Antolin-Diaz, Juan and Petrella, Ivan and Rubio-Ram{\'\i}rez, Juan F.},
  journal = {Journal of Monetary Economics},
  volume  = {117},
  pages   = {798--815},
  year    = {2021}
}

@article{AW17,
  title   = {Policy learning with observational data},
  author  = {Athey, Susan and Wager, Stefan},
  journal = {Econometrica},
  year    = {2021},
  volume  = {89},
  number  = {1},
  pages   = {133--161}
}

@article{barnichon2023sufficient,
  title   = {A sufficient statistics approach for macro policy},
  author  = {Barnichon, R{\'e}gis and Mesters, Geert},
  journal = {American Economic Review},
  volume  = {113},
  number  = {11},
  pages   = {2809--2845},
  year    = {2023}
}

@article{BhattacharyaDupas2012,
  title   = {Inferring welfare maximizing treatment assignment under budget constraints},
  author  = {Bhattacharya, Debopam and Dupas, Pascaline},
  journal = {Journal of Econometrics},
  year    = {2012},
  volume  = {167},
  number  = {1},
  pages   = {168--196},
  doi     = {10.1016/j.jeconom.2011.11.007}
}

@article{berger1976minimax,
  title   = {Minimax estimation of a multivariate normal mean under arbitrary quadratic loss},
  author  = {Berger, James O.},
  journal = {Journal of Multivariate Analysis},
  volume  = {6},
  number  = {2},
  pages   = {256--264},
  year    = {1976}
}

@article{berger1982general,
  author  = {Berger, James O. and Berliner, L. Mark and Zaman, Asad},
  title   = {General admissibility and inadmissibility results for estimation in a control problem},
  journal = {The Annals of Statistics},
  year    = {1982},
  volume  = {10},
  number  = {3},
  pages   = {838--856}
}

@article{bernanke1997systematic,
  title   = {Systematic monetary policy and the effects of oil price shocks},
  author  = {Bernanke, Ben S. and Gertler, Mark and Watson, Mark},
  journal = {Brookings Papers on Economic Activity},
  volume  = {1997},
  number  = {1},
  pages   = {91--157},
  year    = {1997}
}

@book{bertsekas1995dynamic,
  title     = {Dynamic programming and optimal control},
  author    = {Bertsekas, Dimitri P.},
  volume    = {1},
  year      = {1995},
  publisher = {Athena Scientific},
  address   = {Belmont, MA}
}

@article{blanchard1989dynamic,
  title   = {The dynamic effects of aggregate demand and supply disturbances},
  author  = {Blanchard, Olivier and Quah, Danny},
  year    = {1989},
  journal = {American Economic Review},
  volume  = {79},
  number  = {4},
  pages   = {655--673}
}

@article{bojinov2019time,
  title   = {Time series experiments and causal estimands: exact randomization tests and trading},
  author  = {Bojinov, Iavor and Shephard, Neil},
  journal = {Journal of the American Statistical Association},
  volume  = {114},
  number  = {528},
  pages   = {1665--1682},
  year    = {2019}
}

@article{brainard1967uncertainty,
  author  = {Brainard, William C.},
  title   = {Uncertainty and the effectiveness of policy},
  journal = {American Economic Review},
  year    = {1967},
  volume  = {57},
  number  = {2},
  pages   = {411--425}
}

@article{chib1995marginal,
  author  = {Chib, Siddhartha},
  title   = {Marginal likelihood from the {G}ibbs output},
  journal = {Journal of the American Statistical Association},
  year    = {1995},
  volume  = {90},
  number  = {432},
  pages   = {1313--1321},
  doi     = {10.1080/01621459.1995.10476635}
}

@article{chow1976control,
  author  = {Chow, Gregory C.},
  title   = {Control of nonlinear econometric systems with unknown parameters},
  journal = {Econometrica},
  year    = {1976},
  volume  = {44},
  number  = {4},
  pages   = {685--695}
}

@article{christensen2026optimal,
  title   = {Optimal decision rules when payoffs are partially identified},
  author  = {Christensen, Timothy and Moon, Hyungsik Roger and Schorfheide, Frank},
  journal = {Review of Economic Studies},
  year    = {2026},
  note    = {Forthcoming; advance article rdag017}
}

@incollection{christiano1999monetary,
  title     = {Monetary policy shocks: What have we learned and to what end?},
  author    = {Christiano, Lawrence J. and Eichenbaum, Martin and Evans, Charles L.},
  booktitle = {Handbook of Macroeconomics},
  editor    = {Taylor, John B. and Woodford, Michael},
  volume    = {1},
  pages     = {65--148},
  year      = {1999},
  publisher = {Elsevier}
}

@article{cogley2011bayesian,
  title   = {A {B}ayesian approach to optimal monetary policy with parameter and model uncertainty},
  author  = {Cogley, Timothy and De Paoli, Bianca and Matthes, Christian and Nikolov, Kalin and Yates, Tony},
  journal = {Journal of Economic Dynamics and Control},
  volume  = {35},
  number  = {12},
  pages   = {2186--2212},
  year    = {2011}
}

@article{crump2025large,
  title   = {A large {B}ayesian {VAR} of the {U}nited {S}tates economy},
  author  = {Crump, Richard K. and Eusepi, Stefano and Giannone, Domenico and Qian, Eric and Sbordone, Argia M.},
  journal = {International Journal of Central Banking},
  volume  = {21},
  number  = {2},
  pages   = {351--409},
  year    = {2025}
}

@article{dehejia2005program,
  title   = {Program evaluation as a decision problem},
  author  = {Dehejia, Rajeev H.},
  journal = {Journal of Econometrics},
  volume  = {125},
  number  = {1--2},
  pages   = {141--173},
  year    = {2005}
}

@techreport{dieppe2016bear,
  author      = {Dieppe, Alistair and Legrand, Romain and van Roye, Bj{\"o}rn},
  title       = {The {BEAR} toolbox},
  institution = {European Central Bank},
  type        = {Working Paper Series},
  number      = {1934},
  year        = {2016}
}

@article{faust1998robustness,
  title   = {The robustness of identified {VAR} conclusions about money},
  author  = {Faust, Jon},
  journal = {Carnegie-Rochester Conference Series on Public Policy},
  volume  = {49},
  pages   = {207--244},
  year    = {1998}
}

@book{ferguson1967,
  author    = {Ferguson, Thomas S.},
  title     = {Mathematical Statistics: A Decision Theoretic Approach},
  publisher = {Academic Press},
  address   = {New York},
  year      = {1967},
  series    = {Probability and Mathematical Statistics},
  volume    = {1}
}

@book{gali2015monetary,
  title     = {Monetary Policy, Inflation, and the Business Cycle: An Introduction to the New Keynesian Framework and Its Applications},
  author    = {Gal{\'\i}, Jordi},
  edition   = {2},
  year      = {2015},
  publisher = {Princeton University Press},
  address   = {Princeton}
}

@article{gertler2015monetary,
  title   = {Monetary policy surprises, credit costs, and economic activity},
  author  = {Gertler, Mark and Karadi, Peter},
  journal = {American Economic Journal: Macroeconomics},
  volume  = {7},
  number  = {1},
  pages   = {44--76},
  year    = {2015}
}

@article{giacomini2021robust,
  title   = {Robust {B}ayesian inference for set-identified models},
  author  = {Giacomini, Raffaella and Kitagawa, Toru},
  journal = {Econometrica},
  volume  = {89},
  number  = {4},
  pages   = {1519--1556},
  year    = {2021}
}

@incollection{Giacomini_Kitagawa_Read_2026,
  title     = {Robust {B}ayesian analysis for econometrics},
  author    = {Giacomini, Raffaella and Kitagawa, Toru and Read, Matthew},
  booktitle = {Advances in Economics and Econometrics: Twelfth World Congress},
  editor    = {Chernozhukov, Victor and H{\"o}rner, Johannes and La Ferrara, Eliana and Werning, Iv{\'a}n},
  series    = {Econometric Society Monographs},
  publisher = {Cambridge University Press},
  address   = {Cambridge},
  year      = {2026},
  pages     = {117--157}
}

@article{giacomini2022narrative,
  title   = {Narrative restrictions and proxies},
  author  = {Giacomini, Raffaella and Kitagawa, Toru and Read, Matthew},
  journal = {Journal of Business \& Economic Statistics},
  volume  = {40},
  number  = {4},
  pages   = {1415--1425},
  year    = {2022}
}

@article{giacomini2022robust,
  title   = {Robust {B}ayesian inference in proxy {SVARs}},
  author  = {Giacomini, Raffaella and Kitagawa, Toru and Read, Matthew},
  journal = {Journal of Econometrics},
  volume  = {228},
  number  = {1},
  pages   = {107--126},
  year    = {2022}
}

@techreport{giacomini2019ambiguity,
  title       = {Estimation under ambiguity},
  author      = {Giacomini, Raffaella and Kitagawa, Toru and Uhlig, Harald},
  institution = {Centre for Microdata Methods and Practice (cemmap)},
  type        = {cemmap Working Paper},
  number      = {CWP24/19},
  year        = {2019}
}

@article{giannone2015prior,
  author  = {Giannone, Domenico and Lenza, Michele and Primiceri, Giorgio E.},
  title   = {Prior selection for vector autoregressions},
  journal = {Review of Economics and Statistics},
  year    = {2015},
  volume  = {97},
  number  = {2},
  pages   = {436--451},
  doi     = {10.1162/REST_a_00483}
}

@book{hansen2011robustness,
  title     = {Robustness},
  author    = {Hansen, Lars Peter and Sargent, Thomas J.},
  year      = {2008},
  publisher = {Princeton University Press},
  address   = {Princeton}
}

@article{hildreth1963,
  title   = {Bayesian statisticians and remote clients},
  author  = {Hildreth, Clifford},
  journal = {Econometrica},
  volume  = {31},
  number  = {3},
  pages   = {422--438},
  year    = {1963}
}

@article{HiranoPorter2009,
  title   = {Asymptotics for statistical treatment rules},
  author  = {Hirano, Keisuke and Porter, Jack R.},
  journal = {Econometrica},
  year    = {2009},
  volume  = {77},
  number  = {5},
  pages   = {1683--1701},
  doi     = {10.3982/ECTA6630}
}

@unpublished{ishihara2021,
  title  = {Evidence aggregation for treatment choice},
  author = {Ishihara, Takuya and Kitagawa, Toru},
  year   = {2021},
  note   = {arXiv:2108.06473 [econ.EM]},
  url    = {https://doi.org/10.48550/arXiv.2108.06473}
}

@inproceedings{james1961estimation,
  title     = {Estimation with quadratic loss},
  author    = {James, William and Stein, Charles},
  booktitle = {Proceedings of the Fourth Berkeley Symposium on Mathematical Statistics and Probability},
  volume    = {1},
  pages     = {361--379},
  year      = {1961},
  publisher = {University of California Press}
}

@book{kilian2017structural,
  title     = {Structural Vector Autoregressive Analysis},
  author    = {Kilian, Lutz and L{\"u}tkepohl, Helmut},
  year      = {2017},
  publisher = {Cambridge University Press},
  address   = {Cambridge}
}

@article{kitagawa2018should,
  title   = {Who should be treated? {E}mpirical welfare maximization methods for treatment choice},
  author  = {Kitagawa, Toru and Tetenov, Aleksey},
  journal = {Econometrica},
  volume  = {86},
  number  = {2},
  pages   = {591--616},
  year    = {2018}
}

@article{KT21,
  title   = {Equality-minded treatment choice},
  author  = {Kitagawa, Toru and Tetenov, Aleksey},
  journal = {Journal of Business \& Economic Statistics},
  volume  = {39},
  number  = {2},
  pages   = {561--574},
  year    = {2021}
}

@techreport{kitagawa2022policy,
  title       = {Policy choice in time series by empirical welfare maximization},
  author      = {Kitagawa, Toru and Wang, Weining and Xu, Mengshan},
  institution = {Centre for Microdata Methods and Practice (cemmap)},
  type        = {cemmap Working Paper},
  number      = {CWP27/24},
  year        = {2024},
  note        = {arXiv:2205.03970 [econ.EM]}
}

@article{kitagawa2023treatment,
  title   = {Treatment choice, mean square regret and partial identification},
  author  = {Kitagawa, Toru and Lee, Sokbae and Qiu, Chen},
  journal = {The Japanese Economic Review},
  volume  = {74},
  number  = {4},
  pages   = {573--602},
  year    = {2023}
}

@article{kydland1977rules,
  title   = {Rules rather than discretion: The inconsistency of optimal plans},
  author  = {Kydland, Finn E. and Prescott, Edward C.},
  journal = {Journal of Political Economy},
  volume  = {85},
  number  = {3},
  pages   = {473--491},
  year    = {1977}
}

@article{leeper2003modest,
  title   = {Modest policy interventions},
  author  = {Leeper, Eric M. and Zha, Tao},
  journal = {Journal of Monetary Economics},
  volume  = {50},
  number  = {8},
  pages   = {1673--1700},
  year    = {2003}
}

@article{Liao_etal2022_AnnStat,
  title   = {Batch policy learning in average reward {M}arkov decision processes},
  author  = {Liao, Peng and Qi, Zhengling and Wan, Runzhe and Klasnja, Predrag and Murphy, Susan A.},
  journal = {The Annals of Statistics},
  year    = {2022},
  volume  = {50},
  number  = {6},
  pages   = {3364--3387}
}

@article{manski2004statistical,
  title   = {Statistical treatment rules for heterogeneous populations},
  author  = {Manski, Charles F.},
  journal = {Econometrica},
  volume  = {72},
  number  = {4},
  pages   = {1221--1246},
  year    = {2004}
}

@article{manski2021econometrics,
  title   = {Econometrics for decision making: Building foundations sketched by {H}aavelmo and {W}ald},
  author  = {Manski, Charles F.},
  journal = {Econometrica},
  volume  = {89},
  number  = {6},
  pages   = {2827--2853},
  year    = {2021}
}

@article{Manski2000,
  author  = {Manski, Charles F.},
  title   = {Identification problems and decisions under ambiguity: Empirical analysis of treatment response and normative analysis of treatment choice},
  journal = {Journal of Econometrics},
  year    = {2000},
  volume  = {95},
  number  = {2},
  pages   = {415--442}
}

@article{Manski2007,
  author  = {Manski, Charles F.},
  title   = {Minimax-regret treatment choice with missing outcome data},
  journal = {Journal of Econometrics},
  year    = {2007},
  volume  = {139},
  number  = {1},
  pages   = {105--115}
}

@article{MT17,
  title   = {Model selection for treatment choice: Penalized welfare maximization},
  author  = {Mbakop, Eric and Tabord-Meehan, Max},
  journal = {Econometrica},
  volume  = {89},
  number  = {2},
  pages   = {825--848},
  year    = {2021}
}

@article{mckay2023can,
  title   = {What can time-series regressions tell us about policy counterfactuals?},
  author  = {McKay, Alisdair and Wolf, Christian K.},
  journal = {Econometrica},
  volume  = {91},
  number  = {5},
  pages   = {1695--1725},
  year    = {2023}
}

@unpublished{montiel2023decision,
  title  = {Decision theory for treatment choice problems with partial identification},
  author = {Montiel Olea, Jos{\'e} Luis and Qiu, Chen and Stoye, J{\"o}rg},
  year   = {2023},
  note   = {arXiv:2312.17623 [econ.EM]}
}

@article{moon2012bayesian,
  title   = {Bayesian and frequentist inference in partially identified models},
  author  = {Moon, Hyungsik Roger and Schorfheide, Frank},
  journal = {Econometrica},
  volume  = {80},
  number  = {2},
  pages   = {755--782},
  year    = {2012}
}

@article{plagborg2021local,
  title   = {Local projections and {VARs} estimate the same impulse responses},
  author  = {Plagborg-M{\o}ller, Mikkel and Wolf, Christian K.},
  journal = {Econometrica},
  volume  = {89},
  number  = {2},
  pages   = {955--980},
  year    = {2021}
}

@article{plagborg2022instrumental,
  title   = {Instrumental variable identification of dynamic variance decompositions},
  author  = {Plagborg-M{\o}ller, Mikkel and Wolf, Christian K.},
  journal = {Journal of Political Economy},
  volume  = {130},
  number  = {8},
  pages   = {2164--2202},
  year    = {2022}
}

@book{polak1997optimization,
  author    = {Polak, Elijah},
  title     = {Optimization: Algorithms and Consistent Approximations},
  series    = {Applied Mathematical Sciences},
  volume    = {124},
  publisher = {Springer},
  address   = {New York},
  year      = {1997}
}

@article{prescott1972multiperiod,
  author  = {Prescott, Edward C.},
  title   = {The multi-period control problem under uncertainty},
  journal = {Econometrica},
  year    = {1972},
  volume  = {40},
  number  = {6},
  pages   = {1043--1058}
}

@article{romer2004new,
  title   = {A new measure of monetary shocks: Derivation and implications},
  author  = {Romer, Christina D. and Romer, David H.},
  journal = {American Economic Review},
  volume  = {94},
  number  = {4},
  pages   = {1055--1084},
  year    = {2004}
}

@incollection{ramey2016macroeconomic,
  title     = {Macroeconomic shocks and their propagation},
  author    = {Ramey, Valerie A.},
  booktitle = {Handbook of Macroeconomics},
  editor    = {Taylor, John B. and Uhlig, Harald},
  volume    = {2},
  pages     = {71--162},
  year      = {2016},
  publisher = {Elsevier}
}

@article{Savage51,
  author  = {Savage, Leonard J.},
  title   = {The theory of statistical decision},
  journal = {Journal of the American Statistical Association},
  year    = {1951},
  volume  = {46},
  number  = {253},
  pages   = {55--67}
}

@article{simon1956dynamic,
  author  = {Simon, Herbert A.},
  title   = {Dynamic programming under uncertainty with a quadratic criterion function},
  journal = {Econometrica},
  year    = {1956},
  volume  = {24},
  number  = {1},
  pages   = {74--81}
}

@article{sims1980macroeconomics,
  title   = {Macroeconomics and reality},
  author  = {Sims, Christopher A.},
  journal = {Econometrica},
  volume  = {48},
  number  = {1},
  pages   = {1--48},
  year    = {1980}
}

@article{sims1982policy,
  title   = {Policy analysis with econometric models},
  author  = {Sims, Christopher A.},
  journal = {Brookings Papers on Economic Activity},
  volume  = {1982},
  number  = {1},
  pages   = {107--164},
  year    = {1982}
}

@article{sims2006does,
  title   = {Does monetary policy generate recessions?},
  author  = {Sims, Christopher A. and Zha, Tao},
  journal = {Macroeconomic Dynamics},
  volume  = {10},
  number  = {2},
  pages   = {231--272},
  year    = {2006}
}

@article{stein1981estimation,
  title   = {Estimation of the mean of a multivariate normal distribution},
  author  = {Stein, Charles M.},
  journal = {The Annals of Statistics},
  volume  = {9},
  number  = {6},
  pages   = {1135--1151},
  year    = {1981}
}

@article{Stoye2012,
  title   = {Minimax regret treatment choice with covariates or with limited validity of experiments},
  author  = {Stoye, J{\"o}rg},
  journal = {Journal of Econometrics},
  year    = {2012},
  volume  = {166},
  number  = {1},
  pages   = {138--156},
  doi     = {10.1016/j.jeconom.2011.06.012}
}

@article{Stoye2009,
  title   = {Minimax regret treatment choice with finite samples},
  author  = {Stoye, J{\"o}rg},
  journal = {Journal of Econometrics},
  year    = {2009},
  volume  = {151},
  number  = {1},
  pages   = {70--81},
  doi     = {10.1016/j.jeconom.2009.02.013}
}

@incollection{strawderman2003minimax,
  title     = {On minimax estimation of a normal mean vector for general quadratic loss},
  author    = {Strawderman, William E.},
  booktitle = {Mathematical Statistics and Applications: Festschrift for Constance van Eeden},
  series    = {IMS Lecture Notes--Monograph Series},
  volume    = {42},
  pages     = {3--14},
  year      = {2003},
  publisher = {Institute of Mathematical Statistics}
}

@article{Tetenov2012,
  title   = {Statistical treatment choice based on asymmetric minimax regret criteria},
  author  = {Tetenov, Aleksey},
  journal = {Journal of Econometrics},
  year    = {2012},
  volume  = {166},
  number  = {1},
  pages   = {157--165},
  doi     = {10.1016/j.jeconom.2011.06.013}
}

@article{theil1957note,
  author  = {Theil, Henri},
  title   = {A note on certainty equivalence in dynamic planning},
  journal = {Econometrica},
  year    = {1957},
  volume  = {25},
  number  = {2},
  pages   = {346--349}
}

@article{uhlig2005effects,
  title   = {What are the effects of monetary policy on output? {R}esults from an agnostic identification procedure},
  author  = {Uhlig, Harald},
  journal = {Journal of Monetary Economics},
  volume  = {52},
  number  = {2},
  pages   = {381--419},
  year    = {2005}
}

@article{waggoner1999conditional,
  title   = {Conditional forecasts in dynamic multivariate models},
  author  = {Waggoner, Daniel F. and Zha, Tao},
  journal = {Review of Economics and Statistics},
  volume  = {81},
  number  = {4},
  pages   = {639--651},
  year    = {1999}
}

@book{wald_statistical_1950,
  title     = {Statistical Decision Functions},
  author    = {Wald, Abraham},
  series    = {Wiley Publications in Statistics},
  publisher = {Wiley},
  address   = {New York},
  year      = {1950}
}

@article{wieland2000monetary,
  title   = {Monetary policy, parameter uncertainty and optimal learning},
  author  = {Wieland, Volker},
  journal = {Journal of Monetary Economics},
  volume  = {46},
  number  = {1},
  pages   = {199--228},
  year    = {2000}
}

@article{wolf2022can,
  title   = {What can we learn from sign-restricted {VARs}?},
  author  = {Wolf, Christian K.},
  journal = {AEA Papers and Proceedings},
  volume  = {112},
  pages   = {471--475},
  year    = {2022}
}

@unpublished{yata2021optimal,
  title  = {Optimal decision rules under partial identification},
  author = {Yata, Kohei},
  year   = {2023},
  note   = {arXiv:2111.04926 [econ.EM]}
}

@article{zaman1981complete,
  author  = {Zaman, Asad},
  title   = {A complete class theorem for the control problem and further results on admissibility and inadmissibility},
  journal = {The Annals of Statistics},
  year    = {1981},
  volume  = {9},
  number  = {4},
  pages   = {812--821}
}

@book{zellner1971introduction,
  author    = {Zellner, Arnold},
  title     = {An Introduction to Bayesian Inference in Econometrics},
  publisher = {Wiley},
  address   = {New York},
  year      = {1971}
}

\newpage

\appendix

\section{Proofs and additional lemmas}\label{app:proofs}

\subsection{Lemmas and proofs for Section~\ref{sec:svar}}\label{app:SVAR}
\paragraph{Companion objects and conventions.}
Partition the autoregressive matrix as $\Pi=\begin{pmatrix}\bs{\nu} & \Pi_1 & \cdots & \Pi_p\end{pmatrix}$, where $\bs{\nu}$ is the $n\times1$ constant and each $\Pi_\ell$ is $n\times n$. Throughout we adopt the convention
\[
\Pi_r:=0_{n\times n}\quad\text{for }r\le0\text{ or }r>p,\qquad \Phi_i:=0_{n\times n}\ \text{ for }i<0,
\]
which we use freely to keep recursive summation limits uniform. Define the companion matrix and the selector
\[
F=\begin{pmatrix}
\Pi_1 & \Pi_2 & \cdots & \Pi_{p-1} & \Pi_p\\
I_n & 0 & \cdots & 0 & 0\\
0 & I_n & \cdots & 0 & 0\\
\vdots & & \ddots & & \vdots\\
0 & 0 & \cdots & I_n & 0
\end{pmatrix}\in\mathbb{R}^{np\times np},
\qquad
J=\begin{pmatrix}I_n & 0_{n\times n(p-1)}\end{pmatrix},
\]
the \emph{reduced-form impulse matrices} and their cumulative sums
\[
\Phi_i:=JF^iJ',\quad \Phi_0=I_n,\qquad \bar\Phi_i:=\sum_{m=0}^{i}\Phi_m,
\]
and the \emph{structural impulse matrices} 
\[
\Theta_i:=\Phi_i B=JF^iJ'B,\quad \Theta_0=B .
\]
We stack the future path in \emph{period-major} order, $\textbf{y}^{\mathrm{pm}}_{1:h}:=\bigl(\textbf{y}_1',\dots,\textbf{y}_h'\bigr)'$, and likewise $\bs{\varepsilon}^{\mathrm{pm}}_{1:h}$. The initial state $\bs{\chi}_0=\bigl(\textbf{y}_0',\textbf{y}_{-1}',\dots,\textbf{y}_{1-p}'\bigr)'$ satisfies $\textbf{z}_0=\bs{\chi}_0$ for the companion state $\textbf{z}_t:=(\textbf{y}_t',\dots,\textbf{y}_{t-p+1}')'$.

\begin{lemma}[Companion powers and their recursions]\label{lem:blocks}
Denote by $[F^{m}]_{ab}$ the $(a,b)$ block ($n\times n$) of $F^{m}$, for $a,b\in\{1,\dots,p\}$. Then for all $m\ge0$:
\begin{align}
[F^{m}]_{a1}&=\Phi_{m-a+1},\qquad a=1,\dots,p,\label{eq:firstcol}\\
\Phi_{m}&=\textstyle\sum_{j=1}^{p}\Pi_j\,\Phi_{m-j},\qquad m\ge1,\ \ \Phi_0=I_n,\label{eq:wold}\\
\bar\Phi_{m}&=I_n+\textstyle\sum_{j=1}^{p}\Pi_j\,\bar\Phi_{m-j},\qquad m\ge0,\label{eq:barphirec}\\
[F^{m}]_{1\ell}&=\textstyle\sum_{b=1}^{\min(m-1,\,p)}\Pi_b\,[F^{m-b}]_{1\ell}+\Pi_{m+\ell-1},\qquad m\ge1.\label{eq:firstrow}
\end{align}
\end{lemma}

\begin{proof}
We use two structural features of $F$, read off its definition: its first block-row is $[F]_{1b}=\Pi_b$ ($b=1,\dots,p$), and its lower block-rows are pure shifts, $[F]_{ab}=I_n\,\delta_{b,a-1}$ for $a\ge2$.

\smallskip
\noindent\emph{Proof of \eqref{eq:firstcol} and \eqref{eq:wold}.} Induction on $m$. For $m=0$, $[F^{0}]_{a1}=I_n\delta_{a1}=\Phi_{1-a}$ (equal to $I_n$ for $a=1$ and to $0$ for $a\ge2$ by the convention $\Phi_{<0}=0$). For the step, the shift rows give, for $a\ge2$,
\[
[F^{m}]_{a1}=\sum_{c}[F]_{ac}[F^{m-1}]_{c1}=[F^{m-1}]_{a-1,1}=\Phi_{(m-1)-(a-1)+1}=\Phi_{m-a+1},
\]
using the inductive hypothesis. For $a=1$, the first row gives $[F^{m}]_{11}=\sum_{b=1}^{p}\Pi_b[F^{m-1}]_{b1}=\sum_{b=1}^{p}\Pi_b\Phi_{m-b}$; but $[F^{m}]_{11}=JF^{m}J'=\Phi_m$ by definition, which establishes \eqref{eq:wold} and is consistent with \eqref{eq:firstcol} at $a=1$.

\smallskip
\noindent\emph{Proof of \eqref{eq:barphirec}.} Summing \eqref{eq:wold} over $m=1,\dots,i$ and adding $\Phi_0=I_n$,
\[
\bar\Phi_i=I_n+\sum_{m=1}^{i}\sum_{j=1}^{p}\Pi_j\Phi_{m-j}
=I_n+\sum_{j=1}^{p}\Pi_j\sum_{m=1}^{i}\Phi_{m-j}
=I_n+\sum_{j=1}^{p}\Pi_j\,\bar\Phi_{i-j},
\]
where $\sum_{m=1}^{i}\Phi_{m-j}=\bar\Phi_{i-j}$ follows by reindexing and $\Phi_{<0}=0$. The base case $\bar\Phi_0=\Phi_0=I_n$ is immediate.

\smallskip
\noindent\emph{Proof of \eqref{eq:firstrow}.} Expanding along the first block-row,
\[
[F^{m}]_{1\ell}=\sum_{b=1}^{p}\Pi_b\,[F^{m-1}]_{b\ell}.
\]
Iterating the shift $[F^{k}]_{b\ell}=[F^{k-1}]_{b-1,\ell}$ ($b\ge2$) gives, for $b\le m$, $[F^{m-1}]_{b\ell}=[F^{m-b}]_{1\ell}$, whereas for $b>m$ the exponent is exhausted first and $[F^{m-1}]_{b\ell}=[F^{0}]_{b-(m-1),\ell}=I_n\,\delta_{b,\,m+\ell-1}$. We split the sum into $b\le m-1$, $b=m$, and $b>m$. The terms $b\le m-1$ give the first sum in \eqref{eq:firstrow}, with genuinely positive exponent $m-b\ge1$. The term $b=m$, present when $m\le p$, equals $\Pi_m\,[F^{0}]_{1\ell}=\Pi_m\,\delta_{\ell 1}$: it contributes $\Pi_m=\Pi_{m+\ell-1}$ when $\ell=1$ and vanishes otherwise. The residual terms $b>m$ yield $\sum_{b=m+1}^{p}\Pi_b\,\delta_{b,\,m+\ell-1}$, which is nonzero only when $m+\ell-1\in\{m+1,\dots,p\}$, i.e.\ $\ell\ge2$ and $m+\ell-1\le p$, in which case it equals $\Pi_{m+\ell-1}$. The two cases are disjoint in $\ell$ and, under the convention $\Pi_r=0$ for $r>p$, combine into the single term $\Pi_{m+\ell-1}$ for all $\ell$. This proves \eqref{eq:firstrow}; for $m=1$ the sum is empty and the formula reads $[F^{1}]_{1\ell}=\Pi_{\ell}$, which matches $[F]_{1\ell}=\Pi_\ell$ directly.
\end{proof}

The following lemma mirrors the results in Section A of the Online Appendix for \citet{antolin2021structural}. We state the results and prove them again using the companion form representation rather than the recursive formulation used in the aforementioned paper. The recursions in Lemma~\ref{lem:blocks} provide the connection between their notation and ours, as more formally argued in Remark~\ref{rmk:aprr}.

\begin{lemma}[Companion form, reduced-form dependence and recursion for the predetermined path, and variable-major ordering]\label{lem:companion}
With the objects and conventions above:

\begin{enumerate}
\item[(i)] \emph{(Companion representation.)} For every $s\ge1$,
\[
\textbf{y}_s
=\underbrace{JF^{s}\bs{\chi}_0+\bar\Phi_{s-1}\,\bs{\nu}}_{=:\,\textbf{b}_s}
+\sum_{i=0}^{s-1}\Theta_i\,\bs{\varepsilon}_{s-i},
\]
and consequently, in stacked period-major form,
\[
\textbf{y}^{\mathrm{pm}}_{1:h}=\textbf{b}^{\mathrm{pm}}_{1:h}+\mathcal{M}\,\bs{\varepsilon}^{\mathrm{pm}}_{1:h},
\qquad
\mathcal{M}=\begin{pmatrix}
\Theta_0 & & & \\
\Theta_1 & \Theta_0 & & \\
\vdots & & \ddots & \\
\Theta_{h-1} & \Theta_{h-2} & \cdots & \Theta_0
\end{pmatrix}.
\]

\item[(ii)] \emph{(Reduced-form dependence of the predetermined path.)} The predetermined component $\textbf{b}_s$ depends on $\bs{\theta}$ only through the autoregressive coefficients $\bs{\pi}=\operatorname{vec}(\Pi)$; it does not depend on the rotation $\textbf{q}$ nor on the Cholesky parameters $\bs{\sigma}_l$. Equivalently $\textbf{b}_s=\textbf{b}_s(\bs{\phi},\bs{\chi}_0)$. The rotation $\textbf{q}$ enters the decomposition only through $\mathcal{M}$, via $\Theta_i=\Phi_i\,\Sigma_l Q$.

\item[(iii)] \emph{(Conditional expectation, deterministic recursion, and
linearity in the state.)} The predetermined component is the conditional
expectation of the outcome given the initial information, $\textbf{b}_s =
\ev\bigl[\textbf{y}_s \mid \mathcal{I}_0\bigr]$. More generally, for any time
origin $t \geq 0$, writing $\textbf{b}_{s|t} := \ev\bigl[\textbf{y}_s \mid
\mathcal{I}_t\bigr]$ --- so that $\textbf{b}_s = \textbf{b}_{s|0}$ --- the
conditional means admit the closed form and autoregressive recursion
\begin{equation}\label{eq:bsrec}
\textbf{b}_{s|t} = JF^{s-t}\bs{\chi}_t + \bar\Phi_{s-t-1}\,\bs{\nu},
\qquad\qquad
\textbf{b}_{s|t}=\bs{\nu}+\sum_{\ell=1}^{p}\Pi_\ell\,\textbf{b}_{s-\ell|t},
\qquad s\ge t+1,
\end{equation}
the latter with initialization $\textbf{b}_{s|t}=\textbf{y}_{s}$ for $s\le t$.
Equivalently, reparametrizing through the unconditional mean
$\bs{\mu}=(I_n-\sum_{\ell=1}^{p}\Pi_\ell)^{-1}\bs{\nu}$, it admits the
pure-autoregressive form
\begin{equation}\label{eq:bsmean}
\textbf{b}_{s|t}-\bs{\mu}=\sum_{\ell=1}^{p}\Pi_\ell\,(\textbf{b}_{s-\ell|t}-\bs{\mu}),\qquad s\ge t+1.
\end{equation}

\item[(iv)] \emph{(Variable-major ordering via the commutation matrix.)} Let $\textbf{K}:=\textbf{K}_{n,h}$ be the commutation matrix with $\textbf{K}\operatorname{vec}(A)=\operatorname{vec}(A')$ for $A\in\mathbb{R}^{n\times h}$, so that $\textbf{K}$ is an orthogonal permutation with $\textbf{K}^{-1}=\textbf{K}'=\textbf{K}_{h,n}$. With the variable-major convention $\textbf{y}_{1:h}=\operatorname{vec}(Y_h)$ of Section~\ref{sec:trajectories} one has $\textbf{y}_{1:h}=\textbf{K}\,\textbf{y}^{\mathrm{pm}}_{1:h}$, hence the multiplier of \eqref{trajdecomp} is
\[
M(\bs{\theta})=\textbf{K}\,\mathcal{M}\,\textbf{K}',
\qquad
\textbf{b}_{1:h}(\bs{\phi},\bs{\chi}_0)=\textbf{K}\,\textbf{b}^{\mathrm{pm}}_{1:h},
\]
and the variable-major shocks $\bs{\varepsilon}_{1:h}=\textbf{K}\,\bs{\varepsilon}^{\mathrm{pm}}_{1:h}$ are zero mean and the identity as covariance matrix.
\end{enumerate}
\end{lemma}

\begin{proof}
\noindent\textbf{Proof of (i).} The reduced form $\textbf{y}_t=\bs{\nu}+\sum_{\ell=1}^{p}\Pi_\ell\textbf{y}_{t-\ell}+\textbf{u}_t$, $\textbf{u}_t=B\bs{\varepsilon}_t$, is the top block of the first-order companion recursion
\[
\textbf{z}_t=\tilde{\bs{\nu}}+F\,\textbf{z}_{t-1}+J'\textbf{u}_t,\qquad \tilde{\bs{\nu}}:=J'\bs{\nu},
\]
the lower blocks being the identities $\textbf{y}_{t-r}=\textbf{y}_{t-r}$. Initializing at $\textbf{z}_0=\bs{\chi}_0$ and unrolling the recursion,
\[
\textbf{z}_s=F^{s}\bs{\chi}_0+\sum_{i=0}^{s-1}F^{i}\tilde{\bs{\nu}}+\sum_{i=0}^{s-1}F^{i}J'\textbf{u}_{s-i}.
\]
Premultiplying by $J$, using $JF^{i}J'=\Phi_i$, $\sum_{i=0}^{s-1}JF^{i}J'\bs{\nu}=\bar\Phi_{s-1}\bs{\nu}$, and $\textbf{u}_{s-i}=B\bs{\varepsilon}_{s-i}$ with $JF^{i}J'B=\Theta_i$, yields
\[
\textbf{y}_s=JF^{s}\bs{\chi}_0+\bar\Phi_{s-1}\bs{\nu}+\sum_{i=0}^{s-1}\Theta_i\,\bs{\varepsilon}_{s-i},
\]
which is the scalar-period statement. Stacking $s=1,\dots,h$ and reading off the coefficient of $\bs{\varepsilon}_{s-i}$ produces the block-Toeplitz lower-triangular $\mathcal{M}$ with $(r,c)$ block $\Theta_{r-c}$ for $r\ge c$ and $0$ otherwise.

\smallskip
\noindent\textbf{Proof of (ii).} From (i), $\textbf{b}_s=JF^{s}\bs{\chi}_0+\bar\Phi_{s-1}\bs{\nu}$. Both $F$ (through its blocks $\Pi_\ell$) and $\bar\Phi_{s-1}$ (a polynomial in the $\Pi_\ell$ by \eqref{eq:barphirec}) are functions of the autoregressive coefficients only, and $\bs{\chi}_0$ is data. The on-impact matrix $B=\Sigma_lQ$ never enters $\textbf{b}_s$: in the decomposition of (i) it appears solely inside $\Theta_i=\Phi_iB$, hence solely inside $\mathcal{M}$. Therefore $\textbf{b}_s$ is invariant to $\textbf{q}$ and to $\bs{\sigma}_l$, depending on $\bs{\theta}$ only through $\bs{\pi}\subset\bs{\phi}$.

\smallskip
\noindent\textbf{Proof of (iii).} For the conditional-expectation
identification, take $\ev[\,\cdot \mid \mathcal{I}_0]$ in the decomposition of
(i): the shocks $\bs{\varepsilon}_1, \dots, \bs{\varepsilon}_s$ are i.i.d.,
mean-zero, and independent of $\mathcal{I}_0$, so
$\ev[\textbf{y}_s \mid \mathcal{I}_0] = JF^{s}\bs{\chi}_0 +
\bar\Phi_{s-1}\bs{\nu} = \textbf{b}_s$. For a general origin $t$, the
derivation of (i) applies verbatim with the companion recursion initialized at
$\textbf{z}_t = \bs{\chi}_t$: unrolling from $t$ and taking
$\ev[\,\cdot \mid \mathcal{I}_t]$ gives the closed form in \eqref{eq:bsrec},
whose coefficients depend on $(s,t)$ only through the lead $s-t$. For the
recursion, the deterministic state
$\textbf{z}_s^{\det}:=F^{s-t}\bs{\chi}_t+\sum_{i=0}^{s-t-1}F^{i}\tilde{\bs{\nu}}$
satisfies $\textbf{z}_s^{\det}=\tilde{\bs{\nu}}+F\textbf{z}_{s-1}^{\det}$ with
$\textbf{z}_t^{\det}=\bs{\chi}_t$, and its top block is $\textbf{b}_{s|t}$.
The companion structure of $F$, together with $\tilde{\bs{\nu}}=J'\bs{\nu}$
having zero lower blocks, makes the lower block-rows pure shifts:
$[\textbf{z}_s^{\det}]_b=[\textbf{z}_{s-1}^{\det}]_{b-1}$ for $2\leq b\leq p$.
Iterating the shift and matching the initial condition
$[\bs{\chi}_t]_r=\textbf{y}_{t+1-r}$ gives
\[
[\textbf{z}_s^{\det}]_j=\textbf{b}_{s-j+1|t},\qquad s\ge t,\ \ j=1,\dots,p,
\]
with the convention $\textbf{b}_{r|t}:=\textbf{y}_r$ for $r\le t$. Reading the
top block of the state recursion and substituting,
\[
\textbf{b}_{s|t}=[\textbf{z}_s^{\det}]_1=\bs{\nu}+\sum_{\ell=1}^{p}\Pi_\ell\,[\textbf{z}_{s-1}^{\det}]_\ell=\bs{\nu}+\sum_{\ell=1}^{p}\Pi_\ell\,\textbf{b}_{s-\ell|t},
\]
which is the recursion in \eqref{eq:bsrec}. Under the stationarity assumption
$\bs{\mu}=(I_n-\sum_{\ell}\Pi_\ell)^{-1}\bs{\nu}$ is well defined and is the
fixed point of the recursion; subtracting the identity
$\bs{\mu}=\bs{\nu}+\sum_{\ell}\Pi_\ell\bs{\mu}$ delivers \eqref{eq:bsmean}.

\smallskip
\noindent\textbf{Proof of (iv).} With $Y_h=(\textbf{y}_1\ \cdots\ \textbf{y}_h)'\in\mathbb{R}^{h\times n}$, the variable-major stack is $\textbf{y}_{1:h}=\operatorname{vec}(Y_h)$ and the period-major stack is $\textbf{y}^{\mathrm{pm}}_{1:h}=\operatorname{vec}(Y_h')$. By definition of $\textbf{K}=\textbf{K}_{n,h}$, $\operatorname{vec}(Y_h)=\textbf{K}\operatorname{vec}(Y_h')$, i.e. $\textbf{y}_{1:h}=\textbf{K}\,\textbf{y}^{\mathrm{pm}}_{1:h}$. Applying $\textbf{K}$ to the period-major identity of (i) and substituting $\bs{\varepsilon}^{\mathrm{pm}}_{1:h}=\textbf{K}'\bs{\varepsilon}_{1:h}$,
\[
\textbf{y}_{1:h}
=\textbf{K}\,\textbf{b}^{\mathrm{pm}}_{1:h}+\textbf{K}\,\mathcal{M}\,\textbf{K}'\,\bs{\varepsilon}_{1:h},
\]
which is \eqref{trajdecomp} with $\textbf{b}_{1:h}=\textbf{K}\textbf{b}^{\mathrm{pm}}_{1:h}$ and $M(\bs{\theta})=\textbf{K}\mathcal{M}\textbf{K}'$. Since $\textbf{K}$ is an orthogonal permutation, $\bs{\varepsilon}_{1:h}=\textbf{K}\bs{\varepsilon}^{\mathrm{pm}}_{1:h}$ has zero mean and the identity as covariance matrix.
\end{proof}

\begin{remark}[Relation to \citet{antolin2021structural}]\label{rmk:aprr}
The decomposition in Lemma~\ref{lem:companion}(i) coincides with the one derived by \citet{antolin2021structural}, but our derivation proceeds through the companion form whereas theirs is recursive. They adopt the row convention, writing the structural form as $\textbf{y}_t'A_0=\textbf{x}_t'A_+ +\bs{\varepsilon}_t'$ with reduced form $\textbf{y}_t'=\textbf{x}_t'\mathcal{B}+\textbf{u}_t'$, $\textbf{u}_t'=\bs{\varepsilon}_t'A_0^{-1}$; here $\mathcal{B}=A_+A_0^{-1}$ collects the reduced-form autoregressive blocks $\mathcal{B}_\ell$ and the constant $c$, and we write the calligraphic $\mathcal{B}$ to distinguish their reduced-form blocks from our on-impact matrix $B$. Transposing their model maps their objects into ours via
\[
\Pi_\ell=\mathcal{B}_\ell',\qquad \bs{\nu}=c',\qquad B=A_0^{-1\prime}=M_0'.
\]

The link between the recursive and the companion form is provided by the recursions of Lemma~\ref{lem:blocks}. Their Appendix~A defines objects $K_i,N_i^{\ell},M_i$ through mutually coupled recursions; under the map above these are exactly the transposes of our companion objects,
\[
M_i=\Theta_i'=(\Phi_iB)',\qquad K_i=\bar\Phi_i',\qquad N_i^{\ell}=\bigl([F^{i}]_{1\ell}\bigr)',
\]
and transposing our recursions \eqref{eq:barphirec} and \eqref{eq:firstrow}, together with $\Theta_i=\sum_{j=1}^{p}\Pi_j\Theta_{i-j}$ from \eqref{eq:wold}, reproduces their recursions for $K_i$, $N_i^{\ell}$, and $M_i$ term by term (the differing summation limits agreeing under $\mathcal{B}_j=0$ for $j>p$). In particular, since $[\bs{\chi}_0]_\ell=\textbf{y}_{1-\ell}$ and $[F^{s}]_{1\ell}=(N_s^{\ell})'$, the identity $$JF^{s}\bs{\chi}_0=\sum_{\ell=1}^{p}[F^{s}]_{1\ell}\,\textbf{y}_{1-\ell}=\sum_{\ell=1}^{p}(N_s^{\ell})'\textbf{y}_{1-\ell}$$ holds, so our predetermined component equals the transpose of theirs, $$\textbf{b}_s=\bigl(c\,K_{s-1}+\sum_{\ell=1}^{p}\textbf{y}_{1-\ell}'N_s^{\ell}\bigr)'$$
\end{remark}

The following Lemma also mirrors a result from \citet{antolin2021structural}. Once again, we state it and prove it using the companion form representation.

\begin{lemma}\label{lem:MMprime}
The matrix $M(\bs{\theta})M(\bs{\theta})'$ depends on $\bs{\theta}$ only through the reduced-form parameter $\bs{\phi}$.
\end{lemma}

\begin{proof}
By Lemma~\ref{lem:companion}(iv), $M(\bs{\theta})=\textbf{K}\mathcal{M}\textbf{K}'$ with $\textbf{K}$ orthogonal, so $M(\bs{\theta})M(\bs{\theta})'=\textbf{K}\mathcal{M}\mathcal{M}'\textbf{K}'$ and it suffices to show that $\mathcal{M}\mathcal{M}'$ depends only on $\bs{\phi}$. Since the $(s,t)$ block of $\mathcal{M}$ is $\Theta_{s-t}=\Phi_{s-t}B$, the multiplier factors as
\[
\mathcal{M}=\mathcal{G}\,(I_h\otimes B),
\qquad
\mathcal{G}:=\begin{pmatrix}
\Phi_0 & & & \\
\Phi_1 & \Phi_0 & & \\
\vdots & & \ddots & \\
\Phi_{h-1} & \Phi_{h-2} & \cdots & \Phi_0
\end{pmatrix}\in\mathbb{R}^{nh\times nh},
\]
hence
\[
\mathcal{M}\mathcal{M}'
=\mathcal{G}\,(I_h\otimes BB')\,\mathcal{G}'
=\mathcal{G}\,(I_h\otimes \Sigma_u)\,\mathcal{G}',
\]
using $BB'=\Sigma_lQQ'\Sigma_l'=\Sigma_l\Sigma_l'=\Sigma_u$. The factor $\mathcal{G}$ depends only on the autoregressive coefficients $\bs{\pi}$ (via \eqref{eq:wold}) and $\Sigma_u$ only on $\bs{\sigma}_l$, so $\mathcal{M}\mathcal{M}'$, and therefore $M(\bs{\theta})M(\bs{\theta})'$, depends on $\bs{\theta}$ only through $\bs{\phi}=(\bs{\pi}',\bs{\sigma}_l')'$.
\end{proof}

\subsection{Lemmas and proofs for Section~\ref{sec:counterfactuals}}\label{app:counterfactuals}

\subsubsection{Contemporaneous regime}

\begin{proposition}[Block-Lower triangular Toeplitz structure of the multiplier]\label{prop:toeplitz}
Partition the multiplier $M(\bs{\theta})$ of \eqref{trajdecomp} into $n^2$ blocks of dimension $h\times h$, $[M]_{jk}$ ,indexed by a variable $j\in\{1,\dots,n\}$ (rows) and a shock $k\in\{1,\dots,n\}$ (columns), with $(s,t)$ entry $[M]_{jk}^{s,t}$ for variable horizon $s\in\{1,\dots,h\}$ and shock period $t\in\{1,\dots,h\}$. Then
\[
[M]_{jk}^{s,t}=[\Theta_{s-t}]_{jk}\,\mathbf 1\{s\ge t\},
\]
so every block $[M]_{jk}$ is lower-triangular and Toeplitz, with $\ell$-th subdiagonal equal to the scalar impulse response $[\Theta_\ell]_{jk}$ for $\ell=0,\dots,h-1$. Under the contemporaneous-regime renaming $j\in\{i,x_1,\dots,x_{n-1}\}$, $k\in\{v,\eta_1,\dots,\eta_{n-1}\}$, the blocks $M_{v,c}^{i},M_{v,c}^{x_j},M_{\eta_k,c}^{i},M_{\eta_k,c}^{x_j}$ are therefore lower-triangular Toeplitz.
\end{proposition}

\begin{proof}
By Lemma~\ref{lem:companion}(iv), $M(\bs{\theta})=\textbf{K}\mathcal{M}\textbf{K}'$ with $\textbf{K}=\textbf{K}_{n,h}$. In the period-major matrix $\mathcal{M}$ the natural block index is the time pair and the within-block index is the variable pair: by Lemma~\ref{lem:companion}(i), the $(s,t)$ horizon-time block of $\mathcal{M}$ is $\Theta_{s-t}$ (with $\Theta_i=0$ for $i<0$), whose $(j,k)$ entry is $[\mathcal{M}]_{s,t}^{j,k}=[\Theta_{s-t}]_{jk}\,\mathbf 1\{s\ge t\}$. From the defining property $\textbf{K}\operatorname{vec}(A)=\operatorname{vec}(A')$, read entrywise, pre- and post- multiplying by $\textbf{K}$ and $\textbf{K}^\prime$ swaps the block and within block indices of both rows and columns,\footnote{Applying the defining property of $\textbf{K}$ to $ \text{vec}(\textbf{e}_k\textbf{e}_t')\in \mathbb{R}^{n h}$ shows that $\textbf{K}$ acts on $\{1,\dots,nh\}$ as the permutation $\pi\big((t-1)n+k\big)=(k-1)h+t$. Both the rows and the columns of $\mathcal{M}$ are indexed in exactly this stacked format --- rows by $(s-1)n+j$ for the pair (horizon $s$, variable $j$), columns by $(t-1)n+k$ for (period $t$, shock $k$). For any permutation matrix $P$ with $P\textbf{e}_a=\textbf{e}_{\pi(a)}$ and any $A\in\mathbb{R}^{nh\times nh}$, $[PAP']_{\pi(a),\pi(b)}=\textbf{e}_{\pi(a)}'PAP'\textbf{e}_{\pi(b)}=\textbf{e}_a'A\textbf{e}_b=A_{ab}$. Hence pre- and post-multiplying $\mathcal{M}$ by $\textbf{K}$ and $\textbf{K}'$ moves the entry at row $(s-1)n+j$, column $(t-1)n+k$  to row $(j-1)h+s$, column $(k-1)h+t$.} thus yielding:
\[
[M]_{jk}^{s,t}=[\mathcal{M}]_{s,t}^{j,k}=[\Theta_{s-t}]_{jk}\,\mathbf 1\{s\ge t\}.
\]
The entry vanishes for $s<t$, which is lower-triangularity; for $s\ge t$ it depends on $(s,t)$ only through the difference $s-t$, hence is constant along diagonals, which is the Toeplitz property. The four named families of blocks can be realized by the appropriate choice of indexes $j\in\{i,x_1,\dots,x_{n-1}\}$, $k\in\{v,\eta_1,\dots,\eta_{n-1}\}$.
\end{proof}

\begin{remark}[Lower-triangular Toeplitz matrices as polynomials in the lag operator]\label{rmk:LTT}
Let $L$ denote the finite-horizon lag operator: the $h\times h$ matrix with ones on the first lower subdiagonal and zeros elsewhere, so that $L^h=0$ and, on a stacked path, $(L\textbf{v})_s=v_{s-1}$. A matrix is lower-triangular Toeplitz with subdiagonals $a_0,a_1,\dots,a_{h-1}$ if and only if it is the polynomial $\sum_{\ell=0}^{h-1}a_\ell L^\ell$: the lower-triangular Toeplitz matrices are exactly the polynomials in $L$. Products of polynomials in $L$ are polynomials in $L$ (of degree at most $h-1$, since $L^h=0$) and polynomials in a given matrix commute, which gives closure under multiplication --- with the product automatically lower-triangular Toeplitz --- and commutativity. For inversion, given nonzero diagonal $a_0\neq0$, write $A=a_0(I_h-N)$ with $N:=-a_0^{-1}(A-a_0I_h)$ strictly lower triangular, hence nilpotent ($N^h=0$): the Neumann series terminates, $A^{-1}=a_0^{-1}\sum_{m=0}^{h-1}N^m$, and since $N$ is a polynomial in $L$ so is $A^{-1}$, which is therefore lower-triangular Toeplitz. Note that lower triangularity alone already yields an algebra closed under inversion, while the Toeplitz structure adds commutativity, ensures that products and inverses remain Toeplitz and is what makes the kernel of a lower-triangular Toeplitz map a tail coordinate subspace (Corollary~\ref{cor:rank}).

The identification extends from the algebra to its action on paths. Since $L^m\textbf{e}_1=\textbf{e}_{m+1}$ for $m=0,\dots,h-1$, every $\textbf{v}=(v_1,\dots,v_h)'\in\mathbb{R}^h$ can be written as $\textbf{v}=v(L)\,\textbf{e}_1$ for the polynomial $v(L):=\sum_{m=0}^{h-1}v_{m+1}L^m$, and the identification is one-to-one. For a lower-triangular Toeplitz $A$ with symbol $a(L)$, substitution and associativity give $A\textbf{v}=a(L)\,v(L)\,\textbf{e}_1$, and by closure the product is again a polynomial in $L$: expanding and collecting powers, $a(L)v(L)=\sum_{k=0}^{h-1}c_kL^k$ with $c_k=\sum_{\ell+m=k}a_\ell v_{m+1}$, the terms of degree $k\ge h$ being annihilated by $L^h=0$. Applying to $\textbf{e}_1$ and substituting $m=s-1-\ell$,
\[
(A\textbf{v})_s=c_{s-1}=\sum_{\ell=0}^{s-1}a_\ell\,v_{s-\ell},\qquad s=1,\dots,h,
\]
Note that, in particular, $p(L)\textbf{e}_1$ is the vector of coefficients $(p_0,\dots,p_{h-1})'$ of any polynomial $p$, and in particular $p(L)\textbf{e}_1=\textbf{0}$ if and only if all coefficients of degree $0,\dots,h-1$ vanish.
\end{remark}

\begin{lemma}[Sequential implementation of the contemporaneous regime]\label{lem:sequential}
Suppose $[B]_{iv}\neq 0$. Decompose the coefficient matrix on the non-policy shocks in \eqref{vpaths_c} into its $h\times h$ blocks,
\[
-(M_{v,c}^{i})^{-1}M_{\eta,c}^{i}
=\begin{pmatrix} G^{\eta_1} & \cdots & G^{\eta_{n-1}} \end{pmatrix},
\qquad G^{\eta_k}:=-(M_{v,c}^{i})^{-1}M_{\eta_k,c}^{i}\in\mathbb{R}^{h\times h}.
\]
Then each block $G^{\eta_k}$ is lower-triangular and Toeplitz. Hence the period-$t$ policy shock $v_t$ depends on each non-policy shock sequence only through its realizations up to period $t$, $\{\eta^c_{k,s}\}_{s\le t}$, and on the predetermined path $\textbf{b}^i_{1:h}$. 
\end{lemma}

\begin{proof}
By Proposition~\ref{prop:toeplitz}, $M_{v,c}^{i}$ is $h\times h$ lower-triangular Toeplitz with constant diagonal $[\Theta_0]_{iv}=[B]_{iv}\neq 0$. By Remark~\ref{rmk:LTT},  $(M_{v,c}^{i})^{-1}$ is lower-triangular Toeplitz. Each block $M_{\eta_k,c}^{i}$ is also lower-triangular Toeplitz by Proposition~\ref{prop:toeplitz}, so each product $G^{\eta_k}=-(M_{v,c}^{i})^{-1}M_{\eta_k,c}^{i}$ is lower-triangular Toeplitz, being a product within the same algebra. Reading \eqref{vpaths_c} blockwise, the period-$t$ entry of $\textbf{v}_{1:h}$ loads on $\eta^c_{k,s}$ through the $(t,s)$ entry of $G^{\eta_k}$, which vanishes for $s>t$.
\end{proof}

\begin{proof}[Proof of Proposition~\ref{prop:Dstructure}]
\emph{(i) Lower-triangular Toeplitz structure.} The block $M_{v,c}^{i}$ is lower-triangular Toeplitz with nonzero constant diagonal (Proposition~\ref{prop:toeplitz}). By the algebra of lower-triangular Toeplitz matrices in Remark~\ref{rmk:LTT}, $(M_{v,c}^{i})^{-1}$ is lower-triangular Toeplitz, and so is each product $D_{i,c}^{x_j}=M_{v,c}^{x_j}(M_{v,c}^{i})^{-1}$, since each $M_{v,c}^{x_j}$ is lower-triangular Toeplitz by Proposition~\ref{prop:toeplitz}. Both claimed properties follow.

\emph{(ii) Rotation dependence.} In the variable-major ordering, the block $M_{k,c}$ collecting the $h$ columns of $M(\bs{\theta})$ associated with shock $k\in\{1,\dots,n\}$ --- so that $M_{1,c}=M_{v,c}$ and $M_{k,c}=M_{\eta_{k-1},c}$ for $k\ge2$ --- is obtained through the selection matrix $\textbf{e}_k\otimes I_h$. Applying the defining property of $\textbf{K}$ to the matrices $A=\textbf{e}_k\textbf{e}_j'$, for which $\operatorname{vec}(A)=\textbf{e}_j\otimes\textbf{e}_k$ and $\operatorname{vec}(A')=\textbf{e}_k\otimes\textbf{e}_j$, gives $\textbf{K}(\textbf{e}_j\otimes\textbf{e}_k)=\textbf{e}_k\otimes\textbf{e}_j$; reading this column-wise across $j=1,\dots,h$ and transposing, $\textbf{K}'(\textbf{e}_k\otimes I_h)=I_h\otimes\textbf{e}_k$. Then, by Lemma~\ref{lem:companion}(iv), the factorization $\mathcal{M}=\mathcal{G}(I_h\otimes B)$ of the proof of Lemma~\ref{lem:MMprime}, the mixed-product property, and $B\textbf{e}_k=\Sigma_lQ\textbf{e}_k=\Sigma_l\textbf{q}_k$,
\[
M_{k,c}=M(\bs{\theta})(\textbf{e}_k\otimes I_h)
=\textbf{K}\mathcal{M}\textbf{K}'(\textbf{e}_k\otimes I_h)
=\textbf{K}\,\mathcal{G}\,(I_h\otimes B)(I_h\otimes\textbf{e}_k)
=\textbf{K}\,\mathcal{G}\,(I_h\otimes\Sigma_l\textbf{q}_k),
\]
so that
\begin{equation}\label{eq:Mcol_factor}
M_{v,c}=\textbf{K}\,\mathcal{G}\,(I_h\otimes\Sigma_l\textbf{q}_1),
\qquad
M_{\eta_k,c}=\textbf{K}\,\mathcal{G}\,(I_h\otimes\Sigma_l\textbf{q}_{k+1}),\quad k=1,\dots,n-1,
\end{equation}
and each shock block depends on $Q$ only through its own column.

\emph{(iii) Sign invariance.} By \eqref{eq:Mcol_factor}, replacing $\textbf{q}_1$ with $-\textbf{q}_1$ maps $M_{v,c}\mapsto-M_{v,c}$, hence $M_{v,c}^{x}\mapsto-M_{v,c}^{x}$ and $M_{v,c}^{i}\mapsto-M_{v,c}^{i}$, and the sign cancels in the ratio:
\[
D_{i,c}\;\mapsto\;(-M_{v,c}^{x})(-M_{v,c}^{i})^{-1}=D_{i,c}.
\]
Invariance of $D_{b,c}$ is immediate. (The same cancellation shows that $D_{i,c}$ is homogeneous of degree zero in $\textbf{q}_1$; on $\mathcal{O}(n)$, this amounts to sign invariance.)
\end{proof}

\begin{proof}[Proof of Proposition~\ref{prop:loss_c}]
\emph{Quadratic form and certainty equivalence.} By \eqref{contcfact} and the definition of $\tilde{\textbf{b}}_{1:h}^c$,
\[
\textbf{x}_{1:h}^c-\textbf{x}^*=D_{i,c}\bar{\textbf{i}}_{1:h}+\tilde{\textbf{b}}_{1:h}^c+D_{\eta,c}\bs{\eta}^c_{1:h}.
\]
Substituting into \eqref{loss_c} and taking the expectation over $\bs{\eta}^c_{1:h}\sim N(0,I_{(n-1)h})$, the terms linear in $\bs{\eta}^c_{1:h}$ have zero mean and $\ev[\bs{\eta}^c_{1:h}(\bs{\eta}^c_{1:h})']=I_{(n-1)h}$, so
\[
L_c=\big(D_{i,c}\bar{\textbf{i}}_{1:h}+\tilde{\textbf{b}}_{1:h}^c\big)'W\big(D_{i,c}\bar{\textbf{i}}_{1:h}+\tilde{\textbf{b}}_{1:h}^c\big)+\text{trace}(D_{\eta,c}'WD_{\eta,c}).
\]
Expanding the first term into its quadratic, linear, and constant parts in $\bar{\textbf{i}}_{1:h}$ gives \eqref{loss_c_quadratic} with $c_c,H_c,\textbf{g}_c$ as stated. The residual multiplier $D_{\eta,c}$ enters only the trace term of $c_c$; the curvature $H_c$ and the linear term $\textbf{g}_c$ are deterministic. This is certainty equivalence.

\emph{Dependence on $\textbf{q}_1$: the deterministic terms.} By Proposition~\ref{prop:Dstructure}(ii), $D_{i,c}$ and $D_{b,c}$ depend on $\bs{\theta}$ only through $\bs{\phi}$ and $\textbf{q}_1$. Since $\textbf{b}_{1:h}=\textbf{b}_{1:h}(\bs{\phi},\bs{\chi}_0)$ by Lemma~\ref{lem:companion}, the gap $\tilde{\textbf{b}}_{1:h}^c=D_{b,c}\textbf{b}_{1:h}-\textbf{x}^*$ depends on $\bs{\theta}$ only through $(\bs{\phi},\textbf{q}_1)$. Hence $H_c=D_{i,c}'WD_{i,c}$, $\textbf{g}_c=D_{i,c}'W\tilde{\textbf{b}}_{1:h}^c$, and the first summand of $c_c$ depend on $Q$ only through $\textbf{q}_1$.

\emph{Dependence on $\textbf{q}_1$: the trace term.} All that is left to show is that $\operatorname{trace}(D_{\eta,c}'WD_{\eta,c})$ depends on $\textbf{q}$ only through $\textbf{q}_1$. Since $D_{i,c}M_{v,c}^{i}=M_{v,c}^{x}$, the policy columns of $M^x_c-D_{i,c}M^i_c$ vanish and, by \eqref{Deta_c}, the non-policy columns equal the residual multiplier:
\begin{equation}\label{eq:Deta_residual}
M^x_c-D_{i,c}\,M^i_c=\begin{pmatrix} O_{(n-1)h\times h} & D_{\eta,c}\end{pmatrix}.
\end{equation}
Taking the outer product of \eqref{eq:Deta_residual}, the zero block contributes nothing:
\[
\big(M^x_c-D_{i,c}M^i_c\big)\big(M^x_c-D_{i,c}M^i_c\big)'
=D_{\eta,c}D_{\eta,c}',
\]
and expanding the left-hand side,
\begin{equation}\label{eq:DetaDeta}
D_{\eta,c}D_{\eta,c}'
=M^x_c(M^x_c)'-D_{i,c}\,M^i_c(M^x_c)'-M^x_c(M^i_c)'\,D_{i,c}'+D_{i,c}\,M^i_c(M^i_c)'\,D_{i,c}'.
\end{equation}
The terms $M^x_c(M^x_c)'$, $M^x_c(M^i_c)'$, and $M^i_c(M^i_c)'$ are blocks of $M(\bs{\theta})M(\bs{\theta})'$, hence depend only on $\bs{\phi}$ by Lemma~\ref{lem:MMprime}, while $D_{i,c}$ depends only on $(\bs{\phi},\textbf{q}_1)$ by Proposition~\ref{prop:Dstructure}(ii). Therefore $D_{\eta,c}D_{\eta,c}'$, and with it $\operatorname{trace}(D_{\eta,c}'WD_{\eta,c})=\operatorname{trace}(WD_{\eta,c}D_{\eta,c}')$, depends on $Q$ only through $\textbf{q}_1$. Combining with the deterministic terms, $c_c$, $H_c$, $\textbf{g}_c$, and hence $L_c$, depend on $Q$ only through $\textbf{q}_1$. Finally, by Proposition~\ref{prop:Dstructure}(iii), $D_{i,c}$ and $\tilde{\textbf{b}}^c_{1:h}=D_{b,c}\textbf{b}_{1:h}-\textbf{x}^*$ are invariant to $\textbf{q}_1\mapsto-\textbf{q}_1$, and by \eqref{eq:DetaDeta} so is $D_{\eta,c}D_{\eta,c}'$, whose remaining blocks are reduced-form. Hence $c_c$, $H_c$, $\textbf{g}_c$, $L_c$, and the minimizer $-H_c^{-1}\textbf{g}_c$ are invariant to the sign of $\textbf{q}_1$.

\emph{Oracle rule.}  With $H_c=D_{i,c}'WD_{i,c}\succ 0$, the quadratic form \eqref{loss_c_quadratic} is strictly convex, and the first-order condition $2H_c\bar{\textbf{i}}_{1:h}+2\textbf{g}_c=0$ gives the unique minimizer $\bar{\textbf{i}}^{*,c}_{1:h}=-H_c^{-1}\textbf{g}_c$.
\end{proof}

\begin{corollary}[Rank of the policy multiplier and delayed transmission]\label{cor:rank}
Suppose $[B]_{iv}\neq0$. For each non-policy variable $x_j$ define its transmission delay
\[
d_j:=\min\big\{\ell\in\{0,\dots,h-1\}:\ [\Theta_\ell]_{x_jv}\neq0\big\},
\]
with $d_j:=h$ if no such $\ell$ exists, and for $J\subseteq\{1,\dots,n-1\}$ let $d_J:=\min_{j\in J}d_j$. Write $M_{v,c}^{x_J}$ and $D_{i,c}^{x_J}$ for the stacks of the blocks $M_{v,c}^{x_j}$ and $D_{i,c}^{x_j}$ over $j\in J$, and $d:=d_{\{1,\dots,n-1\}}$.
\begin{enumerate}
\item[(i)] \emph{(Delayed transmission implies singularity.)} For every $J$,
\[
\ker(M_{v,c}^{x_J})=\ker(D_{i,c}^{x_J})=\operatorname{span}(\textbf{e}_{h-d_J+1},\dots,\textbf{e}_h),
\qquad
\operatorname{rank}(M_{v,c}^{x_J})=\operatorname{rank}(D_{i,c}^{x_J})=h-d_J .
\]
Consequently, for any symmetric positive semidefinite $W$, $\operatorname{span}(\textbf{e}_{h-d+1},\dots,\textbf{e}_h)\subseteq\ker(H_c)$. In particular, if no non-policy variable responds to the policy shock on impact, $d\ge1$, then $H_c$ is singular.
\item[(ii)] \emph{(Kronecker weights.)} Let $W=\Lambda\otimes W_h$ with $\Lambda=\operatorname{diag}(\lambda_1,\dots,\lambda_{n-1})$, $\lambda_j\ge0$, $W_h\succ0$, and let $J:=\{j:\lambda_j>0\}$. Then $\ker(H_c)=\ker(D_{i,c}^{x_J})$, and $H_c\succ0$ if and only if $d_J=0$, i.e.\ if and only if some variable entering the loss responds to the policy shock on impact.
\end{enumerate}
\end{corollary}

\begin{proof}
By Proposition~\ref{prop:toeplitz}, each block $M_{v,c}^{x_j}$ is lower-triangular Toeplitz, hence by Remark~\ref{rmk:LTT} the polynomial $\theta^{(j)}(L)=\sum_{\ell=0}^{h-1}[\Theta_\ell]_{x_jv}L^\ell$ in the finite-horizon lag operator. Since the coefficients of $\theta^{(j)}$ are exactly the impulse responses $[\Theta_\ell]_{x_jv}$, the first nonzero lag of $\theta^{(j)}$ is the transmission delay $d_j$ of the statement, and $\theta^{(j)}=0$ if and only if $d_j=h$. For $d_j<h$, factor
\[
\theta^{(j)}(L)=L^{d_j}g_j(L),\qquad g_j(0)=[\Theta_{d_j}]_{x_jv}\neq0
\]
by the definition of $d_j$ as the first nonzero lag.

Fix $j$ with $d_j<h$. By Remark~\ref{rmk:LTT}, $M_{v,c}^{x_j}\textbf{v}=\big(L^{d_j}g_j(L)v(L)\big)\textbf{e}_1$, which vanishes if and only if $L^{d_j}g_j(L)v(L)=0$ in the truncated polynomial algebra. Since $g_j(0)\neq0$, $g_j$ is invertible (Remark~\ref{rmk:LTT}), and multiplying by $g_j(L)^{-1}$ the condition is equivalent to
\[
L^{d_j}\,v(L)=0 \iff v_1, \ldots, v_{h-d_j} = 0
\]
 Hence
\[
\ker\big(M_{v,c}^{x_j}\big)=\operatorname{span}\big(\textbf{e}_{h-d_j+1},\dots,\textbf{e}_h\big),
\]
a formula that also covers $d_j=h$, where the block vanishes and the kernel is all of $\mathbb{R}^h$. The kernel of the stack $M_{v,c}^{x_J}$ is the intersection over $j\in J$; the tail coordinate subspaces are nested so the intersection is the shortest tail,
\[
\ker(M_{v,c}^{x_J})=\operatorname{span}(\textbf{e}_{h-d_J+1},\dots,\textbf{e}_h),
\qquad
\operatorname{rank}(M_{v,c}^{x_J})=h-d_J.
\]

Since $M_{v,c}^{i}$ is invertible under $[B]_{iv}\neq0$, the results transfer to $D_{i,c}^{x_J}=M_{v,c}^{x_J}(M_{v,c}^{i})^{-1}$. First, $\operatorname{rank}(D_{i,c}^{x_J})=\operatorname{rank}(M_{v,c}^{x_J})$ by invertibility of $M_{v,c}^{i}$. Second, the kernels correspond through $M_{v,c}^{i}$: $\ker(D_{i,c}^{x_J})=M_{v,c}^{i}\,\ker(M_{v,c}^{x_J})$, as $D_{i,c}^{x_J}\textbf{x}=\textbf{0}$ if and only if $(M_{v,c}^{i})^{-1}\textbf{x}\in\ker(M_{v,c}^{x_J})$. Since $M_{v,c}^{i}$ is lower triangular, it maps the tail coordinate subspace $\ker(M_{v,c}^{x_J})$ into itself, so $\ker(D_{i,c}^{x_J})\subseteq\ker(M_{v,c}^{x_J})$; since the two kernels have equal dimension by rank--nullity, they coincide.

For the second part of (i), $\textbf{v}'H_c\textbf{v}=\|W^{1/2}D_{i,c}\textbf{v}\|^2$ for any $W\succeq0$, so $\ker(D_{i,c})\subseteq\ker(H_c)$. Taking $J=\{1,\dots,n-1\}$ in the first part, $\ker(D_{i,c})=\operatorname{span}(\textbf{e}_{h-d+1},\dots,\textbf{e}_h)$, which is nontrivial whenever $d\ge1$.

For (ii), $W^{1/2}=\Lambda^{1/2}\otimes W_h^{1/2}$, so $W^{1/2}D_{i,c}\textbf{v}$ stacks the vectors $\lambda_j^{1/2}W_h^{1/2}D_{i,c}^{x_j}\textbf{v}$ over $j$. Since $W_h$ is invertible and $\lambda_j>0$ exactly for $j\in J$, this vanishes if and only if $D_{i,c}^{x_j}\textbf{v}=\textbf{0}$ for all $j\in J$, that is $\ker(H_c)=\ker(D_{i,c}^{x_J})$. By (i) this kernel is trivial if and only if $d_J=0$, which by $\Theta_0=B$ is the condition $[B]_{x_jv}\neq0$ for some $j\in J$.
\end{proof}

We develop the extension of the contemporaneous regime of Section~\ref{sec:counterfactuals} in which the loss weights the entire vector of endogenous variables, policy instrument included. Throughout, $D_{i,c},D_{b,c},D_{\eta,c}$ denote the base-case multipliers of the non-policy paths $\textbf{x}^c_{1:h}$ in \eqref{Di_c}--\eqref{Deta_c}, and we build their full-vector counterparts by adjoining the policy block.

\subsubsection{Forward guidance regime}

\begin{proof}[Proof of Proposition~\ref{prop:Dstructure_fg}]
\emph{(i) Rotation dependence.} Recall from Section~\ref{sec:fg} that $M_{fg}=M(\bs{\theta})\textbf{K}$; by Lemma~\ref{lem:companion}(iv) and the orthogonality of $\textbf{K}$, it holds
\[
M_{fg}=M(\bs{\theta})\textbf{K}=\textbf{K}\mathcal{M}\textbf{K}'\textbf{K}=\textbf{K}\mathcal{M}.
\]
Partition $\mathcal{M}$ into its period column blocks, $\mathcal{M}=\begin{pmatrix}\mathcal{C}_1 & \cdots & \mathcal{C}_h\end{pmatrix}$, where $\mathcal{C}_t\in\mathbb{R}^{nh\times n}$ collects the responses to the $n$ structural shocks released at time $t$ and has $s$-th row block $\Theta_{s-t}$, $s=1,\dots,h$. Since $\Theta_{s-t}=\Phi_{s-t}B$ with $B$ common to all row blocks, each period block factors as
\begin{equation}\label{eq:Ct_factor}
\mathcal{C}_t=\begin{pmatrix}\Phi_{1-t}\\ \Phi_{2-t}\\ \vdots\\ \Phi_{h-t}\end{pmatrix}B=\mathcal{G}_{\cdot t}\,B,
\end{equation}
where $\mathcal{G}_{\cdot t}$ is the $t$-th block column of the matrix $\mathcal{G}$ defined in the proof of Lemma~\ref{lem:MMprime}. The block $M_{[1]}:=\begin{pmatrix}M_{1,fg} & M_{\eta,fg}\end{pmatrix}$ collects the first $n$ columns of $M_{fg}$ --- the responses to the $n$ structural shocks released at $t=1$ --- so
\[
M_{[1]}=\textbf{K}\,\mathcal{C}_1=\textbf{K}\,\mathcal{G}_{\cdot1}\,B,
\]
and selecting its first $h$ columns --- the policy shocks $\textbf{e}_1$ --- with $B(\textbf{e}_1\ \cdots\ \textbf{e}_h)=\Sigma_lQ(\textbf{e}_1\ \cdots\ \textbf{e}_h)=\Sigma_lQ_{1:h}$,
\begin{equation}\label{eq:Mcol_factor_fg}
M_{1,fg}=\textbf{K}\,\mathcal{G}_{\cdot1}\,\Sigma_l\,Q_{1:h}.
\end{equation}
Hence $M_{1,fg}$, and with it its row blocks $M^{i}_{1,fg},M^{x}_{1,fg}$, depend on $\bs{\theta}$ only through $\bs{\phi}$  and $\textbf{q}_{1:h}$. Since $D_{i,fg}=M^{x}_{1,fg}(M^{i}_{1,fg})^{-1}$ involves only the policy block, and $D_{b,fg}=\bigl(-D_{i,fg}\ \ I_{(n-1)h}\bigr)$ adjoins a constant block, both depend on $\bs{\theta}$ solely through $(\bs{\phi},\textbf{q}_{1:h})$. In contrast $D_{\eta,fg}$ and $D_{2:h,fg}$ collect responses to the remaining shocks, hence involve the remaining columns of $Q$ as well.

\emph{(ii) Invariance to the choice of basis.} Two orthonormal $n\times h$ frames have the same column span if and only if $\bar{Q}_{1:h}=Q_{1:h}R$ for a (unique) orthogonal $R\in\mathcal{O}(h)$ (Take $R= Q_{1:h}'\bar{Q}_{1:h}$). By \eqref{eq:Mcol_factor_fg}, $\bar{M}_{1,fg}=\textbf{K}\mathcal{G}_{\cdot1}\Sigma_l\bar{Q}_{1:h}=M_{1,fg}R$, hence $\bar{M}^{i}_{1,fg}=M^{i}_{1,fg}R$ --- invertible, since $M^{i}_{1,fg}$ and $R$ are --- and $\bar{M}^{x}_{1,fg}=M^{x}_{1,fg}R$. The change of basis cancels in the ratio:
\[
\bar{D}_{i,fg}=M^{x}_{1,fg}R\,\big(M^{i}_{1,fg}R\big)^{-1}=M^{x}_{1,fg}\big(M^{i}_{1,fg}\big)^{-1}=D_{i,fg},
\]
and $\bar{D}_{b,fg}=\bigl(-\bar{D}_{i,fg}\ \ I_{(n-1)h}\bigr)=D_{b,fg}$.
\end{proof}

\begin{proof}[Proof of Proposition~\ref{prop:loss_fg}]
\emph{Quadratic form and certainty equivalence.} By \eqref{xpaths_fg} and the definition of $\tilde{\textbf{b}}^{fg}_{1:h}$,
\[
\textbf{x}^{fg}_{1:h}-\textbf{x}^*=D_{i,fg}\bar{\textbf{i}}^e_{1:h}+\tilde{\textbf{b}}^{fg}_{1:h}+D_{\eta,fg}\bs{\eta}^{fg}_1+D_{2:h,fg}\tilde{\bs{\varepsilon}}_{2:h}.
\]
The shocks $\bs{\eta}^{fg}_1\sim N(0,I_{n-h})$ and $\tilde{\bs{\varepsilon}}_{2:h}\sim N(0,I_{n(h-1)})$ are independent with mean zero, so taking the expectation in \eqref{loss_fg} the terms linear in the shocks vanish and the two stochastic blocks contribute $\operatorname{trace}(D_{\eta,fg}'WD_{\eta,fg})$ and $\operatorname{trace}(D_{2:h,fg}'WD_{2:h,fg})$:
\[
L_{fg}=\big(D_{i,fg}\bar{\textbf{i}}^e_{1:h}+\tilde{\textbf{b}}^{fg}_{1:h}\big)'W\big(D_{i,fg}\bar{\textbf{i}}^e_{1:h}+\tilde{\textbf{b}}^{fg}_{1:h}\big)+\operatorname{trace}(D_{\eta,fg}'WD_{\eta,fg})+\operatorname{trace}(D_{2:h,fg}'WD_{2:h,fg}).
\]
Expanding the first term into its quadratic, linear, and constant parts in $\bar{\textbf{i}}^e_{1:h}$ gives \eqref{loss_fg_quadratic} with $c_{fg},H_{fg},\textbf{g}_{fg}$ as stated. The shock multipliers $D_{\eta,fg}$ and $D_{2:h,fg}$ enter only the trace terms of $c_{fg}$; the curvature $H_{fg}$ and the linear term $\textbf{g}_{fg}$ are deterministic. This is certainty equivalence.

\emph{Dependence on $\textbf{q}_{1:h}$: the deterministic terms.} By Proposition~\ref{prop:Dstructure_fg}(i), $D_{i,fg}$ and $D_{b,fg}$ depend on $\bs{\theta}$ only through $(\bs{\phi},\textbf{q}_{1:h})$. Since $\textbf{b}_{1:h}=\textbf{b}_{1:h}(\bs{\phi},\bs{\chi}_0)$ by Lemma~\ref{lem:companion}, the gap $\tilde{\textbf{b}}^{fg}_{1:h}=D_{b,fg}\textbf{b}_{1:h}-\textbf{x}^*$ depends on $\bs{\theta}$ only through $(\bs{\phi},\textbf{q}_{1:h})$. Hence $H_{fg}$, $\textbf{g}_{fg}$, and the first summand of $c_{fg}$ depend on $Q$ only through $\textbf{q}_{1:h}$.

\emph{Dependence on $\textbf{q}_{1:h}$: the trace terms.} Recall from the proof of Proposition~\ref{prop:Dstructure_fg} the identities $M_{fg}=\textbf{K}\mathcal{M}$, the partition of $\mathcal{M}$ into its period column blocks $\mathcal{C}_1,\dots,\mathcal{C}_h$, the factorization $\mathcal{C}_t=\mathcal{G}_{\cdot t}B$ of \eqref{eq:Ct_factor}, and $M_{[1]}=\begin{pmatrix} M_{1,fg} & M_{\eta,fg}\end{pmatrix}=\textbf{K}\mathcal{C}_1$. By $BB'=\Sigma_lQQ'\Sigma_l'=\Sigma_u$,
\begin{equation}\label{eq:fg_rf}
\mathcal{C}_t\mathcal{C}_t'=\mathcal{G}_{\cdot t}\,\Sigma_u\,\mathcal{G}_{\cdot t}',\qquad t=1,\dots,h,
\end{equation}
which depends only on $\bs{\phi}$.

For the future block, $D_{2:h,fg}=M^{x}_{2:h,fg}$ collects the bottom $(n-1)h$ rows of $M_{2:h,fg}=\textbf{K}\begin{pmatrix}\mathcal{C}_2&\cdots&\mathcal{C}_h\end{pmatrix}$. Taking the outer product block-column by block-column,
\[
M_{2:h,fg}M_{2:h,fg}'=\textbf{K}\Big(\textstyle\sum_{t=2}^{h}\mathcal{C}_t\mathcal{C}_t'\Big)\textbf{K}',
\]
so $D_{2:h,fg}D_{2:h,fg}'$, its bottom-right $(n-1)h\times(n-1)h$ block, depends only on $\bs{\phi}$ by \eqref{eq:fg_rf}.

For the period-$1$ block, since $D_{i,fg}M_{1,fg}^{i}=M_{1,fg}^{x}$, the policy columns of $M^x_{[1]}-D_{i,fg}M^i_{[1]}$ vanish and, by \eqref{Dmatexpr}, the remaining columns equal the residual multiplier:
\begin{equation}\label{eq:Deta_residual_fg}
M^x_{[1]}-D_{i,fg}\,M^i_{[1]}=\begin{pmatrix} O_{(n-1)h\times h} & D_{\eta,fg}\end{pmatrix}.
\end{equation}
Taking the outer product of \eqref{eq:Deta_residual_fg}, the zero block contributes nothing:
\[
\big(M^x_{[1]}-D_{i,fg}M^i_{[1]}\big)\big(M^x_{[1]}-D_{i,fg}M^i_{[1]}\big)'
=D_{\eta,fg}D_{\eta,fg}',
\]
and expanding the left-hand side,
\begin{equation}\label{eq:DetaDeta_fg}
D_{\eta,fg}D_{\eta,fg}'
=M^x_{[1]}(M^x_{[1]})'-D_{i,fg}\,M^i_{[1]}(M^x_{[1]})'-M^x_{[1]}(M^i_{[1]})'\,D_{i,fg}'+D_{i,fg}\,M^i_{[1]}(M^i_{[1]})'\,D_{i,fg}'.
\end{equation}
The outer product terms in \eqref{eq:DetaDeta_fg} are blocks of $M_{[1]}M_{[1]}'=\textbf{K}\,\mathcal{C}_1\mathcal{C}_1'\,\textbf{K}'$, hence depend only on $\bs{\phi}$ by \eqref{eq:fg_rf}, while $D_{i,fg}$ depends only on $(\bs{\phi},\textbf{q}_{1:h})$ by Proposition~\ref{prop:Dstructure_fg}(i). Therefore $D_{\eta,fg}D_{\eta,fg}'$ depends on $Q$ only through $\textbf{q}_{1:h}$. Since $\operatorname{trace}(D'WD)=\operatorname{trace}(WDD')$ for both stochastic multipliers, both trace terms, and with them $c_{fg}$, $H_{fg}$, $\textbf{g}_{fg}$, and $L_{fg}$, depend on $Q$ only through $\textbf{q}_{1:h}$.

\emph{Invariance to the choice of basis.} By Proposition~\ref{prop:Dstructure_fg}(ii), replacing $Q_{1:h}$ with any orthonormal basis of the same span leaves $D_{i,fg}$ and $D_{b,fg}$ --- hence $\tilde{\textbf{b}}^{fg}_{1:h}$, $H_{fg}$, $\textbf{g}_{fg}$, and the first summand of $c_{fg}$ --- unchanged; by \eqref{eq:DetaDeta_fg} the same holds for $D_{\eta,fg}D_{\eta,fg}'$, whose remaining blocks are reduced-form, and $D_{2:h,fg}D_{2:h,fg}'$ does not depend on $Q$ at all. 

\emph{Oracle rule.} With  $H_{fg}=D_{i,fg}'WD_{i,fg} \succ 0 $, the quadratic form \eqref{loss_fg_quadratic} is strictly convex, and the first-order condition $2H_{fg}\bar{\textbf{i}}^e_{1:h}+2\textbf{g}_{fg}=0$ gives the unique minimizer $\bar{\textbf{i}}^{*,fg}_{1:h}=-H_{fg}^{-1}\textbf{g}_{fg}$.
\end{proof}

\paragraph{Ex-post timing convention and connection to \citet{barnichon2023sufficient}}
\label{rmk:timing}
The formulation of the forward-guidance decision problem is mixed in its timing: the action is chosen \emph{ex ante} --- the loss \eqref{loss_fg} integrates over all shocks dated $t\ge1$, including $\bs{\eta}^{fg}_1$ --- but implemented \emph{ex post}, since the shocks \eqref{e1_solution} enforcing the expected path react to $\bs{\eta}^{fg}_1$: the policymaker commits to the plan at $t=0$ and then steers against the time-$1$ non-policy shocks to keep the expected path aligned with the plan. We adopt this convention mainly to align the timing of the forward-guidance framework with the sequential formulation that arises naturally in the contemporaneous regime (Lemma~\ref{lem:sequential}). 

Under a \emph{fully ex-post} convention, instead, the decision itself is taken after observing the time-$1$ information carried by the non-policy shocks: the loss conditions on $\bs{\eta}^{fg}_1$, integrating only over $\tilde{\bs{\varepsilon}}_{2:h}$, and the oracle becomes state-dependent. Implementing this convention requires the policymaker to observe the shocks. In practice, however, $\bs{\eta}^{fg}_1$ and $D_{\eta,fg}$ enter the conditional loss only through the combination $\tilde{\textbf{b}}^{fg}_{1:h}+D_{\eta,fg}\bs{\eta}^{fg}_1$, which equals the time-$1$ forecast of the target gap net of the policy contribution (Remark~\ref{rem:opp}): a policymaker who updates their forecasts after the non-policy shocks have realized, and feeds the updated forecasts into the decision problem, would solve, at the oracle level, the ex-post problem exactly. If the parameters are not known, the problems may differ. The ex-post convention changes the  decision problem substantially, since the oracle responds to the realized state. We do not tackle this case here. Assuming the policymaker observes the shocks, the problem fits the same decisional framework: the conditional loss is again a quadratic form in the action with the same Hessian $H_{fg}$, the time-$1$ forecast replacing the time-$0$ one in the linear coefficient, so the oracle, Bayes and minimax rules of Sections~\ref{sec:decision-theory} and~\ref{sec:bayes} apply with that substitution.

\begin{remark}[Connection with the optimal path of \citet{barnichon2023sufficient}]\label{rem:opp}
The oracle rule \eqref{istar_fg} maps directly onto the optimal policy path of
\citet{barnichon2023sufficient} (henceforth BM), once their framework is cast
into the SVAR and the timing conventions are aligned. For a stabilization
objective ($\textbf{x}^*=\textbf{0}$), BM characterize the optimal expected
instrument path $\ev_1 P_1^{opt}$ through two sufficient statistics --- the
impulse responses $\mathcal{R}_y^0$ of the policy objectives to the policy news
shocks and their forecasts $\ev_1 Y_1^0$ under a baseline policy --- together
with the policy-response matrix $\mathcal{R}_p^0$, which maps policy news shocks
into the instrument path. BM's policy objectives $Y$ collect only the target
variables --- our $\textbf{x}_{1:h}$ --- while our stacked vector
$\textbf{y}_{1:h}$ also includes the instrument path, which corresponds to their
policy path $P$. The key difference to account for is one of timing: BM solve a
\emph{conditional} decision problem, in which expectations are formed under the
time-$1$ information set $\mathcal{F}_1=\{\textbf{X}_{-1},\Xi_1\}$, which
includes the realized non-policy shocks $\Xi_1$ --- the counterpart of our
$\bs{\eta}^{fg}_1$. Their baseline forecasts therefore embed the time-$1$
realizations. Identifying BM's baseline rule ($\phi^0$ in their notation) with the historical rule
embedded in the SVAR, setting the baseline policy shocks to zero
($\bs{\epsilon}_1^0=\textbf{0}$), truncating their paths at horizon $h$, and
setting $\mathcal{W}=W$, the correspondences read
\[
\mathcal{R}_y^0 \;\leftrightarrow\; M^x_{1,fg}, \qquad
\mathcal{R}_p^0 \;\leftrightarrow\; M^i_{1,fg}, \qquad
\ev_1 Y_1^0 \;\leftrightarrow\; \textbf{b}^x_{1:h}+M^x_{\eta,fg}\bs{\eta}^{fg}_1,
\qquad
\ev_1 P_1^0 \;\leftrightarrow\; \textbf{b}^i_{1:h}+M^i_{\eta,fg}\bs{\eta}^{fg}_1,
\]
where the last two objects are the time-$1$ forecasts of the target and
instrument paths under the historical rule. BM summarize optimality through the
\emph{optimal policy perturbation} (OPP): the adjustment $\bs{\delta}^*$ to the
baseline path such that
$\ev_1P_1^{opt}=\ev_1P_1^0+\mathcal{R}_p^0\bs{\delta}^*$, with closed form
\[
\bs{\delta}^*=-\bigl(\mathcal{R}_y^{0\prime}\mathcal{W}\mathcal{R}_y^0\bigr)^{-1}
\mathcal{R}_y^{0\prime}\mathcal{W}\,\ev_1Y_1^0.
\]
Lemma~\ref{lem:bm-sc} formalizes the equivalence between the state-contingent
oracle of Remark~\ref{rmk:timing} and BM's optimal path. BM's maintained
invertibility of $\mathcal{R}_y^{0\prime}\mathcal{W}\mathcal{R}_y^0$ is the
counterpart of the rank condition on $M^x_{1,fg}$ in
Proposition~\ref{prop:loss_fg}.
\end{remark}

\begin{lemma}[Ex-post oracle and equivalence with the BM optimal path]\label{lem:bm-sc}
Suppose $M^i_{1,fg}$ is invertible, $W$ is symmetric positive definite and
$M^x_{1,fg}$ has full column rank $h$.
Define the state-contingent target gap
\[
\check{\textbf{b}}^{fg}_{1:h}(\bs{\eta}^{fg}_1)
:= D_{b,fg}\,\textbf{b}_{1:h}+D_{\eta,fg}\,\bs{\eta}^{fg}_1 .
\]
Then:
\begin{enumerate}
\item[(i)] the loss conditional on $\bs{\eta}^{fg}_1$ admits a unique minimizer, an optimal state contingent policy $\bar{\textbf{i}}^{*,sc}_{1:h}(\bs{\eta}^{fg}_1)$ given  by
\[
\bar{\textbf{i}}^{*,sc}_{1:h}(\bs{\eta}^{fg}_1)
= -\bigl(D_{i,fg}'WD_{i,fg}\bigr)^{-1}D_{i,fg}'W\,
\left(\check{\textbf{b}}^{fg}_{1:h}(\bs{\eta}^{fg}_1) - \textbf{x}^*\right),
\]
i.e.\ the oracle \eqref{istar_fg} with $\tilde{\textbf{b}}^{fg}_{1:h}$ replaced
by $\check{\textbf{b}}^{fg}_{1:h}(\bs{\eta}^{fg}_1)$;
\item[(ii)] Let $\textbf{x}^*=\textbf{0}$. Then, under the correspondences of Remark~\ref{rem:opp}, with
$\mathcal{W}=W$ and $\bs{\epsilon}^0_1=\textbf{0}$, it coincides with BM's
optimal path:
\[
\bar{\textbf{i}}^{*,sc}_{1:h}
= \ev_1P_1^0+\mathcal{R}_p^0\,\bs{\delta}^*
= \ev_1P_1^{opt}.
\]
\end{enumerate}
\end{lemma}

\begin{proof}[Proof of Lemma~\ref{lem:bm-sc}]
\emph{(i)} Conditional on $\bs{\eta}^{fg}_1$, the counterfactual paths
\eqref{xpaths_fg} read
\[
\textbf{x}^{fg}_{1:h}
= D_{i,fg}\bar{\textbf{i}}^e_{1:h}
+ \check{\textbf{b}}^{fg}_{1:h}(\bs{\eta}^{fg}_1)
+ D_{2:h,fg}\tilde{\bs{\varepsilon}}_{2:h},
\qquad
\tilde{\bs{\varepsilon}}_{2:h}\mid\mathcal{I}_1 \sim N(0,I_{n(h-1)}).
\]
Taking expectations conditional on $\bs{\eta}^{fg}_1$ in the quadratic loss and
using $\ev[\tilde{\bs{\varepsilon}}_{2:h}\mid\mathcal{I}_1]=\textbf{0}$,
\[
\ev\bigl[(\textbf{x}^{fg}_{1:h}- \textbf{x}^*)'W(\textbf{x}^{fg}_{1:h}- \textbf{x}^*)
\,\big|\,\bs{\eta}^{fg}_1\bigr]
= \bigl(D_{i,fg}\bar{\textbf{i}}^e_{1:h}
+\check{\textbf{b}}^{fg}_{1:h} - \textbf{x}^* \bigr )'W
\bigl(D_{i,fg}\bar{\textbf{i}}^e_{1:h}+\check{\textbf{b}}^{fg}_{1:h}- \textbf{x}^*\bigr)
+ \operatorname{trace}\bigl(D_{2:h,fg}'WD_{2:h,fg}\bigr).
\]
By a standard least-squares argument, the unique minimizer is
$\bar{\textbf{i}}^{*,sc}_{1:h}=-H_{fg}^{-1}D_{i,fg}'W(\check{\textbf{b}}^{fg}_{1:h}- \textbf{x}^*)$,
which is \eqref{istar_fg} with $\check{\textbf{b}}^{fg}_{1:h}- \textbf{x}^*$ in place of
$\tilde{\textbf{b}}^{fg}_{1:h}$.

\emph{(ii)} We first express $\check{\textbf{b}}^{fg}_{1:h}$ in terms of BM's
baseline forecasts. From the definitions of $D_{b,fg}$ and $D_{\eta,fg}$ in
\eqref{Dmatexpr} and the correspondences of Remark~\ref{rem:opp},
\[
\check{\textbf{b}}^{fg}_{1:h}
= \bigl(\textbf{b}^x_{1:h}+M^x_{\eta,fg}\bs{\eta}^{fg}_1\bigr)
- D_{i,fg}\bigl(\textbf{b}^i_{1:h}+M^i_{\eta,fg}\bs{\eta}^{fg}_1\bigr)
= \ev_1Y_1^0-D_{i,fg}\,\ev_1P_1^0.
\]
Next, writing $D_{i,fg}=M^x_{1,fg}(M^i_{1,fg})^{-1}$,
\[
H_{fg}=D_{i,fg}'WD_{i,fg}
=(M^i_{1,fg})^{-\prime}\,\bigl(M^{x\prime}_{1,fg}WM^x_{1,fg}\bigr)\,
(M^i_{1,fg})^{-1},
\]
so that
\[
H_{fg}^{-1}D_{i,fg}'W
= M^i_{1,fg}\bigl(M^{x\prime}_{1,fg}WM^x_{1,fg}\bigr)^{-1}
M^{x\prime}_{1,fg}W .
\]
In particular $H_{fg}^{-1}D_{i,fg}'W\,D_{i,fg}=I_h$. Combining the last three
displays,
\begin{align*}
\bar{\textbf{i}}^{*,sc}_{1:h}
&= -H_{fg}^{-1}D_{i,fg}'W
\bigl(\ev_1Y_1^0-D_{i,fg}\,\ev_1P_1^0\bigr)
= \ev_1P_1^0
- M^i_{1,fg}\bigl(M^{x\prime}_{1,fg}WM^x_{1,fg}\bigr)^{-1}
M^{x\prime}_{1,fg}W\,\ev_1Y_1^0 \\
&= \ev_1P_1^0
- \mathcal{R}_p^0\bigl(\mathcal{R}_y^{0\prime}\mathcal{W}\mathcal{R}_y^0\bigr)^{-1}
\mathcal{R}_y^{0\prime}\mathcal{W}\,\ev_1Y_1^0
= \ev_1P_1^0+\mathcal{R}_p^0\,\bs{\delta}^*,
\end{align*}
which is BM's optimal path $\ev_1P_1^{opt}$.
\end{proof}

\subsection{General loss function}\label{app:genloss}

Under the contemporaneous regime the enforced instrument path coincides with the action, $\textbf{i}^c_{1:h}=\bar{\textbf{i}}_{1:h}$, and carries no predetermined or non-policy component. Partitioning the endogenous vector as $\textbf{y}^c_{1:h}=(\textbf{i}^c_{1:h}{}',\,(\textbf{x}^c_{1:h})')'$, the full paths admit the representation
\[
\textbf{y}^c_{1:h}=\tilde{D}_{a,c}\,\bar{\textbf{i}}_{1:h}+\tilde{D}_{b,c}\,\textbf{b}_{1:h}+\tilde{D}_{\eta,c}\,\bs{\eta}^c_{1:h},
\]
with the full-vector multipliers
\[
\tilde{D}_{a,c}=\begin{pmatrix} I_h \\ D_{i,c}\end{pmatrix}\ (nh\times h),
\qquad
\tilde{D}_{b,c}=\begin{pmatrix} O_{h\times nh} \\ D_{b,c}\end{pmatrix}\ (nh\times nh),
\qquad
\tilde{D}_{\eta,c}=\begin{pmatrix} O_{h\times (n-1)h} \\ D_{\eta,c}\end{pmatrix}\ (nh\times (n-1)h).
\]
The top block-rows record that the instrument equals the action and has no predetermined or stochastic component; the bottom block-rows reproduce the base-case multipliers.
 Let $W$ be a symmetric positive semidefinite $nh\times nh$ matrix, partitioned into policy and non-policy blocks,
\[
W=\begin{pmatrix} W_{ii} & W_{ix}\\ W_{xi} & W_{xx}\end{pmatrix},\qquad W_{ix}=W_{xi}',
\]
and let $\textbf{y}^*=(\textbf{i}^*{}',\,\textbf{x}^*{}')'$ be the target. Define the full predetermined target gap $\tilde{\textbf{b}}^{\textbf{y}}=\tilde{D}_{b,c}\textbf{b}_{1:h}-\textbf{y}^*$, whose policy block is $-\textbf{i}^*$ (the instrument has no predetermined component) and whose non-policy block is the base-case gap $\tilde{\textbf{b}}^{c}=D_{b,c}\textbf{b}_{1:h}-\textbf{x}^*$. Substituting $\textbf{y}^c_{1:h}-\textbf{y}^*=\tilde{D}_{a,c}\bar{\textbf{i}}_{1:h}+\tilde{\textbf{b}}^{\textbf{y}}+\tilde{D}_{\eta,c}\bs{\eta}^c_{1:h}$ into $\ev[(\cdot)'W(\cdot)]$ and taking expectations exactly as in the proof of Proposition~\ref{prop:loss_c} gives the quadratic form \eqref{loss_c_quadratic} with
\[
H_c=\tilde{D}_{a,c}'W\tilde{D}_{a,c},\qquad
\textbf{g}_c=\tilde{D}_{a,c}'W\tilde{\textbf{b}}^{\textbf{y}},\qquad
c_c=(\tilde{\textbf{b}}^{\textbf{y}})'W\tilde{\textbf{b}}^{\textbf{y}}+\operatorname{trace}(\tilde{D}_{\eta,c}'W\tilde{D}_{\eta,c}).
\]
Expanding by the policy/non-policy partition and using $\tilde{D}_{a,c}=(I_h;\,D_{i,c})$,
\begin{align}
H_c&=W_{ii}+W_{ix}D_{i,c}+D_{i,c}'W_{xi}+D_{i,c}'W_{xx}D_{i,c},\label{eq:Hc_full}\\
\textbf{g}_c&=-W_{ii}\textbf{i}^*-W_{ix}\textbf{x}^*+W_{ix}D_{b,c}\textbf{b}_{1:h}
+D_{i,c}'\big(W_{xx}\tilde{\textbf{b}}^{c}-W_{xi}\textbf{i}^*\big).\label{eq:gc_full}
\end{align}
Since the top block-row of $\tilde{D}_{\eta,c}$ is zero, $\tilde{D}_{\eta,c}'W\tilde{D}_{\eta,c}=D_{\eta,c}'W_{xx}D_{\eta,c}$, so the trace term reduces to that of the base case with $W$ replaced by $W_{xx}$. Setting $W_{ii}=O$ and $W_{ix}=O$ recovers the base-case coefficients of Proposition~\ref{prop:loss_c} exactly.

The two properties of Proposition~\ref{prop:loss_c} carry over. The added terms in \eqref{eq:Hc_full}--\eqref{eq:gc_full} involving $W_{ii},W_{ix},\textbf{i}^*,\textbf{x}^*$ are deterministic, so certainty equivalence is unaffected: $\tilde{D}_{\eta,c}$ still enters only $c_c$. They also depend on the  rotation matrix $Q$ through $D_{i,c}$ and $D_{b,c}$; since the latter depend on $Q$ only through $\textbf{q}_1$ (Proposition~\ref{prop:Dstructure}) and $\tilde{D}_{\eta,c}'W\tilde{D}_{\eta,c}=D_{\eta,c}'W_{xx}D_{\eta,c}$ depends on $Q$ only through $\textbf{q}_1$ by the argument of \eqref{eq:DetaDeta}, the whole loss continues to depend on $Q$ only through $\textbf{q}_1$.

\paragraph{Interest-rate smoothing.} The leading application of a direct cost on the instrument is \emph{interest-rate smoothing}: a discounted penalty on the period-to-period change of the policy path, anchored to a scalar reference level $i^*$ --- for instance the interest rate prevailing at the time of the decision, or a historical average. It corresponds to no cross terms, $W_{ix}=O$, a target
\[
\textbf{i}^*=i^*\,\textbf{1}_h=(i^*,\dots,i^*)',
\]
and a policy weight
\[
W_{ii}=\rho\,\Delta'\Lambda_\beta\Delta,\qquad
\Delta=\begin{pmatrix}
1 & & & \\ -1 & 1 & & \\ & \ddots & \ddots & \\ & & -1 & 1
\end{pmatrix}\in\mathbb{R}^{h\times h},\qquad
\Lambda_\beta=\operatorname{diag}(\beta,\beta^2,\dots,\beta^h),\qquad \rho>0,
\]
with $\Delta$ the first-difference matrix and $\beta\in(0,1]$ the discount factor of the loss. To see that this reproduces the smoothing penalty, note $\Delta\textbf{1}_h=\textbf{e}_1$, so $\Delta(\bar{\textbf{i}}_{1:h}-i^*\textbf{1}_h)=\Delta\bar{\textbf{i}}_{1:h}-i^*\textbf{e}_1$, and
\[
(\bar{\textbf{i}}_{1:h}-\textbf{i}^*)'W_{ii}(\bar{\textbf{i}}_{1:h}-\textbf{i}^*)
=\rho\big\|\Lambda_\beta^{1/2}\big(\Delta\bar{\textbf{i}}_{1:h}-i^*\textbf{e}_1\big)\big\|^2
=\rho\sum_{s=1}^{h}\beta^s(\bar i_s-\bar i_{s-1})^2,\qquad \bar i_0:=i^*,
\]
the discounted penalty on the successive changes of the instrument, with the first difference $\bar i_1-i^*$ measuring the jump away from the anchor.

With these choices \eqref{eq:Hc_full}--\eqref{eq:gc_full} become
\[
H_c=\rho\,\Delta'\Lambda_\beta\Delta+D_{i,c}'W_{xx}D_{i,c},
\qquad
\textbf{g}_c=D_{i,c}'W_{xx}\tilde{\textbf{b}}^{c}-\rho\beta\,i^*\textbf{e}_1,
\]
using $\Delta'\Lambda_\beta\Delta\,\textbf{1}_h=\Delta'\Lambda_\beta\textbf{e}_1=\beta\,\Delta'\textbf{e}_1=\beta\,\textbf{e}_1$, and the oracle path is
\begin{equation}\label{eq:oracle_smoothing}
\bar{\textbf{i}}^{*,c}_{1:h}(\rho)
=-\big(\rho\,\Delta'\Lambda_\beta\Delta+D_{i,c}'W_{xx}D_{i,c}\big)^{-1}\big(D_{i,c}'W_{xx}\,\tilde{\textbf{b}}^{c}-\rho\beta\,i^*\textbf{e}_1\big).
\end{equation}
Since $\Delta$ is invertible and $\Lambda_\beta\succ0$, $\Delta'\Lambda_\beta\Delta\succ0$ and $H_c\succeq\rho\,\Delta'\Lambda_\beta\Delta\succ0$ for any $\rho>0$: the smoothed oracle is well defined even when the base-case Hessian $D_{i,c}'W_{xx}D_{i,c}$ is singular, without the positive definiteness assumption of Proposition~\ref{prop:loss_c}.As $\rho\to\infty$ the penalty dominates and the oracle tends to the flat path $i^*\textbf{1}_h$. As $\rho\to0$ the oracle converges to the base-case oracle \eqref{istar_c} whenever $D_{i,c}'W_{xx}D_{i,c}\succ0$. When the base-case Hessian is singular the base-case argmin is a set, and the limit selects within it the path with the smallest smoothing penalty, a standard property of Tikhonov regularization. Under delayed transmission this selection has a simple form: by Corollary~\ref{cor:rank} the argmin fixes the first $h-d$ entries and leaves the last $d$ free, and the limit continues the path flat at its last determined value.

\subsection{Lemmas and proofs for Section~\ref{sec:decision-theory}}\label{app:decisions}

\begin{proof}[Proof of Proposition \ref{prop_bayes_quad}]
\textit{Claim 1.} Taking posterior expectations in~(\ref{loss_quad}), the posterior expected loss is
\[
\ev_{\bs{\phi},\textbf{q}\mid\textbf{y}}\bigl[L(\bar{\textbf{i}}_{1:h},\bs{\phi},\textbf{q})\bigr] = \ev[c\mid\textbf{y}] + \bar{\textbf{i}}_{1:h}'\, \ev[H\mid\textbf{y}]\, \bar{\textbf{i}}_{1:h} + 2\, \bar{\textbf{i}}_{1:h}'\, \ev[\textbf{g}\mid\textbf{y}],
\]
which is strictly convex in $\bar{\textbf{i}}_{1:h}$ because $\ev[H\mid\textbf{y}]\succ 0$. Its first-order condition $\ev[H\mid\textbf{y}]\, \bar{\textbf{i}}_{1:h} + \ev[\textbf{g}\mid\textbf{y}] = \textbf{0}$ yields~(\ref{bayes_quad}).

\textit{Claim 2.} The structural shocks enter~(\ref{loss_quad}) only through their second moments, and these appear only in the baseline loss $c(\bs{\phi},\textbf{q})$; the matrix $H$ and the vector $\textbf{g}$ are functions of the multipliers and of the predetermined component alone. Since $c$ enters the posterior expected loss additively and is independent of the action, it leaves~(\ref{bayes_quad}) unchanged. The optimal rule therefore depends on $(\bs{\phi},\textbf{q})$ only through $(H,\textbf{g})$, neither of which involves the shock second moments, establishing certainty equivalence.
\end{proof}

\begin{proof}[Proof of Proposition \ref{prop_regret_ce}]
\textit{Claim 1.} By~(\ref{loss_quad}) the loss is minimized at $\bar{\textbf{i}}^{*}_{1:h} = -H^{-1}\textbf{g}$, with value $\min_{\bar{\textbf{i}}_{1:h}} L = c - \textbf{g}'H^{-1}\textbf{g}$. Hence
\[
Re(\bar{\textbf{i}}_{1:h}, \bs{\phi}, \textbf{q}) = L(\bar{\textbf{i}}_{1:h}, \bs{\phi}, \textbf{q}) - \min_{\bar{\textbf{i}}_{1:h}'} L(\bar{\textbf{i}}_{1:h}', \bs{\phi}, \textbf{q}) = \bar{\textbf{i}}_{1:h}' H \bar{\textbf{i}}_{1:h} + 2\, \bar{\textbf{i}}_{1:h}' \textbf{g} + \textbf{g}'H^{-1}\textbf{g},
\]
in which $c$ cancels; substituting $\textbf{g} = -H\bar{\textbf{i}}^{*}_{1:h}$ and completing the square gives~(\ref{regret_quad}).

\textit{Claim 2.} Since $Re = L - \min_{\bar{\textbf{i}}_{1:h}'} L$, the two integrated objectives satisfy
\[
\ev_{\pi_{\bs{\phi},\textbf{q}}}\!\bigl[\ev_{\textbf{y}\mid \bs{\phi}, \textbf{q}}[Re(\hat{\textbf{i}}_{1:h}(\textbf{y}),\bs{\phi},\textbf{q})]\bigr] = \ev_{\pi_{\bs{\phi},\textbf{q}}}\!\bigl[R(\hat{\textbf{i}}_{1:h},\bs{\phi},\textbf{q})\bigr] - \ev_{\pi_{\bs{\phi},\textbf{q}}}\!\Bigl[\min_{\bar{\textbf{i}}_{1:h}'} L(\bar{\textbf{i}}_{1:h}',\bs{\phi},\textbf{q})\Bigr],
\]
and the subtracted term does not depend on $\hat{\textbf{i}}_{1:h}$. The two objectives therefore share the same minimizer.
\end{proof}

\begin{proof}[Proof of Proposition~\ref{prop:loss_nodes}]
\emph{(i)} As in the proof of Proposition~\ref{prop:Dstructure_fg},
$M_{1,fg}=\textbf{K}\,\mathcal{G}_{\cdot1}\,\Sigma_l\,Q_{1:m}$ --- now with
$m$ columns --- so $M^i_{1,fg}$, $M^x_{1,fg}$, and with them
$\tilde M^i_{1,fg}=S_{\textbf{s}}M^i_{1,fg}$ and
$\tilde D_{i,fg}=M^x_{1,fg}(\tilde M^i_{1,fg})^{-1}$, depend on
$\bs{\theta}$ only through $(\bs{\phi},\textbf{q}_{1:m})$. For the
invariance, two orthonormal frames share the column span if and only if
$\bar{Q}_{1:m}=Q_{1:m}R$ with $R\in\mathcal{O}(m)$; then
$M_{1,fg}\mapsto M_{1,fg}R$, $\tilde M^i_{1,fg}\mapsto \tilde M^i_{1,fg}R$,
and the change of basis cancels:
\[
\tilde D_{i,fg}\;\mapsto\; M^x_{1,fg}R\,\big(\tilde M^i_{1,fg}R\big)^{-1}
= M^x_{1,fg}\big(\tilde M^i_{1,fg}\big)^{-1}=\tilde D_{i,fg}.
\]
Invariance of $\tilde D_{b,fg}$ follows.

\emph{(ii)} The quadratic form and certainty equivalence follow as in the
proof of Proposition~\ref{prop:loss_fg}, with
$\tilde D_{i,fg}\bar{\textbf{i}}^e_{\textbf{s}}$ in place of
$D_{i,fg}\bar{\textbf{i}}^e_{1:h}$. For the trace terms: the future block is
unchanged, and $D_{2:h,fg}D_{2:h,fg}'$ depends only on $\bs{\phi}$ by
\eqref{eq:fg_rf}. For the period-$1$ block, the residual identity
\eqref{eq:Deta_residual_fg} holds with $\tilde D_{i,fg}S_{\textbf{s}}$ in
place of $D_{i,fg}$: the policy columns vanish since
$\tilde D_{i,fg}S_{\textbf{s}}M^i_{1,fg}
=M^x_{1,fg}(\tilde M^i_{1,fg})^{-1}\tilde M^i_{1,fg}=M^x_{1,fg}$, so
\[
\tilde D_{\eta,fg}\tilde D_{\eta,fg}'
=\big(M^x_{[1]}-\tilde D_{i,fg}S_{\textbf{s}}M^i_{[1]}\big)
\big(M^x_{[1]}-\tilde D_{i,fg}S_{\textbf{s}}M^i_{[1]}\big)'.
\]
Expanding, the outer products of $M^x_{[1]}$ and $M^i_{[1]}$ are blocks of
$\textbf{K}\,\mathcal{C}_1\mathcal{C}_1'\,\textbf{K}'$ and depend only on
$\bs{\phi}$ by \eqref{eq:fg_rf}; $S_{\textbf{s}}$ is a fixed selection; and
$\tilde D_{i,fg}$ depends only on $(\bs{\phi},\textbf{q}_{1:m})$ and is
basis-invariant by (i). Hence $\tilde D_{\eta,fg}\tilde D_{\eta,fg}'$ ---
though not $\tilde D_{\eta,fg}$ itself, whose columns depend on the basis of
the non-policy block --- and with it both trace terms, $\tilde c_{fg}$,
$\tilde H_{fg}$, $\tilde{\textbf{g}}_{fg}$, and $L_{fg}$, depend on $Q$ only
through the column span of $Q_{1:m}$.

\emph{(iii)} As in the proof of Proposition~\ref{prop:loss_fg}:
$\tilde H_{fg}=\big((\tilde M^i_{1,fg})^{-1}\big)'\,
\big(M^{x\prime}_{1,fg}WM^x_{1,fg}\big)\,(\tilde M^i_{1,fg})^{-1}$
is positive definite since $M^x_{1,fg}$ has full column rank $m$ and
$\tilde M^i_{1,fg}$ is nonsingular; strict convexity gives the unique
minimizer \eqref{istar_nodes}.
\end{proof}

\begin{proof}[Proof of Lemma~\ref{lemma_js}]
Let $\Sigma^{1/2}H\Sigma^{1/2} = \Gamma'\Lambda\Gamma$ be the spectral decomposition, with $\Gamma$ orthogonal and $\Lambda = \mathrm{diag}(\lambda_1, \ldots, \lambda_k) \succ 0$, and set $A = \Gamma\Sigma^{-1/2}$. Since $A\Sigma A' = I_k$, the transformed observation satisfies $Y = AX \sim N(\bs{\theta}, I_k)$ with $\bs{\theta} = A\bs{\mu}$. To any rule $\bs{\delta}$ associate the rule $\bs{\delta}_2(Y) = A\bs{\delta}(A^{-1}Y)$; since $(A^{-1})'HA^{-1} = \Gamma\Sigma^{1/2}H\Sigma^{1/2}\Gamma' = \Lambda$, Theorem 4.1(a) in \citet{strawderman2003minimax} gives $R(\bs{\delta}, \bs{\mu}) = R_2(\bs{\delta}_2, \bs{\theta})$, where $R_2$ is the risk in the canonical problem with observation $Y \sim N(\bs{\theta}, I_k)$ and loss $L_2(d, \bs{\theta}) = (d - \bs{\theta})'\Lambda(d - \bs{\theta})$. The identity rule maps to the identity rule, while $\bs{\delta}^{JS}_t$ maps to $\bs{\delta}_2^g(Y) = Y + g(Y)$ with the James--Stein-type perturbation
\[
g(y) \equiv g_t(y) = -\,t\,\frac{\Lambda^{-1}y}{\|\Lambda^{-1/2}y\|^2},
\]
since
\begin{align*}
A\,Q^{-1}\Sigma^{-1}A^{-1} &= \Gamma\Sigma^{-1/2}\,Q^{-1}\Sigma^{-1}\,\Sigma^{1/2}\Gamma'
= \Gamma\bigl(\Sigma^{1/2}Q\Sigma^{1/2}\bigr)^{-1}\Gamma'
= \Lambda^{-1}, \\
X'\Sigma^{-1}Q^{-1}\Sigma^{-1}X &= Y'(A^{-1})'\Sigma^{-1}Q^{-1}\Sigma^{-1}A^{-1}Y = Y'\Lambda^{-1}Y = \|\Lambda^{-1/2}Y\|^2.
\end{align*}
Write $s(y) = \|\Lambda^{-1/2}y\|^2 = \sum_{j=1}^k \lambda_j^{-1}y_j^2$, so that $g_i(y) = -t\,\lambda_i^{-1}y_i/s(y)$. By Theorem 3.1(a) in \citet{strawderman2003minimax},
\[
R_2(\bs{\theta}, \bs{\delta}_2^g) - R_2(\bs{\theta}, \bs{\delta}_0)
= \sum_{i=1}^k \lambda_i\, \ev\!\left[g_i^2(Y) + 2\,\frac{\partial g_i(Y)}{\partial Y_i}\right].
\]
Noting $g_i^2 = t^2\lambda_i^{-2}y_i^2/s^2$ and $\partial s/\partial y_i = 2\lambda_i^{-1}y_i$, so that $\partial g_i/\partial y_i = -t\lambda_i^{-1}/s + 2t\lambda_i^{-2}y_i^2/s^2$, yields
\begin{align*}
\sum_{i=1}^k \lambda_i\left[g_i^2 + 2\,\frac{\partial g_i}{\partial y_i}\right]
&= \frac{t^2}{s^2}\sum_{i=1}^k \lambda_i^{-1}y_i^2 \;-\; \frac{2tk}{s} \;+\; \frac{4t}{s^2}\sum_{i=1}^k \lambda_i^{-1}y_i^2 \\
&= \frac{t^2 - 2t(k-2)}{s} = -\,\frac{t\bigl(2(k-2)-t\bigr)}{s}.
\end{align*}
Applying the risk equivalence gives
\[
R(\bs{\delta}^{JS}_t, \bs{\mu}) - R(\bs{\delta}_0, \bs{\mu})
= -\,\ev\!\left[\frac{t\bigl(2(k-2) - t\bigr)}{\|\Lambda^{-1/2}Y\|^2}\right]
= -\,t\bigl(2(k-2) - t\bigr)\,\ev\!\left[\frac{1}{X'\Sigma^{-1}Q^{-1}\Sigma^{-1}X}\right],
\]
which is \eqref{js_riskgain}. For $0 < t < 2(k-2)$ the expectation is strictly positive, and $t \mapsto t\bigl(2(k-2)-t\bigr)$ is maximized at $t = k-2$.
\end{proof}

\subsection{Feasible forward guidance: projection approach}

We consider here an alternative approach for the empirically relevant case where only $m<h$ news shocks are available in the data, which we call the projection approach. Compared to the main proposal adopted in Section \ref{sec:feasible-fg}, here an action remains an
announcement $\bar{\textbf{i}}^e_{1:h}\in\reals^h$, but we now assume that
the SVAR model \eqref{SVAR} only contains $m<h$ distinct policy news shocks
$e_{1t},\ldots,e_{mt}$, so that the matrix $M_{1,fg}$ of
Section~\ref{sec:fg} is $nh\times m$ with full column rank. The policymaker is subject to an implementation constraint: only expected
paths spanned by the impulse responses collected in $M^i_{1,fg}$ can be
enforced, and the policymaker might not be able to solve
\eqref{lcrrestr_fg} exactly for a generic announcement. We resolve this
with the following two assumptions.

\begin{enumerate}
\item At each $\bs{\theta}$, the announcement $\bar{\textbf{i}}^e_{1:h}$ is
understood as the closest \emph{announceable} path, the orthogonal
projection\footnote{Orthogonality is in the Euclidean metric, a convention
for which we claim no optimality. Our approach follows
\citet{mckay2023can}.} of
$\bar{\textbf{i}}^e_{1:h}$ onto the set of expected paths the $m$ shocks
can generate ex ante,
\[
\mathcal{S}_{e}(\bs{\theta}) := \bigl\{\textbf{b}^i_{1:h}
+ M^i_{1,fg}\textbf{e}^m_1 \, : \, \textbf{e}^m_1\in \reals^m\bigr\},
\]
that is, writing
$(M^i_{1,fg})^{+}:=\bigl[(M^i_{1,fg})'M^i_{1,fg}\bigr]^{-1}(M^i_{1,fg})'$
for the Moore--Penrose inverse,
\begin{equation}\label{ifeas_exante}
\bar{\textbf{i}}^{e,feas}_{1:h}(\bar{\textbf{i}}^e_{1:h}, \bs{\theta}, \bs{\chi}_0)
= \textbf{b}^i_{1:h} + M^i_{1,fg}(M^i_{1,fg})^{+}
\bigl(\bar{\textbf{i}}^e_{1:h} - \textbf{b}^i_{1:h}\bigr).
\end{equation}

\item At $t=1$, once the non-policy shocks $\bs{\eta}^{fg}_1$ have
materialized, the policymaker releases the shocks that enforce
$\bar{\textbf{i}}^{e,feas}_{1:h}$ as closely as possible, that is, those
implementing the orthogonal projection of $\bar{\textbf{i}}^{e,feas}_{1:h}$
onto the realized feasible set
\[
\mathcal{S}_{e}(\bs{\theta}, \bs{\eta}_1^{fg}) := \bigl\{\textbf{b}^i_{1:h}
+ M^i_{1,fg}\textbf{e}^m_1 + M^i_{\eta,fg}\bs{\eta}^{fg}_1
\, : \, \textbf{e}^m_1\in \reals^m\bigr\}.
\]
\end{enumerate}

By a standard least squares argument, the shocks selected under the two
assumptions are
\begin{equation}\label{em_solution_feas}
\textbf{e}^m_1(\bar{\textbf{i}}^e_{1:h}, \bs{\eta}^{fg}_1, \bs{\theta}, \bs{\chi}_0)
= (M^i_{1,fg})^{+}
\bigl[\bar{\textbf{i}}^e_{1:h} - (\textbf{b}^i_{1:h}
+ M^i_{\eta,fg}\bs{\eta}^{fg}_1)\bigr].
\end{equation}
Two remarks are in order. First, the two assumptions compose into one
projection: since $(M^i_{1,fg})^{+}M^i_{1,fg}=I_m$, substituting
\eqref{ifeas_exante} for $\bar{\textbf{i}}^{e,feas}_{1:h}$ in the second
assumption returns \eqref{em_solution_feas}, so projecting the announcement
onto $\mathcal{S}_{e}(\bs{\theta})$ first and then onto
$\mathcal{S}_{e}(\bs{\theta},\bs{\eta}^{fg}_1)$ gives the same shocks as
projecting $\bar{\textbf{i}}^e_{1:h}$ directly onto
$\mathcal{S}_{e}(\bs{\theta},\bs{\eta}^{fg}_1)$. The two sets are parallel
--- they differ by the translation $M^i_{\eta,fg}\bs{\eta}^{fg}_1$ and are
generated by the same columns of $M^i_{1,fg}$ --- and the component of the
announcement orthogonal to those columns never plays a role. Second, the
shocks enforce $\bar{\textbf{i}}^{e,feas}_{1:h}$ only up to the part of
$M^i_{\eta,fg}\bs{\eta}^{fg}_1$ that is not spanned by the columns of
$M^i_{1,fg}$. This residual is mean zero: as with the non-policy shocks of
the contemporaneous regime, it enters the loss only through its constant
and leaves the optimal action, and the comparison of decision rules,
unaffected.\footnote{The second assumption adopts the timing convention used in Section \ref{sec:fg} of the paper, as defined in
Footnote~\ref{fn:exante}: the policy shocks are implemented at $t=1$, after
$\bs{\eta}^{fg}_1$ has materialized. Under the fully ex-ante convention
they are selected at $t=0$ and solve the same least squares problem with
$\ev[\bs{\eta}^{fg}_1\mid\mathcal{I}_0]=\textbf{0}$ in place of the
realization, so that the projection is onto $\mathcal{S}_{e}(\bs{\theta})$;
under the hybrid convention only the contemporaneous shock reacts to
$\bs{\eta}^{fg}_1$. With $m<h$ none of these conventions enforces a generic
announcement exactly, and the conclusion of Footnote~\ref{fn:exante}
extends unchanged: the conventions differ only in how much of
$M^i_{\eta,fg}\bs{\eta}^{fg}_1$ the shocks can lean against, the same mean-zero
residual discussed above, which enters the loss through the trace terms
alone. The oracle, plug-in, Bayes and robust Bayes regret rules are therefore the same across
conventions.} When $m=h$,
$\mathcal{S}_{e}(\bs{\theta})=\reals^h$, the projection in the first
assumption is the identity and \eqref{em_solution_feas} reduces to
\eqref{e1_solution}.

Substituting \eqref{em_solution_feas} into \eqref{yh_fg} and retaining the
non-policy block, the counterfactual target paths have the same form as
\eqref{xpaths_fg},
\begin{equation}\label{xpaths_fg_m}
\tilde{\textbf{x}}^{fg}_{1:h} = \tilde D_{i,fg}\,\bar{\textbf{i}}^e_{1:h}
+ \tilde D_{b,fg}\,\textbf{b}_{1:h}
+ \tilde D_{\eta,fg}\,\bs{\eta}^{fg}_1 + D_{2:h,fg}\,\tilde{\bs{\varepsilon}}_{2:h},
\end{equation}
with
\[
\overset{((n-1)h\times h)}{\tilde D_{i,fg}} = M^x_{1,fg}(M^i_{1,fg})^{+}
\]
\[
\overset{((n-1)h\times nh)}{\tilde D_{b,fg}} =
\begin{pmatrix} -\tilde D_{i,fg} & I_{(n-1)h}\end{pmatrix},\qquad
\overset{((n-1)h\times (n-m))}{\tilde D_{\eta,fg}} =
M^x_{\eta,fg} - \tilde D_{i,fg}M^i_{\eta,fg}.
\]
The following proposition shows that the feasible loss function retains the
properties of Propositions~\ref{prop:Dstructure_fg}
and~\ref{prop:loss_fg}, with one difference: the action space has dimension
$h$ while the loss has curvature only in $m$ directions.

\begin{proposition}[Feasible forward-guidance loss and oracle in the projection approach]\label{prop:loss_feas_proj}
Suppose $M^i_{1,fg}$ has full column rank $m\leq h$ and $W$ is symmetric
positive semidefinite. Write
$Q_{1:m}:=(\textbf{q}_1\ \cdots\ \textbf{q}_m)$ and
$\textbf{q}_{1:m}:=\operatorname{vec}(Q_{1:m})$.
\begin{enumerate}
\item[(i)] $\tilde D_{i,fg}$ and $\tilde D_{b,fg}$ depend on $\bs{\theta}$
only through $(\bs{\phi}, \textbf{q}_{1:m})$ and are invariant to rotations
of $\{\textbf{q}_1,\dots,\textbf{q}_m\}$.
\item[(ii)] The loss is the quadratic form
\begin{equation}\label{loss_feas_quadratic}
L_{fg}(\bar{\textbf{i}}^e_{1:h}, \bs{\phi}, Q, \bs{\chi}_0)
= \tilde c_{fg} + (\bar{\textbf{i}}^e_{1:h})'\tilde H_{fg}\,
\bar{\textbf{i}}^e_{1:h}
+ 2\,(\bar{\textbf{i}}^e_{1:h})'\tilde{\textbf{g}}_{fg},
\end{equation}
where, with
$\tilde{\textbf{b}}^{fg}_{1:h} := \tilde D_{b,fg}\textbf{b}_{1:h}-\textbf{x}^*$,
\begin{align*}
\tilde c_{fg} &:= (\tilde{\textbf{b}}^{fg}_{1:h})' W \tilde{\textbf{b}}^{fg}_{1:h}
+ \operatorname{trace}(\tilde D_{\eta,fg}' W \tilde D_{\eta,fg})
+ \operatorname{trace}(D_{2:h,fg}' W D_{2:h,fg}), \\
\tilde H_{fg} &:= \tilde D_{i,fg}' W \tilde D_{i,fg}, \qquad
\tilde{\textbf{g}}_{fg} := \tilde D_{i,fg}' W \tilde{\textbf{b}}^{fg}_{1:h}.
\end{align*}
In particular, the loss depends on $Q$ only through the column span of
$Q_{1:m}$.
\item[(iii)] The loss depends on the announcement only through
$(M^i_{1,fg})^{+}\bar{\textbf{i}}^e_{1:h}$. Assuming $H_{fg}$ is positive
semi-definite with rank $m$, the
minimizers of \eqref{loss_feas_quadratic} form the affine set
$\bar{\textbf{i}}^{*,feas}_{1:h}+\ker (M^i_{1,fg})^{+}$ of dimension $h-m$,
where
\begin{equation}\label{istar_feas}
\bar{\textbf{i}}^{*,feas}_{1:h}(\bs{\phi},\textbf{q}_{1:m},\bs{\chi}_0)
= \textbf{b}^i_{1:h} - M^i_{1,fg}
\bigl[(M^x_{1,fg})'WM^x_{1,fg}\bigr]^{-1}(M^x_{1,fg})'W
\bigl(\textbf{b}^x_{1:h}-\textbf{x}^*\bigr)
\end{equation}
is the unique minimizer lying in $\mathcal{S}_{e}(\bs{\theta})$, and every
other minimizer is enforced as \eqref{istar_feas}. 
\end{enumerate}
\end{proposition}

One difference from the case $m=h$ concerns uniqueness of the oracle path.
By Proposition~\ref{prop:loss_feas_proj}(iii), the loss depends on the
announcement only through $(M^i_{1,fg})^{+}\bar{\textbf{i}}^e_{1:h}$: two
announced paths that differ by an element of $\ker (M^i_{1,fg})^{+}$, the
$(h-m)$-dimensional orthogonal complement of the column space of
$M^i_{1,fg}$, are enforced identically and have the same loss. The oracle
--- and with it the plug-in --- is therefore unique only as an element of
$\mathcal{S}_{e}(\bs{\theta})$: \eqref{istar_feas} is the minimizer that is
itself announceable, and every other minimizer is enforced as
\eqref{istar_feas}. The plug-in rule we report in the numerical simulations
and the empirical application is this representative evaluated at the point
estimate,
\[
\hat{\textbf{i}}^{pi}_{feas}
= \textbf{b}^i_{1:h} - \hat M^i_{1,fg}
\bigl[(\hat M^x_{1,fg})'W\hat M^x_{1,fg}\bigr]^{-1}(\hat M^x_{1,fg})'W
\bigl(\textbf{b}^x_{1:h}-\textbf{x}^*\bigr),
\]
which is by construction announceable under $\hat{\bs{\theta}}$.

The Bayes rule is the minimizer of the posterior expectation of
\eqref{loss_feas_quadratic} and is generally unique over $\reals^h$. While
$\tilde H_{fg}(\bs{\theta})$ has rank $m$ at each parameter point, averaging
it over the posterior generically makes $\ev[\tilde H_{fg}\mid\textbf{y}]$
full rank, because the column space of $M^i_{1,fg}$ varies across draws.
This is not guaranteed, but it holds in our applications. Assuming that
posterior averaging yields an invertible Hessian, the Bayes rule is unique
in $\reals^h$ and satisfies
\[
\hat{\textbf{i}}^{B}_{feas}
:= -\,\ev[\tilde H_{fg}\mid \textbf{y}]^{-1}
\ev[\tilde{\textbf{g}}_{fg}\mid \textbf{y}].
\]
Note that, unlike the feasible plug-in, the Bayes action is not guaranteed
to lie in $\mathcal{S}_e(\bs{\theta})$ for any $\bs{\theta}$ and may
therefore not be announceable. Its risk is nevertheless well defined: like
every announcement, it is evaluated under the two assumptions above, so the
loss it incurs at each parameter value is the loss of the path actually
enforced.

\begin{proof}[Proof of Proposition~\ref{prop:loss_feas_proj}]
\emph{(i)} As in the proof of Proposition~\ref{prop:Dstructure_fg},
$M_{1,fg}=\textbf{K}\,\mathcal{G}_{\cdot1}\,\Sigma_l\,Q_{1:m}$ --- now with
$m$ columns --- so $M^i_{1,fg}$, $M^x_{1,fg}$, and with them
$(M^i_{1,fg})^{+}$ and $\tilde D_{i,fg}=M^x_{1,fg}(M^i_{1,fg})^{+}$, depend
on $\bs{\theta}$ only through $(\bs{\phi},\textbf{q}_{1:m})$. For the
invariance, two orthonormal frames share the column span if and only if
$\bar{Q}_{1:m}=Q_{1:m}R$ with $R\in\mathcal{O}(m)$; then
$M_{1,fg}\mapsto M_{1,fg}R$, $(M^i_{1,fg})^{+}\mapsto R'(M^i_{1,fg})^{+}$,
and the change of basis cancels:
\[
\tilde D_{i,fg}\;\mapsto\; M^x_{1,fg}R\,R'(M^i_{1,fg})^{+}
= M^x_{1,fg}(M^i_{1,fg})^{+}=\tilde D_{i,fg}.
\]
Invariance of $\tilde D_{b,fg}$ follows.

\emph{(ii)} The quadratic form and certainty equivalence follow as in the
proof of Proposition~\ref{prop:loss_fg}, with $\tilde D_{i,fg}$ in place of
$D_{i,fg}$. For the trace terms: the future block is unchanged, and
$D_{2:h,fg}D_{2:h,fg}'$ depends only on $\bs{\phi}$ by \eqref{eq:fg_rf}. For
the period-$1$ block, the residual identity \eqref{eq:Deta_residual_fg}
holds with $\tilde D_{i,fg}$ in place of $D_{i,fg}$: the policy columns
vanish since
$\tilde D_{i,fg}M^i_{1,fg}=M^x_{1,fg}(M^i_{1,fg})^{+}M^i_{1,fg}=M^x_{1,fg}$,
so
\[
\tilde D_{\eta,fg}\tilde D_{\eta,fg}'
=\big(M^x_{[1]}-\tilde D_{i,fg}M^i_{[1]}\big)
\big(M^x_{[1]}-\tilde D_{i,fg}M^i_{[1]}\big)'.
\]
Expanding, the outer products of $M^x_{[1]}$ and $M^i_{[1]}$ are blocks of
$\textbf{K}\,\mathcal{C}_1\mathcal{C}_1'\,\textbf{K}'$ and depend only on
$\bs{\phi}$ by \eqref{eq:fg_rf}, while $\tilde D_{i,fg}$ depends only on
$(\bs{\phi},\textbf{q}_{1:m})$ and is basis-invariant by (i). Hence
$\tilde D_{\eta,fg}\tilde D_{\eta,fg}'$ --- though not $\tilde D_{\eta,fg}$
itself, whose columns depend on the basis of the non-policy block --- and
with it both trace terms, $\tilde c_{fg}$, $\tilde H_{fg}$,
$\tilde{\textbf{g}}_{fg}$, and $L_{fg}$, depend on $Q$ only through the
column span of $Q_{1:m}$.

\emph{(iii)} Since
$\tilde D_{i,fg}\bar{\textbf{i}}^e_{1:h}
=M^x_{1,fg}\bigl((M^i_{1,fg})^{+}\bar{\textbf{i}}^e_{1:h}\bigr)$, the loss
is a function of $(M^i_{1,fg})^{+}\bar{\textbf{i}}^e_{1:h}$ alone, and
adding any element of $\ker (M^i_{1,fg})^{+}$ --- the $(h-m)$-dimensional
orthogonal complement of the column space of $M^i_{1,fg}$ --- leaves it
unchanged. Writing
$\tilde H_{fg}=\bigl((M^i_{1,fg})^{+}\bigr)'
\bigl((M^x_{1,fg})'WM^x_{1,fg}\bigr)(M^i_{1,fg})^{+}$ and using
$\operatorname{rank}\bigl((M^i_{1,fg})^{+}\bigr)=m$ gives
$\operatorname{rank}(\tilde H_{fg})=m$. On $\mathcal{S}_{e}(\bs{\theta})$,
parametrized by
$\bar{\textbf{i}}^e_{1:h}=\textbf{b}^i_{1:h}+M^i_{1,fg}\textbf{e}^m_1$, the
identity $(M^i_{1,fg})^{+}M^i_{1,fg}=I_m$ gives
$\tilde D_{i,fg}\bar{\textbf{i}}^e_{1:h}+\tilde{\textbf{b}}^{fg}_{1:h}
= M^x_{1,fg}\textbf{e}^m_1 + \bigl(\textbf{b}^x_{1:h}-\textbf{x}^*\bigr)$,
so the loss is a strictly convex quadratic in $\textbf{e}^m_1$ with Hessian
$(M^x_{1,fg})'WM^x_{1,fg}$, and the unique minimizer is
\[
\textbf{e}^{m,*}_1
= -\bigl[(M^x_{1,fg})'WM^x_{1,fg}\bigr]^{-1}(M^x_{1,fg})'W
\bigl(\textbf{b}^x_{1:h}-\textbf{x}^*\bigr),
\]
whose image is \eqref{istar_feas}. Any minimizer differs from it by an
element of $\ker (M^i_{1,fg})^{+}$ and is therefore enforced as
\eqref{istar_feas}. When $m=h$, $(M^i_{1,fg})^{+}=(M^i_{1,fg})^{-1}$,
$\mathcal{S}_{e}(\bs{\theta})=\reals^h$ and \eqref{istar_feas} reduces to
\eqref{istar_fg}.
\end{proof}

\subsection{Lemmas and proofs for Section~\ref{sec:sequential}}\label{app:sequential}

This appendix collects the lemmas behind
Proposition~\ref{prop:representativeness} and their proofs.
Lemma~\ref{lem:perperiod} is the Markov representation \eqref{eq:markov}.
We then cast the sequential problem in the standard form of the
linear-quadratic control literature, which requires the explicit
construction of $\textbf{A}$, $\textbf{B}$ and $\textbf{w}_t$ in
\eqref{eq:standardform}. Lemma~\ref{lem:theil} states the certainty
equivalence theorem of \citet{theil1957note} for the class of strategies
of our problem, and Lemma~\ref{lem:timeinv} establishes well-posedness and
the coincidence between the re-optimizing path and the solution of the
dynamic program. The proof of the proposition follows.

\begin{lemma}[Markov representation of the contemporaneous counterfactuals]\label{lem:perperiod}
Suppose $B_{iv} \neq 0$ and let $\mathcal{D}_i$, $\mathcal{D}_\eta$ and
$\mathcal{D}_\chi$ be the impact multipliers defined in
Section~\ref{sec:seq-oracle}. Under the enforced path $\bar{\textbf{i}}_{1:h}$,
the period-$t$ counterfactual outcome of the target variables depends on the
past only through the state, and satisfies \eqref{eq:markov},
\[
\textbf{x}_t = \mathcal{D}_i\, \bar{i}_t + \mathcal{D}_\chi\, \bs{\chi}_{t-1} + \mathcal{D}_\eta\, \bs{\eta}_t,
\qquad t = 1, \dots, h,
\]
where $\bs{\chi}_{t-1}$ is the counterfactual state generated by the enforced
values $(\bar{i}_1, \dots, \bar{i}_{t-1})$, the shocks $(\bs{\eta}_1, \dots,
\bs{\eta}_{t-1})$, and the initial state $\bs{\chi}_0$.
\end{lemma}

\begin{proof}
Fix $t$ and condition on the counterfactual history up to $t-1$, which
determines the state $\bs{\chi}_{t-1}$. The SVAR \eqref{SVAR} gives
$\textbf{y}_t = \Pi\bs{\chi}_{t-1} + B\bs{\varepsilon}_t$, whose instrument
row reads
\[
i_t = b^i_{t|t-1} + B_{iv}v_t + B_{i\eta}\bs{\eta}_t, \qquad \textbf{b}_{t|t-1} = \Pi\bs{\chi}_{t-1}.
\]
Enforcing $i_t = \bar{i}_t$ and solving yields the unique consistent surprise
\[
v_t = B_{iv}^{-1}\bigl(\bar{i}_t - b^i_{t|t-1} - B_{i\eta}\bs{\eta}_t\bigr).
\]
Substituting into the target rows,
\[
\textbf{x}_t = \textbf{b}^x_{t|t-1} + B_{xv}v_t + B_{x\eta}\bs{\eta}_t
= \mathcal{D}_i\,\bar{i}_t + \bigl(\textbf{b}^x_{t|t-1} - \mathcal{D}_i\, b^i_{t|t-1}\bigr) + \mathcal{D}_\eta\,\bs{\eta}_t,
\]
and the middle term equals $\begin{pmatrix} -\mathcal{D}_i &
I_{n-1}\end{pmatrix}\Pi\bs{\chi}_{t-1} = \mathcal{D}_\chi\bs{\chi}_{t-1}$.
Since $\bs{\chi}_{t-1}$ is a function of $(\bs{\chi}_0, \bar{i}_1, \dots,
\bar{i}_{t-1}, \bs{\eta}_1, \dots, \bs{\eta}_{t-1})$, the past enters only
through the state. That the recursive and the stacked constructions coincide
follows from uniqueness of the counterfactual: given the invertibility of
$M^i_{v,c}$, the enforcement constraint \eqref{instrconstr_c} pins down the
policy shocks uniquely, cf.\ \eqref{vpaths_c}. Finally, $\mathcal{D}_i$ and
$\mathcal{D}_\eta$ are the $h=1$ case of \eqref{Di_c} and \eqref{Deta_c},
since at $h=1$ the blocks of $M(\bs{\theta})$ reduce to the impact responses
$M^i_{v,c} = B_{iv}$, $M^x_{v,c} = B_{xv}$, $M^i_{\eta,c} = B_{i\eta}$,
$M^x_{\eta,c} = B_{x\eta}$.
\end{proof}

From Lemma~\ref{lem:perperiod} we obtain a
first-order controlled stochastic recursion for the state vector
$\bs{\chi}_t$. One timing convention requires care. In the dynamic-control
literature the control chosen at a given stage does not affect the same-stage
state, whereas in our conventions the instrument $i_t$ is part of
$\textbf{y}_t$ and hence of $\bs{\chi}_t$. Accordingly, the control chosen at
date $t-1$ is the following period's instrument, $u_{t-1} := i_t$, and we
index system and disturbances by the outcome period $t$; our stage $t$ thus
corresponds to stage $k = t-1$ in \citet{bertsekas1995dynamic}. With this
convention Lemma~\ref{lem:perperiod} yields \eqref{eq:standardform}, that is
\[
\bs{\chi}_{t} = \textbf{A}\,\bs{\chi}_{t-1} + \textbf{B}\, u_{t-1} + \textbf{w}_{t},
\qquad t = 1, \dots, h,
\]
with\footnote{The letters $\textbf{A}$ and $\textbf{B}$ denote the
standard-form transition and control-loading objects and should not be confused
with the structural impact matrix $B$ of \eqref{SVAR}, which only enters through
the impact multipliers $\mathcal{D}_i, \mathcal{D}_\chi, \mathcal{D}_\eta$.}
\begin{equation}\label{eq:AB}
\textbf{A} := \begin{pmatrix}
\mathbf{0}_{1\times np} \\
\mathcal{D}_\chi \\
\begin{pmatrix} I_{n(p-1)} & O_{n(p-1)\times n}\end{pmatrix}
\end{pmatrix} ,
\qquad
\textbf{B} := \begin{pmatrix} 1 \\ \mathcal{D}_i \\ \mathbf{0}_{n(p-1)} \end{pmatrix} ,
\qquad
\textbf{w}_t := \begin{pmatrix} 0 \\ \mathcal{D}_\eta\, \bs{\eta}_{t} \\ \mathbf{0}_{n(p-1)} \end{pmatrix} .
\end{equation}
Here $\textbf{A} \in \reals^{np \times np}$, while $\textbf{B}$ and
$\textbf{w}_t$ are $np$-dimensional column vectors, the control being scalar,
and the disturbances $\textbf{w}_t$ are i.i.d.\ with zero mean and finite
covariance.

\paragraph{Strategies and Theil's theorem.} Along any instrument path,
deterministic or not, the counterfactual is generated period by period as
in Lemma~\ref{lem:perperiod}. Since \eqref{eq:markov} is linear with
coefficients that do not depend on the path, unrolling it gives, pathwise,
\begin{equation}\label{eq:stackedstrategy}
\textbf{x}_{1:h} = D_{i,c}\,\textbf{i} + \textbf{s},
\qquad
\textbf{s} := D_{b,c}\,\textbf{b}_{1:h} + D_{\eta,c}\,\bs{\eta}_{1:h},
\end{equation}
the stacked representation of Section~\ref{sec:contemp}, with
$\ev_0[\textbf{s}] = D_{b,c}\textbf{b}_{1:h}$. By
Lemma~\ref{lem:sequential} the law of $\bs{\eta}_{1:h}$, and hence of
$\textbf{s}$, is the same under every path. Let $\textbf{s}_t$ denote the
$(n-1)$-vector collecting the period-$t$ entries of $\textbf{s}$, and
similarly $(D_{i,c}\textbf{i})_t$; by the lower triangularity of the blocks
of $D_{i,c}$ the latter depends on $i_1, \dots, i_t$ only. Following
\citet{theil1957note}, a \emph{strategy} is an instrument path
$\textbf{i} = (i_1, \dots, i_h)'$ whose entry $i_t$ is a function of
$\textbf{s}_1, \dots, \textbf{s}_{t-1}$, with
$\ev_0[\|\textbf{i}\|^2] < \infty$. We denote by $\mathcal{S}$ the set of
strategies. It is a linear space, since measurability and square
integrability are preserved by linear combinations, and it contains the
deterministic paths $\reals^h$. The expected loss of a strategy is
\begin{equation}\label{eq:lossstrategy}
\mathcal{L}(\textbf{i})
:= \ev_0\Bigl[\,\sum_{t=1}^{h} \ell_t(\bs{\chi}_t)\Bigr]
= \ev_0\bigl[(D_{i,c}\textbf{i} + \textbf{s} - \textbf{x}^*)'\,W\,
(D_{i,c}\textbf{i} + \textbf{s} - \textbf{x}^*)\bigr],
\end{equation}
finite on $\mathcal{S}$ because $D_{i,c}\textbf{i} + \textbf{s} -
\textbf{x}^*$ has finite second moments. On deterministic paths
$\mathcal{L}$ is the loss $L_c(\bar{\textbf{i}}_{1:h}, \bs{\theta})$ of
\eqref{loss_c}.

When $\bs{\theta}$ is known, Theil's class coincides with the class of
instrument paths adapted to the information of the policymaker, that is,
with $i_t$ a function of $\mathcal{I}_{t-1}$. Along any path in either
class, $\textbf{y}_\tau = (i_\tau, \textbf{x}_\tau')'$ with $\textbf{x}_\tau
= (D_{i,c}\textbf{i})_\tau + \textbf{s}_\tau$ by \eqref{eq:stackedstrategy},
and $\mathcal{I}_0$ is generated by $\bs{\chi}_0$. By induction on $\tau$,
the history $\mathcal{I}_\tau$ is a function of $\bs{\chi}_0$,
$\textbf{s}_1, \dots, \textbf{s}_\tau$ and the path, and conversely
$\textbf{s}_\tau = \textbf{x}_\tau - (D_{i,c}\textbf{i})_\tau$ is a function
of $\mathcal{I}_\tau$, so the two information structures generate the same
$\sigma$-fields and the two classes of square integrable paths coincide.

\begin{lemma}[Theil's certainty equivalence]\label{lem:theil}
Suppose $H_c \succ 0$. If $\textbf{f}^* \in \mathcal{S}$ minimizes
$\mathcal{L}$ over $\mathcal{S}$, then
\[
\ev_0[\textbf{f}^*] = \bar{\textbf{i}}^{*,c}_{1:h}(\bs{\chi}_0),
\qquad
f^*_1 = \bar{i}^{*,c}_1 .
\]
\end{lemma}

\begin{proof}
This is the result of \citet[Section~2]{theil1957note}, whose welfare
function is minus our loss and whose matrix $K$ equals $-2H_c$. We
reproduce the argument in our notation to make explicit where the
hypotheses enter. Fix a deterministic direction $\textbf{v} \in \reals^h$
and a scalar $\epsilon$. Since $\mathcal{S}$ is a linear space containing
$\reals^h$, $\textbf{f}^* + \epsilon\textbf{v} \in \mathcal{S}$, and
expanding \eqref{eq:lossstrategy},
\[
\mathcal{L}(\textbf{f}^* + \epsilon\textbf{v}) - \mathcal{L}(\textbf{f}^*)
= 2\epsilon\,\textbf{v}'\Bigl(H_c\,\ev_0[\textbf{f}^*]
+ D_{i,c}'W\bigl(\ev_0[\textbf{s}] - \textbf{x}^*\bigr)\Bigr)
+ \epsilon^2\,\textbf{v}'H_c\textbf{v},
\]
where the deterministic $\textbf{v}$ has been taken out of the expectations
and $\ev_0[D_{i,c}'WD_{i,c}\textbf{f}^*] = H_c\,\ev_0[\textbf{f}^*]$; both
terms are finite by square integrability. The left-hand side is nonnegative
for every $\epsilon$ by optimality of $\textbf{f}^*$, so the coefficient of
$\epsilon$ vanishes, for otherwise a small $\epsilon$ of its sign would make
the right-hand side negative. As $\textbf{v}$ ranges over $\reals^h$,
\[
H_c\,\ev_0[\textbf{f}^*]
= -\,D_{i,c}'W\bigl(D_{b,c}\textbf{b}_{1:h} - \textbf{x}^*\bigr),
\]
and since $H_c \succ 0$ the solution is \eqref{eq:resoracle} at $t = 0$,
the ex-ante oracle path of Proposition~\ref{prop:loss_c}. For the first
entry, $f^*_1$ is a function of no $\textbf{s}_\tau$, hence a constant
given the information at time zero, so
$f^*_1 = \ev_0[f^*_1] = \bar{i}^{*,c}_1$.
\end{proof}

Two uniqueness statements are needed to relate now the optimal strategy above to the solution of a dynamic program and the re-optimizing solution: the shifted
solution \eqref{eq:resoracle} must be uniquely determined at every date, and
the control prescribed by the dynamic program must be unique. The first follows
from the positive definiteness of $H_c$, which part (i) of the following lemma
transfers to every residual horizon. Part (ii) shows that the stage objectives
of the two formulations differ by a constant at every date and state, so that
the strict convexity of the shifted problem transfers to the dynamic program
and delivers the second, part (iii). The lemma thus shows that no assumption
is needed beyond the one guaranteeing uniqueness of the time-zero oracle.

\begin{lemma}[Well-posedness of the sequential problem]\label{lem:timeinv}
Suppose $W$ is time-separable with $W_s \succeq 0$ for every $s$, and that
$H_c$ is positive definite. Then:
\begin{enumerate}
\item[(i)] for every $t = 0, \dots, h-1$, $H^{(t)} \succ 0$ and the shifted
problem \eqref{eq:shiftedproblem} admits the unique solution
\eqref{eq:resoracle};
\item[(ii)] for every $t = 0, \dots, h-1$ and every state $\bs{\chi}_t$, the
optimal continuation values of the shifted problem and of the dynamic program
described by \eqref{eq:perloss} and \eqref{eq:standardform} differ by a
deterministic constant, independent of the state and of the controls, and so
do their stage objectives as functions of the control $u_t = i_{t+1}$;
\item[(iii)] the dynamic program admits a unique sequence of optimal controls,
and for every $t = 0, \dots, h-1$ and every state $\bs{\chi}_t$ the optimal
control at decision date $t$ is the first entry of the shifted solution
\eqref{eq:resoracle}.
\end{enumerate}
\end{lemma}

\begin{proof}
\emph{(i)} Solving the SVAR forward from $\bs{\chi}_t$, the paths of the shifted
problem admit the decomposition \eqref{trajdecomp} over the residual horizon
$h-t$,
\begin{equation}\label{eq:trajdecomp_shifted}
\textbf{y}_{t+1:h} = \textbf{b}_{t+1:h|t}
+ M^{(h-t)}(\bs{\theta})\,\bs{\varepsilon}_{t+1:h},
\end{equation}
with predetermined component $\textbf{b}_{t+1:h|t}$ and future structural shocks
$\bs{\varepsilon}_{t+1:h}$ stacked in the same variable-major ordering as
$\textbf{y}_{t+1:h}$. Partition the matrix $M^{(h-t)}(\bs{\theta})$ as in
Proposition~\ref{prop:toeplitz}, and note that, since the entries of each
variable--shock block of the multiplier depend on $(s,\tau)$ only through the lag
$s-\tau$, the multiplier of the shifted problem is the same function of
$\bs{\theta}$ evaluated at horizon $h-t$, and each of its $n^2$ blocks is a
leading principal sub-block of the corresponding block of $M(\bs{\theta})$. The
construction of Section~\ref{sec:contemp} uses only this decomposition together with
the invertibility of $M^{i,(h-t)}_{v,c}$, which is lower-triangular with constant
diagonal $B_{iv} \neq 0$; it therefore delivers verbatim the multipliers
\eqref{Di_c}--\eqref{Deta_c} and the quadratic representation of
Proposition~\ref{prop:loss_c} at horizon $h-t$, with $\textbf{b}_{t+1:h|t}$ in
place of $\textbf{b}_{1:h}$.

It remains to show that $H^{(t)} \succ 0$. By \eqref{Di_c}, read blockwise, each
of the $n-1$ blocks of $D_{i,c}$ is the product of two lower-triangular matrices,
$D_{i,c}^{x_j} = M_{v,c}^{x_j}\bigl(M_{v,c}^{i}\bigr)^{-1}$, and is therefore
itself lower-triangular and Toeplitz (Propositions~\ref{prop:toeplitz}
and~\ref{prop:Dstructure}(i)). For lower-triangular matrices, extraction of the
leading principal submatrix commutes with products and with inverses; since the
blocks of $M^{(h-t)}(\bs{\theta})$ are the leading principal
$(h-t)\times(h-t)$ sub-blocks of those of $M(\bs{\theta})$, it follows that each
block $D^{(t),x_j}_{i}$ is the leading principal sub-block of $D^{x_j}_{i,c}$.
Those blocks being Toeplitz, their leading and trailing principal submatrices of
equal size coincide, both having $(s,\tau)$ entry equal to the subdiagonal
coefficient of order $s-\tau$; hence $D^{(t),x_j}_{i}$ is equally the trailing
principal sub-block of $D^{x_j}_{i,c}$, that is, its entries of index
$s,\tau > t$.
Let $E_t$ denote the \emph{horizon-selection} matrix and $S_t$ its blockwise
counterpart on the variable-major stack,
\[
E_t := \begin{pmatrix} O_{t \times (h-t)} \\ I_{h-t}\end{pmatrix},
\qquad
S_t := I_{n-1} \otimes E_t ,
\]
so that post-multiplication by $E_t$ selects the last $h-t$ columns of an
$h$-column matrix, and pre-multiplication by $S_t'$ selects the rows of
horizon $s > t$ within each of the $n-1$ blocks. The trailing principal submatrix of $H_c$ obtained by
deleting its first $t$ rows and columns is then $E_t'H_cE_t$, and two identities
deliver the claim. First, by the above, the last $h-t$ columns of each block of
$D_{i,c}$ vanish on the first $t$ rows and reduce to $D^{(t),x_j}_{i}$ below,
so that
\[
D_{i,c}\,E_t = S_t\,D^{(t)}_{i}.
\]
Second, $S_t'\,W\,S_t = W^{(t)}$ by the definition \eqref{eq:Wshifted}, the
left-hand side representing the quadratic form of $W$ restricted to the horizons
$s > t$. Combining,
\[
E_t'\,H_c\,E_t
= \bigl(D_{i,c}E_t\bigr)'\,W\,\bigl(D_{i,c}E_t\bigr)
= \bigl(D^{(t)}_{i}\bigr)'\,S_t'\,W\,S_t\,D^{(t)}_{i}
= \bigl(D^{(t)}_{i}\bigr)'\,W^{(t)}\,D^{(t)}_{i}
= H^{(t)}.
\]
Principal submatrices of positive definite matrices being positive definite,
$H^{(t)} \succ 0$. The shifted loss is then strictly convex and
Proposition~\ref{prop:loss_c}, applied at horizon $h-t$ with
$\textbf{b}_{t+1:h|t}$ in place of $\textbf{b}_{1:h}$, delivers the unique
minimizer \eqref{eq:resoracle}.

\emph{(ii)} For $t = 0, \dots, h-1$, let $V_t(\bs{\chi}_t)$ be the value of
the shifted problem as in \eqref{eq:shiftedproblem} and $J_t(\bs{\chi}_t)$
the continuation value of the dynamic program over the outcome periods
$t+1, \dots, h$, with $J_h \equiv 0$ and
\[
J_t(\bs{\chi}_t) := \min_{u_t \in \reals}\;
\ev_t\bigl[\ell_{t+1}(\bs{\chi}_{t+1}) + J_{t+1}(\bs{\chi}_{t+1})\bigr].
\]
We show by backward induction that the minimum defining $J_t$ is attained at
a unique point, that $J_t$ is a quadratic function of the state, and that
\[
V_t(\bs{\chi}_t) = J_t(\bs{\chi}_t) + \Delta_t,
\qquad \Delta_t \geq 0 \text{ a deterministic constant},
\]
with the two stage objectives in $u_t$ differing by a constant only.

For $t = h-1$ both problems minimize $\ev_{h-1}[\,\ell_h(\bs{\chi}_h)\,]$ over
the scalar $i_h$, where $\ell_t(\bs{\chi}_t) =
\tilde{\bs{\chi}}_{t}'\tilde{Q}_{t}\tilde{\bs{\chi}}_{t}$ is the period loss of
\eqref{eq:perloss}. By part (i) this is a quadratic in $i_h$ with curvature
$H^{(h-1)} > 0$, so the minimizer is unique, $J_{h-1}$ is quadratic in the
state, $V_{h-1} = J_{h-1}$ and $\Delta_{h-1} = 0$.
Suppose the claim holds at $t+1$. Splitting the shifted problem at $t$ into its
first entry $u_t = i_{t+1}$ and its tail $\textbf{a} := \textbf{i}_{t+2:h} \in
\reals^{h-t-1}$, and minimizing iteratively,
\[
V_t(\bs{\chi}_t) = \min_{u_t \in \reals}\;
\Bigl\{ \ev_t\bigl[\ell_{t+1}(\bs{\chi}_{t+1})\bigr]
+ \min_{\textbf{a} \in \reals^{h-t-1}} \ev_t\bigl[\,q_{t+1}(\textbf{a};
\bs{\chi}_{t+1})\,\bigr] \Bigr\},
\]
where $q_{t+1}(\textbf{a}; \bs{\chi}_{t+1}) = \ev_{t+1}\bigl[\sum_{s=t+2}^{h}
\ell_s(\bs{\chi}_s)\bigr]$ is the objective of the shifted problem at decision
date $t+1$, as in \eqref{eq:shiftedproblem}. By part (i), the quadratic
representation of Proposition~\ref{prop:loss_c} applies at horizon $h-t-1$,
giving explicitly
\[
q_{t+1}(\textbf{a}; \bs{\chi}_{t+1})
= \textbf{a}'H^{(t+1)}\textbf{a} + 2\,\textbf{a}'\textbf{g}_{t+1}(\bs{\chi}_{t+1})
+ \tilde{\textbf{b}}^{(t+1)\prime}W^{(t+1)}\tilde{\textbf{b}}^{(t+1)} + c^{(t+1)}(\bs{\chi}_{t+1}),
\]
where
\[
\tilde{\textbf{b}}^{(t+1)} := D^{(t+1)}_{b}\,\textbf{b}_{t+2:h|t+1} - \textbf{x}^*_{t+2:h},
\qquad
\textbf{g}_{t+1}(\bs{\chi}_{t+1}) := \bigl(D^{(t+1)}_{i}\bigr)'W^{(t+1)}
\tilde{\textbf{b}}^{(t+1)},
\]
as in Section~\ref{sec:seq-oracle}, and $c^{(t+1)}(\bs{\chi}_{t+1})$ collects
the terms that do not involve the controls. The coefficient
$\textbf{g}_{t+1}$ is affine in $\bs{\chi}_{t+1}$: by the closed form in
\eqref{eq:bsrec} with time origin $t+1$, each conditional mean satisfies
$\textbf{b}_{s|t+1} = JF^{\,s-t-1}\bs{\chi}_{t+1}$ under the zero-constant
convention. Stacking by periods and rearranging into variable-major ordering
makes $\tilde{\textbf{b}}^{(t+1)}$ affine in $\bs{\chi}_{t+1}$ and
$\textbf{g}_{t+1}$ inherits affinity. Since $H^{(t+1)} \succ 0$ by part (i),
the inner minimizer is $\textbf{a}^* = -\bigl(H^{(t+1)}\bigr)^{-1}\textbf{g}_{t+1}$,
with minimized value
\[
\min_{\textbf{a}}\, q_{t+1}(\textbf{a}; \bs{\chi}_{t+1})
= \tilde{\textbf{b}}^{(t+1)\prime}W^{(t+1)}\tilde{\textbf{b}}^{(t+1)}
+ c^{(t+1)}
- \textbf{g}_{t+1}'\bigl(H^{(t+1)}\bigr)^{-1}\textbf{g}_{t+1}\,.
\]
Repeating the minimization for $\ev_t[q_{t+1}(\textbf{a}; \bs{\chi}_{t+1})]$,
a quadratic in $\textbf{a}$ with the same Hessian $H^{(t+1)}$ and linear term
$\ev_t[\textbf{g}_{t+1}]$, yields the difference
\begin{align*}
\min_{\textbf{a}}\,\ev_t\bigl[q_{t+1}(\textbf{a}; \bs{\chi}_{t+1})\bigr]
- \ev_t\Bigl[\min_{\textbf{a}}\, q_{t+1}(\textbf{a}; \bs{\chi}_{t+1})\Bigr]
&= \ev_t\bigl[\textbf{g}_{t+1}'\bigl(H^{(t+1)}\bigr)^{-1}\textbf{g}_{t+1}\bigr]
- \ev_t[\textbf{g}_{t+1}]'\bigl(H^{(t+1)}\bigr)^{-1}\ev_t[\textbf{g}_{t+1}] \\
&= \operatorname{trace}\Bigl(\bigl(H^{(t+1)}\bigr)^{-1}
\operatorname{Var}_t\bigl(\textbf{g}_{t+1}\bigr)\Bigr) =: \kappa_t \;\geq\; 0,
\end{align*}
the conditional expectations
$\ev_t\bigl[\tilde{\textbf{b}}^{(t+1)\prime}W^{(t+1)}\tilde{\textbf{b}}^{(t+1)}\bigr]$
and $\ev_t\bigl[c^{(t+1)}\bigr]$ appearing identically on both sides and
cancelling. Since $\textbf{g}_{t+1}$ is affine in $\bs{\chi}_{t+1} =
\textbf{A}\bs{\chi}_t + \textbf{B}u_t + \textbf{w}_{t+1}$ and the disturbance
is additive, the state and the control shift only the mean of
$\textbf{g}_{t+1}$, so $\operatorname{Var}_t(\textbf{g}_{t+1})$, and hence
$\kappa_t$, depends on neither. This $\kappa_t$ is the one-step gain of
Remark~\ref{rmk:hierarchy}: at the optimal $u_t$, $V_t = \ev_t[\ell_{t+1}] +
\min_{\textbf{a}}\ev_t[q_{t+1}]$ and $\ev_t[V_{t+1}] = \ev_t[\min_{\textbf{a}}
q_{t+1}]$, so $\kappa_t = V_t(\bs{\chi}_t) - \ev_t[\ell_{t+1}(\bs{\chi}_{t+1}) +
V_{t+1}(\bs{\chi}_{t+1})]$. The inner minimization therefore equals
$\ev_t\bigl[V_{t+1}(\bs{\chi}_{t+1})\bigr] + \kappa_t$, and, by the induction
hypothesis,
\[
V_t(\bs{\chi}_t) = \min_{u_t}\;
\ev_t\bigl[\ell_{t+1}(\bs{\chi}_{t+1}) + J_{t+1}(\bs{\chi}_{t+1})\bigr]
+ \Delta_{t+1} + \kappa_t .
\]
Write
\[
\psi_t(u_t; \bs{\chi}_t) := \ev_t\bigl[\ell_{t+1}(\bs{\chi}_{t+1})\bigr]
+ \ev_t\bigl[J_{t+1}(\bs{\chi}_{t+1})\bigr]
\]
for the stage objective of the dynamic program, the function minimized over
$u_t$ in the display above, and
\[
\varphi_t(u_t; \bs{\chi}_t) := \ev_t\bigl[\ell_{t+1}(\bs{\chi}_{t+1})\bigr]
+ \min_{\textbf{a}}\ev_t\bigl[q_{t+1}(\textbf{a}; \bs{\chi}_{t+1})\bigr]
\]
for the stage objective of the shifted problem, so that $V_t =
\min_{u_t}\varphi_t$. Substituting the expression for the inner minimum
obtained above,
\[
\varphi_t(u_t; \bs{\chi}_t) = \psi_t(u_t; \bs{\chi}_t) + \Delta_{t+1} + \kappa_t ,
\]
and the constant depends neither on $\bs{\chi}_t$ nor on $u_t$: $\Delta_{t+1}$
by the induction hypothesis, $\kappa_t$ by the argument above.

We now show that $\varphi_t$ is a strictly convex quadratic in $u_t$. By the
law of iterated expectations, $\ev_t[\ell_{t+1}] + \ev_t[q_{t+1}(\textbf{a};
\bs{\chi}_{t+1})] = q_t\bigl((u_t, \textbf{a}); \bs{\chi}_t\bigr)$, so
$\varphi_t(u_t; \bs{\chi}_t) = \min_{\textbf{a}} q_t\bigl((u_t, \textbf{a});
\bs{\chi}_t\bigr)$ is the objective of the shifted problem concentrated over
the tail. By part (i), $q_t$ is a quadratic form in $\textbf{i}_{t+1:h} =
(u_t, \textbf{a})$ with Hessian $H^{(t)} \succ 0$. Partition
\[
H^{(t)} = \begin{pmatrix} h_{11} & \textbf{h}_{12}' \\ \textbf{h}_{12} & H_{22}
\end{pmatrix},
\qquad
\textbf{g}^{(t)} = \begin{pmatrix} g_1 \\ \textbf{g}_2 \end{pmatrix},
\]
conformably with $(u_t, \textbf{a})$, where $\textbf{g}^{(t)}$ is the linear
coefficient of $q_t$. For fixed $u_t$ the minimizer over $\textbf{a}$ is
$\textbf{a}^*(u_t) = -H_{22}^{-1}(\textbf{h}_{12}u_t + \textbf{g}_2)$, affine
in $u_t$, and substituting it back gives
\[
\varphi_t(u_t; \bs{\chi}_t) = s_t\,u_t^2
+ 2\,u_t\bigl(g_1 - \textbf{h}_{12}'H_{22}^{-1}\textbf{g}_2\bigr)
+ \text{const},
\qquad
s_t := h_{11} - \textbf{h}_{12}'H_{22}^{-1}\textbf{h}_{12} .
\]
The curvature $s_t$ is the Schur complement of $H_{22}$ in $H^{(t)}$, and it
is positive. Both stage objectives are
therefore strictly convex quadratics in $u_t$ with the same unique minimizer,
$J_t$ is well defined and quadratic in $\bs{\chi}_t$, and
$V_t = J_t + \Delta_t$ with $\Delta_t := \Delta_{t+1} + \kappa_t$. This
completes the induction.

\emph{(iii)} By part (ii) the stage objective of the dynamic program is
strictly convex in $u_t$ at every decision date and state, so the dynamic
program admits a unique sequence of optimal controls. Its minimizer coincides
with that of the shifted stage objective, which by part (i) is the first entry
of \eqref{eq:resoracle}, the first coordinate of the unique joint minimizer. In particular the optimal control at decision date $t$ is affine in the
state, since \eqref{eq:resoracle} is affine in $\textbf{b}_{t+1:h|t}$, which
is linear in $\bs{\chi}_t$ under the zero-constant convention.
\end{proof}

\begin{proof}[Proof of Proposition~\ref{prop:representativeness}]
By Lemma~\ref{lem:timeinv}(iii) the re-optimizing path
$\textbf{i}^{seq}_{1:h}$ coincides, date by date and state by state, with
the optimal control of the dynamic program described by \eqref{eq:perloss}
and \eqref{eq:standardform}. By the principle of optimality
\citep[Section~1.3]{bertsekas1995dynamic} the strategy generated by the
dynamic program minimizes the expected loss over all strategies adapted to
the information sets $\{\mathcal{I}_{t}\}$. The re-optimizing path belongs to
$\mathcal{S}$: each entry is the first component of \eqref{eq:resoracle},
affine in $\bs{\chi}^*_t$ by Lemma~\ref{lem:timeinv}(iii), and the states
have finite second moments by \eqref{eq:standardform}. It therefore
minimizes $\mathcal{L}$ over $\mathcal{S}$, and Lemma~\ref{lem:theil} gives
$\ev_0[\textbf{i}^{seq}_{1:h}] = \bar{\textbf{i}}^{*,c}_{1:h}$ and
$i^{seq}_1 = \bar{i}^{*,c}_1$. Since $H_c \succ 0$ makes $\mathcal{L}$
strictly convex on $\mathcal{S}$, the re-optimizing path is in fact the
unique optimal strategy.
\end{proof}

\begin{corollary}[Representativeness under delayed transmission]\label{cor:repr_delay}
Suppose $W$ is time-separable with $W_t = w_t\Lambda$, $w_t > 0$, and
$\Lambda = \operatorname{diag}(\lambda_1, \dots, \lambda_{n-1}) \succeq 0$.
Let $J := \{j : \lambda_j > 0\}$ be nonempty and let $d := d_J$ be the
transmission delay of Corollary~\ref{cor:rank}, with $1 \leq d < h$. Then:
\begin{enumerate}
\item[(i)] for every $t = 0, \dots, h-1$,
\[
\ker H^{(t)} = \operatorname{span}\bigl(\textbf{e}_{h-t-d+1}, \dots,
\textbf{e}_{h-t}\bigr),
\]
the whole of $\reals^{h-t}$ when $h-t \leq d$. The objective of the shifted
problem \eqref{eq:shiftedproblem} does not depend on the last $\min(d, h-t)$
entries of the residual path, and its minimizer is unique in the remaining
$h-t-\min(d, h-t)$ entries; 
\item[(ii)] the re-optimizing path is well defined in its first $h-d$ entries,
the ex-ante oracle path is unique in its first $h-d$ entries, and
\[
\bar{\textbf{i}}^{*,c}_{1:h-d} = \ev_0\bigl[\textbf{i}^{seq}_{1:h-d}\bigr],
\qquad i^{seq}_1 = \bar{i}^{*,c}_1 .
\]
The last $d$ entries of both paths are undetermined.
\end{enumerate}
\end{corollary}

\begin{proof}
\emph{(i)} By the proof of Lemma~\ref{lem:timeinv}(i), the multiplier
$D^{(t)}_i$ of the shifted problem has blocks $D^{(t),x_j}_i$ equal to the
leading principal $(h-t)\times(h-t)$ sub-blocks of the blocks of $D_{i,c}$,
hence lower-triangular Toeplitz with the same subdiagonal coefficients. The
transmission delay of $x_j$ at the residual horizon is therefore
$\min(d_j, h-t)$, and $\min_{j \in J}\min(d_j, h-t) = \min(d_J, h-t) =
\min(d, h-t)$.
By time-separability, $W^{(t)} = \Lambda \otimes
\operatorname{diag}(w_{t+1}, \dots, w_h)$, which is of the Kronecker form of
Corollary~\ref{cor:rank}(ii) with $W_h \succ 0$. Applying that corollary at
horizon $h-t$ gives the null space claimed in (i). 

The objective of the shifted problem is $q_t(\textbf{i}_{t+1:h}; \bs{\chi}_t)
= \textbf{i}_{t+1:h}'H^{(t)}\textbf{i}_{t+1:h} + 2\,\textbf{i}_{t+1:h}'
\textbf{g}^{(t)} + \text{const}$ with $\textbf{g}^{(t)} = (D^{(t)}_i)'
W^{(t)}\tilde{\textbf{b}}^{(t)}$. For $\textbf{v} \in \ker H^{(t)}$,
$\|(W^{(t)})^{1/2}D^{(t)}_i\textbf{v}\|^2 = \textbf{v}'H^{(t)}\textbf{v} = 0$
gives $W^{(t)}D^{(t)}_i\textbf{v} = \textbf{0}$ and hence
$\textbf{v}'\textbf{g}^{(t)} = 0$. The objective is therefore constant along
$\ker H^{(t)}$, so it does not depend on the last $\min(d, h-t)$ entries. The
set of minimizers is an affine subspace parallel to $\ker H^{(t)}$; since
the kernel is spanned by coordinate vectors, this pins down the remaining
entries uniquely. 

\emph{(ii)} By part (i), for $t \leq h-d-1$ the minimizer of the shifted
problem is unique in its first $h-t-d$ entries. In particular its first entry
is unique, so $i^{seq}_{t+1}$ is well defined for $t+1 \leq h-d$. At $t = 0$
the ex-ante oracle path is unique in its first $h-d$ entries. For
$t \geq h-d$ every residual entry is free and irrelevant for the loss.

Fix the last $d$ entries of every path at an arbitrary value, say zero, and
call \emph{restricted} the resulting one-shot problem, shifted problems at
every $t \leq h-d-1$, and dynamic program, each with action space
$\reals^{h-t-d}$ at decision date $t$. The restricted dynamic program has
decision dates $t = 0, \dots, h-d-1$, the transition \eqref{eq:standardform}, and a terminal cost at the state $\bs{\chi}_{h-d}$ equal to the
expected loss over the remaining periods,
\[
J_{h-d}(\bs{\chi}_{h-d}) := \ev_{h-d}\Bigl[\sum_{s=h-d+1}^{h}
\ell_s(\bs{\chi}_s)\Bigr], \qquad u_s = 0 \text{ for } s \geq h-d ,
\]
with $J_t(\bs{\chi}_t) := \min_{u_t}\ev_t[\ell_{t+1}(\bs{\chi}_{t+1}) +
J_{t+1}(\bs{\chi}_{t+1})]$ for $t = h-d-1, \dots, 0$. Iterating
\eqref{eq:standardform} with $u_s = 0$, the terminal cost is a quadratic
function of $\bs{\chi}_{h-d}$ plus a constant, so this is the
linear-quadratic problem of \citet[Section~4.1]{bertsekas1995dynamic} with
$h-d$ stages and a quadratic terminal cost.

None of the restricted objectives depends on the value at which the last $d$
entries are fixed. For the one-shot and shifted problems this is part (i),
which shows that $q_t$ is constant along $\ker H^{(t)}$. Their restricted
minimizers are therefore the determined entries of the unrestricted ones,
and in particular the re-optimizing path over $t \leq h-d-1$, which is built
from the shifted minimizers, is unchanged by the normalization. For the
dynamic program the fixed entries enter only through the terminal cost, and
$J_{h-d}(\bs{\chi}_{h-d}) = q_{h-d}(\textbf{i}_{h-d+1:h}; \bs{\chi}_{h-d})$ is
the shifted objective at $t = h-d$, where by part (i) the kernel of
$H^{(h-d)}$ is all of $\reals^{d}$; hence $J_{h-d}$, and with it every
$J_t$, does not depend on the normalization either.

The restricted Hessians are positive definite and nested as in
Lemma~\ref{lem:timeinv}(i). Let $E := (I_{h-d} \;\; O_{(h-d)\times d})'$ and
$H^R_c := E'H_cE$, the leading principal $(h-d)\times(h-d)$ block of $H_c$
and the Hessian of the restricted one-shot problem. For
$\textbf{v}_1 \in \reals^{h-d}$, $\textbf{v}_1'H^R_c\textbf{v}_1 = 0$ if and
only if $(\textbf{v}_1', \textbf{0}')' \in \ker H_c$, which by part (i) at
$t = 0$ forces $\textbf{v}_1 = \textbf{0}$; hence $H^R_c \succ 0$. The
Hessian of the restricted shifted problem at $t \leq h-d-1$ is the leading
principal $(h-t-d)\times(h-t-d)$ block of $H^{(t)} = E_t'H_cE_t$, that is
the block of $H_c$ with indices $t+1, \dots, h-d$, which is the trailing
principal block of $H^R_c$ obtained by deleting its first $t$ rows and
columns. It is therefore positive definite, and the restricted problems have
the nested structure of Lemma~\ref{lem:timeinv}(i) with $h-d$ in place of $h$
and $H^R_c$ in place of $H_c$.
Parts (ii) and (iii) of Lemma~\ref{lem:timeinv} then go through verbatim for
the restricted problems, with $h-d$ in place of $h$, $H^R_c$ in place of
$H_c$, the terminal cost $J_{h-d}$ in place of the boundary condition
$J_h \equiv 0$, and the backward induction started at $t = h-d-1$. There, by
the law of iterated expectations, the restricted shifted objective is
$q_{h-d-1} = \ev_{h-d-1}[\ell_{h-d}(\bs{\chi}_{h-d}) +
J_{h-d}(\bs{\chi}_{h-d})]$, the same function the dynamic program minimizes
over the scalar $i_{h-d}$; its curvature is the leading $1\times1$ block of
$H^{(h-d-1)}$, positive by the above, so $\Delta_{h-d-1} = 0$ and the base
case holds. The induction step does not involve the boundary condition.
Lemma~\ref{lem:theil} applies to the restricted problem as well. The
restricted strategies are the elements of $\mathcal{S}$ whose last $d$
entries are zero, a linear subspace $\mathcal{S}^R$ containing the
restricted deterministic paths; along $\textbf{f} \in \mathcal{S}^R$,
\eqref{eq:stackedstrategy} reads $\textbf{x}_{1:h} = D_{i,c}E\,
\textbf{f}_{1:h-d} + \textbf{s}$, and the curvature of the restricted loss
is $E'H_cE = H^R_c \succ 0$. The perturbation argument in the proof of
Lemma~\ref{lem:theil}, with directions in $\reals^{h-d}$, gives for every
minimizer $\textbf{f}^*$ of $\mathcal{L}$ over $\mathcal{S}^R$ that
$\ev_0[\textbf{f}^*_{1:h-d}]$ is the unique minimizer of the restricted
deterministic loss, which by the above is $\bar{\textbf{i}}^{*,c}_{1:h-d}$.
The proof of Proposition~\ref{prop:representativeness} then applies to the
restricted problem: the restricted re-optimizing path is the optimal
strategy of the restricted dynamic program by the restricted part (iii), it
minimizes $\mathcal{L}$ over $\mathcal{S}^R$ by the principle of optimality,
and the restricted Lemma~\ref{lem:theil} yields
$\bar{\textbf{i}}^{*,c}_{1:h-d} = \ev_0[\textbf{i}^{seq}_{1:h-d}]$ and
$i^{seq}_1 = \bar{i}^{*,c}_1$. The last $d$ entries of both paths are
undetermined by part (i).
\end{proof}

\section{Further details on the empirical and numerical specifications}\label{app:specification}

This appendix collects the details of the specification behind the
empirical application of Section~\ref{sec:empirical} and the numerical risk
analysis of Section~\ref{sec:numerical-risk}: the VAR systems and data
(Appendix~\ref{app:spec_data}), the prior and the posterior simulation
(Appendix~\ref{app:spec_prior}), the selection of the prior hyperparameters
(Appendix~\ref{app:prior_selection}), the decision problems
(Appendix~\ref{app:spec_decision}), the set-identified exercise
(Appendix~\ref{app:spec_setid}) and the design of the numerical risk analysis
(Appendix~\ref{app:spec_numerical}). All computations are carried out in
MATLAB R2025a.

\subsection{VAR systems and data}\label{app:spec_data}

All series are quarterly. The macroeconomic block is common to the three
systems: the output gap (HP-filtered real GDP), annualized GDP-deflator
inflation, annualized commodity-price inflation and the federal funds rate.
Each system adds to it the shock series of its identification scheme,
summarized in Table~\ref{tab:spec_systems}. Every system is a VAR(4) with a
constant, estimated on 1970Q1--2005Q4 ($T=144$), with the four quarters of
1969 as initial conditions; the time window 2006--2007 is the horizon over which the policy exercise is carried over. The observations for these periods only matter for the observed path of the
 policy instrument which is compared to plug-in, Bayes and robust Bayes rules. The two high-frequency series are
available from 1988Q4 (Gertler--Karadi) and 1991Q3 (slope factor) and are set to
zero before.

\begin{table}[htbp]
\centering
\small
\caption{VAR systems. $\dim\bs\phi$: autoregressive coefficients, intercepts and
the log-Cholesky factor of $\Sigma$; in brackets, the coordinates left free once
the block-exogeneity restrictions are imposed exactly
(Appendix~\ref{app:spec_numerical}).}
\label{tab:spec_systems}
\begin{tabular}{llcc}
\toprule
System & Endogenous variables (ordering) & $n$ & $\dim\bs\phi$ \\
\midrule
Contemporaneous & $y$, $\pi$, $\pi^{c}$, RR, $i$ & 5 & 120 (104) \\
Forward guidance & GK, SF, $y$, $\pi$, $\pi^{c}$, RR, $i$ & 7 & 231 (183) \\
Set-identified  & $y$, $\pi$, $\pi^{c}$, $i$ & 4 & 78 \\
\bottomrule
\end{tabular}

\smallskip
\parbox{0.92\linewidth}{\footnotesize $y$: output gap; $\pi$: GDP-deflator
inflation; $\pi^c$: commodity-price inflation; $i$: federal funds rate; RR:
\citet{romer2004new} shock; GK: high-frequency surprise of
\citet{gertler2015monetary}; SF: slope-factor surprise
\citep{barnichon2023sufficient}.}
\end{table}

In the two point-identified systems we impose block exogeneity restrictions such that the
equations of the shock series contain only lags of the shock series
themselves. In the contemporaneous system the Romer--Romer equation excludes the
lags of the four macroeconomic variables (16 coefficients); in the forward
guidance system the three instrument equations exclude the lags of the
macroeconomic block, $y$, $\pi$, $\pi^c$ and $i$ (48 coefficients). The shock
series can then be read as exogenous to the macroeconomy, while the
macroeconomic block responds to them freely. In the estimation the
restrictions are imposed through the prior (Appendix~\ref{app:spec_prior}); in
the numerical risk analysis they are imposed exactly.

\subsection{Prior and posterior simulation}\label{app:spec_prior}

Every system is estimated under an independent Normal--inverse-Wishart prior,
as implemented in version 5.2.0 of the BEAR toolbox
\citep{dieppe2016bear}. The VAR
coefficients (autoregressive coefficients and intercepts) have a Gaussian
prior of the Minnesota type, independent of an inverse-Wishart prior on the
reduced-form covariance $\Sigma$. Unlike the conjugate prior, this
specification allows the prior mean on own lags to differ across variables and the tightness to differ
across own and cross-variable coefficients. The prior is centred on a
univariate AR(1) for each variable: the mean of the first own lag is $0.9$ for
the output gap, inflation and the funds rate, $0.3$ for commodity inflation
and $0$ for the shock series, and all other coefficients have mean zero. The
prior variances are governed by five hyperparameters:
\begin{itemize}
\item $\lambda_1$, the overall tightness, scales the prior standard deviation of
all the autoregressive coefficients;
\item $\lambda_2$ tightens the coefficients on the lags of the other variables
relative to those on own lags;
\item $\lambda_3=1$ makes the prior standard deviation decay harmonically with
the lag;
\item $\lambda_4=100$ makes the prior on the intercepts diffuse;
\item $\lambda_5$ shrinks the coefficients excluded by block exogeneity
(Appendix~\ref{app:spec_data}) toward zero.
\end{itemize}
We tie the last one to the first two,
$\lambda_5=10^{-4}/(\lambda_1\lambda_2)$, so that the prior standard deviation
of a restricted coefficient does not depend on $(\lambda_1,\lambda_2)$. The prior on $\Sigma$ is weakly informative.

The posterior is simulated with the Gibbs sampler of the independent
Normal--inverse-Wishart model: $15{,}000$ iterations, of which the first
$5{,}000$ are discarded, leaving $10{,}000$ draws. In the empirical exercises every draw is retained, stationary or not, since all the objects of the decision problem are finite-horizon and remain well defined; the plug-in rule is computed at the posterior mean of the retained draws. In the numerical risk analysis, instead, the parameter points are restricted to stationary draws, since the Gaussian approximation to the sampling distribution of the estimator relies on stationarity (Appendix~\ref{app:spec_numerical}).

\subsection{Hyperparameters selection}\label{app:prior_selection}

We select the overall tightness $\lambda_1$ and the cross-variable weight
$\lambda_2$ by empirical Bayes in the same spirit of \citet{giannone2015prior}, by maximizing the marginal likelihood
$p(Y\mid\lambda_1,\lambda_2)$ over a grid 
with all other hyperparameters held at the values of
Appendix~\ref{app:spec_prior}. Note that our prior is not conjugate, so neither the posterior nor the marginal
likelihood is available in closed form: we estimate the marginal likelihood
from the Gibbs output \citep{chib1995marginal} in $\log_{10}$ units. The maximization is over a discrete grid in
$(\lambda_1,\lambda_2)$.

For the two point-identified systems we evaluate a coarse grid,
$\lambda_1\in\{0.1,0.2,0.3,0.4,0.5\}$ and $\lambda_2\in\{0.25,0.5,1\}$, and then
refine it around its maximum in steps of $0.05$ in $\lambda_1$ and $0.25$ in
$\lambda_2$. For the set-identified system a single grid with the finer
steps, $\lambda_1\in\{0.1,0.15,0.2,0.25,0.3,0.4\}$ and
$\lambda_2\in\{0.25,0.5,0.75,1\}$, already has an interior maximum. The
selected values are $(\lambda_1,\lambda_2)=(0.15,1)$ in the contemporaneous
system, $(0.15,0.75)$ in the forward guidance system and $(0.25,0.5)$ in the
set-identified system.

In the three systems the marginal likelihood surface has a ridge: as
$\lambda_1$ increases, the value of $\lambda_2$ that maximizes the likelihood
falls, so that the tightness of the cross-variable coefficients,
$\lambda_1\lambda_2$, stays between $0.1$ and $0.2$.
Table~\ref{tab:prior_selection} reports the marginal likelihood along the
ridge. Near the maximum the ridge is flat: the $\log_{10}$ marginal
likelihood stays within one unit of its maximum for $\lambda_1$ between $0.15$
and $0.3$ in the contemporaneous and set-identified systems and between $0.1$
and $0.25$ in the forward guidance system. Away from the ridge it falls
quickly: with $\lambda_2$ held at $1$, for instance, raising $\lambda_1$ to
$0.4$ costs about eight $\log_{10}$ units in the contemporaneous system. The
data therefore pin down how much the cross-variable coefficients are shrunk
more sharply than how this shrinkage is split between $\lambda_1$ and
$\lambda_2$. 

\begin{table}[htbp]
\centering
\small
\caption{Prior selection: the ridge of the marginal likelihood. For each value
of $\lambda_1$ on the grid, the value of $\lambda_2$ that maximizes the
marginal likelihood, the $\log_{10}$ marginal likelihood there and its
distance from the maximum over the grid (in bold).}
\label{tab:prior_selection}
\begin{tabular}{ccrr@{\qquad}ccrr@{\qquad}ccrr}
\toprule
\multicolumn{4}{c}{Contemporaneous} & \multicolumn{4}{c}{Forward guidance} & \multicolumn{4}{c}{Set-identified} \\
\cmidrule(lr){1-4}\cmidrule(lr){5-8}\cmidrule(lr){9-12}
$\lambda_1$ & $\lambda_2$ & $\log_{10}p$ & gap & $\lambda_1$ & $\lambda_2$ & $\log_{10}p$ & gap & $\lambda_1$ & $\lambda_2$ & $\log_{10}p$ & gap \\
\midrule
0.10 & 1.00 & $-566.72$ & $-1.60$ & 0.10 & 1.00 & $-256.74$ & $-0.69$ & 0.10 & 1.00 & $-529.61$ & $-1.07$ \\
\textbf{0.15} & \textbf{1.00} & $\mathbf{-565.12}$ & $0.00$ & \textbf{0.15} & \textbf{0.75} & $\mathbf{-256.05}$ & $0.00$ & 0.15 & 0.75 & $-529.06$ & $-0.52$ \\
0.20 & 0.75 & $-565.12$ & $0.00$ & 0.20 & 0.50 & $-256.34$ & $-0.29$ & 0.20 & 0.50 & $-528.80$ & $-0.26$ \\
0.25 & 0.75 & $-565.62$ & $-0.50$ & 0.25 & 0.50 & $-256.87$ & $-0.82$ & \textbf{0.25} & \textbf{0.50} & $\mathbf{-528.54}$ & $0.00$ \\
0.30 & 0.50 & $-565.85$ & $-0.74$ & 0.30 & 0.50 & $-258.33$ & $-2.28$ & 0.30 & 0.50 & $-528.69$ & $-0.15$ \\
0.40 & 0.50 & $-567.33$ & $-2.21$ & 0.40 & 0.25 & $-259.79$ & $-3.73$ & 0.40 & 0.25 & $-529.56$ & $-1.03$ \\
0.50 & 0.25 & $-568.94$ & $-3.83$ & 0.50 & 0.25 & $-261.59$ & $-5.53$ &      &      &           &         \\
\bottomrule
\end{tabular}
\end{table}

\subsection{Decision problems}\label{app:spec_decision}

 The decision is taken at the beginning of 2006
($t=1$ is 2006Q1) over $h=8$ quarters, with the dual-mandate loss of
Section~\ref{sec:emp_setup}: weights one on the squared output gap and on the
squared deviation of inflation from $\pi^\ast=2$, discount factor
$\beta=0.99$. The initial condition $\bs\chi_0$ consists of the four quarters of
2005, and $i_0=3.98$ is the funds rate in 2005Q4. Counterfactual paths are
reported as posterior means, with 90\% equi-tailed pointwise credible regions around Bayes rule counterfactuals. 

In the contemporaneous regime, the action is the path
$\bar{\textbf{i}}_{1:h}$, enforced period by period through the Romer--Romer
shock. Since the impact response of the targets to the policy shock is zero,
the last action has no leverage on the loss. We set it equal to the
second-to-last and solve the $(h-1)$-dimensional problem; figures show the
rules up to $h-1$.

In the forward guidance regime, the announcement is enforced by the $m=3$
shocks GK, SF and RR, each identified as the Cholesky response to its own
series. With $m<h$ a generic expected path cannot be enforced, and the empirical and numerical exercises in the paper implement the feasible forward guidance rules of Section \ref{sec:feasible-fg}. The action is the expected rate at the maturities $\textbf{s}=(1,4,8)$, and
the expected path at the remaining horizons is implied by the shocks that
enforce it (Section~\ref{sec:feasible-fg}). The construction requires the
inverse of $\tilde M^i_{1,fg}=S_{\textbf{s}}M^i_{1,fg}$, and we restrict the
reduced-form parameter space to
\[
\Phi_{\underline\sigma}=\big\{\bs\phi:\ \sigma_{\min}\big(\tilde M^i_{1,fg}(\bs\phi)\big)\geq\underline\sigma\big\},
\qquad \underline\sigma=0.01,
\]
where $\sigma_{\min}$ is the smallest singular value. The restriction is a
regularization: it requires the weakest combination of the announcement
shocks to move the expected funds rate at the announced maturities by a
minimal amount, and it bounds $\|(\tilde M^i_{1,fg})^{-1}\|$ by
$1/\underline\sigma$, so that the posterior expectations entering the Bayes
rule are finite. The value of $\underline\sigma$ is a choice of the
policymaker and could be calibrated based on economic considerations. The posterior is truncated
accordingly: of the $10{,}000$ draws, $2{,}476$ lie outside $\Phi_{\underline\sigma}$, and the Bayes rule averages over the remaining $7{,}524$. The plug-in is the oracle at the
posterior mean of the retained draws; it is not itself restricted to
$\Phi_{\underline\sigma}$. 

\subsubsection{Interest-rate smoothing}\label{app:spec_smoothing}

The smoothing penalty of Appendix~\ref{app:genloss},
$\rho\sum_{s=1}^{h}\beta^{s}(\bar i_s-\bar i_{s-1})^2
=\rho\,\|\Lambda_\beta^{1/2}(\Delta\bar{\textbf{i}}_{1:h}-i^\ast\textbf{e}_1)\|^2$,
is anchored at the prevailing rate in $2005q4$,   $i^\ast=i_0=3.98$. In the contemporaneous regime the penalty applies to
the instrument path itself, and adds the deterministic terms
$\rho\,\Delta'\Lambda_\beta\Delta$ to the curvature and
$-\rho\beta i^\ast\textbf{e}_1$ to the linear term of the loss. The loss is linear in $\rho$, and so are its posterior moments:
writing $H(\rho)=H+\rho H_\Delta$ and $\textbf{g}(\rho)=\textbf{g}+\rho\,\textbf{g}_\Delta$
for the curvature and linear term with the penalty terms just described, the
Bayes path at weight $\rho$ is
\[
\hat{\textbf{i}}_B(\rho)=-\big(\bar H+\rho\,\bar H_\Delta\big)^{-1}
\big(\bar{\textbf{g}}+\rho\,\bar{\textbf{g}}_\Delta\big),
\]
where bars denote posterior means, and $H_\Delta, \textbf{g}_\Delta$ are taken to be the appropriate terms as described above. The posterior moments are computed once, at $\rho=0$, and the rule at
any weight follows in closed form.

The calibrated weight $\rho^\ast$ is the smallest $\rho\geq0$ such that the
Bayes path is no
rougher than the observed path, where the roughness of a path is the
discounted sum the penalty weighs,
$r(\textbf{i})=\|\Lambda_\beta^{1/2}(\Delta\textbf{i}-i^\ast\textbf{e}_1)\|^2$,
and $r(\textbf{i}^{obs})=0.868$. We obtain $\rho^\ast=4.95$.

\subsection{Set-identified empirical exercise}\label{app:spec_setid}

 The system is the four-variable VAR of
Table~\ref{tab:spec_systems}, with the prior selected in
Appendix~\ref{app:prior_selection} ($\lambda_1=0.25$, $\lambda_2=0.5$), and
every posterior draw is retained. For each draw $\bs\phi_g$ we sample
candidate rotations $\textbf{q}=\textbf{z}/\|\textbf{z}\|$ with
$\textbf{z}\sim N(0,I_4)$, which is uniform on the unit sphere. We accept
those satisfying the sign restrictions of the robust Bayes exercise of
Section~\ref{sec:empirical} (funds rate up, output gap and inflation down, on
impact and in the following three quarters), together with the relevance
restriction $(\Sigma^{1/2}_{tr}\textbf{q})_i^2\geq\omega\,\Sigma_{ii}$,
$\omega=0.5$, where $\Sigma^{1/2}_{tr}$ is the lower Cholesky factor of
$\Sigma$ and $i$ indexes the funds rate. The sampler stops at $R=5{,}000$
accepted rotations or after $2\times10^7$ candidates; draws with no accepted
rotation have an empty identified set and are dropped. 

We approximate the worst-case plug-in rule \eqref{wcplugin}, the robust Bayes rule \eqref{rb_rule} and the robust Bayes regret rule \eqref{rbre_rule} numerically, in three steps.
\begin{enumerate}
    \item For each of the $G = 9{,}998$ posterior draws $\bs{\phi}_g$ we sample $R = 5{,}000$ rotations $q_{gr}$ satisfying the sign and relevance restrictions, and we replace the supremum over $\mathcal{I}_{q_1}(\bs{\phi}_g)$ by the maximum over the sampled rotations. The worst-case plug-in rule uses the rotations sampled at the posterior mean $\hat{\bs{\phi}}$ only.
    \item For each pair $(g,r)$ the loss of a path is the quadratic $\bar{\textbf{i}}'H_{gr}\bar{\textbf{i}}+2\textbf{g}_{gr}'\bar{\textbf{i}}+c_{gr}$, and the regret replaces $c_{gr}$ with $\textbf{g}_{gr}'H_{gr}^{-1}\textbf{g}_{gr}$. The sampled criterion,
    \[
    \frac1G\sum_{g=1}^G\max_{r\leq R}\Big(\bar{\textbf{i}}'H_{gr}\bar{\textbf{i}}+2\textbf{g}_{gr}'\bar{\textbf{i}}+c_{gr}\Big),
    \]
    is the posterior average of the maximum of $R$ quadratics: a convex, piecewise-quadratic function of the path, which we minimize over the box $[0,20]^h$.
    \item We minimize it by iterating on the worst case, a Newton-type variant of the Pshenichnyi--Pironneau--Polak method for finite minimax problems \citep[Section 2.4]{polak1997optimization}. At the current path the algorithm selects, for every draw, the rotation attaining the maximum. With these rotations fixed, the criterion is a single quadratic with a closed-form minimizer, and the algorithm moves toward it by the largest step that lowers the criterion. It then re-selects the worst rotations at the new path, and repeats until the selection stops changing and the step vanishes.
\end{enumerate}
The solution is certified by weak duality. For any weights $\mu_{gr}\geq0$ with $\sum_r\mu_{gr}=1$ for each $g$, the combination $\frac1G\sum_g\sum_r\mu_{gr}(\cdot)$ of the quadratics is a minorant of the criterion, so its minimum over the box bounds the optimum from below. We choose the weights that maximize this bound among the pairs within a relative $10^{-8}$ of their draw's maximum, and stop when the relative gap between the attained value and the bound falls below $10^{-7}$; the gaps attained by the reported rules are below $10^{-10}$. The certificate refers to the sampled problem: it does not account for the Monte Carlo error of the posterior average or for the approximation of the supremum over the identified set by the maximum over $R$ sampled rotations.

\subsection{Numerical risk analysis}\label{app:spec_numerical}
We work in the
coordinates $\bs\phi$ of Section~\ref{sec:emp_setup}: the autoregressive
coefficients stacked lag by lag, the intercepts and the log-Cholesky factor of
$\Sigma$. The coefficients restricted by block exogeneity are set to exactly
zero, in the parameter space, the signals and the posterior, and the
experiment takes place on the remaining free coordinates. The parameter points at which we conduct the Monte Carlo experiments are the posterior draws of
the main specification that imply a stationary VAR (spectral radius below
$1-10^{-3}$). Points at which the oracle path
at the true parameter has Euclidean norm above $300$ are excluded, and under the feasible forward guidance regime
so are the points outside $\Phi_{\underline\sigma}$. From
the remaining points we draw at random, with a fixed seed, $I=5{,}000$ points
$\bs\phi_1,\ldots,\bs\phi_I$ at which the risks are computed. Since the points
are posterior draws, statistics over them approximate the posterior
distribution of the risk comparisons.

 At each point the signal is drawn from the Gaussian
limit experiment $\hat{\bs\phi}\mid\bs\phi_i\sim N(\bs\phi_i,\hat\Sigma_0)$.
The covariance $\hat\Sigma_0$ is the posterior covariance of the same system
estimated under a normal-diffuse prior: a Normal prior on the coefficients
with $\lambda_1=100$, which is flat for practical purposes, the
block-exogeneity restrictions imposed through $\lambda_5=10^{-6}$, and the
Jeffreys prior on $\Sigma$. It is computed over the stationary draws of that
estimation.  Taking the signal covariance from the posterior of the
estimation prior would let the prior inform both the precision of the signal
and the Bayes rule that processes it, using the prior information twice.

The prior of the
benchmark rule is meant to replicate the estimation prior of Appendix~\ref{app:spec_prior}, at the
selected hyperparameters, expressed in the coordinates $\bs\phi$. To avoid having to use a Gibbs sampler at each signal draw, we modify the prior so that we have an analytical posterior. On the
autoregressive coefficients and the intercepts it is Gaussian and independent
across coordinates, with mean $\textbf{m}$ and diagonal precision $D$ given by
the Minnesota prior moments. On the log-Cholesky coordinates it is flat. The prior variance is scaled by $\kappa>0$: the prior density is
$\pi(\bs\phi)\propto\exp\{-\tfrac{1}{2\kappa}(\bs\phi-\textbf{m})'D(\bs\phi-\textbf{m})\}$,
with $D$ diagonal and equal to zero on the log-Cholesky coordinates, so that
the prior is Gaussian on the coefficients and flat, hence improper, on
$\Sigma$; $\kappa=1$ is the baseline. Since the signal is Gaussian in all the
free coordinates, the posterior remains Gaussian and proper, with precision
$\hat\Sigma_0^{-1}+\kappa^{-1}D\succ0$:
$\bs\phi\mid\hat{\bs\phi}\sim N(\bs\mu(\hat{\bs\phi}),\Omega)$ with
\[
\Omega=\big(\hat\Sigma_0^{-1}+\kappa^{-1}D\big)^{-1}
=\big(I+\kappa^{-1}\hat\Sigma_0D\big)^{-1}\hat\Sigma_0,
\]
\[
\bs\mu(\hat{\bs\phi})=\Omega\big(\hat\Sigma_0^{-1}\hat{\bs\phi}+\kappa^{-1}D\textbf{m}\big)
=\big(I+\kappa^{-1}\hat\Sigma_0D\big)^{-1}\big(\hat{\bs\phi}+\kappa^{-1}\hat\Sigma_0D\textbf{m}\big),
\]
where the second expressions, which avoid inverting the ill-conditioned
$\hat\Sigma_0$, are the ones we compute.  The Bayes action at
$\hat{\bs\phi}$ is
$\hat{\textbf{i}}_{B}(\hat{\bs\phi})=-\ev[H\mid\hat{\bs\phi}]^{-1}\ev[\textbf{g}\mid\hat{\bs\phi}]$
(Proposition~\ref{prop_bayes_quad}), with the expectations approximated by the
average over $200$ draws from $N(\bs\mu(\hat{\bs\phi}),\Omega)$. Draws
  outside $\Phi_{\underline\sigma}$ in the FG nodes regime are rejected: the posterior of the benchmark rule
is truncated to the restricted parameter space. Table \ref{tab:num_quartiles_sigma} reports some statistics suggestive of the sensitivity of the decision problem on the choice of $\underline\sigma$. The gains of the Bayes rule are largest for values of $\sigma_{\min}$ close to the threshold, where the posterior uncertainty about the weakest combination of announcement shocks is large relative to its size. The Bayes rule still outperforms the plug-in in every quartile, both in the share of parameter points at which it has lower risk and in the median and mean share of regret removed. As one would expect, the performance of the plug-in improves the higher the threshold $\underline\sigma$.

\begin{table}[tbp]
\centering
\label{tab:num_quartiles_sigma}
\begin{tabular}{lcrrrr}
\toprule
 & & Plug-in has & \multicolumn{2}{c}{Regret ratio} & Plug-in regret \\
\cmidrule(lr){4-5}
Quartile & $\sigma_{\min}$ & lower risk (\%) & Median & Geom.\ mean & (median) \\
\midrule
1 & [0.010, 0.019] & 0.0 & 5.89 & 5.76 & 213.4 \\
2 & [0.019, 0.029] & 1.0 & 3.54 & 3.58 & 92.0 \\
3 & [0.029, 0.044] & 3.1 & 2.38 & 2.45 & 52.4 \\
4 & [0.044, 0.155] & 15.5 & 1.49 & 1.58 & 30.6 \\
\midrule
\multicolumn{6}{l}{\emph{Higher truncation threshold $\underline\sigma$ (approximation)}} \\
\multicolumn{2}{l}{$\underline\sigma=0.01$ (run), 5000 points} & 4.9 & 2.99 & 2.99 & 68.9 \\
\multicolumn{2}{l}{$\underline\sigma=0.02$, 3578 points} & 6.8 & 2.26 & 2.34 & 49.2 \\
\multicolumn{2}{l}{$\underline\sigma=0.05$, 943 points} & 17.6 & 1.42 & 1.48 & 29.0 \\
\bottomrule
\end{tabular}
\caption{\small Forward guidance: the Bayes rule against the plug-in, by quartiles of $\sigma_{\min}$  Regret ratio: $(R_{PI}-L^*)/(R_B-L^*)$, median and geometric mean over the points. }
\end{table}

At each point $\bs\phi_i$ we draw $S=400$ signals $\hat{\bs\phi}_s\sim N(\bs\phi_i,\hat\Sigma_0)$. Signals implying a non-stationary VAR are kept (about $5\%$ of the signals in both regimes); signals at which every posterior draw is rejected are discarded for all the rules (12 and 14 of the 2 million signals in the contemporaneous and forward guidance regimes). The same signals and the same posterior draws are used for every rule (common random numbers).

The plug-in action at a signal is the oracle rule evaluated at $\hat{\bs\phi}_s$, where the oracle is the solution of the $(h-1)$-dimensional reduced problem in the contemporaneous regime and the unique oracle at the selected maturities in feasible forward guidance. The plug-in is not restricted to $\Phi_{\underline\sigma}$: signals outside it are kept, and the plug-in is computed there as at any other signal. The feasible James--Stein rule \eqref{eq:shrunk_plugin_feasible} shrinks the plug-in defined above, in the coordinates in which the loss weight is positive definite (the reduced path in the contemporaneous regime, the announced rates in feasible forward guidance). $\hat H$ is the loss Hessian at the signal. $\hat S$ is estimated by a parametric bootstrap at the signal: for $200$ draws $\bs\theta_r\sim N(\hat{\bs\phi}_s,\hat\Sigma_0)$ we compute the plug-in at $\bs\theta_r$, discarding draws whose path has Euclidean norm above $c$, and take the sample covariance of the plug-in across the retained draws. The reference path is the status-quo path at $i_0$, and we compute the rule at a grid of values of $t$ over the range $0<t\leq2(k-2)$ of Lemma~\ref{lemma_js}, in unit steps in the contemporaneous regime and in half steps in feasible forward guidance.

Each pairwise comparison discards the signals at which either rule prescribes a path of Euclidean norm above $c=500$, so that the two risks are computed on the same signals. The risk of each rule is the average of its losses at the true $\bs\phi_i$ over the retained signals.

\end{document}